\documentclass[12pt,letterpaper,oneside,onecolumn,portrait]{book} 
\usepackage[Sonny]{fncychap} 
\usepackage{amssymb,amsfonts,amsmath,amscd,amsthm}
\usepackage{graphicx}
\usepackage[dvips]{epsfig}
\usepackage{rotating}
\usepackage{color}
\usepackage{url}
\usepackage{bm}
\usepackage{multirow}
\usepackage{booktabs}
\usepackage{algorithm}
\usepackage{algorithmicx}
\usepackage{algpseudocode}
\usepackage{amsmath,amssymb}
\usepackage{fancyvrb}
\usepackage{subfig}
\usepackage{wrapfig}
\usepackage{framed}    
\usepackage{booktabs}  
\usepackage{array} 
\usepackage{textcomp} 
\usepackage{geometry}
\usepackage{enumerate}

\errorcontextlines\maxdimen
\makeatletter

\newcommand*{\algrule}[1][\algorithmicindent]{\makebox[#1][l]{\hspace*{.5em}\thealgruleextra\vrule height
\thealgruleheight depth \thealgruledepth}}%

\newcommand*{\thealgruleextra}{}
\newcommand*{\thealgruleheight}{.75\baselineskip}
\newcommand*{\thealgruledepth}{.25\baselineskip}
\algdef{SE}[DOWHILE]{Do}{doWhile}{\algorithmicdo}[1]{\algorithmicwhile\ #1}%

\newcount\ALG@printindent@tempcnta
\def\ALG@printindent{%
    \ifnum \theALG@nested>0
        \ifx\ALG@text\ALG@x@notext
        \else
            \unskip
            \addvspace{-1pt}
            \ALG@printindent@tempcnta=1
            \loop
                \algrule[\csname ALG@ind@\the\ALG@printindent@tempcnta\endcsname]%
                \advance \ALG@printindent@tempcnta 1
            \ifnum \ALG@printindent@tempcnta<\numexpr\theALG@nested+1\relax
            \repeat
        \fi
    \fi
    }%

\usepackage{etoolbox}
\patchcmd{\ALG@doentity}{\noindent\hskip\ALG@tlm}{\ALG@printindent}{}{\errmessage{failed to patch}}
\makeatother

\newbox\statebox
\newcommand{\myState}[1]{%
    \setbox\statebox=\vbox{#1}%
    \edef\thealgruleheight{\dimexpr \the\ht\statebox+1pt\relax}%
    \edef\thealgruledepth{\dimexpr \the\dp\statebox+1pt\relax}%
    \ifdim\thealgruleheight<.75\baselineskip
        \def\thealgruleheight{\dimexpr .75\baselineskip+1pt\relax}%
    \fi
    \ifdim\thealgruledepth<.25\baselineskip
        \def\thealgruledepth{\dimexpr .25\baselineskip+1pt\relax}%
    \fi
    \State #1%
    \def\thealgruleheight{\dimexpr .75\baselineskip+1pt\relax}%
    \def\thealgruledepth{\dimexpr .25\baselineskip+1pt\relax}%
}

\DeclareMathOperator{\E}{\mathbb{E}}
\DeclareMathOperator{\V}{\mathbb{V}}

\usepackage{titlesec}
\titleformat{\section}{\LARGE\sffamily}{\thesection}{1em}{}
\titleformat{\subsection}{\Large\sffamily}{\thesubsection}{1em}{}
\titleformat{\subsubsection}{\large\sffamily}{\thesubsubsection}{1em}{}

 \ChTitleVar{\huge\sf}

\usepackage{hyperref}                       
\hypersetup{colorlinks=true,linkcolor=blue,citecolor=blue}

\hypersetup{
   pdftitle = {QUESO user's manual},
   pdfsubject = {The QUESO Library -- Quantification of Uncertainty for Estimation, Simulation, and Optimization},
   pdfkeywords = {research, uncertainty quantification, statistical inverse and statistical forward problems, validation.},
   pdfauthor = {Kemelli C. Estacio-Hiroms}
   }

\newcommand{\Quesoweb}{\url{https://github.com/libqueso}}
\newcommand{\Queso}{QUESO}

\newcommand{\bv}[1]{\ensuremath{\mbox{\boldmath$ #1 $}}}
\newcommand{\post}{\text{posterior}}
\newcommand{\prior}{\text{prior}}
\newcommand{\D}{ {\bf D}}
\newcommand{\be}{\begin{equation}}
\newcommand{\ee}{\end{equation}}

\newcommand{\myverb}[1]{ \indent{ \begin{verbatim} #1 \end{verbatim} } }
\newcommand{\QUESOversion}{0.51.0}

\usepackage{color}
\definecolor{dkgreen}{rgb}{0,0.6,0}
\definecolor{gray}{rgb}{0.5,0.5,0.5}
\definecolor{mauve}{rgb}{0.58,0,0.82}

\usepackage{listings} 
\newcommand{\chainsizeresults}{20000}
\newcommand{\new}[1]{#1}

\makeatletter
\def\BState{\State\hskip-\ALG@thistlm}
\makeatother

\begin{document}

\setlength{\unitlength}{1.0in}
\pagestyle{headings}
\markright{}
\pagenumbering{roman}
\numberwithin{equation}{section}
\numberwithin{figure}{section}
\numberwithin{table}{section}


\thispagestyle{empty}
{\setlength{\parindent}{0cm}\bf{\sf The QUESO Library}}\hfill $~$\\
\begin{picture}(8,0.1)
\linethickness{3pt}
\put(0,0.1){\line(1,0){6.6}}
\end{picture}

\begin{flushright}
\sf
User's Manual\\
Version \QUESOversion\\
\end{flushright}

\vfill

\begin{center}
\begin{LARGE}
\sf\bf
Quantification of Uncertainty for Estimation,\\
Simulation, and Optimization (QUESO)\\
\end{LARGE}
\end{center}

\vfill
$~$\\

\noindent
{\bf\sf Editors:}\hfill \\
{\sf Kemelli C. Estacio-Hiroms}  \\
{\sf Ernesto E. Prudencio} \\
{\sf Nicholas P. Malaya} \\
{\sf Manav Vohra} \\
{\sf Damon McDougall} \\

\vfill

\begin{minipage}[b]{0.20\linewidth}
\includegraphics[height=4\baselineskip]{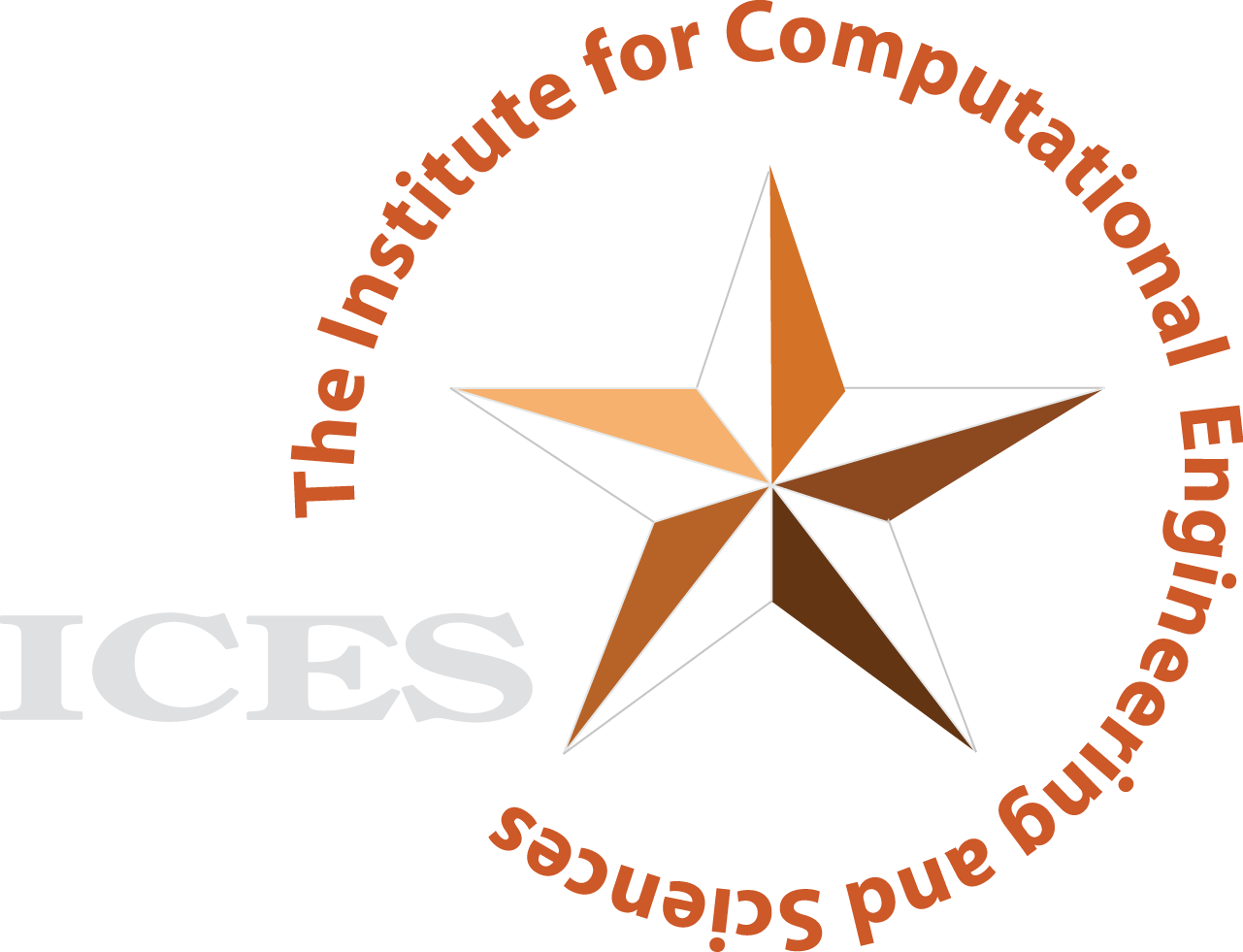}
\end{minipage}
\hfill
\begin{minipage}[b]{0.80\linewidth}
\small\sf
Center for Predictive Engineering and Computational Sciences (PECOS) \hfill\\
Institute for Computational and Engineering Sciences (ICES) \hfill\\
The University of Texas at Austin\hfill\\
Austin, TX 78712, USA
\end{minipage}
$~$\\
\begin{picture}(8,0.1)
\linethickness{1.5pt}
\put(0,0.1){\line(1,0){6.6}}
\end{picture}

\clearpage

\thispagestyle{empty}
$~$\\
\vfill
Copyright \copyright\ 2008-2013 The PECOS Development Team, \texttt{http://pecos.ices.utexas.edu}\\
Permission is granted to copy, distribute and/or modify this document under the terms of
the GNU Free Documentation License, Version 1.2 or any later version published by the Free
Software Foundation; with the Invariant Sections being ``GNU General Public License'' and
``Free Software Needs Free Documentation'', the Front-Cover text being ``A GNU Manual'',
and with the Back-Cover text being ``You have the freedom to copy and modify this GNU Manual''.
A copy of the license is included in the section entitled ``GNU Free Documentation License''.

\clearpage
\addcontentsline{toc}{chapter}{Abstract}
\centerline{\LARGE\sffamily Abstract}
$~$\\

QUESO stands for Quantification of Uncertainty for Estimation, Simulation and Optimization and consists of
 a collection of algorithms and C++ classes intended for
research in uncertainty quantification,
including
the solution of statistical inverse and statistical forward problems,
the validation of mathematical models under uncertainty, and
the prediction of quantities of interest from such models along with
the quantification of their uncertainties.

QUESO is designed for flexibility, portability, easy of use and
easy of extension. Its software design follows an object-oriented
approach and its code is written on C++ and over MPI. It can run over
uniprocessor or multiprocessor environments.

QUESO contains two forms of documentation:
a user's manual available in PDF format
and
a lower-level code documentation available in web based/HTML format.

This is the user's manual: it gives an overview of the QUESO capabilities,
provides procedures for software execution, and includes example studies.

\clearpage
$~$\\

\clearpage
\addcontentsline{toc}{chapter}{Disclaimer}
\centerline{\LARGE\sffamily Disclaimer}
$~$\\
    This document was prepared
    by The University of Texas at Austin.
    Neither the University of Texas
    at Austin, nor any of its institutes, departments and employees, make any warranty, express or implied,
    or assume any legal liability or responsibility for the accuracy, completeness, or
    usefulness of any information, apparatus, product, or process disclosed, or represent
    that its use would not infringe privately owned rights. Reference herein to any specific
    commercial product, process, or service by trade name, trademark, manufacturer, or otherwise,
    does not necessarily constitute or imply its endorsement, recommendation, or favoring by
    The University of Texas at Austin or any of its institutes, departments and employees thereof.
    The views and opinions expressed herein do not necessarily state or reflect
    those of The University of Texas at Austin or any institute or department
    thereof.

    \new{
	QUESO library as well as this material are provided as is, with absolutely no warranty
    expressed or implied.  Any use is at your own risk.}

\clearpage
$~$\\

\clearpage 
{\markboth{}{}
\addtocontents{toc}{\protect\markboth{}{}}
}
\tableofcontents


\clearpage
\addcontentsline{toc}{chapter}{Preface}
\thispagestyle{empty}
\chapter*{Preface}
$~$\\
The QUESO project started in 2008 as part
of the efforts of the recently established Center for Predictive Engineering and Computational Sciences (PECOS)
at the Institute for Computational and Engineering Sciences (ICES) at The University of Texas at Austin.

The PECOS Center was selected by the National Nuclear Security Administration (NNSA) as one of its new five centers of excellence
under the Predictive Science Academic Alliance Program (PSAAP).
The goal of the PECOS Center is
to advance predictive science and to develop the next generation of advanced computational methods and tools
for the calculation of reliable predictions on the behavior of complex phenomena and systems (multiscale, multidisciplinary).
This objective demands a systematic, comprehensive treatment of the calibration and validation of the mathematical models involved,
as well as the quantification of the uncertainties inherent in such models.
The advancement of predictive science is essential for the application of Computational Science to the solution of realistic problems of national interest.

The QUESO library is released as open source under Version 2.1 of the GNU
Lesser General Public License and is available for free download world-wide.
See https://www.gnu.org/licenses/lgpl-2.1.html for more information on the
LGPLv2.1 software use agreement.\\

\noindent
{\bf Contact Information:}\\
Paul T. Bauman,
Kemelli C. Estacio-Hiroms or
Damon McDougall,
\\
Institute for Computational and Engineering Sciences\\
1 University Station C0200\\
Austin, Texas 78712\\
email: pecos-dev@ices.utexas.edu\\
web: http://pecos.ices.utexas.edu\\
$~$\\

\section*{Referencing the QUESO Library}
When referencing the QUESO library in a publication, please cite the following:
\begin{verbatim}
@incollection{QUESO,
  author    = "Ernesto Prudencio and Karl W. Schulz",
  title     = {The Parallel C++ Statistical Library `QUESO':
               Quantification of Uncertainty for Estimation,
               Simulation and Optimization},
  booktitle = {Euro-Par 2011: Parallel Processing Workshops},
  series    = {Lecture Notes in Computer Science},
  publisher = {Springer Berlin / Heidelberg},
  isbn      = {978-3-642-29736-6},
  keyword   = {Computer Science},
  pages     = {398-407},
  volume    = {7155},
  url       = {http://dx.doi.org/10.1007/978-3-642-29737-3_44},
  year      = {2012}
}

@TechReport{queso-user-ref,
   Author      = {K.C. Estacio-Hiroms and E.E. Prudencio and N.P. Malaya
                  and M. Vohra and D. McDougall},
   Title       = {{T}he {QUESO} {L}ibrary, {U}ser's {M}anual},
   Institution = {Center for Predictive Engineering and Computational Sciences
                  (PECOS), at the Institute for Computational and Engineering
                  Sciences (ICES), The University of Texas at Austin},
   Note        = {Consistent with QUESO version 0.56.0},
   Year        = {2016}
}


@Misc{queso-web-page,
  Author = {{QUESO} Development Team},
  Title  = {{T}he {QUESO} {L}ibrary: {Q}uantification of {U}ncertainty
            for {E}stimation, {S}imulation and {O}ptimization},
  Note   = \url{https://github.com/libqueso/},
  Year   = {2008-2013}
}

\end{verbatim}
%
%
\section*{\Queso{} Development Team}

The QUESO development team currently consists of
Paul T. Bauman,
Sai Hung Cheung,
Kemelli C. Estacio-Hiroms,
Nicholas Malaya,
Damon McDougall,
Manav Vohra,
Kenji Miki,
Todd A. Oliver,
Ernesto E. Prudencio,
Karl W. Schulz,
Chris Simmons, and
Rhys Ulerich.

%
\section*{Acknowledgments}

This work has been supported by the United States Department of Energy,
under the National Nuclear Security Administration Predictive Science Academic Alliance Program (PSAAP) award number [DE-FC52-08NA28615], and
under the Office of Science Scientific Discovery through Advanced Computing (SciDAC) award number [DE-SC0006656].

We would also like to thank
James Martin,
Roy Stogner and
Lucas Wilcox
for interesting discussions and constructive feedback.

%
%
\section*{Target Audience}


\vspace{10pt}
\new{

QUESO is a collection of statistical algorithms and programming constructs supporting research into the uncertainty quantification (UQ) of models and their predictions. UQ may be a very complex and time consuming task, involving many steps:
decide which physical model(s) to use;
decide which reference or experimental data to use;
decide which discrepancy models to use;
decide which quantity(ies) of interest (QoI) to compute;
decide which parameters to calibrate;
perform computational runs and collect results;
analyze computational results, and eventually reiterate;
predict QoI(s) with uncertainty.

The purpose of this manual is \underline{not} to teach UQ and its methods, but rather to introduce QUESO library so it can be used as a tool to assist and facilitate the uncertainty quantification of the user's application.
Thus, the target audience of this manual is researchers who have solid background in Bayesian methods, are  comfortable with UNIX concepts and the command line, and have  knowledge of a programming language, preferably
C/C++. Bellow we suggest some useful literature:
\vspace{-4pt}
\begin{enumerate}
\item Probability, statistics, random variables \cite{DasGupta2008,Durret2005,JacodProtter2004};\vspace{-9pt}
\item Bayes' formula \cite{CarlinLouis2009,GelmanEtAl2004,Jaynes2003,Ro04};\vspace{-9pt}
\item Markov chain Monte Carlo (MCMC) methods \cite{CaSo07,GrMi01,HaLaMiSa06,HaSaTa01,Hast_1970,KaSo05,Laine08,Metr_1953,Mira01};\vspace{-9pt}
\item Monte Carlo methods \cite{RoCa04};\vspace{-9pt}
\item Kernel density estimation \cite{Silverman1986};\vspace{-9pt}
\item C++ \cite{StlJosuttis1999,CppLipman2005};\vspace{-9pt}
\item Message Passing Interface (MPI) \cite{Openmpi,Mpich};\vspace{-9pt}
\item UNIX/Linux (installation of packages, compilation, linking);\vspace{-9pt}
\item MATLAB/GNU Octave (for dealing with output files generated by QUESO); and \vspace{-9pt}
\item UQ issues in general \cite{VvUqReport2012}.\vspace{-9pt}
\end{enumerate}
}


\pagenumbering{arabic}


\chapter{Introduction}\label{ch-introduction}
\thispagestyle{headings}
\markboth{Chapter \ref{ch-introduction}: Introduction}{Chapter \ref{ch-introduction}: Introduction}

QUESO is a parallel object-oriented statistical library dedicated to the research of statistically robust, scalable, load balanced, and fault-tolerant mathematical algorithms for the quantification of uncertainty (UQ) of mathematical models and their predictions. 

The purpose of this chapter is to introduce relevant terminology, mathematical and statistical concepts, statistical algorithms, together with an overall description of how the user's application may be linked with the QUESO library.

\section{Preliminaries}

Statistical inverse theory reformulates inverse problems as problems of statistical inference by means of Bayesian statistics: all quantities are modeled as random variables, and probability distribution of the quantities encapsulates the uncertainty observed in their values. The solution to the inverse problem is then the probability distribution of the quantity of interest when all information available has been incorporated in the model. This (posterior) distribution describes the degree of confidence about the quantity after the measurement has been performed \cite{KaSo05}.

Thus, the solution to the statistical inverse problem may be given by Bayes' formula, which express the posterior distribution as a function of the prior distribution and the data represented through the likelihood function.

The likelihood function has an open form and its evaluation is highly computationally expensive.  Moreover, simulation-based posterior inference requires a large number of forward calculations to be performed, therefore fast and efficient sampling techniques are required for posterior inference.

It is often not straightforward to obtain explicit posterior point estimates of the solution, since it usually involves the evaluation of a high-dimensional integral with respect to a possibly non-smooth posterior distribution. In such cases, an alternative integration technique is the Markov chain Monte Carlo method: posterior means may be estimated using the sample mean from a series of random draws from the posterior distribution.

QUESO is designed in an abstract way so that it can be used by any computational model, as long as a likelihood function (in the case of statistical inverse problems) and a quantity of interest (QoI) function (in the case of statistical forward problems) is provided by the user application.

QUESO provides tools for both sampling algorithms for statistical inverse problems, following Bayes' formula, and statistical forward problems. It contains Monte Carlo solvers (for autocorrelation, kernel density estimation and accuracy assessment), MCMC (e.g. Metropolis Hastings \cite{Metr_1953,Hast_1970}) as well as the DRAM \cite{HaLaMiSa06} (for sampling from probability distributions); it also has the capacity to handle many chains or sequences in parallel, each chain or sequence itself demanding many computing nodes because of the computational model being statistically explored \cite{PrSc12}.

\section{Key Statistical Concepts}\label{sec:statistical_concepts}

A computational model is a combination of a
mathematical model and a discretization that enables the approximate
solution of the mathematical model using computer algorithms and  might be used in two different types of problems:
forward or inverse. 

Any computational model is composed of a vector $\boldsymbol{\theta}$ of $n$ {\it parameters}, {\it state variables} $\mathbf{u}$, and {\it state equations} $\mathbf{r}(\boldsymbol{\theta},\mathbf{u}) = \mathbf{0}$.
Once the solution $\mathbf{u}$ is available, the computational model also includes extra functions for e.g.
the calculation of {\it model output data} $\mathbf{y} = \mathbf{y}(\boldsymbol{\theta},\mathbf{u})$, and the {\it prediction} of a
vector $\mathbf{q} = \mathbf{q}(\boldsymbol{\theta},\mathbf{u})$ of $m$~quantities~of~interest\text{ (QoI)},

Parameters designate all model variables that are neither state variables
nor further quantities computed by the model, such as: material properties, coefficients, constitutive parameters, boundary conditions, initial conditions,
external forces, parameters for modeling the model error, characteristics of an experimental apparatus (collection of devices and procedures),
discretization choices and numerical algorithm options.
%

In the case of a forward problem, the parameters $\boldsymbol{\theta}$ are given and
one then needs to compute $\mathbf{u}$, $\mathbf{y}$ and/or $\mathbf{q}$.
In the case of an inverse problem, however, experimental data $\mathbf{d}$ is given and
one then needs to {\it estimate} the values of the parameters $\boldsymbol{\theta}$ that
cause $\mathbf{y}$ to best fit  $\mathbf{d}$.

The process of parameter estimation is also referred to as model calibration or model update, and it usually precedes the computation of a QoI, a process called model prediction. 

\begin{figure*}[htb]
\begin{minipage}[b]{0.5\textwidth}
\setlength{\unitlength}{3500sp}%
\begingroup\makeatletter\ifx\SetFigFont\undefined%
\gdef\SetFigFont#1#2#3#4#5{%
  \reset@font\fontsize{#1}{#2pt}%
  \fontfamily{#3}\fontseries{#4}\fontshape{#5}%
  \selectfont}%
\fi\endgroup%
\begin{picture}(3714,1284)(799,-703)
\thicklines
{\color[rgb]{0,0,0}\put(3241,-61){\vector( 1, 0){1260}}
}%
{\color[rgb]{0,0,0}\put(2161,-691){\framebox(1080,1260){}}
}%
{\color[rgb]{0,0,0}\put(811,-61){\vector( 1, 0){1350}}
}%
{\color[rgb]{0,0,0}\put(3241,299){\vector( 1, 0){1260}}
}%
{\color[rgb]{0,0,0}\put(3241,-421){\vector( 1, 0){1260}}
}%
\put(2341, 74){\makebox(0,0)[lb]{\smash{{\SetFigFont{10}{14.4}{\rmdefault}{\mddefault}{\updefault}{\color[rgb]{0,0,0}Forward}%
}}}}
\put(2341,-286){\makebox(0,0)[lb]{\smash{{\SetFigFont{10}{14.4}{\rmdefault}{\mddefault}{\updefault}{\color[rgb]{0,0,0}Problem}%
}}}}
\put(3286,344){\makebox(0,0)[lb]{\smash{{\SetFigFont{10}{14.4}{\rmdefault}{\mddefault}{\updefault}{\color[rgb]{0,0,0}State $\mathbf{u}=?$}%
}}}}
\put(3286,-16){\makebox(0,0)[lb]{\smash{{\SetFigFont{10}{14.4}{\rmdefault}{\mddefault}{\updefault}{\color[rgb]{0,0,0}Output $\mathbf{y}=?$}%
}}}}
\put(3286,-376){\makebox(0,0)[lb]{\smash{{\SetFigFont{10}{14.4}{\rmdefault}{\mddefault}{\updefault}{\color[rgb]{0,0,0}Prediction $\mathbf{q}=?$}%
}}}}
\put(1396, 29){\makebox(0,0)[lb]{\smash{{\SetFigFont{10}{14.4}{\rmdefault}{\mddefault}{\updefault}{\color[rgb]{0,0,0}Input $\boldsymbol{\theta}$}%
}}}}
\end{picture}%
\\
\centering
(a)
\end{minipage}
\begin{minipage}[b]{0.5\textwidth}
\setlength{\unitlength}{3500sp}%
\begingroup\makeatletter\ifx\SetFigFont\undefined%
\gdef\SetFigFont#1#2#3#4#5{%
  \reset@font\fontsize{#1}{#2pt}%
  \fontfamily{#3}\fontseries{#4}\fontshape{#5}%
  \selectfont}%
\fi\endgroup%
\begin{picture}(3714,1284)(799,-703)
\thicklines
{\color[rgb]{0,0,0}\put(3241,-61){\vector( 1, 0){1260}}
}%
{\color[rgb]{0,0,0}\put(2161,-691){\framebox(1080,1260){}}
}%
{\color[rgb]{0,0,0}\put(811,299){\vector( 1, 0){1350}}
}%
{\color[rgb]{0,0,0}\put(811,-61){\vector( 1, 0){1350}}
}%
{\color[rgb]{0,0,0}\put(811,-421){\vector( 1, 0){1350}}
}%
\put(2341, 74){\makebox(0,0)[lb]{\smash{{\SetFigFont{10}{14.4}{\rmdefault}{\mddefault}{\updefault}{\color[rgb]{0,0,0}Inverse}%
}}}}
\put(2341,-286){\makebox(0,0)[lb]{\smash{{\SetFigFont{10}{14.4}{\rmdefault}{\mddefault}{\updefault}{\color[rgb]{0,0,0}Problem}%
}}}}
\put(1550,-331){\makebox(0,0)[lb]{\smash{{\SetFigFont{10}{14.4}{\rmdefault}{\mddefault}{\updefault}{\color[rgb]{0,0,0}$\mathbf{y}(\boldsymbol{\theta},\mathbf{u})$}%
}}}}
\put(3250, 29){\makebox(0,0)[lb]{\smash{{\SetFigFont{10}{14.4}{\rmdefault}{\mddefault}{\updefault}{\color[rgb]{0,0,0}Parameters $\boldsymbol{\theta}=?$}%
}}}}
\put(1200, 29){\makebox(0,0)[lb]{\smash{{\SetFigFont{10}{14.4}{\rmdefault}{\mddefault}{\updefault}{\color[rgb]{0,0,0}$\mathbf{r}(\boldsymbol{\theta},\mathbf{u})=\mathbf{0}$}%
}}}}
\put(801,389){\makebox(0,0)[lb]{\smash{{\SetFigFont{10}{14.4}{\rmdefault}{\mddefault}{\updefault}{\color[rgb]{0,0,0}Experimental $\mathbf{d}$}%
}}}}
\end{picture}%
\\
\centering 
(b)
\end{minipage}
\vspace{-20pt}
\caption{The representation of (a) a generic forward problem and (b) a generic inverse problem.}
\label{fig-generic-problems}
\end{figure*}
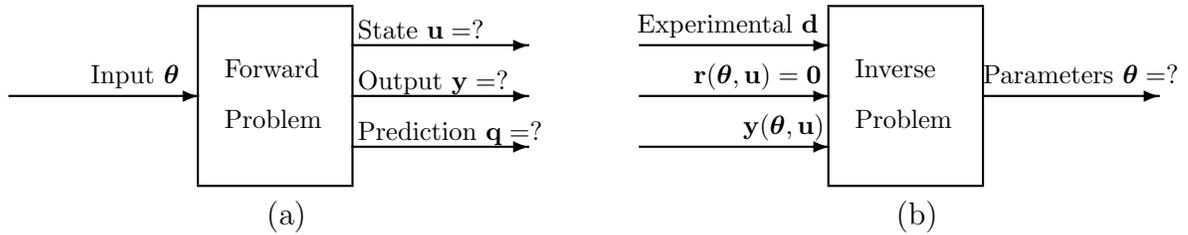

Figure~\ref{fig-generic-problems} represents general inverse and forward problems respectively.
There are many possible sources of uncertainty on a computational model. 
First, $\mathbf{d}$ need not be equal to the actual values of observables because of errors in the measurement process. Second, the values of the input parameters to the phenomenon might not be precisely known. Third, the appropriate set of
equations governing the phenomenon might not be well understood. 

Computational models can be classified as either deterministic or stochastic -- which are the ones of interest here.  In deterministic models, all parameters are assigned numbers, and no parameter is related to the parametrization of a random variable (RV) or field. As a
consequence, a deterministic model assigns a number to each of the components of quantities $\mathbf{u}$, $\mathbf{y}$ and $\mathbf{q}$. In stochastic models, however, at least one parameter is assigned a probability density function (PDF) or is related to the parametrization of a RV or field, causing $\mathbf{u}$, $\mathbf{y}$ and $\mathbf{q}$ to become random variables.  Note that not all components of $\boldsymbol{\theta}$ need to be treated as random. As long as at least one component is random, $\boldsymbol{\theta}$ is a random vector, and the problem is stochastic.

\begin{figure}[h!]
\centerline{
\setlength{\unitlength}{3500sp}%
\begingroup\makeatletter\ifx\SetFigFont\undefined%
\gdef\SetFigFont#1#2#3#4#5{%
  \reset@font\fontsize{#1}{#2pt}%
  \fontfamily{#3}\fontseries{#4}\fontshape{#5}%
  \selectfont}%
\fi\endgroup%
\begin{picture}(3807,1284)(796,-703)
\thicklines
{\color[rgb]{0,0,0}\put(2161,-691){\framebox(1080,1260){}}
}%
{\color[rgb]{0,0,0}\put(811,209){\vector( 1, 0){1350}}
}%
{\color[rgb]{0,0,0}\put(811,-331){\vector( 1, 0){1350}}
}%
{\color[rgb]{0,0,0}\put(3241,-61){\vector( 1, 0){1350}}
}%
\put(2300,119){\makebox(0,0)[lb]{\smash{{\SetFigFont{10}{14.4}{\rmdefault}{\mddefault}{\updefault}{\color[rgb]{0,0,0}Statistical}%
}}}}
\put(2340,-421){\makebox(0,0)[lb]{\smash{{\SetFigFont{10}{14.4}{\rmdefault}{\mddefault}{\updefault}{\color[rgb]{0,0,0}Problem}%
}}}}
\put(2340,-151){\makebox(0,0)[lb]{\smash{{\SetFigFont{10}{14.4}{\rmdefault}{\mddefault}{\updefault}{\color[rgb]{0,0,0}Forward}%
}}}}
\put(811,-241){\makebox(0,0)[lb]{\smash{{\SetFigFont{10}{14.4}{\rmdefault}{\mddefault}{\updefault}{\color[rgb]{0,0,0}$\mathbf{q}(\boldsymbol{\theta})$}%
}}}}
\put(811,299){\makebox(0,0)[lb]{\smash{{\SetFigFont{10}{14.4}{\rmdefault}{\mddefault}{\updefault}{\color[rgb]{0,0,0}Input RV $\boldsymbol{\Theta}$}%
}}}}
\put(3321, 29){\makebox(0,0)[lb]{\smash{{\SetFigFont{10}{14.4}{\rmdefault}{\mddefault}{\updefault}{\color[rgb]{0,0,0}Output RV $\mathbf{Q}$}%
}}}}
\end{picture}%
\\
}
\caption{
The representation of a statistical forward problem.
$\boldsymbol{\Theta}$ denotes a random variable related to parameters,
$\boldsymbol{\theta}$ denotes a realization of $\boldsymbol{\Theta}$ and
$\mathbf{Q}$ denotes a random variable related to quantities of interest.
}
\label{fig-sfp-queso}
\end{figure}
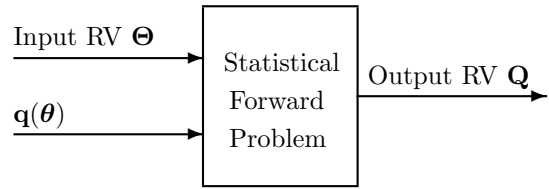

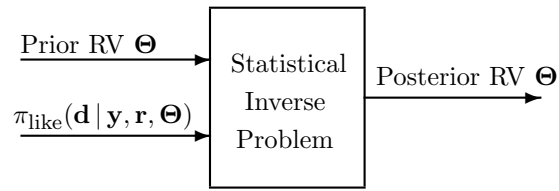
\begin{figure}[h!]
\centerline{
\setlength{\unitlength}{3500sp}%
\begingroup\makeatletter\ifx\SetFigFont\undefined%
\gdef\SetFigFont#1#2#3#4#5{%
  \reset@font\fontsize{#1}{#2pt}%
  \fontfamily{#3}\fontseries{#4}\fontshape{#5}%
  \selectfont}%
\fi\endgroup%
\begin{picture}(3807,1284)(796,-703)
\thicklines
{\color[rgb]{0,0,0}\put(3241,-61){\vector( 1, 0){1260}}
}%
{\color[rgb]{0,0,0}\put(2161,-691){\framebox(1080,1260){}}
}%
{\color[rgb]{0,0,0}\put(811,-331){\vector( 1, 0){1350}}
}%
{\color[rgb]{0,0,0}\put(811,209){\vector( 1, 0){1350}}
}%
\put(3331, 29){\makebox(0,0)[lb]{\smash{{\SetFigFont{10}{14.4}{\rmdefault}{\mddefault}{\updefault}{\color[rgb]{0,0,0}Posterior RV $\boldsymbol{\Theta}$}%
}}}}
\put(2300,119){\makebox(0,0)[lb]{\smash{{\SetFigFont{10}{14.4}{\rmdefault}{\mddefault}{\updefault}{\color[rgb]{0,0,0}Statistical}%
}}}}
\put(2340,-421){\makebox(0,0)[lb]{\smash{{\SetFigFont{10}{14.4}{\rmdefault}{\mddefault}{\updefault}{\color[rgb]{0,0,0}Problem}%
}}}}
\put(2400,-151){\makebox(0,0)[lb]{\smash{{\SetFigFont{10}{14.4}{\rmdefault}{\mddefault}{\updefault}{\color[rgb]{0,0,0}Inverse}%
}}}}
\put(811,254){\makebox(0,0)[lb]{\smash{{\SetFigFont{10}{14.4}{\rmdefault}{\mddefault}{\updefault}{\color[rgb]{0,0,0}Prior RV $\boldsymbol{\Theta}$}%
}}}}
\put(760,-241){\makebox(0,0)[lb]{\smash{{\SetFigFont{10}{14.4}{\rmdefault}{\mddefault}{\updefault}{\color[rgb]{0,0,0}$\pi_{\mbox{\scriptsize{like}}}(\mathbf{d}\,|\,\mathbf{y},\mathbf{r},\boldsymbol{\Theta})$}%
}}}}
\end{picture}%
\\
}
\caption{
The representation of a statistical inverse problem.
$\boldsymbol{\Theta}$ denotes a random variable related to parameters,
$\boldsymbol{\theta}$ denotes a realization of $\boldsymbol{\Theta}$ and
$\mathbf{r}$ denotes model equations,
$\mathbf{y}$ denotes some model output data and
$\mathbf{d}$ denotes experimental data.
}
\label{fig-sip-queso}
\end{figure}

In the case of forward problems, statistical forward problems can be represented very similarly to deterministic forward problems,
as seen in Figure \ref{fig-sfp-queso}.
In the case of inverse problems, as depicted in Figure \ref{fig-sip-queso}, however, the conceptual connection between deterministic and statistical problems
is not as straightforward.

QUESO adopts a Bayesian analysis \cite{KaSo05, Ro04} for statistical inverse problems, interpreting the posterior PDF
\begin{equation}\label{eq-Bayes-solution}
\pi_{\text{posterior}}(\boldsymbol{\theta}|\mathbf{d})=\frac{\pi_{\text{prior}}(\boldsymbol{\theta})\pi_{\text{likelihood}}(\mathbf{d}|\boldsymbol{\theta})}{\pi(\mathbf{d})}
\end{equation}
as the solution. Such solutions combine the prior information $\pi_{\text{prior}}(\boldsymbol{\theta})$ of the parameters,
the information $\pi(\mathbf{d})$ on the data, and the likelihood $\pi_{\text{likelihood}}(\mathbf{d}|\boldsymbol{\theta})$ that the model computes certain data values with a given set of input parameters.

This semantic interpretation of achieving a posterior knowledge on the parameters (on the model)
after combining some prior model knowledge with experimental information provides an important mechanism for dealing with uncertainty.
Although mathematically simple, is not computationally trivial. 

\section{The Software Stack of an Application Using QUESO}


An application using QUESO falls into three categories: a statistical inverse problem (IP), a statistical forward problem (FP), or combinations of both.
In each problem the user might deal with up to five vectors of potentially very different sizes:
parameters $\boldsymbol{\theta}$, state $\mathbf{u}$, output $\mathbf{y}$, data $\mathbf{d}$ and QoIs $\mathbf{q}$.

\begin{figure}[!htbp]
\centerline{
\includegraphics[scale=0.4,clip=true]{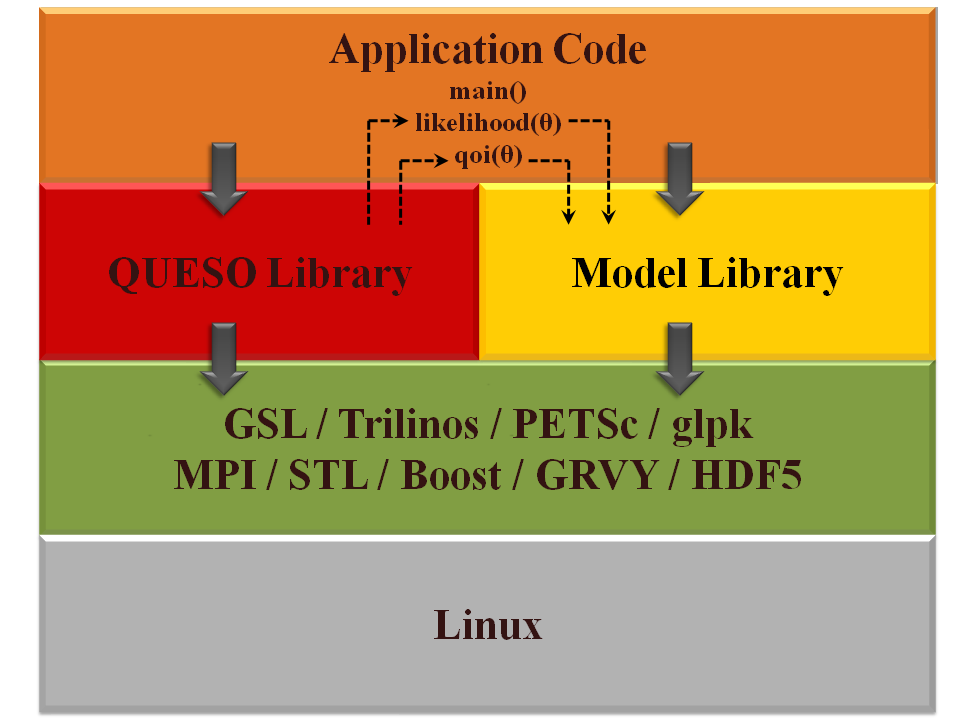}
}
\caption{
An application software stack.
QUESO requires the input
of a likelihood routine $\pi_{\text{like}}:\mathbb{R}^n\rightarrow\mathbb{R}_+$ for IPs and 
of a QoI routine $\mathbf{q}:\mathbb{R}^n\rightarrow\mathbb{R}^m$ for FPs.
These application level routines provide the bridge between
the statistical algorithms in QUESO,
physics 
knowledge in the model library, and relevant 
experimental (calibration
    and validation) data.
}
\label{fig-sw-stack}
\end{figure}

Algorithms in the QUESO library require the supply
of a likelihood routine $\pi_{\text{like}}:\mathbb{R}^n\rightarrow\mathbb{R}_+$ for statistical inverse problems and 
of a QoI routine $\mathbf{q}:\mathbb{R}^n\rightarrow\mathbb{R}^m$ for statistical forward problems. These routines
exist at the application level and provide the necessary bridge between the statistical algorithms in QUESO,
model knowledge in the model library and scenario and experimental data in the disk space.
%
Figure~\ref{fig-sw-stack} shows the software stack of a typical application that uses QUESO. In the figure, the symbol $\boldsymbol{\theta}$ represents a vector of $n\geqslant 1$ parameters. 

Even though QUESO deals directly with $\boldsymbol{\theta}$ and $\mathbf{q}$ only,
it is usually the case the one of the other three vectors ($\mathbf{u}$, $\mathbf{y}$ and $\mathbf{d}$) will have the biggest number of components and will therefore
dictate the size of the minimum parallel environment to be used in a problem.
So, for example, even though one processor might be sufficient for handling $\boldsymbol{\theta}$, $\mathbf{y}$, $\mathbf{d}$ and $\mathbf{q}$,
eight processors at least might be necessary to solve for $\mathbf{u}$.
QUESO currently only requires that the amounts $n$ and $m$ can be handled by the memory available to one processor,
which allows the analysis of problems with thousands of parameters and QoIs, a large amount even for state of the art UQ algorithms.

QUESO currently supports three modes of parallel execution:
an application user may simultaneously run:
\begin{description}
\item[(a)] multiple instances of a problem where the physical model requires a single processor, or
\item[(b)] multiple instances of a problem where the physical model requires multiple processors, or
\item[(c)] independent sets of types (a) and (b).
\end{description}

For example, suppose an user wants to use the Metropolis-Hastings (MH) algorithm to solve a statistical IP, and that 1,024 processors are available.
If the physical model is simple enough to be handled efficiently by a single processor, then the user can run 1,024 chains simultaneously, as in case (a).
If the model is more complex and requires, say, 16 processors, then the user can run 64 chains simultaneously, as in case (b), with 16 processors per chain.
QUESO treats this situation by using only 1 of the 16 processors to handle the chain.
When a likelihood evaluation is required, all 16 processors call the likelihood routine simultaneously.
Once the likelihood returns its value, QUESO puts  15 processors into idle state until the routine is called again or the chain completes.
Case (c) is useful, for instance, in the case of a computational procedure involving two models,
where a group of processors can be split into two groups, each handling one model.
Once the two-model analysis end, the combined model can use the full set of processors.\footnote{The parallel capabilities of QUESO have been exercised on the Ranger system of the TACC \cite{tacc} with up to 16k processors.}

\section{Algorithms for solving Statistical Inverse Problems}

The goal of inference is to characterize the posterior PDF, or to evaluate
point or interval estimates based on the posterior~\cite{HuMa01}.  Samples from
posterior can be obtained using Markov chain Monte Carlo (MCMC) which require
only pointwise evaluations of the unnormalized posterior.  The resulting
samples can then be used to either visually present the posterior or its
marginals, or to construct sample estimates of posterior expectations.
Examples of MCMC are: the Metropolis-Hastings (MH)
algorithm~\cite{Metr_1953,Hast_1970}, the Delayed Rejection (DR)
algorithm~\cite{GrMi01,Mira01}, and Adaptive Metropolis (AM)~\cite{HaSaTa01}
which are combined together in the Delayed Rejection Adaptive Metropolis, DRAM,
algorithm~\cite{HaLaMiSa06}. The DRAM is implemented in QUESO and available for
the solution of SIP. MCMC methods are well-established and
documented~\cite{CaSo07,GrMi01,HaLaMiSa06,HaSaTa01,Hast_1970,KaSo05,Laine08,Metr_1953,Mira01};
thus only brief description of the DRAM algorithm is presented in Section
\ref{sec:DRAM}.

During model construction, errors arising from imperfect modeling and
uncertainties due to incomplete information about the system and its
environment always exist; thus, there has been a crescent interest in Bayesian
model class updating  and selection
\cite{ChingChen2007,ChOlPr10,CheungPrudencio2012}. 

Model updating refers to the methodology that determines the most plausible
model for a system, given a prior PDF. One stochastic method that handles model
updating successfully is the multilevel method. Throughout the years, sereveral
versions of the same method have been implemented as improvements of its
predecessors~\cite{BeckAu2002,ChingChen2007,CheungPrudencio2012}. QUESO hosts
the novel Adaptive Multilevel Stochastic Simulation Algorithm
(AMSSA)~\cite{CheungPrudencio2012}, which is described in Section
\ref{sec:ML:intro}. For details about the method, please refer to
\cite{CheungPrudencio2012}.


\subsection{DRAM Algorithm}\label{sec:DRAM}

DRAM is a combination of two ideas for improving the efficiency of
Metropolis-Hastings type Markov chain Monte Carlo (MCMC) algorithms, Delayed
Rejection and Adaptive Metropolis~\cite{DRAMtool}. 

Random walk Metropolis-Hasting algorithm with Gaussian proposal distribution is
useful in simulating from the posterior distribution in many Bayesian data
analysis situations.
%
In order for the chain to be efficient, the proposal covariance must somehow be
tuned to the shape and size of the target distribution. This is important in
highly nonlinear situations, when there are correlation between the components
of the posterior, or when the dimension of the parameter is high. The problem
of adapting the proposal distribution using the chain simulated so far is that
when the accepted values depend on the history of the chain, it is no longer
Markovian and standard convergence results do not apply. One solution is to use
adaptation only for the burn-in period and discard the part of the chain where
adaptation has been used. In that respect, the adaptation can be thought as
automatic burn-in. The idea of diminishing adaptation is that when adaptation
works well, its effect gets smaller and we might be able to prove the
ergodicity properties of the chain even when adaptation is used throughout the
whole simulation. This is the ideology behind AM adaptation. On the other hand,
the DR method allows the use of the the current rejected values without losing
the Markovian property and thus allows to adapt locally to the current location
of the target distribution.

In Adaptive Metropolis~\cite{HaSaTa01} the covariance matrix of the Gaussian
proposal distribution is adapted on the fly using the past chain. This
adaptation destroys the Markovian property of the chain, however, it can be
shown that the ergodicity properties of the generated sample remain. How well
this works on finite samples and on high dimension is not obvious and must be
verified by simulations.

Starting from initial covariance $C^{(0)}$, the target covariance is updated at
given intervals from the chain generated so far.
$$
C^{(i)} = s_d \, cov(\text{chain}_1:\text{chain}_i) + s_d \varepsilon I_d,
$$
the small number $\varepsilon$ prevents the sample covariance matrix from
becoming singular. For the scaling factor, the value $s_d = 2.4^2/d$ is
standard optimal choice for Gaussian targets, $d$ being the dimension of the
target~\cite{GelmanEtAl2004}. A standard updating formula for the sample
covariance matrix can be used, so that the whole chain does not need to reside
in the computer memory.

With the Delayed rejection method~\cite{Mira01}, it becomes possible to make
use of several tries after rejecting a value by using different proposals while
keep the reversibility of the chain. Delayed rejection method (DR) works in the
following way. Upon rejection a proposed candidate point, instead of advancing
time and retaining the same position, a second stage move is proposed. The
acceptance probability of the second stage candidate is computed so that
reversibility of the Markov chain relative to the distribution of interest is
preserved. The process of delaying rejection can be iterated for a fixed or
random number of stages, let's say $n_\text{stages}$. The higher stage
proposals are allowed to depend on the candidates so far proposed and rejected.
Thus DR allows partial local adaptation of the proposal within each time step
of the Markov chain still retaining the Markovian property and reversibility.

The first stage acceptance probability in DR is the standard MH acceptance and
it can be written as
\begin{equation*}
\alpha_1(\mathbf{a},\mathbf{x}^{(1)}) = \text{ min}
\left\{ 1,
\frac{\pi(\mathbf{x}^{(1)})}{\pi(\mathbf{a})} \cdot
\frac{q_1(\mathbf{x}^{(1)},\mathbf{a})}{q_1(\mathbf{a},\mathbf{x}^{(1)})}
\right\},
\end{equation*}

Here $\mathbf{a}$ is the current point, $\mathbf{x}^{(1)}$ is the proposed new
value drawn from the distribution $q_1(\mathbf{a}, \cdot)$, and $\pi$ is the
target distribution.  If $\mathbf{x}^{(1)}$ is rejected, a second candidate
$\mathbf{x}^{(2)}$ is drawn from $q_2(\mathbf{a}, \mathbf{x}^{(1)} , \cdot)$
using the acceptance probability
\begin{equation*}
\alpha_2( \mathbf{a}, \mathbf{x}^{(1)}, \mathbf{x}^{(2)}) =\min \left\{1,
\dfrac{\pi( \mathbf{x}^{(2)}) q_1( \mathbf{x}^{(2)}, \mathbf{x}^{(1)}) q_2( \mathbf{x}^{(2)}, \mathbf{x}^{(1)}, \mathbf{a})[1 - \alpha_1( \mathbf{x}^{(2)} , \mathbf{x}^{(1)} )]}{\pi( \mathbf{a}) q_1( \mathbf{a}, \mathbf{x}^{(1)}) q_2( \mathbf{a}, \mathbf{x}^{(1)}, \mathbf{x}^{(2)})[1 - \alpha_1 ( \mathbf{a}, \mathbf{x}^{(1)} )]}
\right\}
\end{equation*}
i.e., it depends not only on the current position of the chain but also on what
we have just proposed and rejected.

As the reversibility property is preserved, this method also leads to the same
stationary distribution $\pi$ as the standard MH algorithm. The procedure can
be iterated further for higher-stage proposals. 
The Gaussian proposal at each stage $i$ is defined as:
\begin{equation} 
\label{eq:qi}
q_i(\underbrace{\mathbf{a},\mathbf{x}^{(1)},\ldots,\mathbf{x}^{(i-1)}}_{i\text{ terms}},\mathbf{z})
=
e^{-\dfrac{1}{2}{\displaystyle \left\{[\mathbf{z}-\mathbf{a}]^T\cdot \left[\mathbf{C}\right]^{-1}\cdot[\mathbf{z}-\mathbf{a}]\right\}}} 
\end{equation}

where the covariance matrix $\mathbf{C}$ and the scalings for the higher-stage
proposal covariances
$1=\gamma_1\leqslant\gamma_2\leqslant\ldots\leqslant\gamma_{n_{\text{stages}}}$
are given.

If $q_i$ denotes the proposal at the $i$-th stage, the acceptance probability
at that stage is:
\begin{equation}\label{eq-alphas}
\alpha_i(\mathbf{a},\mathbf{x}^{(1)},\ldots,\mathbf{x}^{(i)}) = \text{ min}
\left\{1,
\frac {\pi(\mathbf{x}^{(i)})}{\pi(\mathbf{a})} \cdot q_{\text{fraction}} \cdot \alpha_{\text{fraction}}
\right\}.
\end{equation}
where the expressions $q_{\text{fraction}}$ and $\alpha_{\text{fraction}}$ are
given by
\begin{equation*}
q_{\text{fraction}}=
\frac{q_1(\mathbf{x}^{(i)},\mathbf{x}^{(i-1)})}{q_1(\mathbf{a},\mathbf{x}^{(1)})}
\frac{q_2(\mathbf{x}^{(i)},\mathbf{x}^{(i-1)},\mathbf{x}^{(i-2)})}{q_2(\mathbf{a},\mathbf{x}^{(1)},\mathbf{x}^{(2)})}
\ldots
\frac{q_i(\mathbf{x}^{(i)},\mathbf{x}^{(i-1)},\ldots,\mathbf{x}^{(1)},\mathbf{a})}{q_i(\mathbf{a},\mathbf{x}^{(1)},\ldots,\mathbf{x}^{(i-1)},\mathbf{x}^{(i)})}
\end{equation*}
and
\begin{equation*}
\alpha_{\text{fraction}}=
\frac{[1-\alpha_1(\mathbf{x}^{(i)},\mathbf{x}^{(i-1)})]}{[1-\alpha_1(\mathbf{a},\mathbf{x}^{(1)})]}
\frac{[1-\alpha_2(\mathbf{x}^{(i)},\mathbf{x}^{(i-1)},\mathbf{x}^{(i-2)})]}{[1-\alpha_2(\mathbf{a},\mathbf{x}^{(1)},\mathbf{x}^{(2)})]}
\ldots
\frac{[1-\alpha_{i-1}(\mathbf{x}^{(i)},\mathbf{x}^{(i-1)},\ldots,\mathbf{x}^{(1)})]}{[1-\alpha_{i-1}(\mathbf{a},\mathbf{x}^{(1)},\ldots,\mathbf{x}^{(i-1)})]}.
\end{equation*}

Since all acceptance probabilities are computed so that reversibility with
respect to $\pi$ is preserved separately at each stage, the process of delaying
rejection can be interrupted at any stage that is, we can, in advance, decide
to try at most, say, 3 times to move away from the current position, otherwise
we let the chain stay where it is. Alternatively, upon each rejection, we can
toss a p-coin (i.e., a coin with head probability equal to p), and if the
outcome is head we move to a higher stage proposal, otherwise we stay
put~\cite{HaLaMiSa06}.

The smaller overall rejection rate of DR guarantees smaller asymptotic variance
of the estimates based on the chain. The DR chain can be shown to be
asymptotically more efficient that MH chain in the sense of Peskun ordering
(Mira, 2001a). 

%
%
%

Haario, et al. 2006 \cite{HaLaMiSa06} combine AM and DR into a method called
DRAM, in what they claim to be a straightforward possibility amongst the
possible different implementations of the idea, and which is described in this
section.

In order to be able to adapt the proposal, all you need some accepted points to
start with. 

One ``master" proposal is tried first -- i.e., the proposal at the first stage
of DR is adapted just as in AM: the covariance $C^{(1)}$  is computed from the
points of the sampled chain, no matter at which stage these points have been
accepted in the sample path.  After rejection, a try with modified version of
the first proposal is done according to DR. A second proposal can be one with a
smaller covariance, or with different orientation of the principal axes. The
most common choice is to always compute the covariance $C^{(i)}$  of the
proposal for the $i$-th stage ($i=2,\ldots, n_\text{stages}$) simply as a
scaled version of the proposal of the first stage, $$C^{(i)} = \gamma_i
C^{(1)}$$ where the scale factors $\gamma_i$ can be somewhat freely chosen.
Then, the master proposal is adapted using the chain generated so far, and the
second stage proposal follows the adaptation in obvious manner.


The requirements for the DRAM algorithm are:
\begin{itemize}
 \item Number $n_{\text{pos}}\geqslant 2$ of positions in the chain;
 \item Initial guess $\mathbf{m}^{(0)}$;
 \item Number of stages for the DR method: $n_{\text{stages}}\geqslant 1$;
 \item For $1\leqslant i\leqslant n_{\text{stages}}$, functions $q_i:\underbrace{\mathbb{R}^N\times\ldots\times\mathbb{R}^N}_{(i+1)\text{ times}}\rightarrow\mathbb{R}_{+}$, such that $q_i(\mathbf{a},\mathbf{x}^{(1)},\ldots,\mathbf{x}^{(i-1)},\cdot)$ is a PDF for any $(\mathbf{a},\mathbf{x}^{(1)},\ldots,\mathbf{x}^{(i-1)})\in\underbrace{\mathbb{R}^N\times\ldots\times\mathbb{R}^N}_{i\text{ times}}$; i.e., choose $q_i$ as in Equation~\eqref{eq:qi};
 \item Recursively define $\alpha_i:\underbrace{\mathbb{R}^n\times\ldots\times\mathbb{R}^n}_{(i+1)\text{ times}}\rightarrow [0,1],\quad 1\leqslant i\leqslant n_{\text{stages}}$ according to Equation~\eqref{eq-alphas}.
\end{itemize}


Recalling that a sample is defined as:
\begin{equation*} 
\text{a sample } = \mathbf{a}+\mathbf{C}^{1/2}\mathcal{N}(0,I).
\end{equation*}
a simple, but useful, implementation of DRAM is described in Algorithm \ref{alg:DRAM}.
 
\begin{algorithm}[!hp]
\caption{DRAM algorithm \cite{Laine08}.}\label{alg:DRAM}
  \begin{algorithmic}[1]
     \Statex{ \textbf{Input}: Number of positions in the chain $n_{\text{pos}}\geqslant 2$; initial guess $\mathbf{m}^{(0)}$; initial first stage proposal covariance $C^{(0)}$; $n_{\text{stages}}\geqslant 1$; and functions $q_i:\underbrace{\mathbb{R}^N\times\ldots\times\mathbb{R}^N}_{(i+1)\text{ times}}\rightarrow\mathbb{R}_{+}$}

      \myState{Select $s_d$}\Comment{\texttt{scaling factor}}            
      \myState{Select $\varepsilon$}\Comment{\texttt{covariance regularization factor}}            
      \myState{Select $n_0$}\Comment{\texttt{initial non-adaptation period}}            
       
      \For{$i\leftarrow 1$ \textbf{to} $n_{\text{stages}}$} \Comment{\texttt{$n_{\text{stages}}$ is the number of tries allowed}}
	\myState{Select $\gamma_i$}\Comment{\texttt{scalings for the higher-stage proposal covariances}}
      \EndFor
      
      \Do
	\myState{Set $ACCEPT \leftarrow false$~Set $i \leftarrow 1$} \Comment{\texttt{After an initial period of
simulation, adapt the master proposal (target) covariance using the chain generated so far:}}  
	\If {$k \geqslant n_0$}
	  \myState{$C^{(1)} = s_d Cov(\mathbf{m}^{(0)} ,\ldots, \mathbf{m}^{(k-1)}) + s_d \varepsilon I_d$}
	\EndIf
	\Comment{$n_{\text{stages}}$-DR loop: }
	  \Do
  	    \myState{Generate candidate $\mathbf{c}^{(i)}\in\mathbb{R}^N$ by sampling $q_i(\mathbf{m}^{(k)},\mathbf{c}^{(1)},\ldots,\mathbf{c}^{(i-1)},\cdot)$} \Comment{\texttt{$q_i$ is the proposal probability density}}
	    \If{$\mathbf{c}^{(i)}\notin supp(\pi)$}
		\myState{$i \leftarrow i+1 $}
	    \EndIf
	    
	    \If{$\mathbf{c}^{(i)}\in supp(\pi)$}
		\myState{Compute $\alpha_i(\mathbf{m}^{(k)},\mathbf{c}^{(1)},\ldots,\mathbf{c}^{(i-1)},\mathbf{c}^{(i)})$}\Comment{\texttt{acceptance probability}}
		\myState{Generate a sample $\tau \sim \mathcal{U}\left((0,1]\right ) $}
		\myState{\textbf{if} ($\alpha_i < \tau$) \textbf{then} $i\leftarrow i+1$}
		\myState{\textbf{if} ($\alpha_i \geqslant \tau$) \textbf{then} ACCEPT$\leftarrow$true}
	    \EndIf
	    
	    \myState{$C^{(i)} = \gamma_i C^{(1)}$}\Comment{\texttt{Calculate the higher-stage proposal as scaled versions of~$C^{(1)}$, according to the chosen rule}}

	  \doWhile{(ACCEPT=false) and ($i \leqslant n_{\text{stages}}$)}

	    \If{(\text{ACCEPT=true})}
		\myState{Set $\mathbf{m}^{(k+1)}\leftarrow \mathbf{c}^{(i)}$}
	    \EndIf
	    \If{(\text{ACCEPT=false})}
		\myState{Set $\mathbf{m}^{(k+1)}\leftarrow \mathbf{m}^{(k)}$}
	    \EndIf

	  \myState{Set $k \leftarrow k+1$}

      \doWhile{($k+1 < n_{\text{pos}}$)}
	
  \end{algorithmic}
\end{algorithm}

There are six variables in the QUESO input file used to set available options
for the DRAM algorithm, which are described in \ref{sec:MH}. Here, they are
presented presented bellow together with their respective definition in
Algorithm \ref{alg:DRAM}.
\begin{description}

 \item[\texttt{ip\_mh\_dr\_maxNumExtraStages}:] defines how many extra stages
   should be considered in the DR loop ($n_\text{stages}$);
 
 \item[\texttt{ip\_mh\_dr\_listOfScalesForExtraStages}:] defines the list $s$
   of scaling factors that will multiply the covariance matrix (values of
   $\gamma_i$ );

 \item[\texttt{ip\_mh\_am\_adaptInterval}:] defines whether or not there will
   be a period of adaptation;
   
 
 \item[\texttt{ip\_mh\_am\_initialNonAdaptInterval}:] defines the initial
   interval where the proposal covariance matrix will not be changed ($n_0$);
  
 \item[\texttt{ip\_mh\_am\_eta}:] is a factor used to scale the proposal
   covariance matrix, usually set to be $2.4^2/d$, where $d$ is the dimension
   of the problem~\cite{Laine08,HaLaMiSa06} ($s_d$);
 
 \item[\texttt{ip\_mh\_am\_epsilon}:] is the covariance regularization factor
   ($\varepsilon$).

\end{description}
%

\subsection{Adaptive Multilevel Stochastic Simulation Algorithm}
\label{sec:ML:intro}

In this section we rewrite the Bayesian formula \eqref{eq-Bayes-solution} by
making explicit all the implicit model assumptions. Such explication demands
the use of probability logic and the concept of a stochastic system model class
(''model class'' for short); as these concepts enable the comparison of
competing model classes. 

Let $M_j$ be one model class; the choice of $\bv{\theta} $ specifies a
particular predictive model in $M_j$, and, for brevity, we do not explicitly
write $\bv{\theta}_j $ to indicate that the parameters of different model
classes may be different, but this should be understood.  Based on $M_j$, one
can use data $D$ to compute the updated relative plausibility of each
predictive model in the set defined by $M_j$.  This relative plausibility is
quantified by the \textit{posterior} PDF $\pi(\boldsymbol{\theta}|\D,M_j)$.

Bayes' theorem allows the update of the probability of each predictive model
$M_j$ by combining measured data $D$ with the prior PDF into the posterior PDF:
\begin{equation}
\begin{split}
\pi_\post (\bv{\theta}|\D, M_j) &= \dfrac{f(\D|\bv{\theta}, M_j) \cdot \pi_\prior (\bv{\theta} | M_j)}{\pi(\D, M_j)} 
\\
&= \dfrac{f(\D|\bv{\theta}, M_j) \cdot \pi_\prior (\bv{\theta} | M_j)}{\int f(\D|\bv{\theta}, M_j) \cdot \pi_\prior (\bv{\theta} | M_j)\, d\bv{\theta}} 
\end{split}
\end{equation}
where the denominator expresses the probability of getting the data $\D$ based
on $M_j$ and is called the evidence for $M_j$ provided by $\D$;
$\pi_\prior (\bv{\theta} | M_j)$ is the prior PDF of the predictive model
$\bv{\theta}$ within $M_j$; and the likelihood function $f(\D|\bv{\theta}, M_j)$
expresses the probability of getting $\D$ given the predictive model
$\bv{\theta}$ within $M_j$ -- and this allows stochastic models inside a model
class $M_j$ to be compared.


When generating samples of posterior PDF $\pi_\post(\bv{\theta}|D,M_j) $ in
order to forward propagate uncertainty and compute QoI RV's, it is important to
take into account potential multiple modes in the posterior. One simple idea is
to sample increasingly difficult intermediate distributions, accumulating
``knowledge'' from one intermediate distribution to the next, until the target
posterior distribution is sampled.  In \cite{CheungPrudencio2012}, an advanced
stochastic simulation method, referred to as Adaptive Multi Level Algorithms,
is proposed which can generate posterior samples from
$\pi_\post (\bv{\theta}|\D, M_j)$ and compute the log of the evidence
$p(\D | \boldsymbol{\theta},M_j)$ at the same time by adaptively moving samples
from the prior to the posterior through an adaptively selected sequence of
intermediate distributions~\cite{ChOlPr10}.  

Specifically, the intermediate distributions are given by:
\begin{equation}
\label{eq:intermediate_dist}
\pi_\text{int}^{(\ell)} (\bv{\theta}|\D) = f(\bv{\theta}|\D, M_j)^{\tau_\ell} \cdot \pi_\prior (\bv{\theta} | M_j), \quad \ell=0,1,\ldots,L,
\end{equation}
for a given $L > 0$ and a given sequence $0 = \tau_0 < \tau_1 < \ldots < \tau_L = 1$
of exponents.

In order to compute the model evidence $\pi( \D |M_j)$ where:
\begin{equation}
 \label{eq:evidence}
 \pi( \D |M_j)=\int f(\bv{\theta}|\D, M_j) \cdot \pi_\prior (\bv{\theta}|M_j) \, d\bv{\theta}, 
\end{equation}
the use of intermediate distribution is also beneficial.
For that, recall that
\begin{equation}
\label{eq:cl}
\begin{split}
 \pi( \D |M_j) &= \int f(\bv{\theta})\pi (\bv{\theta}) \, d\bv{\theta} \\
  &= \int f \; \pi \, d\bv{\theta} \\
  &= \int f^{1-\tau_{L-1}} f^{\tau_{L-1}-\tau_{L-2}}\ldots f^{\tau_2-\tau_1} f^{\tau_1}\; \pi \, d\bv{\theta} \\
  &= c_1 \int f^{1-\tau_{L-1}} f^{\tau_{L-1}-\tau_{L-2}}\ldots f^{\tau_2-\tau_1} \dfrac{f^{\tau_1}\; \pi}{c_1} \, d\bv{\theta} \\
  &= c_2 c_1 \int f^{1-\tau_{L-1}} f^{\tau_{L-1}-\tau_{L-2}}\ldots  \dfrac{f^{\tau_2-\tau_1} f^{\tau_1}\; \pi}{c_2 c_1} \, d\bv{\theta} \\
  &= c_L c_{L-1} \cdots c_2 c_1.
\end{split}
\end{equation}
Assuming that the prior PDF is normalized (it integrates to one) and if
$\tau_{\ell}$ is small enough, then Monte Carlo method can be efficiently
applied to calculate $c_{\ell}$ in Equation \eqref{eq:cl}. Due to numerical
(in)stability, it is more appropriate to calculate the estimators:
\begin{equation}
 \label{eq:log-cl}
 \tilde{c_i} = \ln c_i, \quad i=1,\ldots, L.
\end{equation}

Combining Equations \eqref{eq:cl} and \eqref{eq:log-cl}, we have:
\begin{equation*}
 \ln[\pi( \D |M_j)] = \tilde{c}_{L}+\tilde{c}_{L-1}+\ldots+\tilde{c}_2+\tilde{c}_1.
\end{equation*}

Computing the log of the evidence instead of calculating the evidence directly
is attractive because the evidence is often too large or too small relative to
the computer precision.
The posterior probability can be calculated readily in terms of the log
evidence, allowing overflow and underflow errors to be avoided
automatically~\cite{ChOlPr10}.  


Now let's define some auxiliary variables for
$k=1,\ldots,n_\text{total}^{(\ell)}$:

\begin{itemize}
 \item $k$-th sample at the $\ell$-th level: 
    \begin{equation}\label{eq:samples}
    \bv{\theta}^{(\ell) [k]},   \quad \ell=0,1,\ldots, L \\ 
    \end{equation} 

 \item Plausibility weight:
    \begin{equation}
    \begin{split}\label{eq:w}
    w^{(\ell) [k]} &= \dfrac{f(\bv{\theta}^{(\ell) [k]}|\D, M_j)^{\tau_\ell} \cdot \pi_\prior (\bv{\theta}^{(\ell) [k]}, M_j)}{f(\bv{\theta}^{(\ell) [k]}|\D, M_j)^{\tau_\ell-1} \cdot \pi_\prior (\bv{\theta}^{(\ell) [k]}, M_j)}  
		=\dfrac{f^{(\tau_{\ell})}(\D | \bv{\theta}^{(\ell) [k]}, M_j) }{f^{(\tau_{\ell-1})}(\D | \bv{\theta}^{(\ell) [k]}, M_j)}, \\
		&= f^{(\tau_{\ell}-\tau_{\ell-1})}(\D | \bv{\theta}^{(\ell) [k]}, M_j), \quad \ell=0,1,\ldots, L \\ 
    \end{split}
    \end{equation}
    
\item Normalized plausibility weight:
    \begin{equation}\label{eq:w-tilde}
    \tilde{w}^{(\ell) [k]} = \dfrac{w^{(\ell) [k]}}{\sum_{s=1}^{n_\text{total}^{(\ell)}}  w^{(\ell) [s]} }, \quad \ell=0,1,\ldots,L 
    \end{equation}

\item Effective sample size:
    \begin{equation}\label{eq:neff}
    n_\text{eff}^{(\ell)} = \dfrac{1}{\sum_{s=1}^{n_\text{total}^{(\ell)}} \left( \tilde{w}^{(\ell) [s]}\right)^2}
    \end{equation}
    
\item Estimate for the sample covariance matrix for $\pi_\text{int}^{(\ell)}$:
    \begin{equation}\label{eq:est_cov}
     \Sigma = \sum_{m=1}^{n_\text{total}^{(\ell-1)}} \tilde{w}_{m} (\bv{\theta}^{(\ell-1) [m]} - \overline{\bv{\theta}}) (\bv{\theta}^{(\ell-1) [m]} - \overline{\bv{\theta}})^{t}, \quad \text{where} \quad
     \overline{\bv{\theta}} = \sum_{m=1}^{n_\text{total}^{(\ell-1)}}  \tilde{w}_{m} \bv{\theta}^{(\ell-1) [m]}
    \end{equation}

\end{itemize}

so we can define the discrete distribution:
\begin{equation}
 \label{eq:distribution}
 P^{(\ell)}(k) = \tilde{w}^{(\ell) [k]} , \quad k=1,2,\ldots, n_\text{total}^{(\ell)}.
\end{equation}

The ML algorithm consists of a series of resampling stages, with each stage
doing the following: given $n_\text{total}^{(\ell)}$ samples from
$\pi_\text{int}^{(\ell)} (\bv{\theta}|\D)$, denoted by
$\bv{\theta}^{(\ell)[k]}, k=1...n_\text{total}^{(\ell)}$ obtain samples from
$\pi_\text{int}^{(\ell+1)} (\bv{\theta}|\D)$, denoted by
$\bv{\theta}^{(\ell+1)[k]}, k=1...n_\text{total}^{(\ell+1)}$. 

This is accomplished by: given the samples
$\bv{\theta}^{(\ell)[k]}, k=1...n_\text{total}^{(\ell)}$, in Equation \eqref{eq:samples}, from
$\pi_\text{int}^{(\ell)} (\bv{\theta}|\D)$, we compute the plausibility weights
$w^{(\ell) [k]}$ given in Equation \eqref{eq:w} with respect to
$\pi_\text{int}^{(\ell+1)} (\bv{\theta}|\D)$. Then we re-sample the uncertain
parameters according to the normalized weights $\tilde{w}^{(\ell) [k]}$, given
in Equation \eqref{eq:w-tilde}, through the distribution in Equation
\eqref{eq:distribution}. This is possible due to the fact that for large
$n_\text{total}^{(\ell)}$ and $n_\text{total}^{(\ell+1)}$, then
$\bv{\theta}^{(\ell+1)[k]}, k=1...n_\text{total}^{(\ell+1)}$ will be
distributed as $\pi_\text{int}^{(\ell+1)}
(\bv{\theta}|\D)$~\cite{ChingChen2007}.

The choice of $\tau_\ell, \ell=1,\ldots,L-1$ is essential. It is desirable to
increase the $\tau$ values slowly so that the transition between adjacent PDFs
is smooth, but if the increase of the $\tau$ values is too slow, the required
number of intermediate stages ($L$ value) will be too
large~\cite{ChingChen2007}. More intermediate stages mean more computational
cost.  In the ML method proposed by \cite{CheungPrudencio2012} and implemented
in QUESO, $\tau_\ell$ is computed through a bissection method so that:
\begin{equation}\label{eq:tau}
\beta_\text{min}^{(\ell)}<\dfrac{n_\text{eff}^{(\ell)}}{n_\text{total}^{(\ell)}} <\beta_\text{max}^{(\ell)}
\end{equation}

\subsubsection{AMSSA Algorithm}

Based on the above results, and recalling that the series of intermediate PDFs,
$\pi_\text{int}^{(\ell)} (\bv{\theta}|\D)$, start from the prior PDF and ends
with the posterior PDF,  Algorithm \ref{alg:ML} can be applied both to draw
samples from the posterior PDF, $\pi_\post (\bv{\theta}|\D, M_j)$, and to
estimate the evidence $\pi( \D ,M_j)$.

\begin{algorithm}[!htb]
\caption{AMSSA Algorithm proposed by \cite{CheungPrudencio2012}.}\label{alg:ML}
\begin{algorithmic}[1]

\Statex{ \textbf{Input}: for each $\ell=0,\ldots,L$: the total amount of samples to be generated at $\ell$-th level ($n_\text{total}^{(\ell)}>0$) and the thresholds ($0<\beta_\text{min}^{(\ell)}<\beta_\text{max}^{(\ell)}<1$) on the effective sample size of the $\ell$-th level}

\Statex{\textbf{Output}: $\bv{\theta}^{(m) [k]}, k=1,\ldots,n_\text{total}^{(m)}$; which are  asymptotically distributed as  $\pi_\post(\bv{\theta}|\D, M_j)$}

\Statex{\textbf{Output}: $\prod_{\ell} c_{\ell}$; which is asymptotically unbiased for $\pi( \D ,M_j)$ }
\setcounter{ALG@line}{32}
\myState{Set $\ell=0$}
\myState{Set $\tau_\ell =0$}
\myState{Sample prior distribution, $\pi_\prior (\bv{\theta} | M_j)$, $n_\text{total}^{(0)}$ times}
\Comment{\texttt{i.e, obtain $\bv{\theta}^{(0) [k]}, k=1,\ldots,n_\text{total}^{(0)}$}}

\While{$\tau_\ell < 1$} 
\Comment{\texttt{At the beginning of the $\ell$-th level, we have the samples $\bv{\theta}^{(\ell-1)[k]}, k=1...n_\text{total}^{(\ell-1)}$ from  $\pi_\text{int}^{(\ell-1)} (\bv{\theta}|\D)$, Equation \eqref{eq:intermediate_dist}.}}

\myState{Set $\ell \leftarrow \ell + 1 $ }\Comment{\texttt{begin next level}}
\myState{Compute plausibility weights $w^{(\ell) [k]}$ via Equation \eqref{eq:w}
 Compute normalized  weights $\tilde{w}^{(\ell) [k]}$ via Equation \eqref{eq:w-tilde}
 Compute $n_\text{eff}^{(\ell)}$ via Equation \eqref{eq:neff}}
\myState{Compute $\tau_\ell$ so that Equation \eqref{eq:tau} is satisfied} 
\If{$\tau_\ell>1$}
\myState{$\tau_\ell \leftarrow 1$ Recompute  $w^{(\ell) [k]}$ and  $\tilde{w}^{(\ell) [k]}$}
\EndIf

\myState{Compute an estimate for the sample covariance matrix for $\pi_\text{int}^{(\ell)}$ via Equation \eqref{eq:est_cov}}
\myState{Select, from previous level, the initial positions for the Markov chains}\label{alg:ML:initialpos}
\myState{Compute sizes of the chains}\Comment{\texttt{the sum of the sizes $=n_\text{total}^{(\ell)}$}} \label{alg:ML:computechainsize}

\myState{Redistribute chain initial positions among processors} \label{alg:ML:redist}
\Comment{\texttt{Then the $n_\text{total}^{(\ell)}$ samples $\bv{\theta}^{(\ell)[k]}$, from $\pi_\text{int}^{(\ell)}(\bv{\theta})$ are generated by doing the following for  $k=1,\ldots,n_\text{total}^{(\ell)}$:}}

\myState{Generate chains: draw a number $k'$ from a discrete distribution $P^{(\ell)}(k)$ in Equation \eqref{eq:distribution} via Metropolis-Hastings}\Comment{\texttt{i.e., obtain $\bv{\theta}^{(\ell) [k]}= P^{(l)[k]}$}}
\myState{Compute $c_{\ell} =\frac{1}{ n_\text{total}^{(\ell-1)}}  \left( \sum_{s=1}^{n_\text{total}^{(\ell-1)}} w_{s} \right)$}\Comment{\texttt{recall that $\pi( \D |M_j) = \prod_{\ell} c_{\ell}$, Equation \eqref{eq:evidence}}}

\EndWhile
\end{algorithmic}

\end{algorithm}

Steps \ref{alg:ML:initialpos} and \ref{alg:ML:computechainsize} in Algorithm
\ref{alg:ML} are accomplished by sampling the distribution in
Equation~\eqref{eq:distribution} a total of $n_\text{total}^{(\ell)} $ times.
The selected indices $k$ determine the samples $\bv{\theta}^{(\ell) [k]}$ to be
used as initial positions, and the number of times an index $k$ is selected
determines the size of the chain beginning at $\bv{\theta}^{(\ell) [k]}$.

At each level $\ell$, many computing nodes can be used to sample the parameter
space collectively. Beginning with $\ell = 0$, the computing nodes:
(a) sample $\pi_\text{int}^{(\ell)}(\bv{\theta}|\bv{D}, M_j)$; 
(b) select some of the generated samples (``knowledge'') to serve as initial positions of Markov chains for the next distribution $\pi_\text{int}^{(\ell+1)}(\bv{\theta}|\bv{D}, M_j)$; and 
(c) generate the Markov chains for $\pi_\text{int}^{(\ell+1)}(\bv{\theta}|\bv{D}, M_j)$. 

The process (a)--(b)--(c) continues until the final posterior distribution is
sampled.  As $\ell$ increases, the selection process tends to value samples
that are located in the regions of high probability content, which gradually
``appear''as  $\tau_\ell$ increases. So, as $\ell$ increases, if the ``good''
samples selected from the  $\ell$-th level to the ($\ell$+1)-th level are not
redistributed among computing nodes before the Markov chains for the (
$\ell$+1)-th level are generated, the ``lucky'' computing nodes (that is, the
ones that had, already at the initial levels, samples in the final posterior
regions of high probability content) will tend to accumulate increasingly more
samples in the next levels. This possible issue is avoided maintaining a
balanced computational load among all computing nodes, which is handled in the
ML by the step in Line \ref{alg:ML:redist}. 

Running the step in Line \ref{alg:ML:redist} of Algorithm \ref{alg:ML} is then
equivalent of solving the following problem: given the number of processors
$N_p$, the total number of runs $n_\text{total}$ and the number of runs $n_j$
(to be) handled by the $j$-th processor; distribute $N_t$ tasks among the $N_p$
processors so that each processor gets its total number $n_j$ of program runs,
$j = 1, \ldots, N_p$, the closest possible to the mean
$\bar{n}=n_\text{total}/N_p$. This parallel implementation of the algorithm is
proposed in \cite{CheungPrudencio2012}, and it has been implemented in QUESO by
the same authors/researchers.

%

\section{Algorithms for solving the Statistical Forward Problem}

The Monte Carlo method is commonly used for analyzing uncertainty propagation,
where the goal is to determine how random variation, lack of knowledge, or
error affects the sensitivity, performance, or reliability of the system that
is being modeled \cite{RoCa04}.

Monte Carlo works by using random numbers to sample, according to a PDF, the
`solution space' of the problem to be solved.  Then, it iteratively evaluates a
deterministic model using such sets of random numbers as inputs.

Suppose we wish to generate random numbers distributed according to a positive
definite function in one dimension $P(x)$.  The function need not be normalized
for the algorithm to work, and the same algorithm works just as easily in a
many dimensional space. The random number sequence $x_i$, $i=0,1,2,\ldots$ is
generated by a random walk as follows:

\begin{enumerate}
\item Choose a starting point $x_0$
\item Choose a fixed maximum step size $\delta$.
\item Given a $x_i$,  generate the next random number as follows: 
  \begin{enumerate}
  \item Choose $x_\text{trial}$  uniformly and randomly in the interval $[x_i-\delta, x_i+\delta]$.
  \item Compute the ratio $w=\dfrac{P(x_\text{trial})}{P(x_i)}$.
 
   Note that $P$ need not be normalized to compute this ratio.
  
  \item If $w >1$ the trial step is in the  right direction, i.e., towards a region of higher probability. 
  
  Accept the step $x_{i+1} =x_\text{trial}$.
  
  \item  If $w <1$ the trial step is in the wrong direction, i.e., towards a region of lower probability.  We should not unconditionally reject this step! So accept the step conditionally if the decrease in probability is smaller than a random amount:
     \begin{enumerate}
     \item Generate a random number $r$ in the interval $[0,1]$.
     \item If $r < w$ accept the trial step $x_{i+1} = x_\text{trial}$.
     \item If $w \leq r $ reject the step $x_{i+1}=x_i$. Note that we don't discard this step! The two steps have the same value.
     \end{enumerate}
  \end{enumerate}
\end{enumerate}

There are essentially two important choices to be made.  First, the initial
point $x_0$ must be chosen carefully. A good choice is close to the maximum of
the desired probability distribution. If this maximum is not known (as is
usually the case in multi-dimensional problems), then the random walker must be
allowed to thermalize i.e., to find a good starting configuration: the
algorithm is run for some large number of steps which are then discarded.
Second, the step size must be carefully chosen. If it is too small, then most
of the trial steps will be accepted, which will tend to give a uniform
distribution that converges very slowly to $P(x)$. If it is too large the
random walker will step right over and may not ` `see" important peaks in the
probability distribution. If the walker is at a peak, too many steps will be
rejected. A rough criterion for choosing the step size is for the
$$ \text{Acceptance ratio} = \dfrac{\text{Number of steps accepted}}{\text{Total number of trial steps}}$$
to be around 0.5.

An implementation of Monte Carlo algorithm is described in Algorithm
\ref{alg:MC}.

\begin{algorithm}[!htb]
\caption{Monte Carlo Algorithm proposed by \cite{Metr_1953}.}\label{alg:MC}
\begin{algorithmic}[1]
\Statex{\textbf{Input}: Starting point $x_0$, step size $\delta$, number of trials $M$, number of steps per trial $N$, unnormalized density or probability function  $P(x)$ for the target distribution.}
\Statex{\textbf{Output}: Random number sequence $x_i$, $i=0,1,2,\ldots$}
\setcounter{ALG@line}{48}
\For{$i=0\ldots M$}
 \For{$j=0\ldots N$}
  \myState{Set $ x_\text{trial} \leftarrow  x_i + (2 \, \text{RAND([0,1])} - 1)  \delta$}
  \myState{Set $ x_\text{trial} \leftarrow  x_i + (2 \, \text{RAND([0,1])} - 1)  \delta$
  Set $ w = P(x_\text{trial}) / P(x) $
  Set $accepts \leftarrow 0$}
  \If{$w \geq 1$} \Comment{\texttt{uphill}}
    \myState{$x_{i+1} \leftarrow x_\text{trial} $}\Comment{\texttt{accept the step}}
    \myState{accepts $\leftarrow$ accepts+1}
  \Else \Comment{\texttt{downhill}}
    \myState{Set $r \leftarrow$ RAND([0,1])}\Comment{\texttt{but not too far}}
    \If{$ r < w$}
      \myState{$ x_{i+1} \leftarrow x_\text{trial} $ }\Comment{accept the step}
      \myState{accepts $\leftarrow$ accepts+1}
    \EndIf
  \EndIf
 \EndFor
\EndFor
\end{algorithmic}
\end{algorithm}

Monte Carlo is implemented in QUESO and it is the chosen algorithm to compute a
sample of the output RV (the QoI) of the SFP for each given sample of the input
RV.


\chapter{Installation}\label{ch-install}
\thispagestyle{headings}
\markboth{Chapter \ref{ch-install}: Installation}{Chapter \ref{ch-install}: Installation}


This chapter covers the basic steps that a user will need follow when beginning to use QUESO: 
how to obtain, configure, compile, build, install, and test the library.  It also presents both QUESO source and installed directory structure, some simple examples and finally,  introduces the user on how to use QUESO together with the user's  application.

This manual is current at the time of printing; however, QUESO library  is under active development.
For the most up-to-date, accurate and complete information, please visit the online \Queso{} Home Page\footnote{\Quesoweb}.
        
\section{Getting started}\label{sec:Pre_Queso}

In operating systems which have the concept of a superuser, it is generally recommended that most application work be done 
using an ordinary account which does not have the ability to make system-wide changes (and eventually break the system via 
(ab)use of superuser privileges).

Thus, suppose you want to install QUESO and its dependencies on the following directory:
\begin{lstlisting}[{basicstyle=\footnotesize\ttfamily}]
$HOME/LIBRARIES/
\end{lstlisting}
so that you will not need superuser access rights. The directory above is referred to as the \Queso{} installation directory (tree).

There are two main steps to prepare your LINUX computing system  for QUESO library: obtain and install \Queso{} dependencies, and define a number of environmental variables. These steps are discussed bellow.

\subsection{Obtain and Install \Queso{} Dependencies}

\Queso{} interfaces to a number of high-quality software packages to provide certain functionalities. While some of them are required for the successful installation of \Queso{}, other may be used for enhancing its performance. 
\Queso{} dependencies are:
\begin{enumerate}

  \item C and C++ compilers. Either \texttt{gcc} or \texttt{icc} are recommended \cite{GCC,ICC}.
  
  \item \textbf{Autotools}: The GNU build system, also known as the Autotools, is a suite of programming tools (Automake, Autoconf, Libtool) designed to assist in making source-code packages portable to many Unix-like systems~\cite{Autotools}.
  
  \item \textbf{STL}: The Standard Template Library is a C++ library of container classes, algorithms, and iterators; it provides many of the basic algorithms and data structures of computer science~\cite{STL}. The STL usually comes packaged with your compiler.

  \item \textbf{GSL}: The GNU Scientific Library is a numerical library for C and C++ programmers. It provides a wide range of mathematical routines such as random number generators, special functions and least-squares fitting~\cite{Gsl}. The lowest version of GSL required by QUESO is GSL 1.10.

  \item \textbf{Boost}: Boost provides free peer-reviewed portable C++ source libraries, which can be used with the C++ Standard Library~\cite{Boost}. QUESO requires Boost 1.35.0 or newer.

  \item \textbf{MPI}: The Message Passing Interface is a standard for parallel programming using the message passing model. E.g. Open MPI~\cite{Openmpi} or MPICH~\cite{Mpich}. \Queso{} requires MPI during the compilation step; however, you may run it in serial mode (e.g. in one single processor) if you wish. 

\end{enumerate}

%

\Queso{} also works with the following optional libraries:

\begin{enumerate}

\item \textbf{GRVY}: The Groovy Toolkit (GRVY) is a library used to house various support functions often required for application development of high-performance, scientific applications. The library is written in C++, but provides an API for development in C and Fortran~\cite{grvy}. QUESO requires GRVY 0.29 or newer.

\item \textbf{HDF5}: The Hierarchical Data Format 5 is a technology suite that makes possible the management of extremely large and complex data collections~\cite{HDF5}. The lowest version required by QUESO is HDF5 1.8.0.

\item \textbf{GLPK}: The GNU Linear Programming Kit package is is a set of routines written in ANSI C and organized in the form of a callable library for solving large-scale linear programming, mixed integer programming, and other related problems~\cite{GLPK}. QUESO works with GLPK versions newer than or equal to  GLPK 4.35.
%
%
 \item{ \textbf{Trilinos}: The Trilinos Project is an effort to develop and implement robust algorithms and enabling technologies using modern object-oriented software design, while still leveraging the value of established libraries. It emphasizes abstract interfaces for maximum flexibility of component interchanging, and provides a full-featured set of concrete classes that implement all abstract interfaces~\cite{Trilinos,TrilinosPage}. QUESO requires Trilinos release to be newer than or equal to  11.0.0.
 {\bf Remark:} An additional requirement for QUESO work with Trilinos is that the latter must have enabled both Epetra and Teuchos libraries.}

\end{enumerate}
%

\new{The majority of QUESO output files is MATLAB$^\circledR$/GNU Octave compatible ~\cite{Matlab,Octave}. Thus, for results visualization purposes, it is recommended that the user have available either one of these tools.
}             

\subsection{Prepare your LINUX Environment}\label{sec:prepare}


This section presents the steps to prepared the environment
considering the user LINUX environment runs a BASH-shell. For other types of shell, such as C-shell, some adaptations may be required.

Before using QUESO, the user must first set a number of environmental variables, and indicate the full path
of the QUESO's dependencies (GSL and Boost) and optional libraries. 

For example, supposing the user wants to install QUESO with two additional libraries: HDF5 and Trilinos. 
Add the following lines to append the location of QUESO's dependencies and optional libraries to the \verb+LD_LIBRARY_PATH+ environment variable:
\begin{lstlisting}[{basicstyle=\footnotesize\ttfamily}]
$ export LD_LIBRARY_PATH=$LD_LIBRARY_PATH:\
  $HOME/LIBRARIES/gsl-1.15/lib/:\
  $HOME/LIBRARIES/boost-1.53.0/lib/:\
  $HOME/LIBRARIES/hdf5-1.8.10/lib/:\
  $HOME/LIBRARIES/trilinos-11.2.4/lib:
\end{lstlisting}
which can be placed in the user's \verb+.bashrc+ or other startup file. 

In addition, the user must set the following environmental
variables:
\begin{lstlisting}[{basicstyle=\footnotesize\ttfamily}]
$ export CC=gcc
$ export CXX=g++
$ export MPICC=mpicc
$ export MPICXX=mpic++
$ export F77=fort77
$ export FC=gfortran
\end{lstlisting}


\section{Obtaining a Copy of \Queso{}}

The latest supported public release of \Queso{} is available in the form of a tarball (tar format compressed with gzip) from \Queso{} Home Page: \Quesoweb.

Suppose you have copied the file `\verb+queso-0.47.1.tar.gz+' into \texttt{\$HOME/queso\_download/}.
Then just follow these commands to expand the tarball:
\begin{lstlisting}[{basicstyle=\footnotesize\ttfamily}]
$ cd $HOME/queso_download/
$ tar xvf queso-0.47.1.tar.gz
$ cd queso-0.47.1   	#enter the directory 
\end{lstlisting}

Naturally, for versions of \Queso{} other than \QUESOversion, the file names in the above commands must be adjusted.

\subsection{Recommended Build Directory Structure}\label{sec:Queso_tree}

Via Autoconf and Automake, \Queso{} configuration facilities provide a great deal 
of flexibility for configuring and building the existing \Queso{} packages. However,
unless a user has prior experience with Autotools, we strongly recommend
the following process to build and maintain local builds of \Queso{} (as an example, see note on Section \ref{sec:summary}).
To start, we defined three useful terms:

\begin{description}
 \item [Source tree] - The directory structure where the \Queso{} source files are located. A source
tree is is typically the result of expanding an \Queso{} distribution source code bundle, such as a tarball.
 \item [Build tree] 
- The tree where \Queso{} is built. It is always related to a specific source tree, and it is the directory structure where object and library files are located. Specifically, this is the tree where you invoke \texttt{configure, make}, etc. to build and install \Queso{}. 
 \item [Installation tree] - The tree where \Queso{} is installed. It is typically the \texttt{prefix} argument given to \Queso{}'s configure script; it is the directory from which you run installed \Queso{} executables.
\end{description}

Although it is possible to run \verb+./configure+ from the source tree (in the directory where the configure file is located), we recommend separate build trees. The greatest advantage to having a separate build tree is that multiple builds of the library
can be maintained from the same source tree~\cite{Trilinos}. 

\section{Configure QUESO Building Environment}\label{sec:Queso_configure}

\Queso{} uses the GNU Autoconf system for configuration, which detects various features of the host system and creates Makefiles. 
The configuration process can be controlled through environment variables, command-line switches, and host configuration files.
For a complete list of switches type:
\begin{lstlisting}[{basicstyle=\footnotesize\ttfamily}]
$ ./configure  --help  
\end{lstlisting}
from the top level of the source tree (exemplified as \verb+$HOME/queso_download/queso-0.47.1+ in this report). 

This command will also display the help page for \Queso{} options.  Many of the \Queso{} configure options are used to describe 
the details of the build. For instance, to include HDF5, a package that is not currently built by default, append \texttt{--with-hdf5=DIR}, 
where \texttt{DIR} is the root directory of HDF5 installation,  to the configure invocation line. 

\Queso{} default installation location is `\texttt{/usr/local}', which requires superuser privileges. To use a path
 other than `\texttt{/usr/local}', specify the path with the `\texttt{--prefix=PATH}' switch. For instance, to follow the suggestion
 given in Section \ref{sec:Pre_Queso}, the user should append `\verb+--prefix=$HOME/LIBRARIES+'.

Therefore, the basic steps to configure QUESO using Boost, GSL (required), HDF5 and Trilinos (optional) for installation at `\verb+$HOME/LIBRARIES/QUESO-0.51.0+' are:
\begin{lstlisting}[{basicstyle=\footnotesize\ttfamily}]
$  ./configure --prefix=$HOME/LIBRARIES/QUESO-0.51.0 \
  --with-boost=$HOME/LIBRARIES/boost-1.53.0 \
  --with-gsl=$HOME/LIBRARIES/gsl-1.15 \
  --with-hdf5=$HOME/LIBRARIES/hdf5-1.8.10 \
  --with-trilinos=$HOME/LIBRARIES/trilinos-11.2.4
  \end{lstlisting}

Note: the directory `\verb+$HOME/LIBRARIES/QUESO-0.51.0+' does not need to exist in advance, since it will be created by the command \verb+make install+ described in Section \ref{sec:install_Queso_make}.

\section{Compile, Check and Install \Queso{}}\label{sec:install_Queso_make}
In order to build, check and install the library, the user must enter the following three commands sequentially:
\begin{lstlisting}[{basicstyle=\footnotesize\ttfamily}]
$ make
$ make check       # optional
$ make install 
\end{lstlisting}

Here, \verb+make+ builds the library, confidence tests, and programs;  \verb+ make check+ conducts various test suites in order to check the compiled source; and \verb+make install+ installs \Queso{} library, include files, and support programs.

The files are installed under the installation tree (refer to Section \ref{sec:Queso_tree}), e.g. the directory specified with `\texttt{--prefix=DIR}' in Section \ref{sec:Queso_configure}. The directory, if not existing, will be created automatically.

%
%
%
%


By running \texttt{make check}, several printouts appear in the screen and you should see messages such as:
\begin{lstlisting}[{basicstyle=\footnotesize\ttfamily}]
--------------------------------------------------------------------
(rtest): PASSED: Test 1 (TGA Validation Cycle)
--------------------------------------------------------------------
\end{lstlisting}

The tests printed in  the screen are tests under your QUESO build tree, i.e., they are located at the  directory \verb+$HOME/queso_download/queso-0.47.1/test+ (see Section \ref{sc-source-dir-structure} for the complete list of the directories under QUESO build tree).    
These tests are used as part of the periodic QUESO regression tests, conducted to ensure that more recent program/code changes have not adversely affected existing features of the library.

\section{\Queso{} Developer's Documentation}\label{sec:Queso_docs}

\Queso{} code documentation is written using Doxygen~\cite{Doxygen}, and can be generated by typing in the build tree:
\begin{lstlisting}[{basicstyle=\footnotesize\ttfamily}]
$ make docs
\end{lstlisting}

A directory named \verb+docs+ will be created in \verb+$HOME/queso_download/queso-0.47.1+ (the build tree; your current path) and you may access the code documentation in two different ways:
\begin{enumerate}
 \item HyperText Markup Language (HTML)  format: \verb+docs/html/index.html+, and the browser of your choice can be used to walk through the HTML documentation.

\item Portable Document Format (PDF) format: \verb+docs/queso.pdf+, which can be accessed thought any PDF viewer.
\end{enumerate}
%

\section{Summary of Installation Steps}\label{sec:summary}

Supposing you have downloaded the file `queso-0.47.1.tar.gz' into \texttt{\$HOME/queso\_download/}.
In a BASH shell, the basic steps to configure QUESO using GRVY, Boost and GSL for installation at 
`\verb+$HOME/LIBRARIES/QUESO-0.51.0+'  are:

\begin{lstlisting}[{basicstyle=\footnotesize\ttfamily}]
$ export LD_LIBRARY_PATH=$LD_LIBRARY_PATH:\
  $HOME/LIBRARIES/gsl-1.15/lib/:\
  $HOME/LIBRARIES/boost-1.53.0/lib/:\
  $HOME/LIBRARIES/hdf5-1.8.10/lib/:\
  $HOME/LIBRARIES/trilinos-11.2.4/lib:
$ export CC=gcc
$ export CXX=g++
$ export MPICC=mpicc
$ export MPICXX=mpic++
$ export F77=fort77
$ export FC=gfortran
$ cd $HOME/queso_download/               #enter source dir
$ gunzip < queso-0.47.1.tar.gz  | tar xf -
$ cd $HOME/queso_download/queso-0.47.1   #enter the build dir
$ ./configure --prefix=$HOME/LIBRARIES/QUESO-0.51.0 \
  --with-boost=$HOME/LIBRARIES/boost-1.53.0 \
  --with-gsl=$HOME/LIBRARIES/gsl-1.15 \
  --with-hdf5=$HOME/LIBRARIES/hdf5-1.8.10 \
  --with-trilinos=$HOME/LIBRARIES/trilinos-11.2.4
$ make 
$ make check
$ make install 
$ make docs
$ ls $HOME/LIBRARIES/QUESO-0.51.0 #listing QUESO installation dir
>>  bin  include  lib  examples
\end{lstlisting}

%
%

\section{The Build Directory Structure} \label{sc-source-dir-structure}

The QUESO build directory contains three main directories, \texttt{src}, \texttt{examples} and \texttt{test}. They are listed below and more specific
information about them can be obtained with the Developer's documentation from Section \ref{sec:Queso_docs} above.
\begin{enumerate}
\item \texttt{src}: this directory contains the QUESO library itself, and its main subdirectories are:
  \begin{enumerate}
  \item \texttt{basic/}: contain classes for dealing with vector sets, subsets and spaces, scalar and vector functions and scalar and vector sequences
  \item \texttt{core/}: contain classes that handle \Queso{} environment, and vector/matrix operations
  \item \texttt{stats/}: contain classes that implement vector realizers, vector random variables, statistical inverse and forward problems; and the Monte Carlo and the Metropolis-Hasting solvers
  \end{enumerate}
  
Details of \Queso{}  classes are presented in Chapter \ref{ch-classes}.

\item \texttt{examples}:  examples of different applications, with distinct levels of difficulty, using \Queso{}. The following examples have been thoroughly documented and are included in Chapter \ref{chap:Queso-examples}:
\begin{enumerate}
\item \texttt{gravity/}: inference of the acceleration of gravity via experiments and a solution of a SIP; and forward propagation of uncertainty in the calculation of the distance traveled by a projectile. It is presented in detail in Section~\ref{sec:example_gravity}.


\item \texttt{simpleStatisticalForwardProblem/}: simplest example of how to use \Queso{} to solve a SFP, described in detail in Section \ref{sec:example_sfp}.

\item \texttt{simpleStatisticalInverseProblem/}: simplest example of how to use \Queso{} to solve a SIP, thoroughly described in Section \ref{sec:example_sip}.

\item \texttt{validationCycle/}: presents a combination of SIP and SFP to solve a thermogravimetric analysis problem; this problem has the majority of its code in \verb+*.h+ files, with templated routines. This example is described in Section \ref{sec:example_tga}.

\item \texttt{validationCycle2/}: also presents a combination of SIP and SFP to solve a thermogravimetric analysis problem; but the majority of its code is in \verb+*.C+ files. 
\end{enumerate}

All the examples presented in Chapter \ref{chap:Queso-examples} come with the mathematical formulation, their translation into code, the options input file required by \Queso{} and auxiliary Matlab (GNU Octave compatible) files for data visualization.

The build directory contains only the source files. The executables are available under the QUESO installation directory, together with example of Makefiles that may be used to re-build the examples without the need of re-building the library.

\item  \texttt{test}: a set of tests used as part of the periodic QUESO regression tests, conduct to ensure that more recent program/code changes have not adversely affected existing features of the library, as described in Section \ref{sec:install_Queso_make}. 
\begin{enumerate}
\item \texttt{gsl\_tests}
\item \texttt{t01\_valid\_cycle/}
\item \texttt{t02\_sip\_sfp/}
\item \texttt{t03\_sequence/} 
\item \texttt{t04\_bimodal/} 
\item \texttt{test\_Environment/}
\item \texttt{test\_GaussianVectorRVClass/}
\item \texttt{test\_GslMatrix/}
\item \texttt{test\_GslVector/}
\item \texttt{test\_uqEnvironmentOptions/}
\end{enumerate}

These tests can optionally be called during QUESO installation steps by entering the instruction: \verb+make check+.

\end{enumerate}

\section{The Installed Directory Structure} \label{sc-installed-dir-structure}

After having successfully executed steps described in Sections \ref{sec:Pre_Queso}--\ref{sec:install_Queso_make}, the QUESO installed directory will contain four subdirectories:
\begin{enumerate}
 \item \verb+bin+: contains the executable \verb+queso_version+, which provides information about the installed library. The code bellow presents a sample output:

\begin{lstlisting}[{basicstyle=\footnotesize\ttfamily}]
kemelli@margarida:~/LIBRARIES/QUESO-0.51.0/bin$ ./queso_version 
---------------------------------------------------------------
QUESO Library: Version = 0.47.1 (47.1)

Development Build

Build Date   = 2013-07-12 12:36
Build Host   = margarida
Build User   = kemelli
Build Arch   = i686-pc-linux-gnu
Build Rev    = 40392

C++ Config   = mpic++ -g -O2 -Wall

Trilinos DIR = /home/kemelli/LIBRARIES/trilinos-11.2.4
GSL Libs     = -L/home/kemelli/LIBRARIES/gsl-1.15/lib -lgsl -lgslcblas -lm
GRVY DIR     = 
GLPK DIR     = 
HDF5 DIR     = 
---------------------------------------------------------------
kemelli@margarida:~/LIBRARIES/QUESO-0.51.0/bin$ 
\end{lstlisting}

 \item \verb+lib+: contains the static and dynamic versions of the library. The full to path to this directory, e.g., \verb+$HOME/LIBRARIES/QUESO-0.51.0/lib+ should be added to the user's \verb+LD_LIBRARY_PATH+ environmental variable in order to use QUESO library with his/her application code:
\begin{lstlisting}[{basicstyle=\footnotesize\ttfamily}]
$ export LD_LIBRARY_PATH=$LD_LIBRARY_PATH:\
 $HOME/LIBRARIES/QUESO-0.51.0/lib
\end{lstlisting}

Note that due to \Queso{} being compiled/built with other libraries (GSL, Boost, Trilinos and HDF5), \verb+LD_LIBRARY_PATH+ had already some values set in Section \ref{sec:prepare}.

 \item \verb+include+: contains the library \verb+.h+ files.

 \item \verb+examples+: contains the same examples of QUESO build directory, and listed in Section~\ref{sc-source-dir-structure}, together with their executables and Matlab files that may be used for visualization purposes. A selection of examples are described in details in Chapter \ref{chap:Queso-examples}; the user is invited understand their formulation, to run them and understand their purpose.

\end{enumerate}

\section{Create your Application with the Installed QUESO} \label{sc-use-queso}

Prepare your environment by either running or saving the following command in
your \verb+.bashrc+ (supposing you have a BASH-shell):
\begin{lstlisting}[{basicstyle=\footnotesize\ttfamily}]
$ export LD_LIBRARY_PATH=$LD_LIBRARY_PATH:\
 $HOME/LIBRARIES/QUESO-0.51.0/lib
\end{lstlisting}


Suppose your application code consists of the files:  \verb+example_main.C+,
\verb+example_qoi.C+,  \verb+example_likelihood.C, example_compute.C+ and
respective \verb+.h+ files. Your application code may be linked with QUESO
library through a Makefile such as the one displayed as follows:

\begin{lstlisting}[basicstyle={\footnotesize\ttfamily},deletekeywords={export,rm}]
QUESO_DIR = $HOME/LIBRARIES/QUESO-0.51.0/
BOOST_DIR = $HOME/LIBRARIES/boost-1.53.0/
GSL_DIR = $HOME/LIBRARIES/gsl-1.15/
GRVY_DIR = $HOME/LIBRARIES/grvy-0.32.0
TRILINOS_DIR = $HOME/LIBRARIES/trilinos-11.2.4/

INC_PATHS = \
	-I. \
	-I$(QUESO_DIR)/include \
	-I$(BOOST_DIR)/include/boost-1.53.0 \
	-I$(GSL_DIR)/include \
	-I$(GRVY_DIR)/include \
	-I$(TRILINOS_DIR)/include \

LIBS = \
	-L$(QUESO_DIR)/lib -lqueso \
	-L$(BOOST_DIR)/lib -lboost_program_options \
	-L$(GSL_DIR)/lib -lgsl \
	-L$(GRVY_DIR)/lib -lgrvy \
	-L$(TRILINOS_DIR)/lib -lteuchoscore -lteuchoscomm -lteuchosnumerics \
	-lteuchosparameterlist -lteuchosremainder -lepetra

CXX = mpic++
CXXFLAGS += -O3 -Wall -c

default: all

.SUFFIXES: .o .C

all:	ex_gsl

clean:
	rm -f *~
	rm -f *.o
% 	rm -f example_gsl

ex_gsl: example_main.o example_likelihood.o example_qoi.o example_compute.o
	$(CXX) example_main.o example_likelihood.o example_qoi.o \
	       example_compute.o -o example_gsl $(LIBS)

%.o: %.C
	$(CXX) $(INC_PATHS) $(CXXFLAGS) $<
\end{lstlisting}
%


\chapter{C++ Classes in the Library}\label{ch-classes}
\thispagestyle{headings}
\markboth{Chapter \ref{ch-classes}: C++ Classes in the Library}{Chapter \ref{ch-classes}: C++ Classes in the Library}

QUESO is is a parallel object-oriented statistical library dedicated to the research of   statistically robust, scalable, load balanced, and fault-tolerant mathematical algorithms for the  quantification of uncertainty in realistic computational models and predictions related to natural and engineering systems.

Classes in QUESO can be divided in four main groups: core, templated basic, templated statistical and miscellaneous.
The classed that handle environment (and options), vector and matrix classes are considered \textit{core} classes. Classes implementing vector sets and subsets, vector spaces,  scalar functions, vector functions, scalar sequences and vector sequences are \textit{templated basic} classes; they are necessary for the definition and description of other entities, such as RVs, Bayesian solutions of IPs, sampling algorithms and chains.  Vector realizer, vector RV, statistical IP (and options), MH solver (and options), statistical FP (and options), MC solver (and options) and sequence statistical options are part of \textit{templated statistical} classes. And finally, the \textit{miscellaneous} classes consist of C and FORTRAN interfaces.



\section{Core Classes}

QUESO core classes are the classes responsible for handling the environment (and options), vector
and matrix operations. They are described in the following sections.

%
%
%
%
%
%

\subsection{Environment Class (and Options)}\label{sec:environment_class}

The \texttt{Environment} class sets up the environment underlying the use of
the QUESO library by an executable.  This class is virtual. It is inherited by
\verb+EmptyEnvironment+ and \verb+FullEnvironment+.

The QUESO environment class is instantiated at the application level, right
after \linebreak\verb+MPI_Init(&argc,&argv)+.  The QUESO environment is
required by reference by many constructors in the QUESO library, and is
available by reference from many classes as well.

The constructor of the environment class requires a communicator, the name of
an options input file, and the eventual prefix of the environment in order for
the proper options to be read (multiple environments can coexist, as explained
further below).

The environment class has four primary tasks:
\begin{enumerate}
\item Assigns rank numbers, other than the world rank, to nodes participating in a parallel job,
\item Provides MPI communicators for generating a sequence of vectors in a distributed way,
\item Provides functionality to read options from the options input file (whose name is passed in the constructor of this environment class), and
\item Opens output files for messages that would otherwise be written to the screen (one output file per allowed rank is opened and allowed ranks can be specified through the options input file).
\end{enumerate}

Let $S \geqslant 1$ be the number of problems a QUESO environment will be handling at the same time, in parallel.
$S$ has default value of $1$ and is an option read by QUESO from the input file provided by the user.
The QUESO environment class manages five types of communicators, referred to as:

\begin{enumerate}

\item {\it world}: MPI\_WORLD\_COMM;
\item {\it full}: communicator passed to the environment constructor, of size $F$ and usually equal to the world communicator;
\item {\it sub}: communicator of size $F/S$ that contains the number of MPI nodes necessary to solve a statistical IP or a statistical FP;
\item {\it self}: MPI\_SELF\_COMM, of size 1; and
\item {\it inter0}: communicator of size $S$ formed by all MPI nodes that have subrank 0 in their respective subcommunicators.
 
\end{enumerate}

A {\it subenvironment} in QUESO is the smallest collection of processors
necessary for the proper run of the model code.  An {\it environment} in QUESO
is the collection of all subenvironments, if there is more than one
subenvironment.


Each subenvironment is able to generate a statistical inverse problem and/or a
statistical forward problem; that is, each subenvironment is able to handle a
``sub'' Markov chain (a sequence) of vectors and/or a ``sub'' Monte Carlo
sequence of output vectors.  The ``sub'' sequences can be seen as forming a
``unified'' sequence in a distributed way.  Indeed, the virtual class
\verb+VectorSequence+ provides ``sub'' and ``unified'' statistical operations. 

Thus, if the model code requires 16 processors to run and the user decides to run 64 Markov chains in parallel,
then the environment will consist of a total of $F=1024$ processors and $S=64$ subenvironments, each subenvironment with $F/S=16$ processors.
Any given computing node in a QUESO run has potentially five different ranks.
Each subenvironment is assigned a subid varying from $0$ (zero) to $S-1$, and is able to handle a statistical IP and/or a statistical FP.
That is, each subenvironment is able to handle a {\it sub} Markov chain (a sequence) of vectors and/or a {\it sub} MC sequence of output vectors.
The {\it sub} sequences form an unified sequence in a distributed way.
QUESO takes care of the unification of results for the application programming and for output files.  Of course, if the user is solving just one statistical problem with just one MPI node, then all ranks are equal to zero.

A QUESO subenvironment eventually prints messages to its own output file. In order for that to happen, the requirements are:
\begin{enumerate}
 \item option \verb+m_subDisplayFileName+, a string, must be different than the default value \verb+"."+;
\item  option \verb+m_subDisplayAllowedSet+, a set of sub ids, must contain the id of the sub environment wanting to write a message to the output file;
\item  the previous requirement is automatically satisfied if the option \verb+m_subDisplayAllowAll+, a boolean, is set to 1 (the default value is 0);
\item  the processor wanting to write a message to the output file must have sub rank 0 (zero).
\end{enumerate}

If all requirements are satisfied, then QUESO will generate a file with name \linebreak 
\verb+<m_subDisplayFileName>_sub<sub id>.txt+.   For instance, if \verb+m_subDisplayFileName+ is `\verb+pROblem_775_+' then a node of sub rank 0 in sub environment 17 will write a message to the file `\verb+pROblem_775_sub17.txt+'. The class responsible for reading options one can pass to a QUESO environment through an input file is the \verb+EnvironmentOptions+ class.

\begin{figure}[!hp]
\centering
\includegraphics[scale=0.80,clip=true]{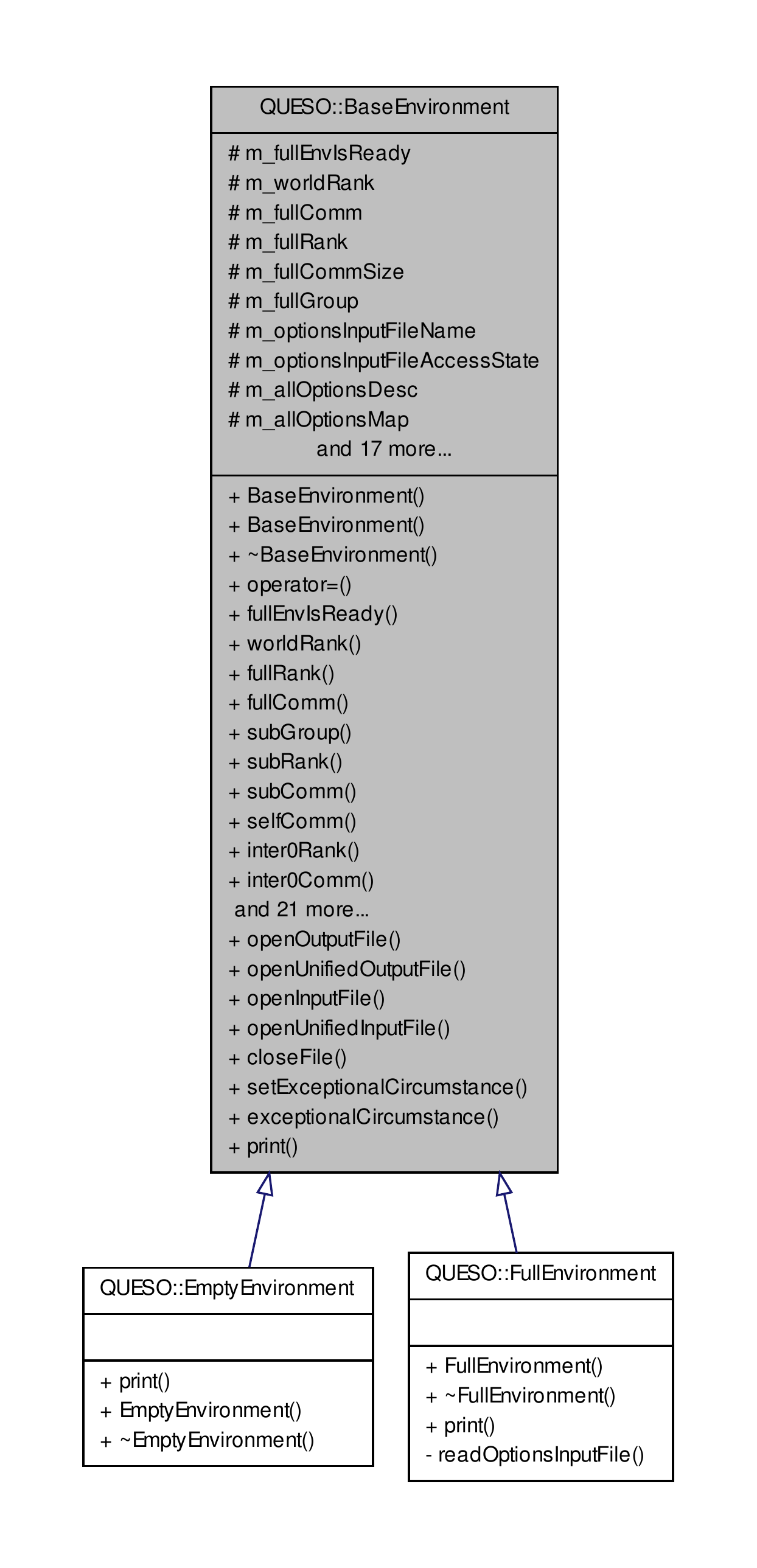}
\vspace*{-1.2cm}
\caption{The class diagram for the {Environment} class described in Section \ref{sec:environment_class}.}
\label{fig-env-class}
\end{figure}

\begin{figure}[!hp]
\centering
\includegraphics[scale=0.60,clip=true]{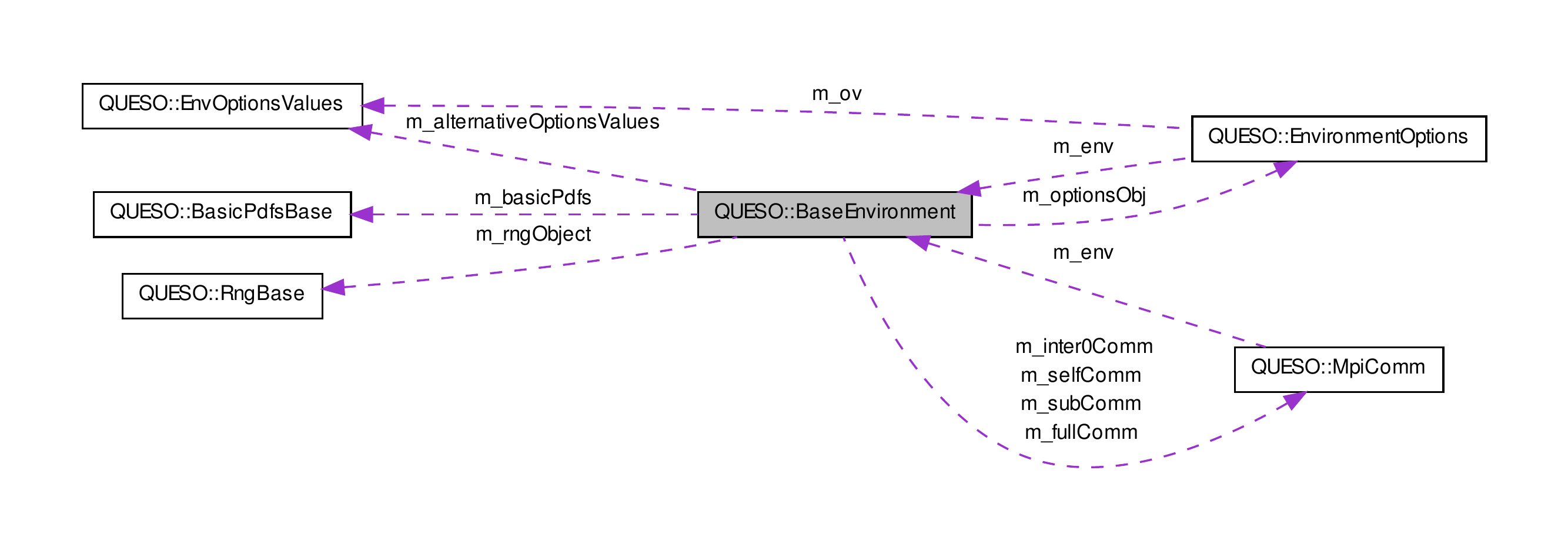}
\vspace*{-1.cm}
\caption{Collaboration graph for the environment class described in Section \ref{sec:environment_class}.}
\label{fig-env-coll}
\end{figure}

\begin{figure}[htpb]
\centering
\includegraphics[scale=0.8,clip=true]{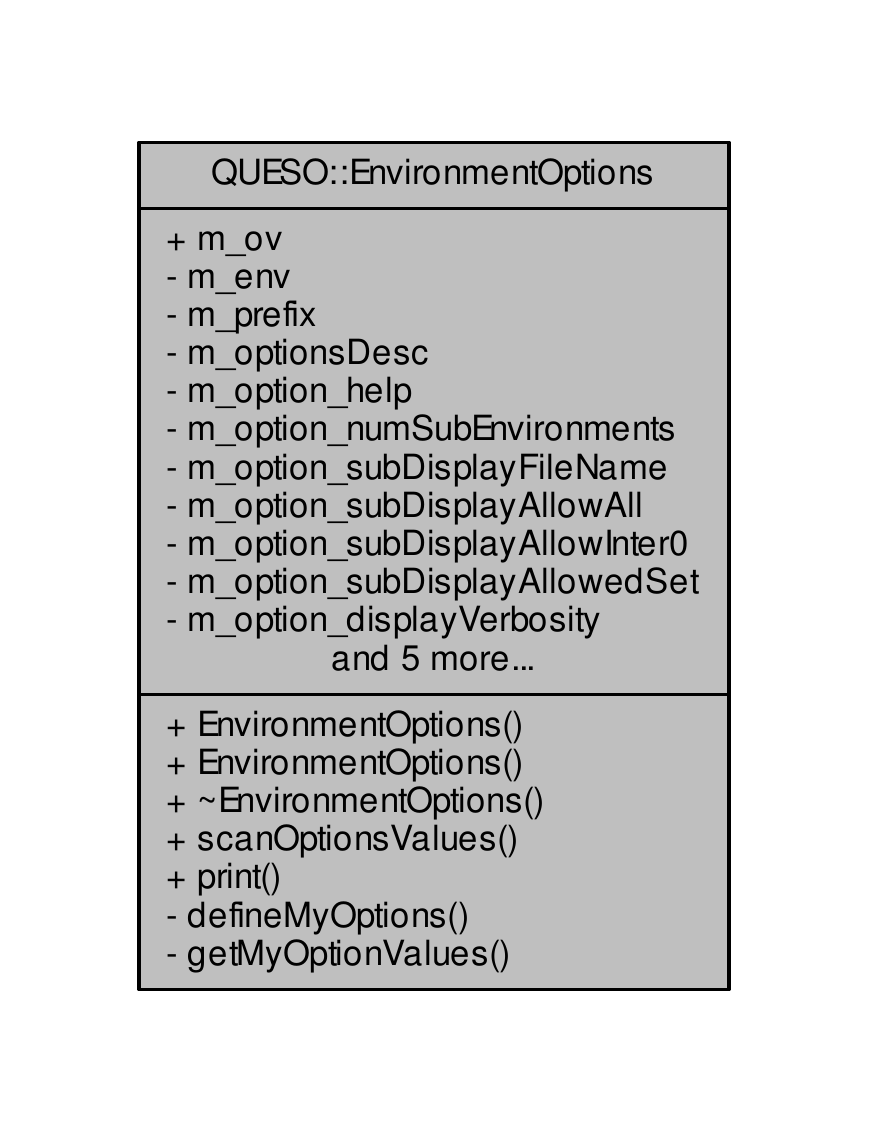}
\vspace*{-1.2cm}
\caption{The environment options class with its attributes and methods.}
\label{fig-env-options-class}
\end{figure}

Figure \ref{fig-env-class} depicts class diagram for the environment class and Figure \ref{fig-env-coll} display its collaboration graph; and Figure  \ref{fig-env-options-class} displays environment options class. 
 Finally, the input file options for a QUESO environment, i.e., the options the user may set in his/her input file when using QUESO together with the application of interest, is presented in Table \ref{tab-env-options}.

%

\begin{table}[htpb]
\begin{center}
\caption{Input file options for a QUESO environment.}
\vspace{-8pt}
\label{tab-env-options}
\footnotesize
\begin{tabular}{l c  m{7cm}}
\toprule
Option name                      &  Default  value & Description \\
\midrule\midrule
\ttfamily \textlangle PREFIX\textrangle env\_help                &     & Produces help message for environment class            \\
\ttfamily\textlangle PREFIX\textrangle env\_numSubEnvironments   &  1  &  Number of subenvironments                \\ 
\ttfamily\textlangle PREFIX\textrangle env\_subDisplayFileName   & \ttfamily"." & Output filename for sub-screen writing     \\ 
\ttfamily\textlangle PREFIX\textrangle env\_subDisplayAllowAll   &  0  & Allows all subenvironments to write to output file \\ 
\ttfamily\textlangle PREFIX\textrangle env\_subDisplayAllowedSet & \ttfamily""  & Subenvironments that will write to output file \\ 
\ttfamily\textlangle PREFIX\textrangle env\_displayVerbosity     &  0  & Sets verbosity				         \\ 
\ttfamily\textlangle PREFIX\textrangle env\_syncVerbosity        &  0  & Sets syncronized verbosity             \\ 
\ttfamily\textlangle PREFIX\textrangle env\_seed                 &  0  & Set seed                             \\ 
%
\bottomrule
\end{tabular}
\end{center}
\end{table}

\subsection{Vector}\label{sec:vector_class}

The Vector class handles all the vector operations carried out in QUESO, and currently has two derived classes: \verb+GslVector+ and \verb+TeuchosVector+. \verb+GslVector+ is based on the GSL vector structure; whereas \verb+TeuchosVector+ is based on Trilinos Teuchos vector structure~\cite{Trilinos}, and therefore, it is only available if QUESO was compiled with Trilinos.  

\begin{figure}[!htpb]
\centering
\vspace*{-.5cm}
\includegraphics[scale=0.50,clip=true]{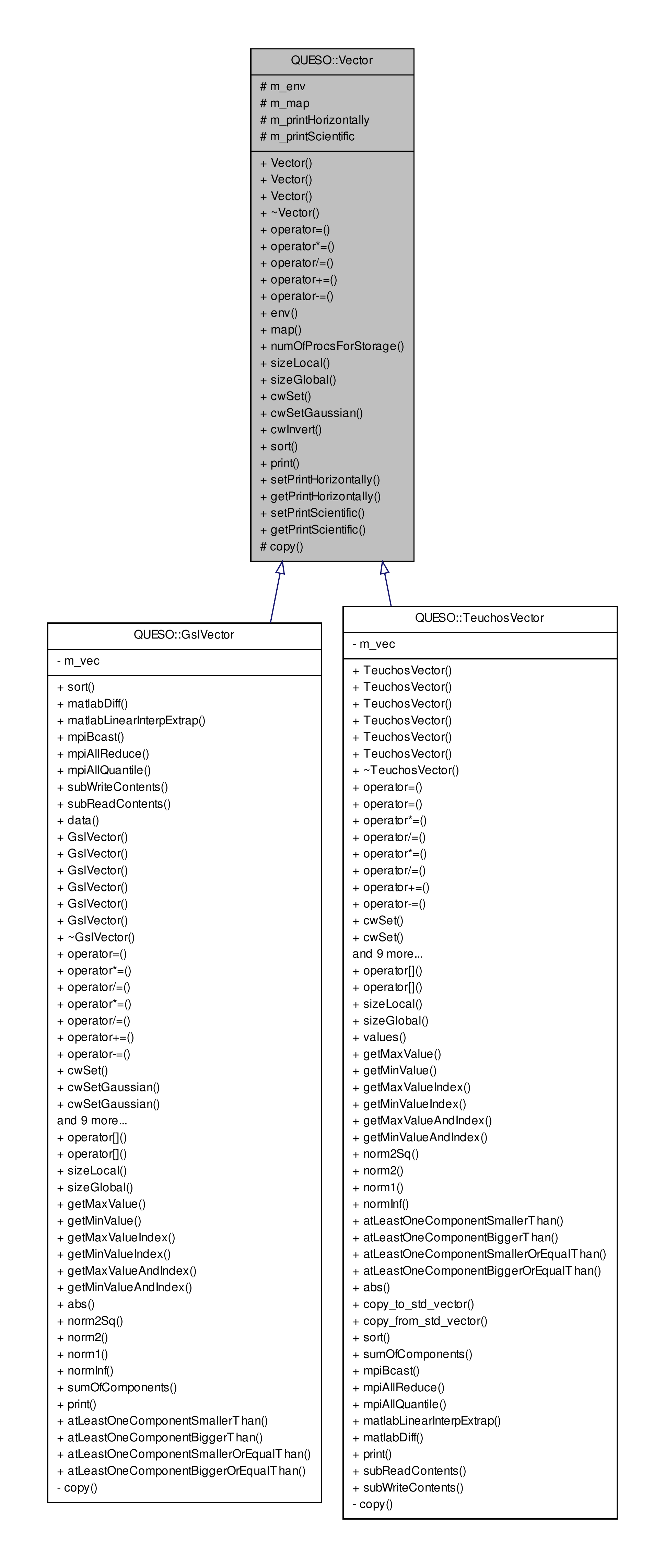}
\vspace*{-.8cm}
\caption{ The class diagram for the vector class described in Section \ref{sec:vector_class}.}
\label{fig-vector-class}
\end{figure}

A class diagram for \verb+Vector+ class is presented in Figure \ref{fig-vector-class}.

\subsection{Matrix}\label{sec:matrix_class}


The Matrix class handles all the matrix operations carried out in QUESO.  Analogously to the vector class case described in the previous section,
matrix class currently has two derived classes: \verb+GslMatrix+ and \verb+TeuchosMatrix+. \verb+GslMatrix+ is based on the GSL matrix structure; whereas \verb+TeuchosMatrix+ is based on Trilinos Epetra matrix structure.

\begin{figure}[!hp]
\centering
\vspace*{-.8cm}
\includegraphics[scale=0.50,clip=true]{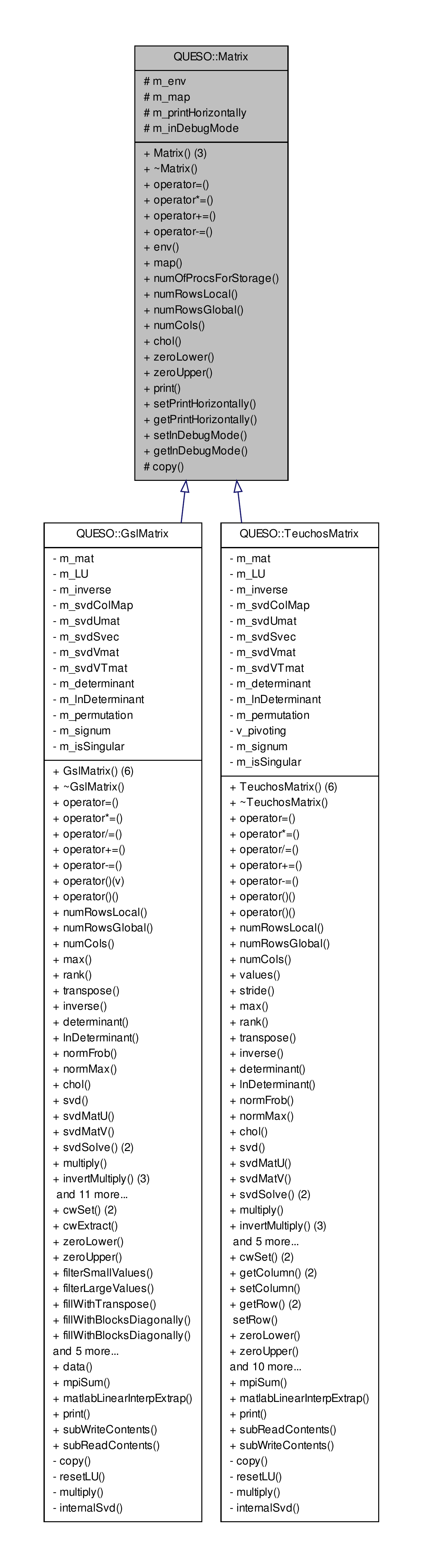}
\vspace*{-.8cm}
\caption{The class diagram for the matrix class.}
\label{fig-matrix-class}
\end{figure}

A class diagram for \verb+Matrix+  is presented in Figure \ref{fig-matrix-class}; it displays its protected attributes together with its member functions. Again, the diagram displays in some detail the inherited classes \verb+GslMatrix+ and \verb+TeuchosMatrix+.

\section{Templated Basic Classes}
The classes in this group are: vector sets, subsets and spaces (Section \ref{sec:vector-set-space}), scalar and vector function classes (Section \ref{sec:scalar-vector-function}), and scalar and vector sequences (Section \ref{sec:scalar-vector-sequence}).

These classes constitute the core entities necessary for the formal
mathematical definition and description of other entities, such as
random variables, Bayesian solutions of inverse problems, sampling algorithms and chains.


\subsection{Vector Set, Subset  and Vector Space Classes}\label{sec:vector-set-space}
The vector set class is fundamental for the proper handling of many mathematical entities.
Indeed, the definition of a scalar function such as $\pi:\mathbf{B}\subset\mathbb{R}^n\rightarrow\mathbb{R}$ requires the
specification of the domain $\mathbf{B}$, which is a {\it subset} of the {\it vector space} $\mathbb{R}^n$, which is itself a {\it set}. Additionally, 
 SIPs need a likelihood routine $\pi_{\text{like}}:\mathbb{R}^n\rightarrow\mathbb{R}_+$,
and SFPs need a QoI routine $\mathbf{q}:\mathbb{R}^n\rightarrow\mathbb{R}^m$; the \textit{sets} $\mathbb{R}^n$, $\mathbb{R}^m$, etc., are {\it vector spaces}.

\begin{figure}[htpb]
\hspace{-1cm}
\includegraphics[scale=0.65,clip=true]{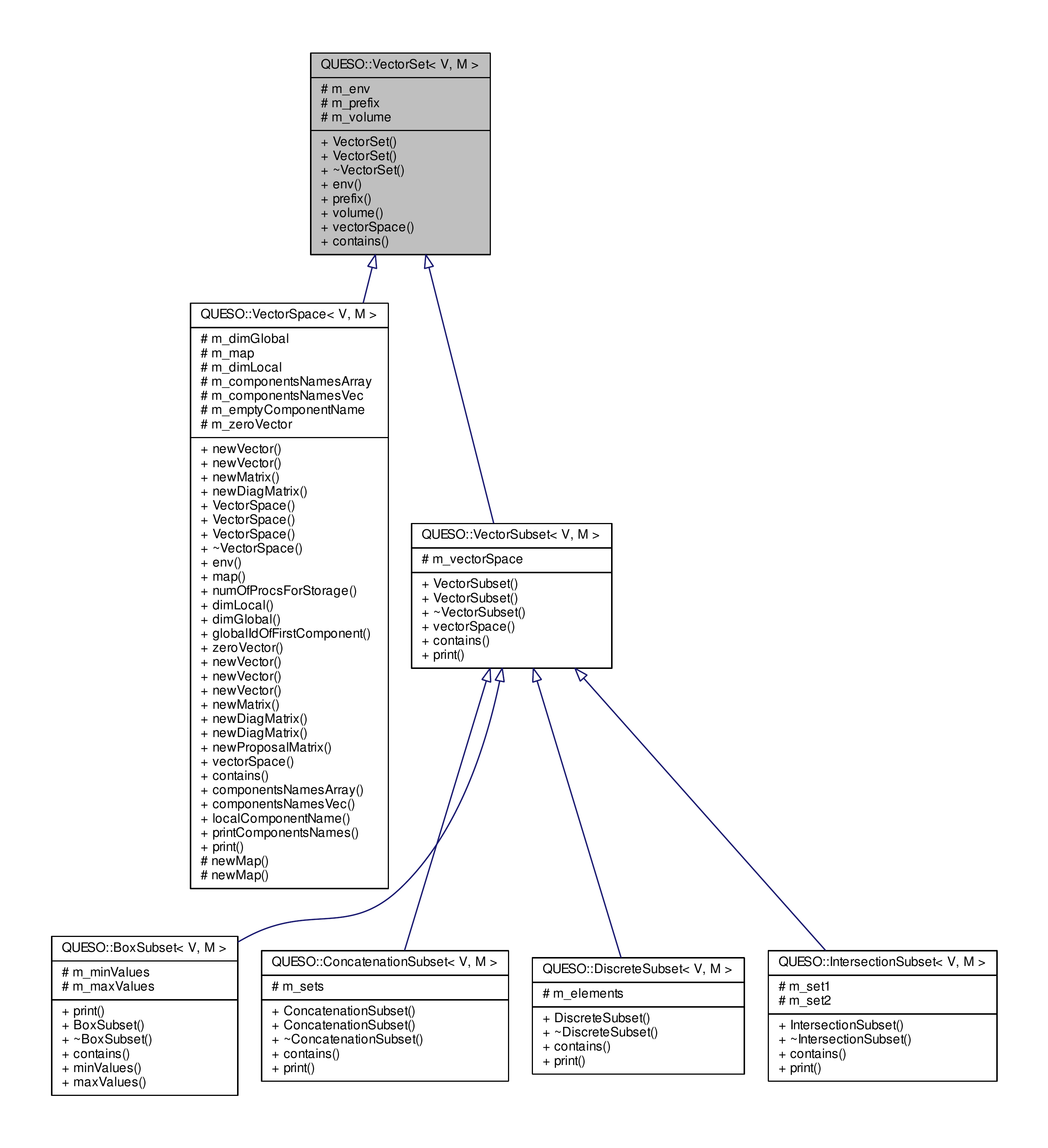}
\vspace*{-1.5cm}
\caption{The class diagram for vector set, vector subset and vector space classes, described in Section \ref{sec:vector-set-space}.}
\label{fig-vector-space-subset-classes}
\end{figure}

The relationship amongst QUESO classes for handling sets, namely \verb+VectorSet+; subsets, namely \verb+VectorSubset+;  and vector spaces, namely \verb+VectorSpace+ is sketched in Figure \ref{fig-vector-space-subset-classes}.
An attribute of the {\it subset} class is the {\it vector space} which it belongs to, and in fact a reference to a vector space is required by the constructor of the subset class. An example of this case is the definition of a scalar function such as $\pi:\mathbf{B}\subset\mathbb{R}^n\rightarrow\mathbb{R}$ above. 

The power of an object-oriented design is clearly featured here.
The intersection subset derived class \verb+IntersectionSubset+ is useful for handling a posterior PDF  on Equation~\eqref{eq-Bayes-solution},
since its domain is the intersection of the domain of the prior PDF with the domain of the likelihood function.

%

\subsection{Scalar Function and Vector Function Classes}\label{sec:scalar-vector-function}

\begin{figure}[htpb]
\includegraphics[scale=0.65,clip=true]{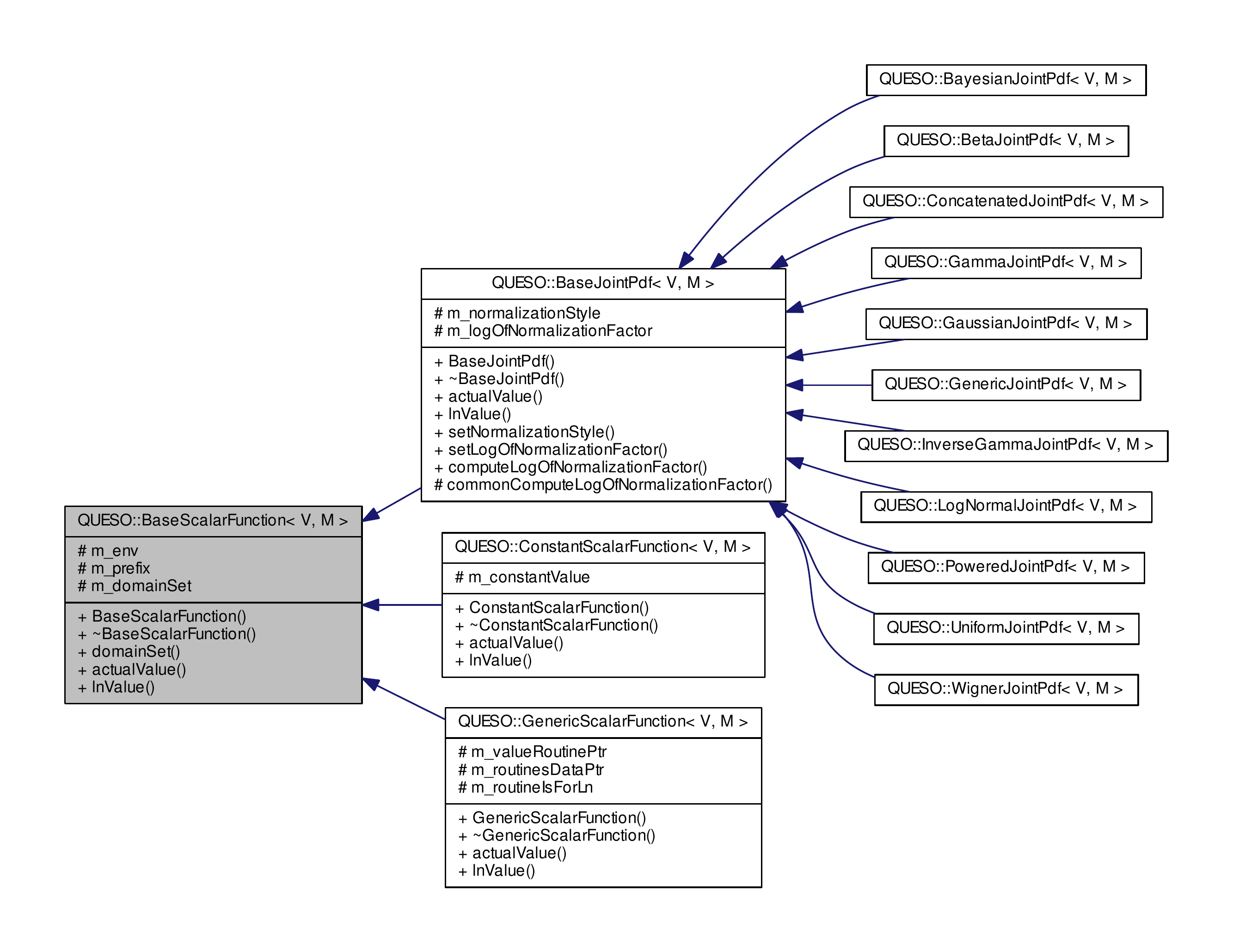}
\vspace*{-1.5cm}
\caption{The class diagram for the scalar function class.}
\label{fig-scalar-function-class}
\end{figure}

Joint PDF, marginal PDF, and CDF are all examples of scalar functions present in statistical problems. 
QUESO currently supports basic PDFs such as uniform and Gaussian and also more complex PDFs, such as the ones coming from a Bayesian analysis. They are implemented in the classes \verb+UniformJointPdf+, \verb+GaussianJointPdf+, and \verb+BayesianJointPdf+, respectively. The posterior PDF may be represented within QUESO by \verb+GenericJointPdf+.
See Diagram~\ref{fig-scalar-function-class} for the scalar function class.

\begin{figure}[htpb]
\centering
\hspace{-40pt}
\includegraphics[scale=0.65,clip=true]{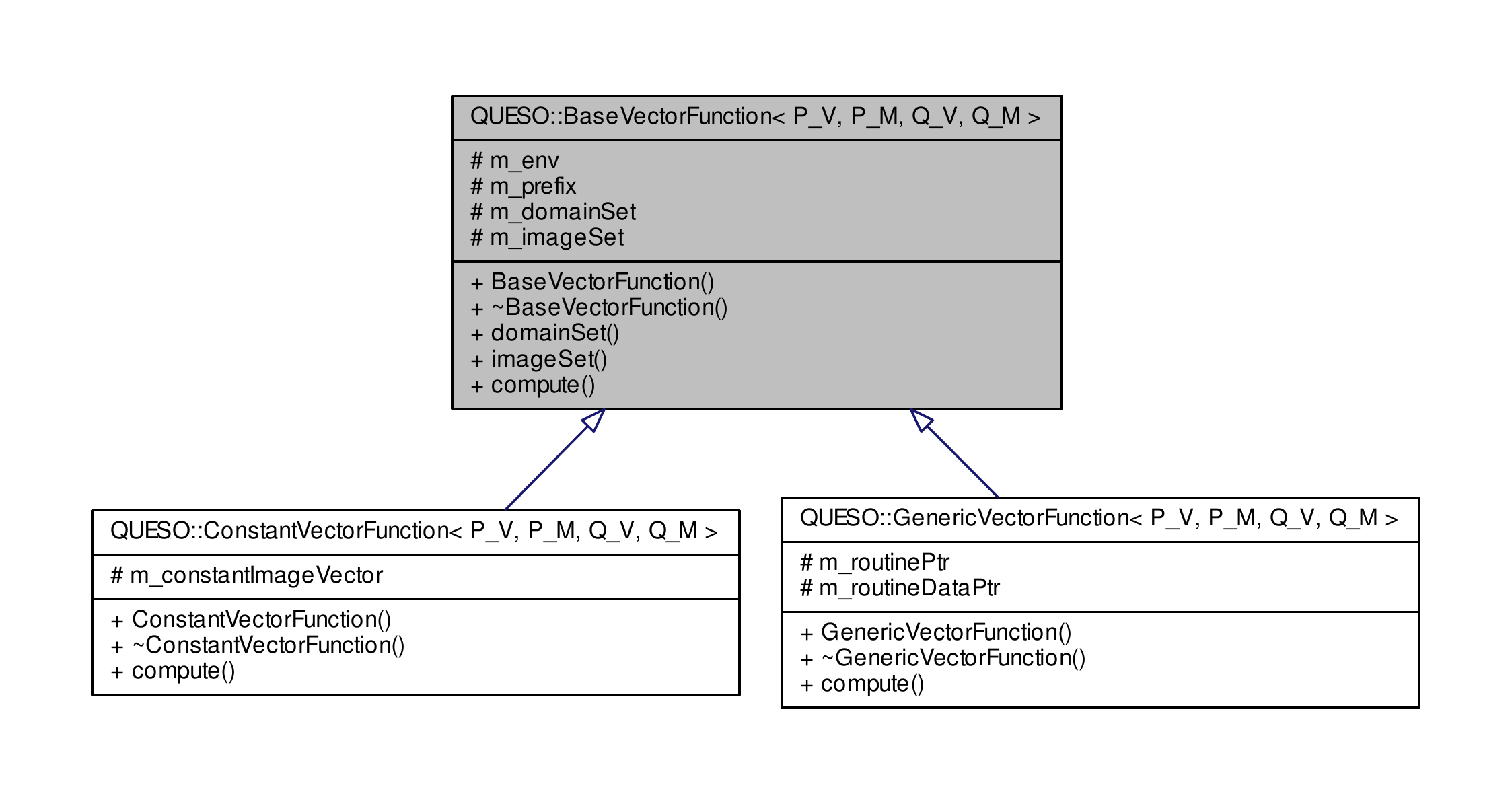}
\vspace{-1.2cm}
\caption{The class diagram for the vector function class described in Section \ref{sec:scalar-vector-function}.} 
\label{fig-vector-function-class}
\end{figure}

The handling of vector functions within QUESO is also quite straightforward. Indeed, the definition of a vector function $\mathbf{q}:\mathbf{B}\subset\mathbb{R}^n\rightarrow\mathbb{R}^m$ requires only the extra specification of the image vector space $\mathbb{R}^m$. The classes representing the vector function class \verb+GenericVectorFunction+ and \verb+ConstantVectorFunction+ are derived  from \verb+BaseVectorFunction+ and are presented in Diagram \ref{fig-vector-function-class} 

\subsection{Scalar Sequence and Vector Sequence Classes}\label{sec:scalar-vector-sequence}
The scalar sequence class contemplates {\it scalar} samples generated by an algorithm, as well as operations that can
be done over them, e.g., calculation of means, variances, and convergence indices.
Similarly, the vector sequence class contemplates {\it vector} samples and operations such as means, correlation matrices and covariance matrices.

\begin{figure}[htpb]
\centering
\includegraphics[scale=0.65,clip=true]{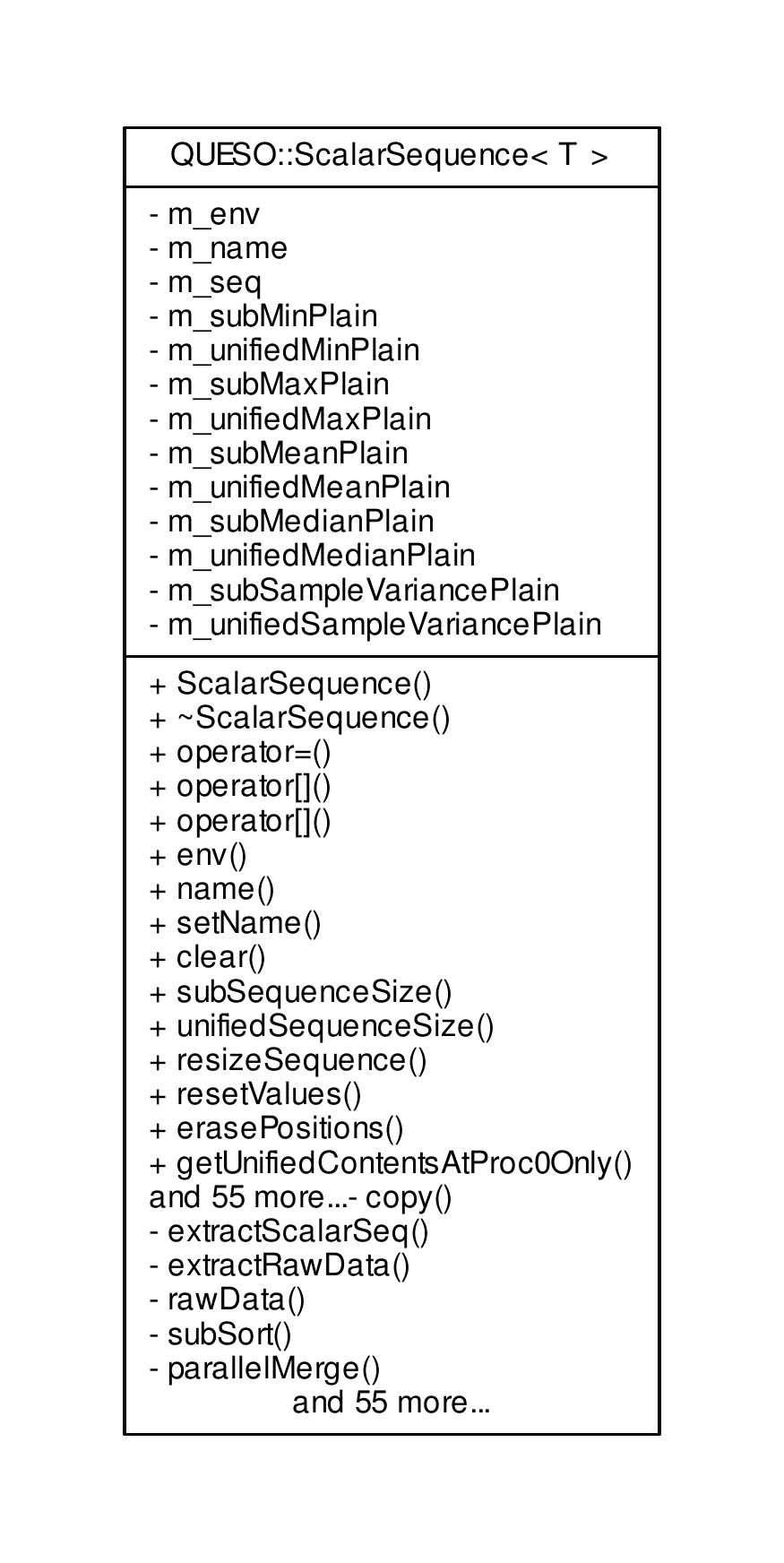}
\vspace{-1.cm}
\caption{The class diagram for the scalar sequence class.}
\label{fig-scalar-sequence-class}
\end{figure}

\begin{figure}[htpb]
\centering
\includegraphics[scale=0.65,clip=true]{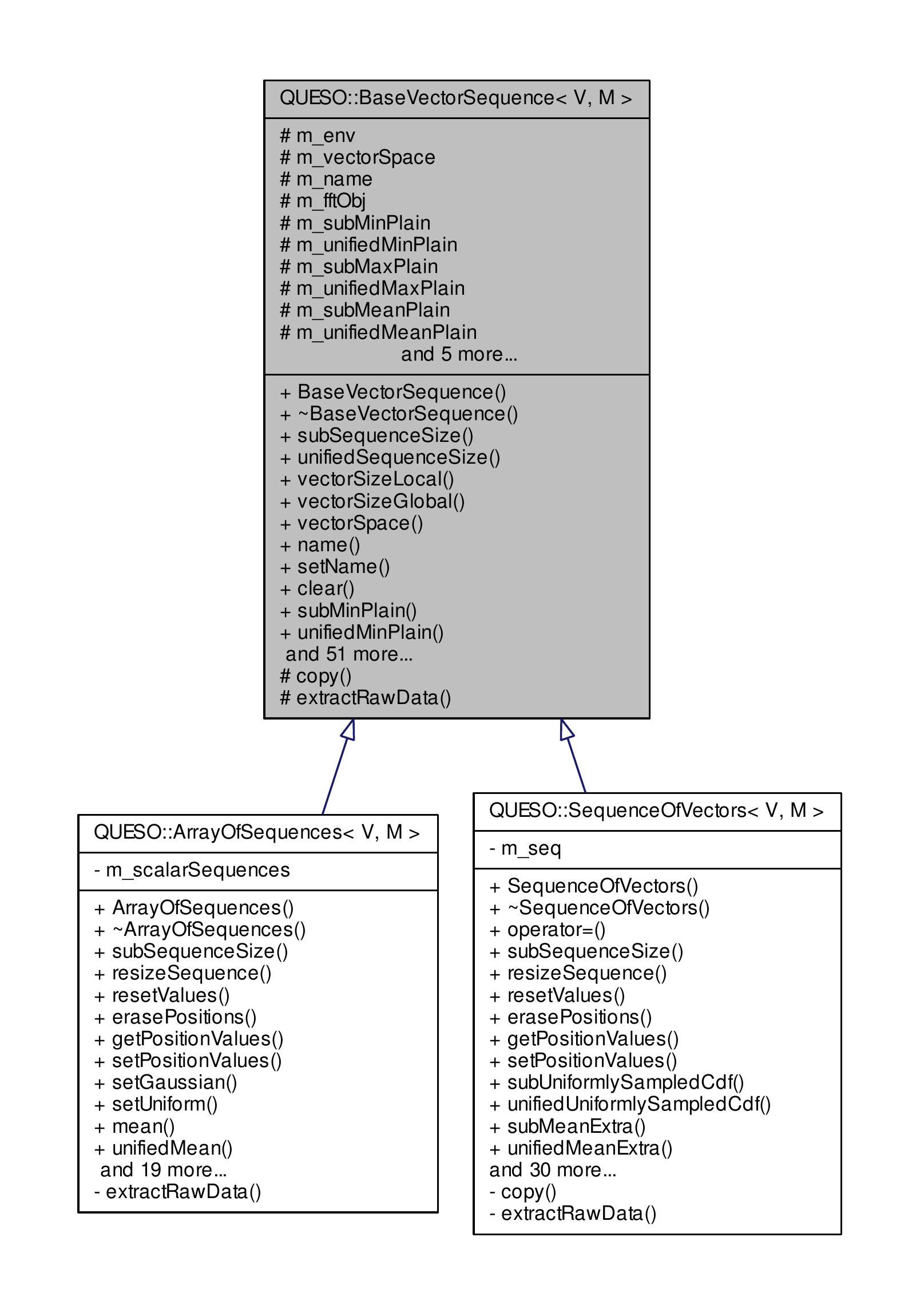}
\vspace{-1.cm}
\caption{The class diagram for the vector sequence class.}
\label{fig-vector-sequence-class}
\end{figure}

Figures \ref{fig-scalar-sequence-class} and \ref{fig-vector-sequence-class} display the class diagram for the scalar sequence  and vector sequence classes, respectively.

\section{Templated Statistical Classes}

The classes in this group are: vector realizer, vector random variable, statistical inverse problem (and options), Metropolis-Hastings solver (and options), statistical forward problem (and options), Monte Carlo solver (and options), and Sequence statistical options.

For QUESO, a SIP has two input entities, a prior RV and
a likelihood routine, and one output entity, the posterior RV, as shown in Chapter \ref{ch-introduction}, Figure~\ref{fig-sip-queso}.
Similarly, a SFP has two input entities, a input RV and
a QoI routine, and one output entity, the output RV, as shown in Figure \ref{fig-sfp-queso}.

\subsection{Vector Realizer Class}\label{sec:vector-realizer-class}
A {\it realizer} is an object that, simply put, contains a \verb+realization()+ operation that returns a sample of a vector RV.
QUESO currently supports several realizers: 
\begin{itemize}
 \item uniform, implemented in \verb+UniformVectorRealizer+,                \vspace{-8pt}
\item Gaussian, implemented in \verb+GaussianVectorRealizer+,               \vspace{-8pt}
\item Log Normal, implemented in \verb+LogNormalVectorRealizer+,            \vspace{-8pt}
\item Gamma,  implemented in \verb+GammaVectorRealizer+,                    \vspace{-8pt}
\item Inverse Gamma, implemented in \verb+InverseGammaVectorRealizer+, and  \vspace{-8pt}
\item Beta, , implemented in \verb+BetaVectorRealizer+,                     \vspace{-8pt}
\end{itemize}
which are all derived from the base class \verb+BaseVectorRealizer+. 
 
QUESO conveniently provides the class \verb+ConcatenatedVectorRealizer+, which allows two distinct realizers to be concatenated.
It also contains a {\it sequence realizer} class for storing samples of a MH algorithm.

\subsection{Vector Random Variable Class}
Vector RVs are expected to have two basic functionalities:
compute the value of its PDF at a point, and generate realizations following such PDF.
The joint PDF (\verb+BaseJointPdf+ and derived classes, see Section \ref{sec:scalar-vector-function}) and vector realizer  (\verb+BaseVectorRealizer+ and derived classes, see Section \ref{sec:vector-realizer-class}) classes allow a straightforward definition and manipulation of vector RVs. Similarly to the vector realizer class above, QUESO also allows users to form new RVs through the concatenation of existing RVs (class \verb+ConcatenatedVectorRV+).

\begin{figure}[htpb]
\centering
\includegraphics[scale=0.7,clip=true]{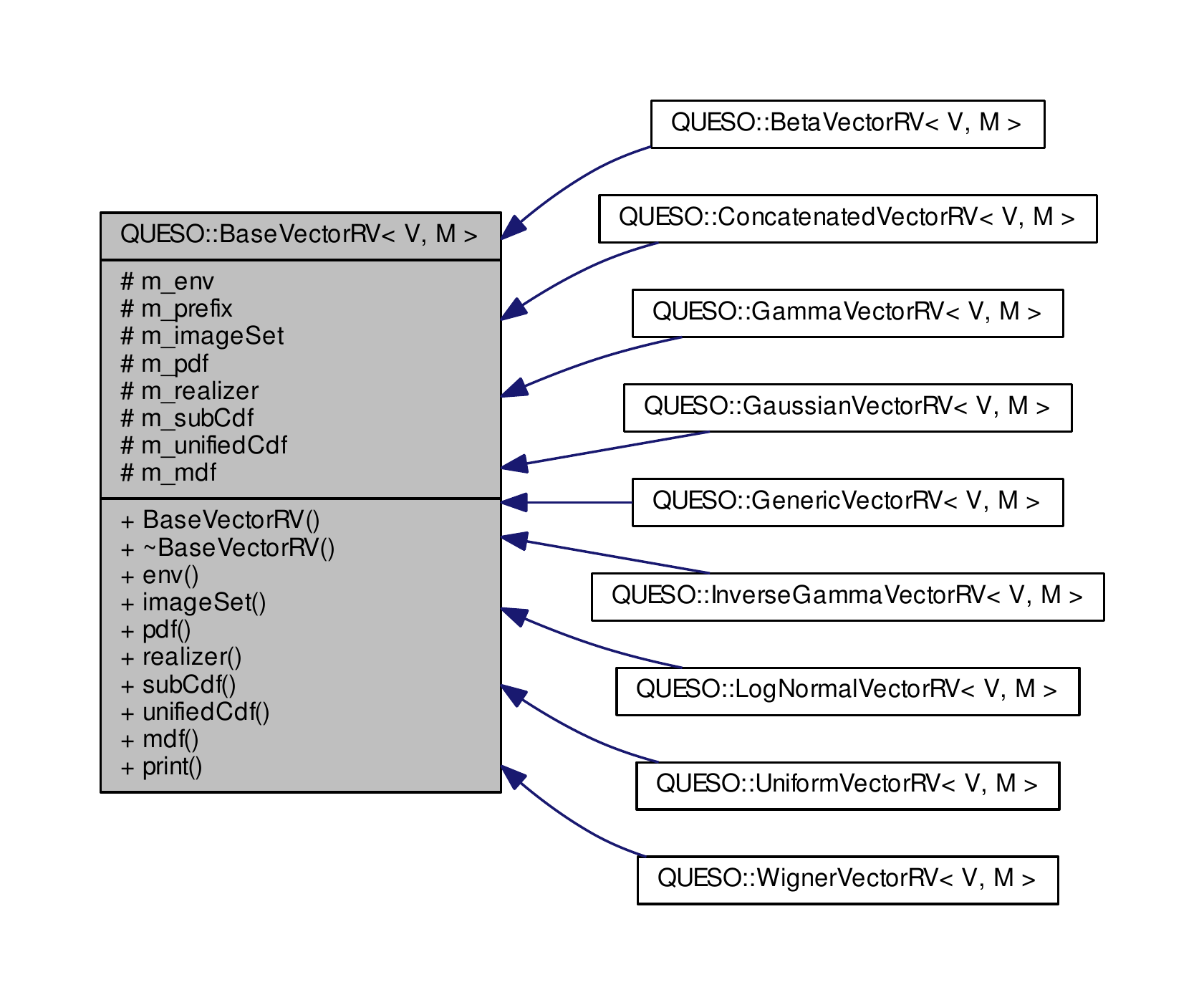}
\vspace{-1.cm}
\caption{The class diagram for the vector random variable class.}
\label{fig-vector-rv-class}
\end{figure}

QUESO currently supports a few vector RVs such as uniform, Gaussian, Gamma and Beta, as depicted in Diagram \ref{fig-vector-rv-class}.
A derived class called {\it generic vector RV} allows QUESO to store the solution of an statistical IP:
a {\it Bayesian joint PDF} becomes the PDF of the posterior RV, while a {\it sequence vector realizer} becomes the realizer of the same posterior RV.

\subsection{Statistical Inverse Problem (and Options)}
\begin{figure}[htpb]
\centering
\subfloat[StatisticalInverseProblem]{
 \includegraphics[trim={0 1.3cm 0 0},clip,scale=0.8]{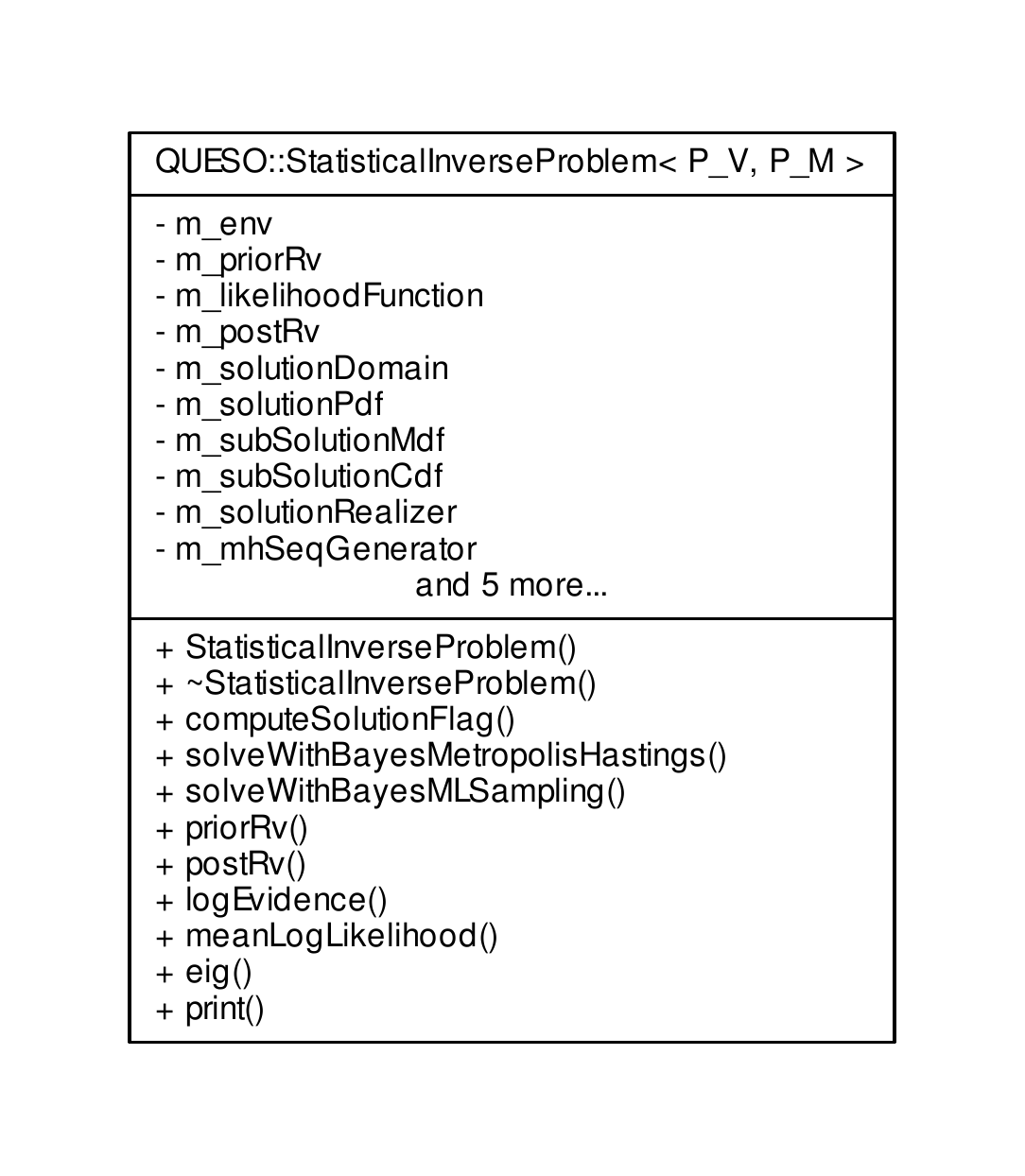}\label{fig-sip-class}}
\subfloat[StatisticalInverseProblemOptions]{
 \includegraphics[trim={0 1.3cm 0 0},clip,scale=0.8]{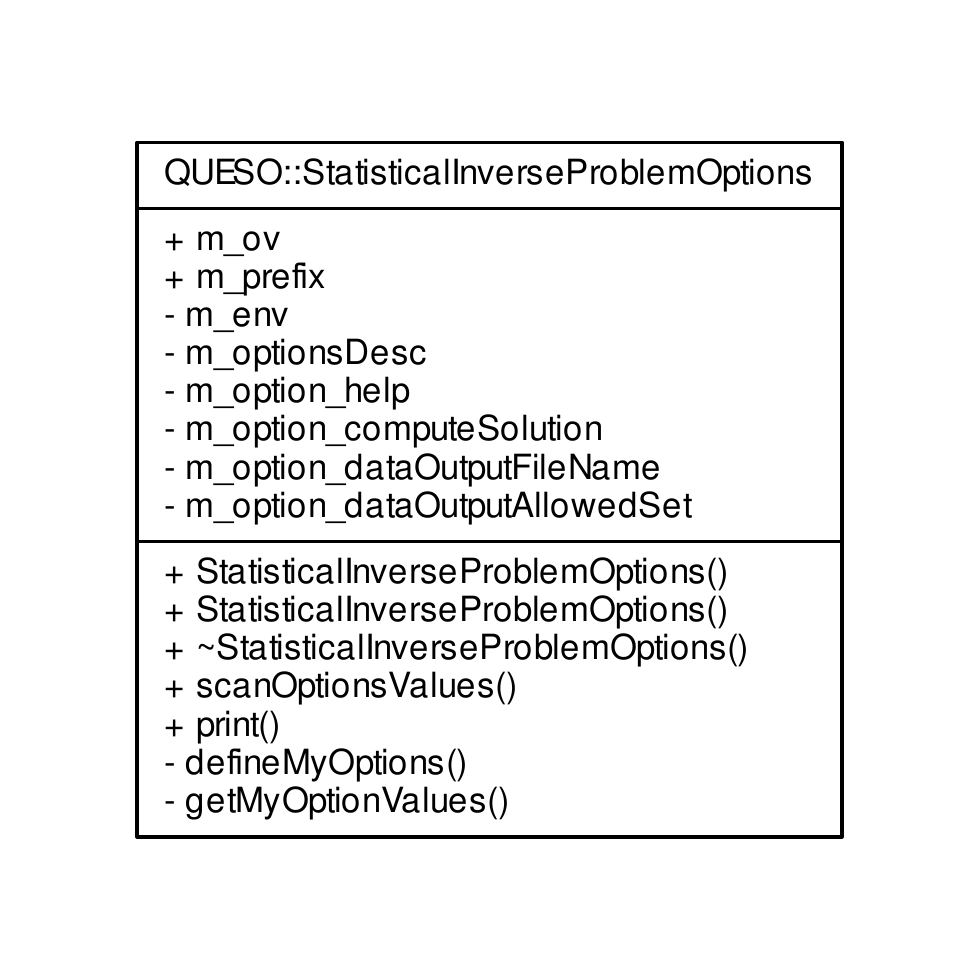}\label{fig-sip-options-class}}
\vspace{-.2cm}
\caption{The statistical inverse problem class, which implements the representation in Figure~\ref{fig-sip-queso}, and statistical inverse problem options class.}
\end{figure}

Similarly to its mathematical concepts, a SIP in QUESO also expects two input entities, a prior RV and a likelihood routine, and one output entity, the posterior RV.
The SIP is represented in QUESO through the templated class \verb+StatisticalInverseProblem<P_V,P_M>+, which is illustrated in Figure \ref{fig-sip-class}.
One important characteristic of the QUESO design is that it  separates `what the problem is' from `how the problem is solved'.
The prior and the posterior RV are instances of the \verb+BaseVectorRv<V,M>+ class, while
the likelihood function is an instance of the \verb+BaseScalarFunction<V,M>+ class.

The solution of a SIP is computed by calling either \verb+solveWithBayesMetropolisHastings()+ or \verb+solveWithBayesMLSampling()+, which are member functions of the class\linebreak\verb+StatisticalInverseProblem<P_V,P_M>+ class.
Upon return from a solution operation, the posterior RV is available through the \verb+postRv()+ member function.
More details are provided about \verb+solveWithBayesMetropolisHastings()+ and \verb+solveWithBayesMLSampling()+ in Sections \ref{sec:MH} and \ref{sec:ML}, respectively.

Figure \ref{fig-sip-options-class} displays the  \verb+StatisticalInverseProblemOptions+ class, i.e. that class that handles a variety of options for solving the SIP. Such options may be provided to QUESO by the user's input file; and they are listed in Table \ref{tab-sip-options}.

\begin{table}[htpb]
\begin{center}
\caption{Input file options for a QUESO statistical inverse problem.}
\vspace{-8pt}
\label{tab-sip-options}
\ttfamily\footnotesize
\begin{tabular}{l c  m{7cm}}
\toprule
\rmfamily Option name                    & \rmfamily Default  Value & \rmfamily Description \\
\midrule\midrule
\textlangle PREFIX\textrangle ip\_help                 &     &  \rmfamily Produces help message for statistical inverse problem   \\
\textlangle PREFIX\textrangle ip\_computeSolution      &  1  &  \rmfamily Computes solution process \\
\textlangle PREFIX\textrangle ip\_dataOutputFileName   & "." &  \rmfamily Name of data output file \\
\textlangle PREFIX\textrangle ip\_dataOutputAllowedSet & ""  &  \rmfamily Subenvironments that will write to data output file  \\
\bottomrule
\end{tabular}
\end{center}
\end{table}

\subsection{Metropolis-Hastings Solver (and Options)}\label{sec:MH}

The templated class that represents a Metropolis-Hastings generator of samples in QUESO is \verb+MetropolisHastingsSG<P_V,P_M>+, where SG stands for 'Sequence Generator'. This class implements the DRAM algorithm of Haario, Laine, Mira and Saksman~\cite{HaLaMiSa06} together with an operation named \verb+generateSequence()+ 
based on the core routine at the MCMC toolbox for MATLAB~\cite{Mcmctool}. 

\begin{figure}[htpb]
\centering
\subfloat[MetropolisHastingsSG]{\label{fig-metropolis-hastings-solver-class}\includegraphics[trim={0 1.3cm 0 0},clip, scale=0.65]{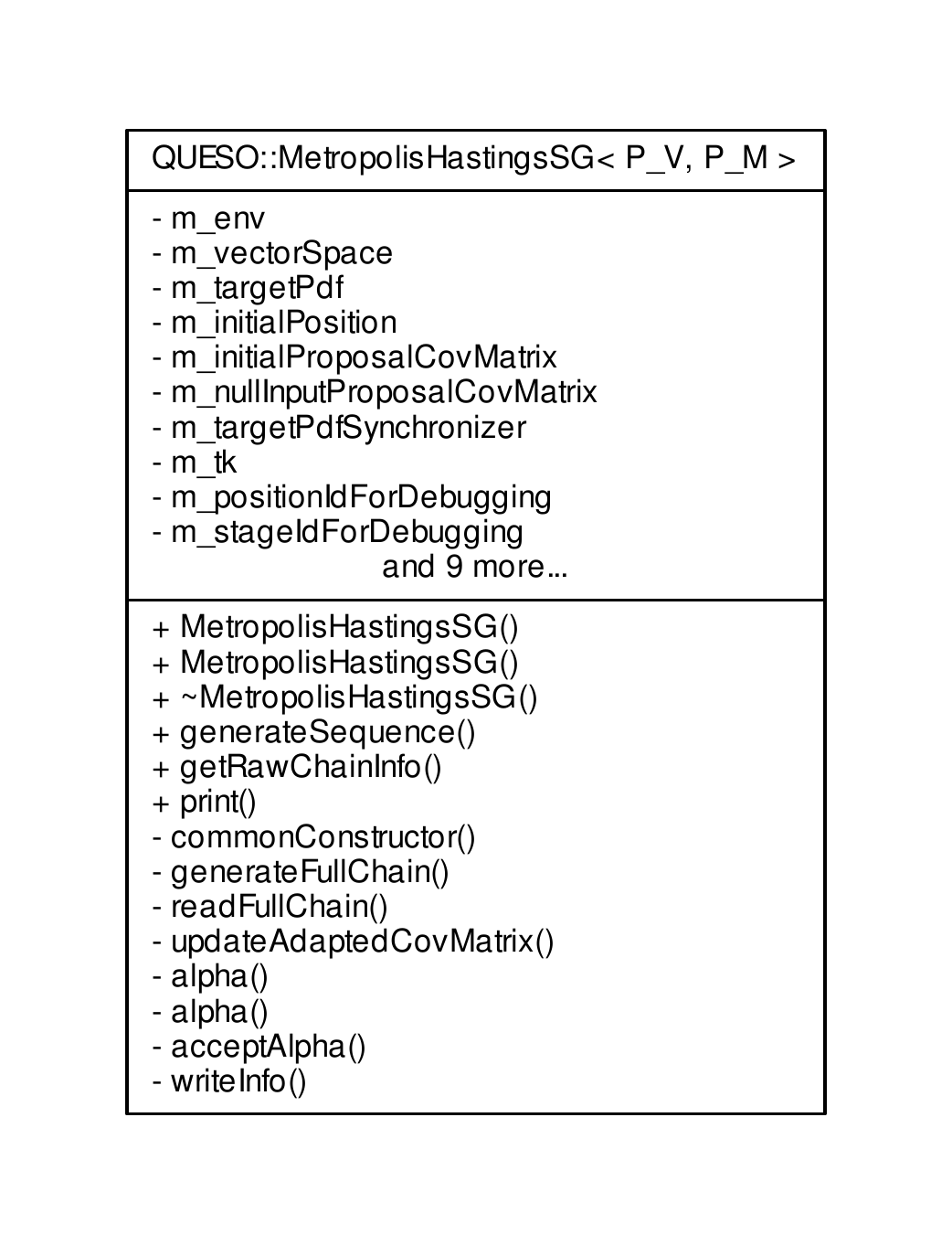}}
\subfloat[MetropolisHastingsSGOptions]{\label{fig-metropolis-hastings-options-class}\includegraphics[trim={0 1.3cm 0 0},clip, scale=0.65]{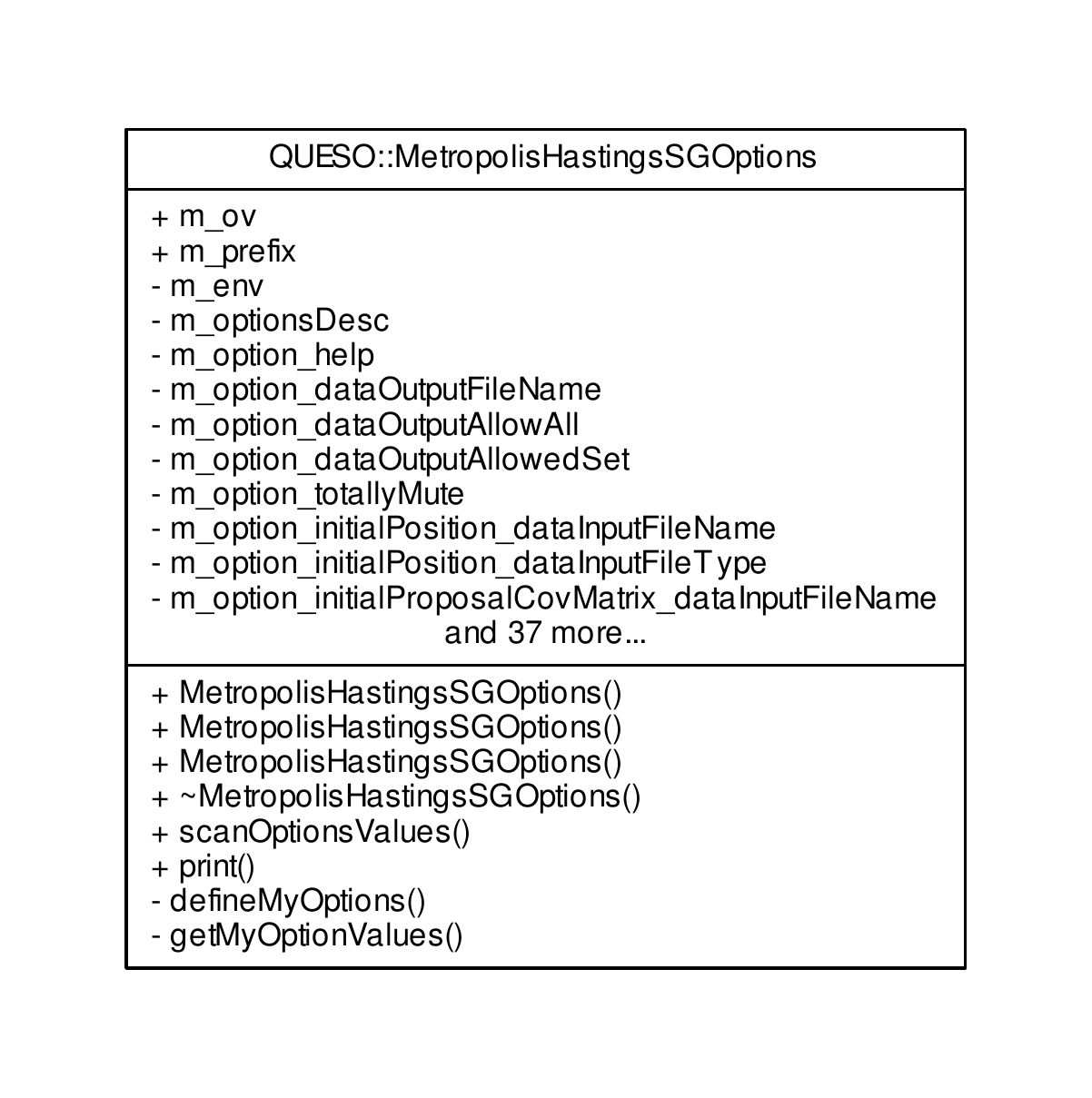}}
\vspace{-.2cm}
\caption{The Metropolis-Hastings sequence generator class and the Metropolis-Hastings sequence generator options class.}
\end{figure}

\begin{figure}[p]
\centering
\includegraphics[scale=.35,clip=true,angle=90]{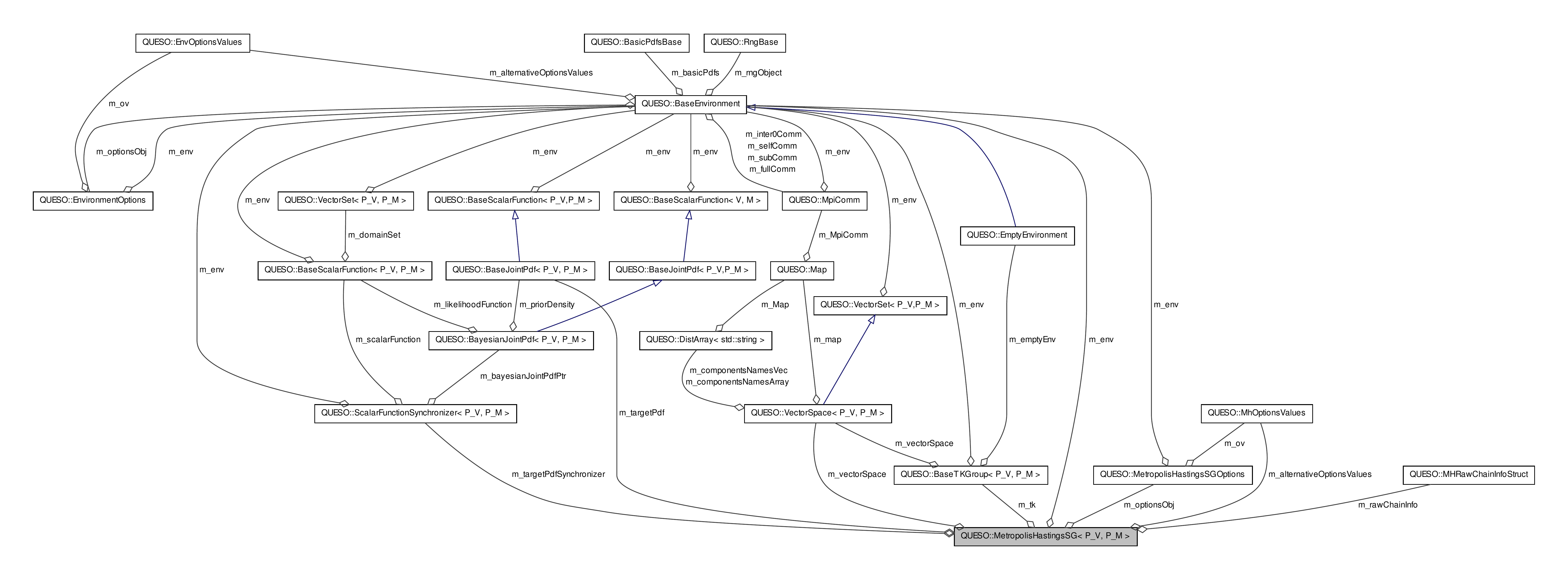}
 \vspace{-.8cm}
\caption{Collaboration graph of the  Metropolis-Hastings sequence generator class.}
\label{fig-metropolis-hastings-coll}
\end{figure}

The Metropolis-Hastings sequence generator class is depicted in Figure \ref{fig-metropolis-hastings-solver-class}; the Metropolis-Hastings sequence generator options class is depicted in Figure \ref{fig-metropolis-hastings-options-class}. A collaboration graph for the Metropolis-Hastings class is presented in Figure \ref{fig-metropolis-hastings-coll}; and the options are presented in Table \ref{tab-metropolis-hastings-options}.


%

\begin{table}[htpb]
\begin{center}
\caption{Input file options for a QUESO Metropolis-Hastings solver.}\label{tab-metropolis-hastings-options}
\vspace*{-8pt}
\ttfamily\footnotesize
\begin{tabular}{l c} 
\toprule
\rmfamily Option Name                                    & \rmfamily Default Value \\
\midrule\midrule
 \textlangle PREFIX\textrangle mh\_dataOutputFileName                       & "."   \\ 
 \textlangle PREFIX\textrangle mh\_dataOutputAllowAll                       & 0     \\ 
 \textlangle PREFIX\textrangle mh\_initialPositionDataInputFileName         & "."   \\ 
 \textlangle PREFIX\textrangle mh\_initialPositionDataInputFileType         & "m"   \\ 
 \textlangle PREFIX\textrangle mh\_initialProposalCovMatrixDataInputFileName& "."   \\ 
 \textlangle PREFIX\textrangle mh\_initialProposalCovMatrixDataInputFileType& "m"   \\ 
 \textlangle PREFIX\textrangle mh\_rawChainDataInputFileName                & "."   \\ 
 \textlangle PREFIX\textrangle mh\_rawChainDataInputFileType                & "m"   \\ 
 \textlangle PREFIX\textrangle mh\_rawChainSize                             & 100   \\ 
 \textlangle PREFIX\textrangle mh\_rawChainGenerateExtra                    &  0    \\ 
 \textlangle PREFIX\textrangle mh\_rawChainDisplayPeriod                    & 500   \\ 
 \textlangle PREFIX\textrangle mh\_rawChainMeasureRunTimes                  &  1    \\ 
 \textlangle PREFIX\textrangle mh\_rawChainDataOutputPeriod                 &  0    \\ 
 \textlangle PREFIX\textrangle mh\_rawChainDataOutputFileName               & "."   \\ 
 \textlangle PREFIX\textrangle mh\_rawChainDataOutputFileType               & "m"   \\ 
 \textlangle PREFIX\textrangle mh\_rawChainDataOutputAllowAll               &  0    \\ 
 \textlangle PREFIX\textrangle mh\_filteredChainGenerate                    &  0    \\ 
 \textlangle PREFIX\textrangle mh\_filteredChainDiscardedPortion            &  0.   \\ 
 \textlangle PREFIX\textrangle mh\_filteredChainLag                         &  1    \\ 
 \textlangle PREFIX\textrangle mh\_filteredChainDataOutputFileName          & "."   \\ 
 \textlangle PREFIX\textrangle mh\_filteredChainDataOutputFileType          & "m"   \\ 
 \textlangle PREFIX\textrangle mh\_filteredChainDataOutputAllowAll          &  0   \\ 
 \textlangle PREFIX\textrangle mh\_displayCandidates                        &  0    \\ 
 \textlangle PREFIX\textrangle mh\_putOutOfBoundsInChain                    &  1    \\ 
 \textlangle PREFIX\textrangle mh\_tkUseLocalHessian                        &  0    \\ 
 \textlangle PREFIX\textrangle mh\_tkUseNewtonComponent                     &  1    \\ 
 \textlangle PREFIX\textrangle mh\_drMaxNumExtraStages                      &  0    \\ 
 \textlangle PREFIX\textrangle mh\_drDuringAmNonAdaptiveInt                 &  1    \\ 
 \textlangle PREFIX\textrangle mh\_amKeepInitialMatrix                      &  0    \\ 
 \textlangle PREFIX\textrangle mh\_amInitialNonAdaptInterval                &  0    \\ 
 \textlangle PREFIX\textrangle mh\_amAdaptInterval                          &  0    \\ 
 \textlangle PREFIX\textrangle mh\_amAdaptedMatricesDataOutputPeriod        &  0    \\ 
 \textlangle PREFIX\textrangle mh\_amAdaptedMatricesDataOutputFileName      & "."   \\ 
 \textlangle PREFIX\textrangle mh\_amAdaptedMatricesDataOutputFileType      & "m"   \\ 
 \textlangle PREFIX\textrangle mh\_amAdaptedMatricesDataOutputAllowAll      &  0    \\ 
 \textlangle PREFIX\textrangle mh\_amEta                                    & 1.    \\ 
 \textlangle PREFIX\textrangle mh\_amEpsilon                                & $1\times 10^{-5}$   \\ 
 \textlangle PREFIX\textrangle mh\_enableBrooksGelmanConvMonitor            & 0    \\ 
 \textlangle PREFIX\textrangle mh\_BrooksGelmanLag                          & 100   \\ 
\bottomrule
\end{tabular}
\end{center}
\end{table}

\subsection{Multilevel Solver (and Options)}\label{sec:ML}

The templated class that represents a Multilevel generator of samples in QUESO is \linebreak \verb+MLSampling<P_V,P_M>+. This class implements the Adaptive Multilevel Stochastic Simulation Algorithm of Cheung and Prudencio~\cite{CheungPrudencio2012}.
The Multilevel sequence generator class is assisted by two extra classes, \verb+MLSamplingOptions+ and \verb+MLSamplingLevelOptions+, for handling the options to be used.

\begin{figure}[htpb]
\centering
\subfloat[MLSampling]{\includegraphics[trim={0 1.3cm 0 0},clip,scale=0.65]{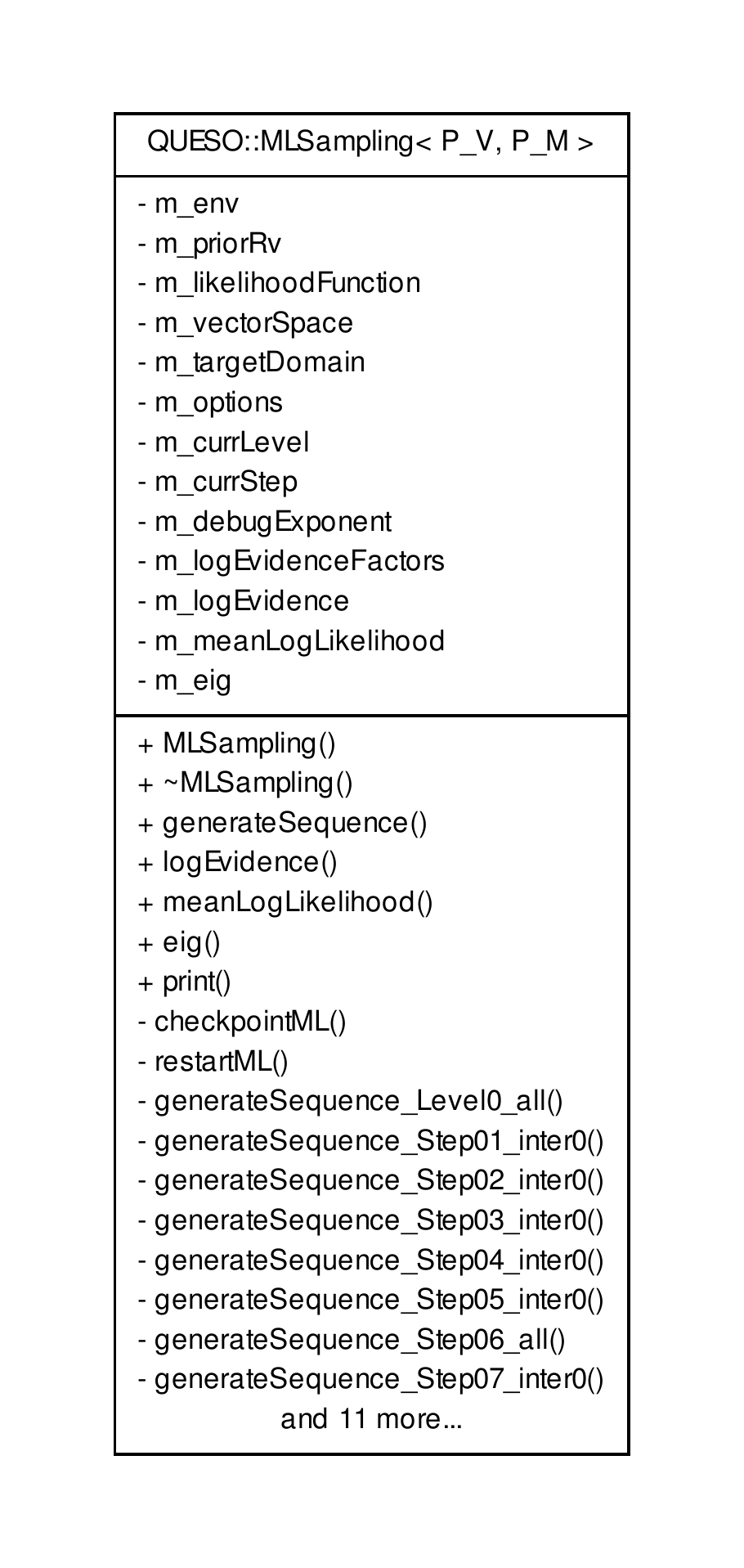}\label{fig-Multilevel-solver-class}}\hspace{-1.5cm}
\subfloat[MLSamplingOptions]{\includegraphics[trim={0 1.3cm 0 0},clip,scale=0.65]{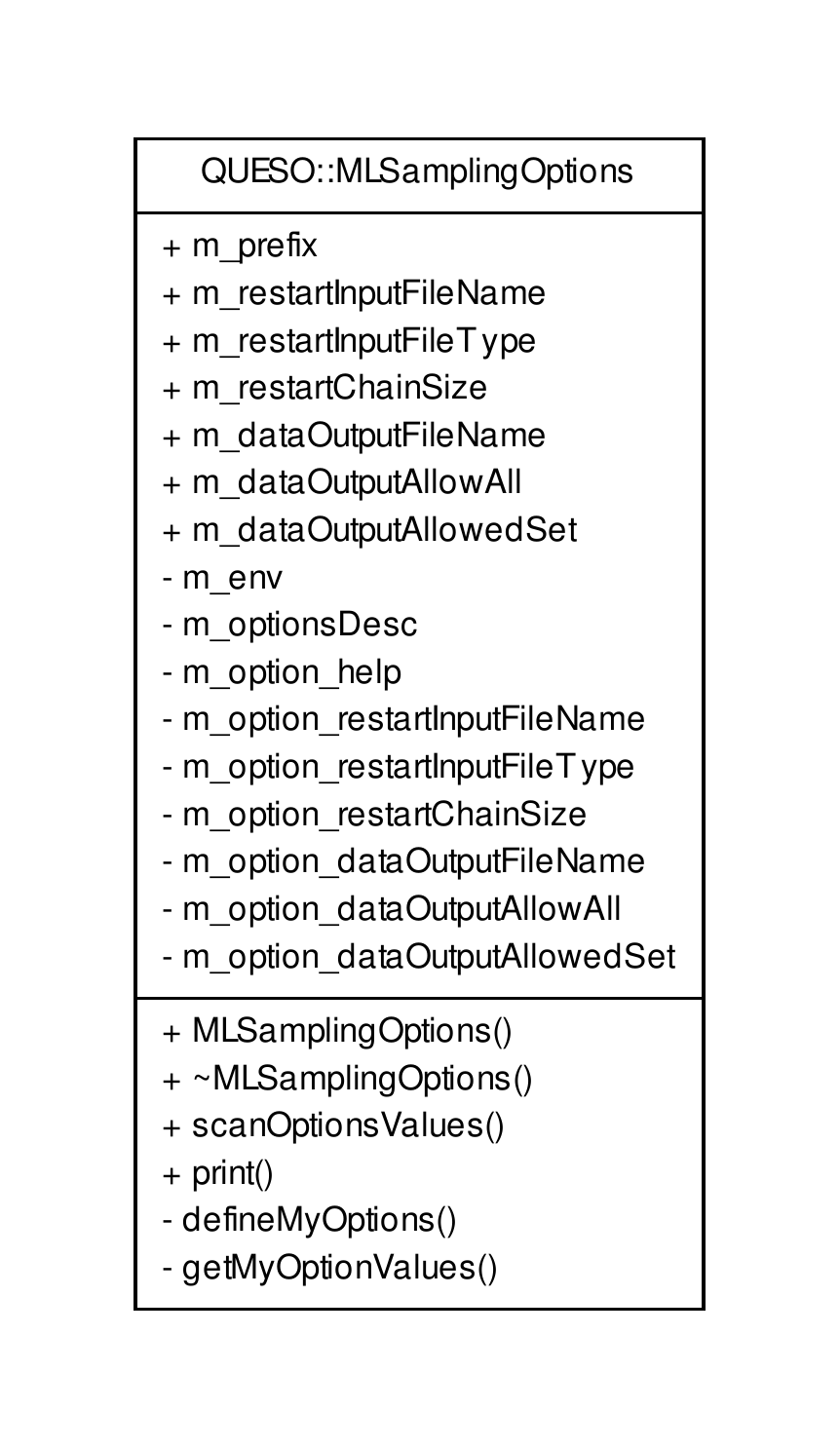}}\hspace{-1.5cm}
\subfloat[MLSamplingLevelOptions]{\includegraphics[trim={0 1.3cm 0 0},clip,scale=0.65]{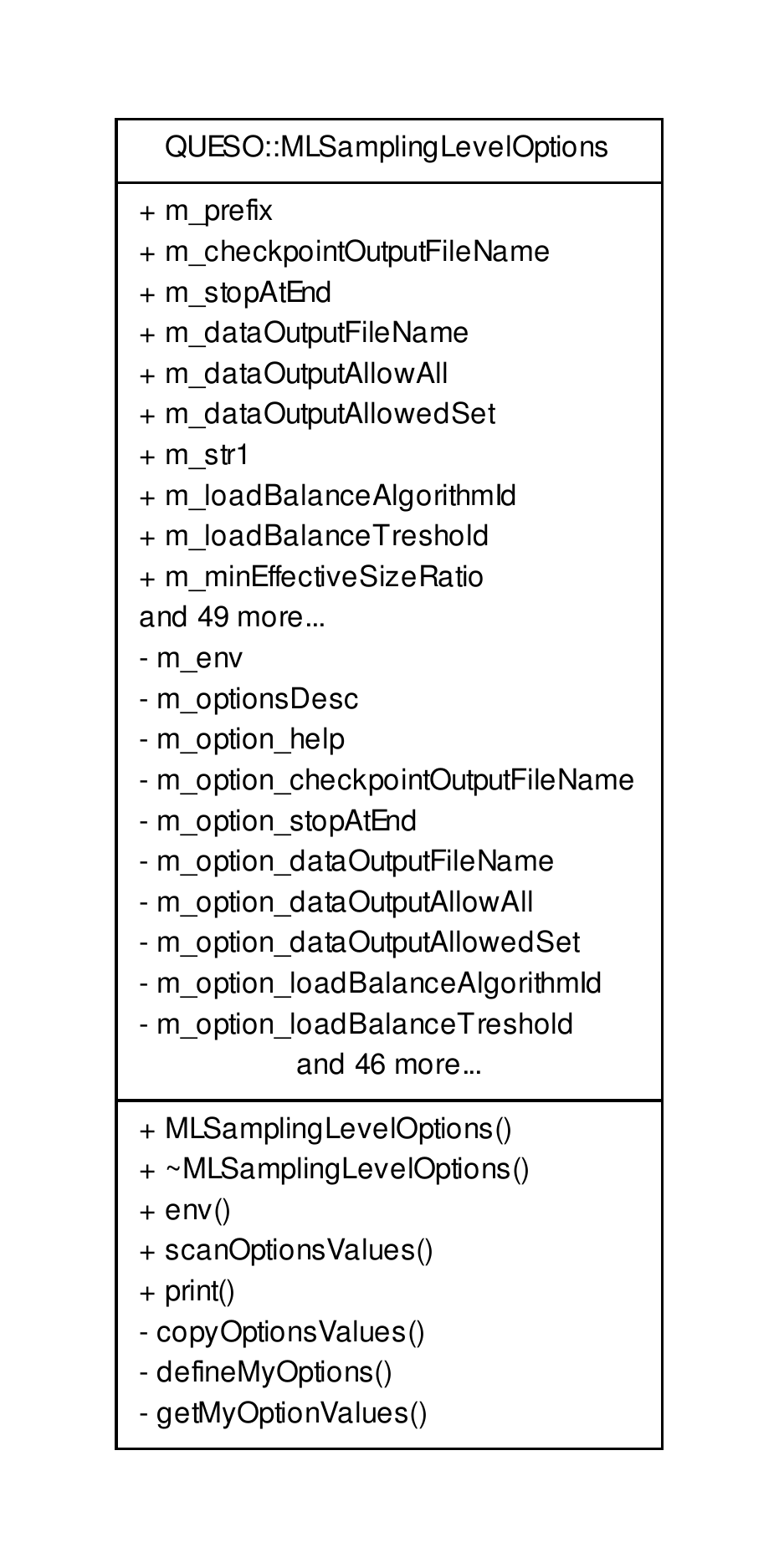}}
 \vspace{-.2cm}
\caption{The Multilevel sequence generator options class (\ref{fig-Multilevel-solver-class}) and its associated classes for handling options.}
\label{fig-Multilevel-options-class}
\end{figure}

\begin{figure}[p]
\centering
\includegraphics[scale=.4,clip=true,angle=90]{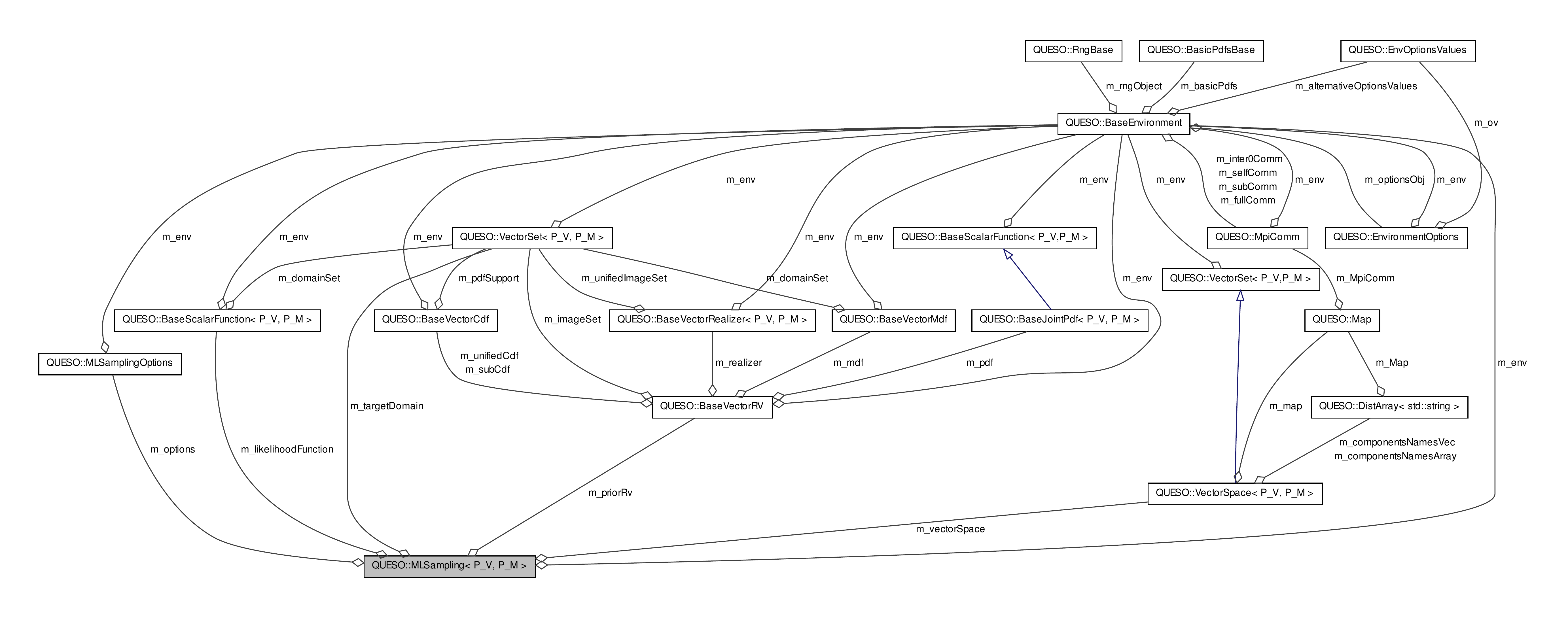}
 \vspace{-.8cm}
\caption{Collaboration graph of the  Multilevel sampling class.}
\label{fig-Multilevel-coll}
\end{figure}

The Multilevel class, the Multilevel options and level options classes are depicted in Figure~\ref{fig-Multilevel-options-class}.  A collaboration graph for the Multilevel class is presented in Figure \ref{fig-Multilevel-coll}; whereas its associated options are presented in Table \ref{tab-Multilevel-options}.
%

\begin{table}[p]
\begin{center}
\caption{Input file options for a QUESO Multilevel solver (to be continued).}\label{tab-Multilevel-options}
\vspace*{-8pt}
\ttfamily\footnotesize
\begin{tabular}{l c} 
\toprule
\rmfamily Option Name                                    & \rmfamily Default Value \\
\midrule\midrule

\textlangle PREFIX\textrangle
ml\_restartOutput\_levelPeriod       & 0    \\ 

\textlangle PREFIX\textrangle 
ml\_restartOutput\_baseNameForFiles  & "."  \\ 

\textlangle PREFIX\textrangle 
ml\_restartOutput\_fileType          & "m"  \\ 

\textlangle PREFIX\textrangle 
ml\_restartInput\_baseNameForFiles   & "."  \\ 

\textlangle PREFIX\textrangle 
ml\_restartInput\_fileType           & "m"  \\ 


 \textlangle PREFIX\textrangle ml\_stopAtEnd                                 & 0    \\
 \textlangle PREFIX\textrangle ml\_dataOutputFileName                        & "."  \\
 \textlangle PREFIX\textrangle ml\_dataOutputAllowAll                        & 0    \\
 \textlangle PREFIX\textrangle ml\_loadBalanceAlgorithmId                    & 2    \\
 \textlangle PREFIX\textrangle ml\_loadBalanceTreshold                       & 1.0  \\
 \textlangle PREFIX\textrangle ml\_minEffectiveSizeRatio                     & 0.85 \\
 \textlangle PREFIX\textrangle ml\_maxEffectiveSizeRatio                     & 0.91 \\
 \textlangle PREFIX\textrangle ml\_scaleCovMatrix                            & 1    \\
 \textlangle PREFIX\textrangle ml\_minRejectionRate                          & 0.50 \\
 \textlangle PREFIX\textrangle ml\_maxRejectionRate                          & 0.75 \\
 \textlangle PREFIX\textrangle ml\_covRejectionRate                          & 0.25 \\
 \textlangle PREFIX\textrangle ml\_minAcceptableEta                          & 0.   \\
 \textlangle PREFIX\textrangle ml\_totallyMute                               & 1    \\
 \textlangle PREFIX\textrangle ml\_initialPositionDataInputFileName          & "."  \\
 \textlangle PREFIX\textrangle ml\_initialPositionDataInputFileType          & "m"  \\
 \textlangle PREFIX\textrangle ml\_initialProposalCovMatrixDataInputFileName & "."  \\
 \textlangle PREFIX\textrangle ml\_initialProposalCovMatrixDataInputFileType & "m"  \\
 \textlangle PREFIX\textrangle ml\_rawChainDataInputFileName                 & "."  \\
 \textlangle PREFIX\textrangle ml\_rawChainDataInputFileType                 & "m"  \\
 \textlangle PREFIX\textrangle ml\_rawChainSize                              & 100  \\
 \textlangle PREFIX\textrangle ml\_rawChainGenerateExtra                     & 0    \\
 \textlangle PREFIX\textrangle ml\_rawChainDisplayPeriod                     & 500  \\
 \textlangle PREFIX\textrangle ml\_rawChainMeasureRunTimes                   & 1    \\
 \textlangle PREFIX\textrangle ml\_rawChainDataOutputPeriod                  & 0    \\
 \textlangle PREFIX\textrangle ml\_rawChainDataOutputFileName                & "."  \\
 \textlangle PREFIX\textrangle ml\_rawChainDataOutputFileType                & "m"  \\
 \textlangle PREFIX\textrangle ml\_rawChainDataOutputAllowAll                & 0    \\
 \textlangle PREFIX\textrangle ml\_filteredChainGenerate                     & 0    \\
 \textlangle PREFIX\textrangle ml\_filteredChainDiscardedPortion             & 0.   \\
 \textlangle PREFIX\textrangle ml\_filteredChainLag                          & 1    \\
 \textlangle PREFIX\textrangle ml\_filteredChainDataOutputFileName           & "."  \\
 \textlangle PREFIX\textrangle ml\_filteredChainDataOutputFileType           & "m"  \\
 \textlangle PREFIX\textrangle ml\_filteredChainDataOutputAllowAll           & 0    \\
 \textlangle PREFIX\textrangle ml\_displayCandidates                         & 0    \\
 \textlangle PREFIX\textrangle ml\_putOutOfBoundsInChain                     & 1    \\

 \textlangle PREFIX\textrangle ml\_tkUseLocalHessian                         & 0    \\
 \textlangle PREFIX\textrangle ml\_tkUseNewtonComponent                      & 1    \\
 \textlangle PREFIX\textrangle ml\_drMaxNumExtraStages                       & 0    \\
 \textlangle PREFIX\textrangle ml\_drScalesForExtraStages                    & 0    \\
 \textlangle PREFIX\textrangle ml\_drDuringAmNonAdaptiveInt                  & 1    \\
 \textlangle PREFIX\textrangle ml\_amKeepInitialMatrix                       & 0    \\
 \textlangle PREFIX\textrangle ml\_amInitialNonAdaptInterval                 & 0    \\
 \textlangle PREFIX\textrangle ml\_amAdaptInterval                           & 0    \\
 \textlangle PREFIX\textrangle ml\_amAdaptedMatricesDataOutputPeriod         & 0    \\
 \textlangle PREFIX\textrangle ml\_amAdaptedMatricesDataOutputFileName       & "."  \\ 
 \textlangle PREFIX\textrangle ml\_amAdaptedMatricesDataOutputFileType       & "m"  \\ 
 \textlangle PREFIX\textrangle ml\_amAdaptedMatricesDataOutputAllowAll       & 0    \\ 
 \textlangle PREFIX\textrangle ml\_amEta                                     & 1.   \\ 
 \textlangle PREFIX\textrangle ml\_amEpsilon                                 & 1.e-5\\ 
\bottomrule
\end{tabular}
\end{center}
\end{table}

%
%
%

\subsection{Statistical Forward Problem (and Options)}
\label{sub:SFP}

\begin{figure}[p]
\centering
\centering
\subfloat[StatisticalForwardProblem]{\includegraphics[trim={0.5cm 1.3cm 0 0},clip,scale=0.7]{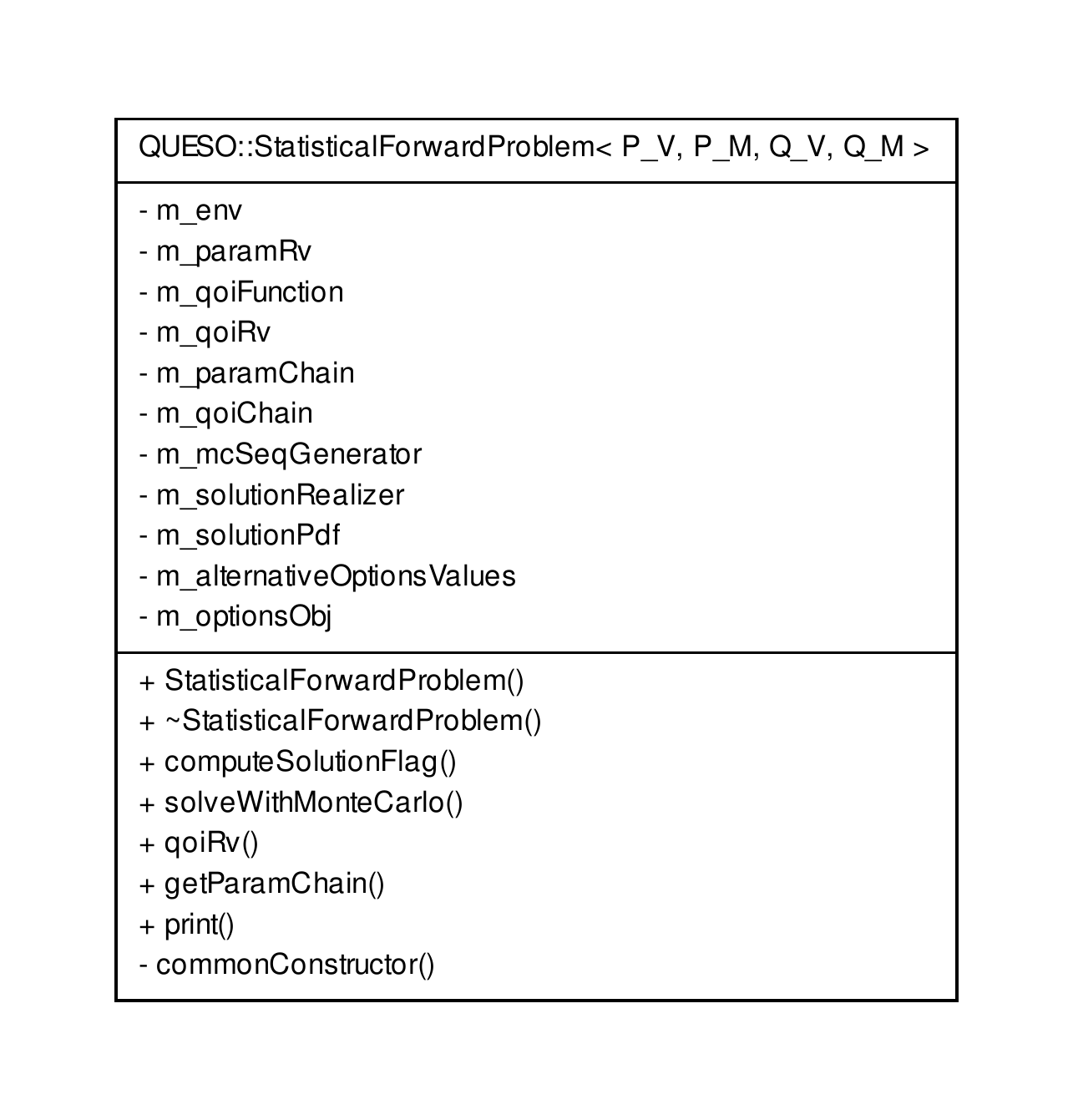}\label{fig-sfp-class}}
\subfloat[StatisticalForwardProblemOptions]{\includegraphics[trim={0.5cm 1.3cm 0 0},clip,scale=0.7]{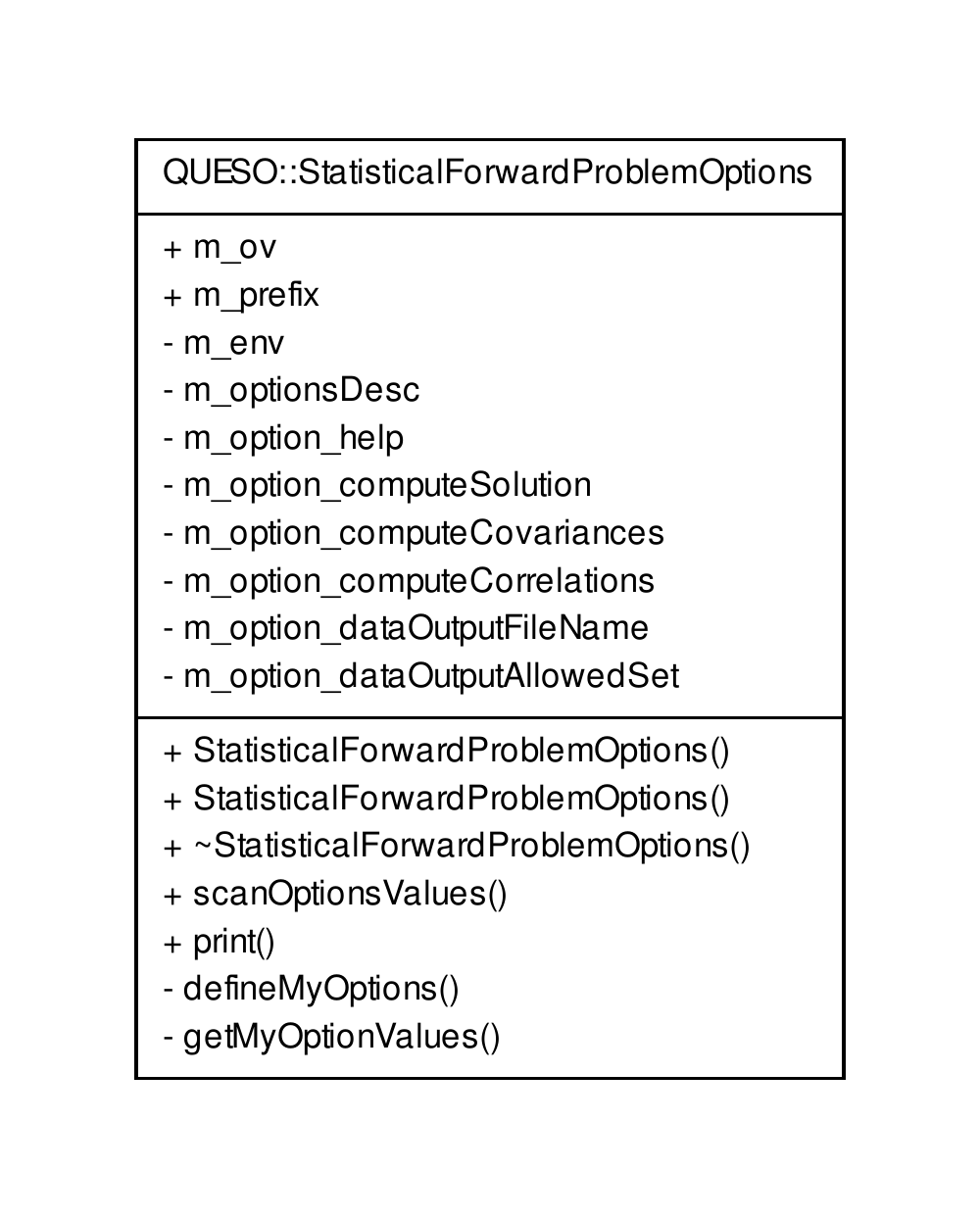}\label{fig-sfp-options-class}}
\vspace{-.2cm}
\caption{The statistical forward problem class, which implements the representation in Figure~\ref{fig-sfp-queso}, and the statistical forward problem options class.}
\end{figure}

A SFP in QUESO also has two input entities, the input (parameter) RV and a QoI function, and one output entity, the QoI RV. 
The SIP is represented through the templated class \verb+StatisticalForwardProblem<P_V,P_M,Q_V,Q_M >+, which diagram is presented in Figure \ref{fig-sfp-class}. Again, the types \verb+P_V+ and \verb+Q_V+ of vectors and types \verb+P_M+ and \verb+Q_M+ of matrices, where \verb+P_+ stands for 'parameter' and \verb+Q_+ stands for 'quantities of interest'.

The input RV and the output QoI RV are instances of the \verb+BaseVectorRv<P_V,P_M>+ class, while
the QoI function is an instance of \verb+BaseVectorFunction<P_V,P_M,Q_V,Q_M>+.
In the template parameters, the prefix \verb+P_+ refers to the parameters, whereas the prefix \verb+Q_+ refers to the QoIs.

In order to find the solution of a SFP, one must call the \verb+solveWithMonteCarlo()+ member function of the \verb+StatisticalForwardProblem<P_V,P_M>+ class.
Upon return from a solution operation, the QoI RV is available through the \verb+qoiRv()+ member function. Such QoI RV  is able to provide:
a vector realizer through the operation \verb+'qoiRv().realizer()'+, which returns an instance of the class \verb+'uqBaseVectorRealizer<Q_V,Q_M>'+.

Figure \ref{fig-sfp-options-class} displays the  statistical forward problem options class, i.e. that class that handles a variety of options for solving the SFP. Such options may be provided to QUESO at the user's input file; and they are listed in Table \ref{tab-sfp-options}. In the table, \texttt{p-q} stands for parameter--quantity of interest.


\begin{table}[htpb]
\caption{Input file options for a QUESO statistical forward problem.}\label{tab-sfp-options}
\vspace{-8pt}
\ttfamily\footnotesize
\begin{center}
\begin{tabular}{l c  m{6cm}}
\toprule
 \rmfamily Option Name                     & \rmfamily Default Value& \rmfamily Description \\
\midrule
\textlangle PREFIX\textrangle fp\_computeSolution      &   1  &\rmfamily Computes the solution process   \\
\textlangle PREFIX\textrangle fp\_computeCovariances   &   1  &\rmfamily Compute \verb+p-q+ covariances    \\ 
\textlangle PREFIX\textrangle fp\_computeCorrelations  &   1  &\rmfamily Compute \verb+p-q+ correlations   \\ 
\textlangle PREFIX\textrangle fp\_dataOutputFileName   &  "." &\rmfamily Name of data output file  \\ 
\textlangle PREFIX\textrangle fp\_dataOutputAllowedSet &  ""  &\rmfamily Subenvironments that will write to data output file   \\ 
\bottomrule
\end{tabular}
\end{center}
\end{table}

\subsection{Monte Carlo Solver (and Options)}

\begin{figure}[p]
\centering
\subfloat[MonteCarloSG]{\includegraphics[trim={0.5cm 1.3cm 0 0},clip,scale=0.7]{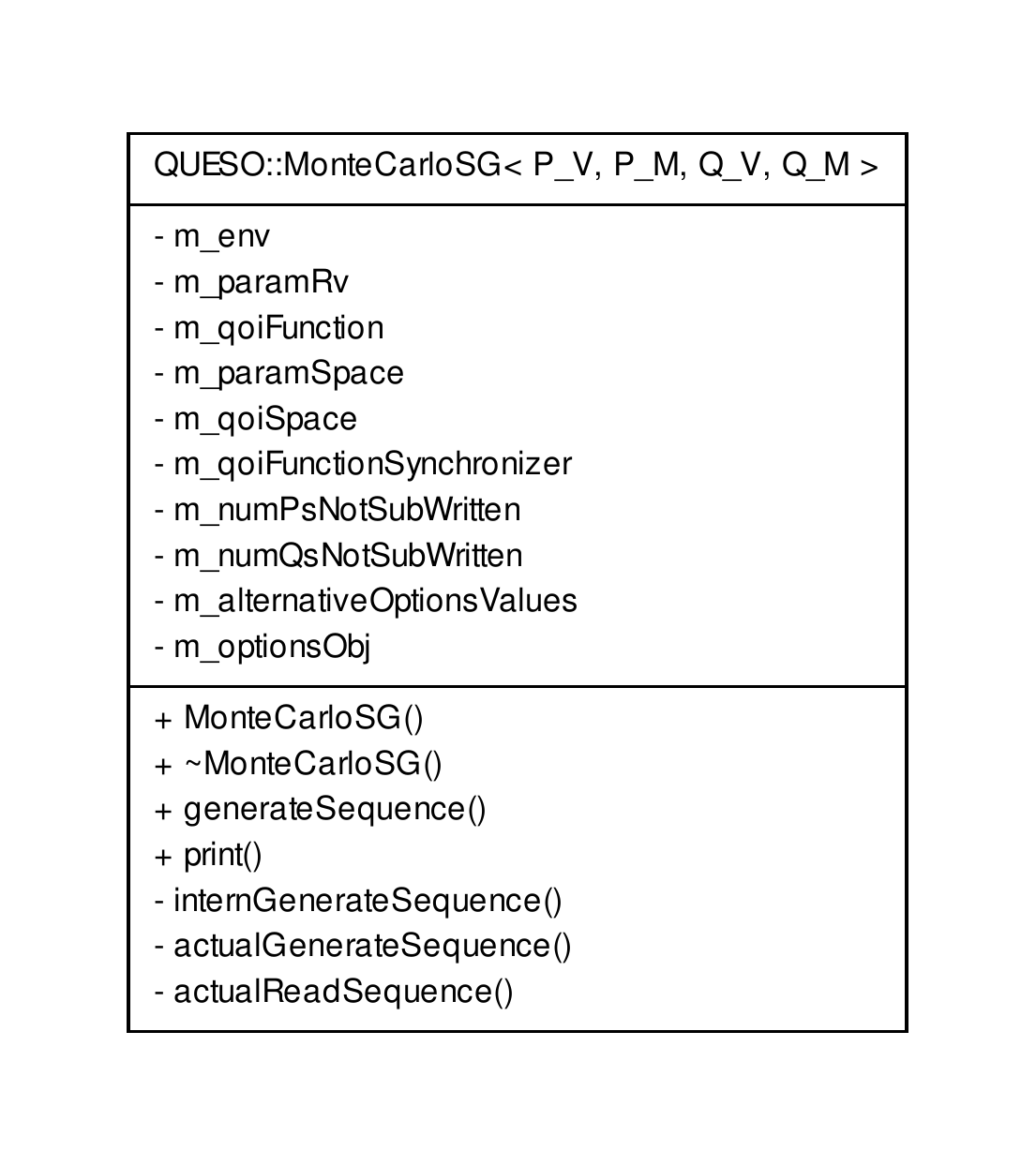}\label{fig-monte-carlo-solver-class}}
\subfloat[MonteCarloSGOptions]{\includegraphics[trim={0.5cm 1.3cm 0 0},clip,scale=0.7]{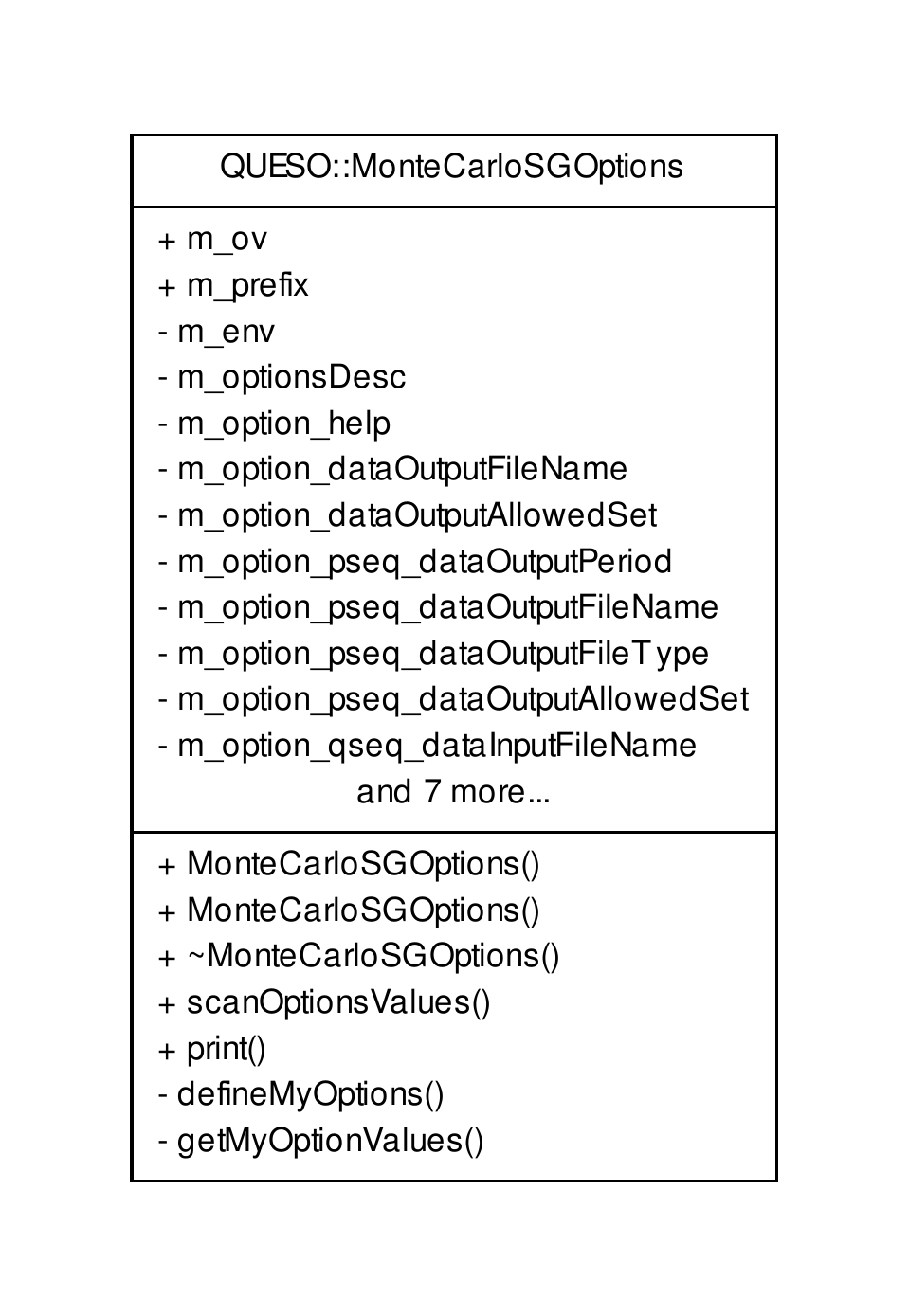}\label{fig-monte-carlo-options-class}}
\vspace*{-.2cm}
\caption{The Monte Carlo sequence generator class and the Monte Carlo sequence generator options class.}
\end{figure}

The templated class that implements a Monte Carlo generator of samples within QUESO is \verb+MonteCarloSG<P_V,P_M,Q_V,Q_M>+, as illustrated in Figure \ref{fig-monte-carlo-solver-class}.
This class has the requirement that the image set of the vector random variable  and the domain set of the QoI function belong to vector spaces of equal dimensions. If the requirements are satisfied, the class constructor reads input options that begin with the string `\verb+<PREFIX>_mc_+' (See Table~\ref{tab-monte-carlo-options}). Options reading is handled by class \verb+MonteCarloOptions+, which is illustrated in Figure~\ref{fig-monte-carlo-options-class}.

\begin{table}[htpb]
\begin{center}
\caption{Input file options for a QUESO statistical forward problem solved via Monte Carlo algorithm.}
\vspace{-8pt}
\label{tab-monte-carlo-options}
\ttfamily\footnotesize
\begin{tabular}{l c  m{6cm}}
\toprule
\rmfamily Option Name     & \rmfamily Default Value \\
\midrule\midrule
\textlangle PREFIX\textrangle mc\_dataOutputFileName           &  "."  \\
\textlangle PREFIX\textrangle mc\_dataOutputAllowedSet         &       \\
\textlangle PREFIX\textrangle mc\_pseq\_dataOutputFileName     &  "."  \\
\textlangle PREFIX\textrangle mc\_pseq\_dataOutputAllowedSet   &       \\
\textlangle PREFIX\textrangle mc\_qseq\_dataInputFileName      &  "."  \\
\textlangle PREFIX\textrangle mc\_qseq\_size                   &  100  \\
\textlangle PREFIX\textrangle mc\_qseq\_displayPeriod          &  500  \\  
\textlangle PREFIX\textrangle mc\_qseq\_measureRunTimes        &    0  \\  
\textlangle PREFIX\textrangle mc\_qseq\_dataOutputFileName     &  "."  \\ 
\textlangle PREFIX\textrangle mc\_qseq\_dataOutputAllowedSet   &       \\  
\bottomrule
\end{tabular}
\end{center}
\end{table}

\section{Miscellaneous Classes and Routines}
 
As the name suggests, QUESO miscellaneous classes and routines have a variety of routines.
For instance, the function \verb+MiscReadDoublesFromString+ is used for reading the options input files and assigning the values to the respective variables, in \texttt{uqMonteCarloSGOptions::} \texttt{getMyOptionValues} and in \verb+MetropolisHastingsSGOptions::getMyOptionValues+. 

QUESO class \verb+BaseOneDGrid+ generates grids necessary for calculating the CDF of a RV; it is required by class \verb+ArrayOfOneDGrids+, which, in turn, is used in both classes: \linebreak \verb+StatisticalForwardProblem+ and \verb+StatisticalInverseProblem+.


\chapter{Important Remarks}

\new{
At this point, the user may feel comfortable and ready to start his/her validation and calibration exercises using QUESO. There are, however, a few quite important remarks that will make the linkage of the QUESO Library with the user application code possible. They are addressed in the following sections.
}

\section{Revisiting Input Options}\label{sec:revisiting_input}

Input options are read from the QUESO input file, whose name is required by the constructor of the QUESO environment class. Herein, suppose that no prefix is defined, i.e., nothing will precede the input variables names (\verb+PREFIX = ""+ in Tables \ref{tab-env-options} -- \ref{tab-monte-carlo-options}). An example of the use of prefixes may be found in the input file \texttt{tgaCycle.inp} under the subdirectory  \texttt{/examples/validationCycle/} of QUESO installation tree.

The first part of a input file commonly handles the environment options. he variable assignment \verb+env_numSubEnvironments = 1+ indicates to QUESO that only one subenvironment should be used. The variable assignment \texttt{env\_subDisplayFileName} \texttt{=} \texttt{outputData/} \texttt{display} create both the subdirectory \verb+outputData/+ and a file named \verb+display_sub0.txt+ that contains all the options listed in the input file together with more specific information, such as the chain run time and the number of delayed rejections. The existence of file  \verb+display_sub0.txt+  allows, for instance, the user in verifying the actual parameters read by QUESO.

For an SIP, the user may set up variables related to the DRAM algorithm. Six important variables are:  
\texttt{ip\_mh\_dr\_maxNumExtraStages} defines how many extra candidates will be generated; 
\texttt{ip\_mh\_dr\_listOfScalesForExtraStages} defines the list $s$ of scaling factors that will multiply the covariance matrix.
The variable \texttt{ip\_mh\_am\_initialNonAdaptInterval} defines the initial interval in which the proposal covariance matrix will not be changed;
whereas \texttt{ip\_mh\_am\_adaptInterval} defines the size of the interval in which each adapted proposal covariance matrix will be used. 
\texttt{ip\_mh\_am\_eta} is a factor used to scale the proposal covariance matrix, usually set to be $2.4^2/d$, where $d$ is the dimension of the problem~\cite{Laine08,HaLaMiSa06}. 
Finally, \texttt{ip\_mh\_am\_epsilon} is the covariance regularization factor used in the DRAM algorithm. 

For a SFP, the variable assignment \verb+fp_computeSolution = 1+ tells QUESO to compute the solution process; the assignment \verb+fp_computeCovariances = 1+,  instructs QUESO to compute parameter--QoI covariances, and analogously, \verb+fp_computeCorrelations = 1+ inform QUESO to compute  parameter--QoI correlations. The name of the data output file can be set with variable \verb+fp_dataOutputFileName arg+;  and \verb+fp_dataOutputAllowedSet+ defines which subenvironments will write to data output file.

An example a complete input file used by QUESO to solve a SIP--SFP is presented in Section \ref{sec:gravity-input-file}; however every application example included in \Queso{} build and installation directories \verb+examples+ has an options input file and the user is invited to familiarize him/herself with them.

\section{Revisiting Priors}
\new{
QUESO offers a variety of prior distributions: uniform, Gaussian, Beta, Gamma, Inverse Gamma, and Log Normal. Also, QUESO presents the option of concatenating any of those priors, through the Concatenated prior. 

Concatenated priors are employed in problems with multiple random parameters. They allow one random parameter to have a different prior distribution then other; i.e., one variable may have a uniform prior distribution whereas other may have a Gaussian prior distribution.

It is important to notice that, in order to set a Gaussian prior, besides providing the mean, the user must also supply the \underline{variance}, not the  standard deviation. 
}

\section{Running with Multiple Chains or Monte Carlo Sequences}

 
As presented in the previous section, the variable \verb+env_numSubEnvironments+ determines how many subenvironments QUESO will work with. Thus, if \verb+env_numSubEnvironments=1+, then   only one subenvironment will be used, and QUESO will use only one set on Monte Carlo chains of size defined by ones of the variables \verb+ip_mh_rawChain_size+ or \verb+fp_mc_qseq_size+, depending either the user is solving a SIP or a SFP.

If the user wants to run QUESO with multiple chains or Monte Carlo sequences, then two variables have to be set in QUESO input file: \verb+env_numSubEnvironments = +$N_s$, with $N_s>1$ is the number of chains and/or Monte Carlo sequences of samples; and \verb+env_seed = +$-z$, with $z\geqslant 1$, so that each processor sets the seed to value MPI\_RANK+$z$.
It is crucial that \verb+env_seed+ takes a \underline{negative} value, otherwise all chain samples are going to be the same.

Also, the total number $N_p$ of processors in the full communicator, usually named \linebreak MPI\_COMM\_WORLD, needs to be a multiple of $N_s$.

\section{Running with Models that Require Parallel Computing}
\new{
It is possible to run QUESO with models that require parallel computing as long as total number of processors $N_p$ is multiple of the number of subenvironments $N_s$. QUESO will internally create $N_s$ subcommunicators, each of size $N_p/N_s$, and make sure that the likelihood and QoI routines are called for all processors in these subcommunicators -- the
likelihood/QoI routine will have available a communicator of size $N_p/N_s$. For instance, if $N_p = 2048$ and $N_s = 16$, then each likelihood/QoI will have available a communicator of size 128. s
Each subcommunicator is accessible through \texttt{env.subComm()}. At the end of the simulation, there will be a total of $N_s$ chains.
%

The user, however, must keep in mind the possible occurrence of race condition, especially in the case where the application is a black box and files are being accessed constantly (e.g. data is being written and read).
}

\section{A Requirement for the DRAM  Algorithm}
\new{
Besides setting up the variables related to the DRAM algorithm in the input file, as described in Section \ref{sec:revisiting_input} above, the user must also provide an initialized \underline{covariance matrix} before calling the DRAM solver, \texttt{solveWithBayesMetropolisHastings(...)}, in his/her application code. 

It is worth to note that this is rather a DRAM requirement~\cite{Laine08}, not a QUESO limitation. An example of the instantiation and initialization of a proposal covariance matrix and its subsequent use is presented in lines 145-147 of Listings \ref{code:gravity_compute_C}, Section \ref{sec:gravity_code}.

}


\chapter{Global Sensitivity Analysis}

\new{
Global sensitivity analysis (GSA) involves a quantitative assessment of variability in the
model output or quantity of interest (QoI) due to uncertain model parameters. Variance based
approaches relying on pseudo-random sampling of prior distributions of the parameters
have been used effectively~\cite{Cukier:1973, Sobol:1990, Saltelli:2008}. However, it can
be understood that estimating the sensitivity indices (`first order effect' and `total
effect') can
be computationally intensive especially in situations where a complex multiphysics model
is simulated for a considered set of parameter samples. In order to mitigate such computational
costs, alternative strategies involving construction and application of cheap surrogates for the
models have been developed. Examples include polynomial chaos
expansions~\cite{Xiu:2002, Ghanem:2003} which have been used extensively for physics-based
models and admit simple analytical expressions for computing the sensitivity
indices~\cite{Vohra:2014, Vohra:2016}.
Other examples include
response surfaces based on Kriging, and radial basis functions~\cite{Gutmann:2001}.

Surrogate models, however, are not the central theme of this chapter. Instead, we focus
our attention on exploiting QUESO to perform a prior based, parametric GSA. As mentioned
earlier, the analysis helps determine relative contribution to the variance of the QoI and
thus the relative importance of the stochastic model parameters. Potentially, such an
analysis could help reduce the dimensionality of an inverse problem. In the following section,
we provide a mathematical framework for the first order effect and total effect sensitivity indices
as well as a brief survey of existing estimators for these indices. In section~\ref{sec:app},
we provide an algorithm based on~\cite{Saltelli:2010}
for using QUESO to perform a prior-based parametric GSA and
further demonstrate its implementation using a simple exercise involving sensitivity analysis
of the slope and intercept of a straight line.
}

\section{Sensitivity Indices}

Consider a model, $\mathcal{G}(\bm \theta)$, where $\bm \theta$ denotes a vector of model
parameters. Variance based measures for the first order effect and total effect sensitivity
indices can be computed as discussed below.

\subsection{First Order Effect}

The first order effect sensitivity index $\mathcal{S}(\theta_{i})$ for a specific model
parameter $(\theta_{i})$ quantifies relative contribution to the variance of the QoI
strictly due to $(\theta_{i})$ and does not consider its interactions with other parameters.
Mathematically, this is expressed as follows:

\be
\mathcal{S}(\theta_{i}) = \frac{\V_{\theta_{i}}(\E_{\bm \theta_{\sim i}}[\mathcal{G}\vert\theta_{i}])}{\V(\mathcal{G})}
\ee

\noindent where $\theta_{i}$ is the $i^{th}$ parameter for which the first order effect index is to be
computed and $\bm \theta_{\sim i}$ denotes a vector of all parameters except $\theta_{i}$. The quantity,
$\E_{\bm \theta_{\sim i}}[\mathcal{G}\vert\theta_{i}]$ denotes the mean estimate of the model output
taken over all possible values of $\bm \theta_{\sim i}$ while using a fixed value of $\theta_{i}$. The
outer variance of this expectation is hence computed over all possible values of $\theta_{i}$.
The quantity, $\V_{\theta_{i}}(\E_{\bm \theta_{\sim i}}[\mathcal{G}\vert\theta_{i}])$ can also be
understood as the expected reduction in variance due to fixed $\theta_{i}$.
It is normalized by, $\V(\mathcal{G})$, i.e. the total variance of of the model output.

\subsection{Total Effect}

The total effect sensitivity index $\mathcal{T}(\theta_{i})$ for a specific model
parameter $(\theta_{i})$ quantifies relative contribution to the variance of the QoI
due to $(\theta_{i})$ and accounts for its interactions with other parameters.
Mathematically, this is expressed as follows:

\begin{eqnarray}
\mathcal{T}(\theta_{i}) &=& \frac{\E_{\bm \theta_{\sim i}}[\V_{\theta_{i}}(\mathcal{G}\vert\bm{\theta}_{\sim i})]}{\V(\mathcal{G})} \\ \vspace{1mm}
			&=& 1 - \frac{\V_{\bm \theta_{\sim i}}(\E_{\theta_{i}}[\mathcal{G}\vert\bm\theta_{\sim i}])}{\V(\mathcal{G})}
\end{eqnarray}

\noindent where $\E_{\bm \theta_{\sim i}}[\V_{\theta_{i}}(\mathcal{G}\vert\bm{\theta}_{\sim i}))$ is the expected variance when all parameters
except $\theta_{i}$ could be fixed. We can also interpret $\mathcal{T}(\theta_{i})$ using the quantity,
$\V_{\bm \theta_{\sim i}}(\E_{\theta_{i}}[\mathcal{G}\vert\bm\theta_{\sim i}])$ which denotes the expected reduction in variance
when all parameters except $\theta_{i}$ could be fixed.

\subsection{Estimation of $\mathcal{S}(\theta_{i})$ and $\mathcal{T}(\theta_{i})$}

In order to estimate the first order effect and the total effect sensitivity indices, we need to numerically estimate
the quantities, $\V_{\theta_{i}}(\E_{\bm \theta_{\sim i}}[\mathcal{G}\vert\theta_{i}])$ and
$\E_{\bm \theta_{\sim i}}[\V_{\theta_{i}}(\mathcal{G}\vert\bm{\theta}_{\sim i})]$ respectively. Tabulated below are commonly used estimators. As discussed in~\cite{Saltelli:2010}, we consider two independent set of samples denoted by the matrices,
$\bm A$ and $\bm B$. Additionally, we consider derived sets of samples denoted by the  matrices,
$\bm A_{\bm B}^{(i)}$ where all columns
are from $\bm A$ except the $i^{th}$ column which is from $\bm B$. Similarly, we can construct the matrix,
$\bm B_{\bm A}^{(i)}$ as well.

\begin{table}[htbp]
\centering
\begin{tabular}{@{}*{6}{c}@{}}
\toprule
& \textsc{Estimator} &  \textsc{Reference} \\
\cmidrule(r){2-3}\cmidrule(l){4-6}
\multirow{4}{*}{$\V_{\theta_{i}}(\E_{\bm \theta_{\sim i}}[\mathcal{G}\vert\theta_{i}])$} & $\frac{1}{N}\sum_{k=1}^{N} f(\bm A)_{k}f(\bm B_{\bm A}^{(i)})_{k} - f_{0}^{2}$ &  Sobol 1990~\cite{Sobol:1990}  \\
\cmidrule(lr){2-2}\cmidrule(lr){3-3}
& $\frac{1}{N}\sum_{k=1}^{N} f(\bm B)_{k}(f(\bm A_{\bm B}^{(i)})_{k} - f(\bm A)_{k})$ &  Saltelli 2010~\cite{Saltelli:2010} \\
\cmidrule(lr){2-2}\cmidrule(lr){3-3}
& $\V(\mathcal{G}) - \frac{1}{2N}\sum_{k=1}^{N} (f(\bm B)_{k} - f(\bm A_{\bm B}^{(i)})_{k})^{2}$ & Jansen 1999~\cite{Jansen:1999}   \\
\cmidrule{1-3}
\multirow{3}{*}{$\E_{\bm \theta_{\sim i}}[\V_{\theta_{i}}(\mathcal{G}\vert\bm{\theta}_{\sim i})]$} & $\V(\mathcal{G}) - \frac{1}{N}\sum_{k=1}^{N} f(\bm A)_{k}f(\bm A_{\bm B}^{(i)})_{k} + f_{0}^{2}$ &  Homma 1996~\cite{Homma:1996}  \\
\cmidrule(lr){2-2}\cmidrule(lr){3-3}
& $\frac{1}{N}\sum_{k=1}^{N} f(\bm A)_{k}(f(\bm A)_{k} - f(\bm A_{\bm B}^{(i)})_{k})$ &  Sobol 2007~\cite{Sobol:2007} \\
\cmidrule(lr){2-2}\cmidrule(lr){3-3}
& $\frac{1}{2N}\sum_{k=1}^{N} (f(\bm A)_{k} - f(\bm A_{\bm B}^{(i)})_{k})^{2}$ & Jansen 1999~\cite{Jansen:1999}   \\
\bottomrule
\end{tabular}
\caption{Commonly used estimators and corresponding references for $\V_{\theta_{i}}(\E_{\bm \theta_{\sim i}}[\mathcal{G}\vert\theta_{i}])$ and $\E_{\bm \theta_{\sim i}}[\V_{\theta_{i}}(\mathcal{G}\vert\bm{\theta}_{\sim i})]$.}
\label{tab:estimators}
\end{table}

Statistical forward problem (SFP) can be solved with QUESO by computing the QoI for pseudo-random samples drawn
from the posterior distribution as discussed in~\ref{sub:SFP}. However, for GSA, we need to generate two independent
data sets comprising pseudo-random samples for the model parameters, drawn from their individual prior distributions.
QoIs are estimated for both sets of samples as well derived matrices for the model parameters, as discussed earlier.

In the following section, we present a simple application involving sensitivity analysis of the slope and y-intercept
of a straight line. SFP on samples from prior distributions of the model parameters is solved to generate the data
which can further be used to compute the first order effect and total effect sensitivity indices.

\section{Application}
\label{sec:app}

We consider the following equation for a straight line:

\be y = mx + c \ee

The slope, $m$ and the y-intercept, $c$ are considered to be uniformly distributed in the intervals,
[2, 5] and [3, 7] respectively. For reference purposes, we provide an algorithm followed by the
C++ code which interfaces with QUESO to generate the required data for GSA as follows.
In order to estimate the first order effect and the total effect sensitivity indices, we solve
the forward problem in QUESO to generate the required set of data. Specifically, we need to generate
(2$n$+2) data files for $n$ model parameters. Hence, in the present case, we need 6 data files as
listed and described below. Note that the pseudorandom samples pertaining to the  individual model parameters
are given by their respective columns. In this case, column 1 corresponds to the slope, $m$ and column 2
corresponds to the y-intercept, $c$. Moreover, $y$ in the above equation is regarded as the QoI.

\begin{enumerate}
\item \texttt{qoi\_samplesA.txt}: Pseudo-random samples and corresponding QoI estimates, regarded as set $\bm A$.
\item \texttt{qoi\_samplesB.txt}: Pseudo-random samples and corresponding QoI estimates, regarded as set $\bm B$.
\item \texttt{m\_qoi\_samplesAi.txt}: All columns from set $\bm A$ except the $i^{th}$ ($i$=1) column which is from
set $\bm B$ and corresponding QoI estimates.
\item \texttt{m\_qoi\_samplesBi.txt}: All columns from set $\bm B$ except the $i^{th}$ ($i$=1) column which is from
set $\bm A$ and corresponding QoI estimates.
\item \texttt{c\_qoi\_samplesAi.txt}: All columns from set $\bm A$ except the $i^{th}$ ($i$=2) column which is from
set $\bm B$ and corresponding QoI estimates.
\item \texttt{c\_qoi\_samplesBi.txt}: All columns from set $\bm B$ except the $i^{th}$ ($i$=2) column which is from
set $\bm A$ and corresponding QoI estimates.
\end{enumerate}

In the above list, let us denote files in 3--6 as the set of derived files.  The following algorithm provides
a sequence of steps as well as snippets of code which could be used to
generate the set of data files to compute the sensitivity indices. Relevant source
files have also been included later in this section.

\begin{algorithm}
\caption*{\textbf{Algorithm}:~Generating data for GSA}
\begin{algorithmic}[1]
\Procedure{Solving SFP with QUESO}{}
\BState Instantiate a QoI object (\texttt{qoi\_mc}):
\Statex \texttt{Qoi\_mc<> qoi\_mc("qoi\_", paramDomain, qoiSpace);}
\BState Instantiate the forward problem (\texttt{fp\_mc}):
\Statex \texttt{QUESO::StatisticalForwardProblem<> fp\_mc("", NULL, priorRv, qoi\_mc, qoiRv);}
\BState Solve the forward problem to generate the data file, \texttt{qoi\_samplesA.txt} i.e. set $\bm A$.
\Statex \texttt{fp\_mc.solveWithMonteCarlo(NULL);}
\BState Repeat steps 2--4 to generate the data file, \texttt{qoi\_samplesB.txt} i.e. set $\bm B$.
\BState Use the two sets of data, $\bm A$ and $\bm B$, to generate intermediate data files comprising samples,
$\bm{A}^{(i)}_{\bm B}$ and $\bm{B}^{(i)}_{\bm A}$ for both $m$ and $c$.
\BState Repeat steps 2--4 four times to generate the derived files: \texttt{m\_qoi\_samplesAi.txt},
\texttt{m\_qoi\_samplesBi.txt}, \texttt{c\_qoi\_samplesAi.txt} and \texttt{c\_qoi\_samplesAi.txt}. (Instead
of estimating the QoI for pseudorandom samples from the prior, the QoI is now computed for the set of samples in
corresponding intermediate data files generated in the previous step.)
\BState Compute $\mathcal{S}(\theta_{i})$ and $\mathcal{T}(\theta_{i})$ using the set of 6 data files generated
in previous steps.
\EndProcedure
\end{algorithmic}

\end{algorithm}

The source code for generating the required set of data files is provided by the header file,
\texttt{sensitivity\_mc.h} and the corresponding source file, \texttt{sensitivity\_mc.C} as follows.

\begin{lstlisting}[caption=File \texttt{sensitivity\_mc.h.}, label={code:sensitivity-h},
linerange={1-29}]
#ifndef QUESO_EXAMPLE_SENSITIVITY_MC_H
#define QUESO_EXAMPLE_SENSITIVITY_MC_H

#include <queso/VectorFunction.h>
#include <queso/DistArray.h>
#include <fstream>

template<class P_V = QUESO::GslVector, class P_M = QUESO::GslMatrix,
	 class Q_V = QUESO::GslVector, class Q_M = QUESO::GslMatrix>
class Qoi_mc : public QUESO::BaseVectorFunction<P_V, P_M, Q_V, Q_M>
{
public:
	Qoi_mc(const char * prefix, const QUESO::VectorSet<P_V, P_M> & domainSet,
	    const QUESO::VectorSet<Q_V, Q_M> & imageSet);
	virtual ~Qoi_mc();
	virtual void compute(const P_V & domainVector, const P_V * domainDirection,
		Q_V & imageVector, QUESO::DistArray<P_V *> * gradVectors,
		QUESO::DistArray<P_M *> * hessianMatrices,
		QUESO::DistArray<P_V *> * hessianEffects) const;

private:
	double x_loc;
	mutable double m,c,mf,cf,count;
	mutable std::fstream qoi_samples;
	mutable std::fstream samples;

};

#endif
\end{lstlisting}

\begin{lstlisting}[caption=File \texttt{sensitivity\_mc.C.}, label={code:sensitivity-c},
linerange={1-70}, numbers=left,stepnumber=1]
#include <cmath>
#include <fstream>
#include <iomanip> // for setprecision

#include <queso/GslVector.h>
#include <queso/GslMatrix.h>
#include <sensitivity_mc.h>

template<class P_V, class P_M, class Q_V, class Q_M>
Qoi_mc<P_V, P_M, Q_V, Q_M>::Qoi_mc(const char * prefix,
    const QUESO::VectorSet<P_V, P_M> & domainSet,
    const QUESO::VectorSet<Q_V, Q_M> & imageSet)
  : QUESO::BaseVectorFunction<P_V, P_M, Q_V, Q_M>(prefix, domainSet, imageSet),
    x_loc(3)
{
}

template<class P_V, class P_M, class Q_V, class Q_M>
Qoi_mc<P_V, P_M, Q_V, Q_M>::~Qoi_mc()
{
  // Deconstruct here
}

template<class P_V, class P_M, class Q_V, class Q_M>
void
Qoi_mc<P_V, P_M, Q_V, Q_M>::compute(const P_V & domainVector,
    const P_V * domainDirection,
    Q_V & imageVector, QUESO::DistArray<P_V *> * gradVectors,
    QUESO::DistArray<P_M *> * hessianMatrices,
    QUESO::DistArray<P_V *> * hessianEffects) const
{
  if (domainVector.sizeLocal() != 2) {
    queso_error_msg("domainVector does not have size 2");
  }
  if (imageVector.sizeLocal() != 1) {
    queso_error_msg("imageVector does not have size 1");
  }

  qoi_samples.open ("c_qoi_samplesAi.txt", std::fstream::in | std::fstream::out | std::fstream::app);

 // ----Generate Qoi using samples from a text files -------------
 count++;
 samples.open ("./files_sense/c_samples_Ai.txt", std::fstream::in | std::fstream::out | std::fstream::app);
 
 int cou = 1;

 while (samples >> mf >> cf){
       if (cou == count){
       		m = mf;
 		c = cf;
       	break;
       }
       cou++;
 }

// ------- Generate Qoi using pseudo-random MC samples ------------
//  std::cout << "m = " << domainVector[0] << std::endl; 
//  m = domainVector[0];  // Sample of the RV 'line slope'
//  c = domainVector[1];  // Sample of the RV 'y-intercept'
  double y_obs = 0.0;
  y_obs = m*x_loc + c;

  qoi_samples << std::setprecision(4) << y_obs << "\t\t" << m << "\t\t" << c << std::endl;

  imageVector[0] = y_obs;
  qoi_samples.close();
  samples.close();
}

template class Qoi_mc<QUESO::GslVector, QUESO::GslMatrix, QUESO::GslVector,
\end{lstlisting}
As shown in lines, 42--54 in the  above listing for \texttt{sensitivity\_mc.C}, in order to generate the set of derived files,
we compute the QoI by reading samples from corresponding intermediate files (such as \texttt{c\_samples\_Ai.txt} in this case).
Whereas, for generating the pair of files, \texttt{qoi\_samplesA.txt} and \texttt{qoi\_samplesB.txt}, we compute the QoI
for pseudo-random samples drawn from the prior distributions for $m$ and $c$ as shown in lines, 57--59 which are commented in
the present case.

\section{Results}

In this section, we provide results for the first order effect sensitivity index as computed using approximations
from Sobol~\cite{Sobol:1990} and Saltelli~\emph{et al.}~\cite{Saltelli:2010} for the quantity,
$\V_{\theta_{i}}(\E_{\bm \theta_{\sim i}}[\mathcal{G}\vert\theta_{i}])$ as provided in Table~\ref{tab:estimators}.

\begin{figure}[htbp]
\begin{center}
\includegraphics[width=1.0\textwidth]{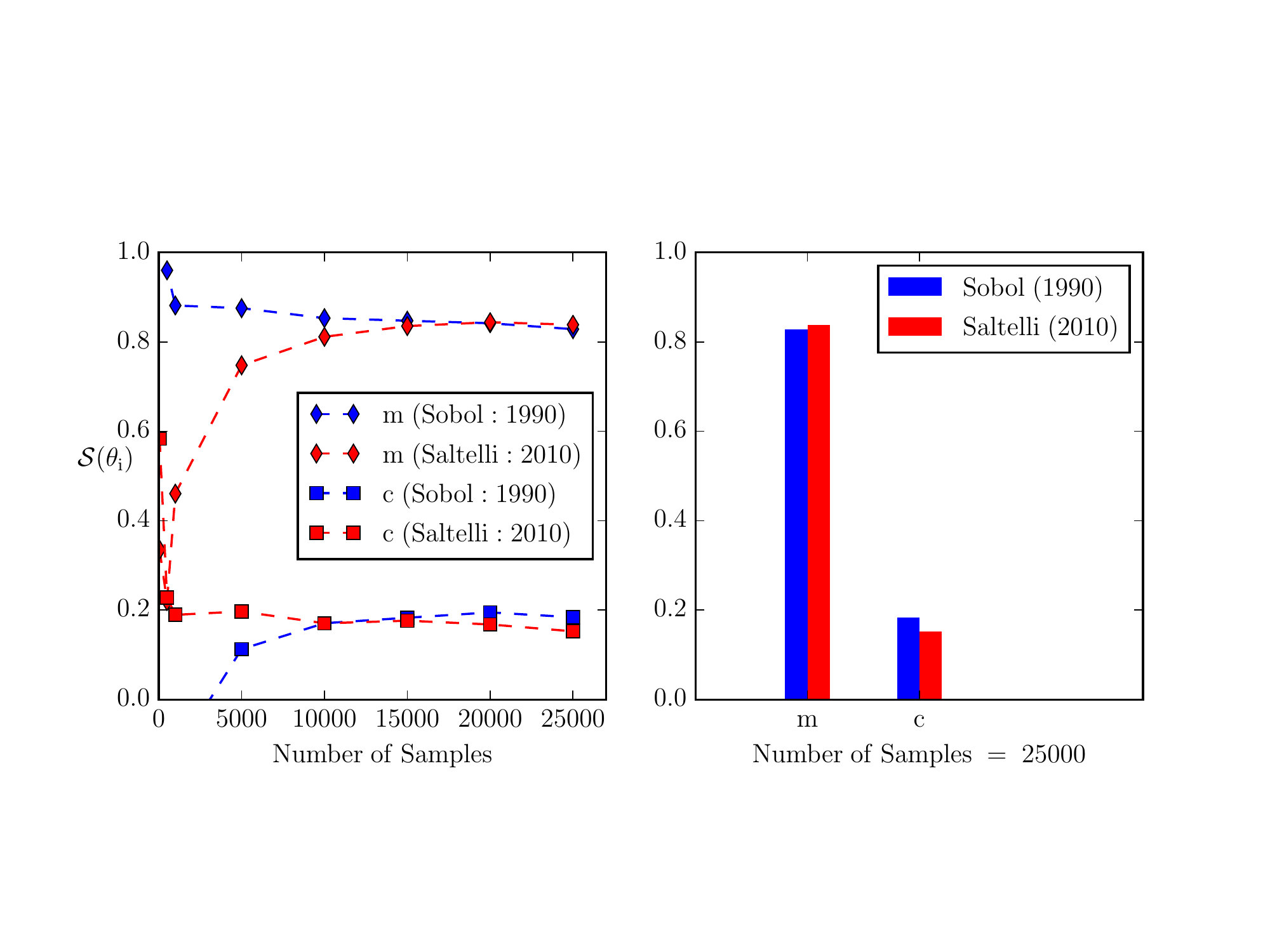}
\end{center}
\caption{(a) Analysis of convergence for the first order sensitivity indices for slope, $m$ and
y-intercept, $c$ with increasing sample size. (b) Bar-graph representation of estimates for $\mathcal{S}(\theta_{i})$
based on estimators suggested by Sobol~\cite{Sobol:1990} and Saltelli~\emph{et al.}~\cite{Saltelli:2010} using 25000
pseudo-random samples.}
\label{fig:sensitivity}
\end{figure}

In Figure~\ref{fig:sensitivity}(a), we perform a convergence study for the first order effect sensitivity index,
$\mathcal{S}(\theta_{i})$. It is observed that for a small number of samples ($<$ 5000), estimates from both,
Sobol and Saltelli estimators exhibit large oscillations with increase in sample size indicating that the
estimates have not yet converged to a stable value. Moreover, we observe large discrepancies are observed between
estimates obtained from the two estimators in this regime. However, as we increase the sample size above 10000, it
the two estimators seem to converge to stable values that are in close agreement. This phenomenon underscores the
need for a large enough sample size for computing the sensitivity indices using pseudo-random sampling techniques.
Optimizing the required number of samples in a way that the sensitivity indices are estimated within  reasonable
accuracy with the least possible sample size is a challenging task and depends on the map from the uncertain
model parameters to the quantity of interest.

Figure~\ref{fig:sensitivity}(b), illustrates estimates for $\mathcal{S}(\theta_{i})$ for the slope, $m$ and the
y-intercept, $c$ based on 25000 samples. For both parameters, estimates from Sobol~\cite{Sobol:1990} and
Saltelli~\emph{et al.}~\cite{Saltelli:2010} are in close agreement. Moreover, the QoI ($y$) is observed
to be much more sensitive to the uncertainty in the slope as compared to the y-intercept.

\section{Concluding Remarks}

Global Sensitivity Analysis can be a computationally challenging task especially  if it involves model
estimates for a complex multiphysics problem. However, in case the forward solve of the model is inexpensive, one
can exploit the SFP machinery in QUESO to compute the sensitivity indices as demonstrated with the
help of a simple example in this chapter. Moreover, when stochastic formulations are proposed to
capture the inadequacy in a model, parametric sensitivity analysis based on prior distributions of
the stochastic parameters can potentially reduce the dimensionality of an inverse problem.
Depending upon the nature of the problem, one can benefit from valuable insight into relative
importance of the parameters with much fewer samples than required for convergence of the estimates as
observed in the case of the straight line problem discussed in this chapter.


\chapter{QUESO Examples}\label{chap:Queso-examples}

This chapter assumes that the user has successfully installed QUESO and its dependencies.
It presents a variety of  examples of how to use QUESO in order to develop applications.

There are examples that solve a statistical inverse problem (Sections \ref{sec:example_sip}, \ref{sec:example_modal}, \ref{sec:example_bimodal} and \ref{sec:example_hysteretic}), a statistical forward problem (Section \ref{sec:example_sfp}) or a combination of both, where the solution of the former serves as input to the later (Sections \ref{sec:example_gravity} and \ref{sec:example_tga}). Three of the first four examples (Sections \ref{sec:example_sip}, \ref{sec:example_gravity} --\ref{sec:example_tga}) use the DRAM algorithm for solving the SIP, and the last three examples use the Multilevel algorithm. 
Each section presents: 
\begin{itemize}
 \item the mathematical models for the SIP and/or the SFP; \vspace*{-6pt}
 \item the application codes  that translate the mathematical language into C++ using the QUESO classes and algorithms;\vspace*{-6pt}
 \item  the input file that contains a list of options for either the Markov chain Monte Carlo algorithm or the Multilevel algorithm (in case of SIPs) and the Monte Carlo algorithm (in case of SFPs) which will be used by QUESO classes and algorithms;\vspace*{-6pt}
 \item examples of Makefiles which may be used to link the code  with QUESO library;\vspace*{-6pt}
 \item how to plot figures using Matlab/GNU Octave  and the output data generated by each application. 
\end{itemize}


All the examples presented in this chapter may be found under the directory \texttt{examples} in both \Queso{} installation and build directories and are compatible with QUESO \QUESOversion{}.

\paragraph*{Note:} Even though the Multilevel method is a methodology very useful for  stochastic system model class comparison (model updating, model selection, model validation)~\cite{Cheung_2009A}, such tasks are not discussed in this manual. Thus the explicit dependency of the statistical variables on the predictive model in the set  $M_j$ as presented in Section \ref{sec:ML:intro} are omitted herein.


\section{\texttt{simpleStatisticalInverseProblem}}\label{sec:example_sip}

According to the Bayesian paradigm, the unobservable parameters
in a statistical model are treated as random. When no data is available,
a prior distribution is used to quantify our knowledge about the parameter.
When data are available, we can update our prior knowledge using the conditional distribution of parameters, given the data. 
The transition from the prior to the posterior is possible via the Bayes theorem:
\begin{equation*}
\pi_{\text{posterior}}(\boldsymbol{\theta}|\mathbf{d})=\frac{\pi_{\text{prior}}(\boldsymbol{\theta})\pi_{\text{likelihood}}(\mathbf{d}|\boldsymbol{\theta})}{\pi(\mathbf{d})}
\end{equation*}

In this example, suppose a random variable of interest with two parameters $\bv{\theta} \in \mathbb{R}^2$ has a uniform prior distribution, and suppose that a suitable likelihood has normal distribution with mean $\bv{\mu}$ and covariance matrix $\bf{C}$, given by:
\begin{equation}\label{eq-example-mu}
\boldsymbol{\mu} = 
\left(\begin{array}{c}
-1 \\
2
\end{array}\right)
\quad
\text{and}
\quad
\mathbf{C} = 
\left[\begin{array}{cc}
4 & 0 \\
0 & 1
\end{array}\right].
\end{equation}

Therefore, we have: 
\begin{equation*}
\pi_{\text{prior}}(\boldsymbol{\theta}) \varpropto 1
\end{equation*}
and
\begin{equation*}
\pi_{\text{like}}(\boldsymbol{\theta}) \varpropto \exp \left(-\frac{1}{2}\left[(\boldsymbol{\theta}-\boldsymbol{\mu})^T[\mathbf{C}^{-1}](\boldsymbol{\theta}-\boldsymbol{\mu})\right] \right),
\end{equation*}
where
\begin{equation*}
\boldsymbol{\theta} = 
\left(
\begin{array}{c}
\theta_1 \\
\theta_2
\end{array}
\right)\in \mathbb{R}^2.
\end{equation*}

Therefore,  posterior PDF is given by:
\begin{equation}\label{eq-example-post}
\pi_{\text{post}}(\boldsymbol{\theta}) \varpropto e^{-\frac{1}{2}\left\{(\boldsymbol{\theta}-\boldsymbol{\mu})^T[\mathbf{C}^{-1}](\boldsymbol{\theta}-\boldsymbol{\mu})\right\}}.
\end{equation}

In this example, we can replace the values for the mean and covariance matrix given in Equation (\ref{eq-example-mu}) into Equation (\ref{eq-example-post}), 
in order to analytically compute both the posterior PDF:
\begin{eqnarray*}\label{eq-example-exact-post}
\pi_{\text{post}}(\boldsymbol{\theta}) & = & \frac{1}{4\pi} \exp\left(-\frac{1}{2}(\boldsymbol{\theta}-\boldsymbol{\mu})^T[\mathbf{C}^{-1}](\boldsymbol{\theta}-\boldsymbol{\mu})\right) \\
                                       & = & \frac{1}{4\pi} \exp\left( -\frac{1}{8}(\theta_1+1)^2 - \frac{1}{2}(\theta_2-2)^2\right), \label{eq-example-exact-joint}
\end{eqnarray*}
and the marginal results for $\theta_1$ and $\theta_2$:
\begin{equation}\label{eq-example-exact-marginal}
\begin{split}
\pi_{\text{post}}(\theta_1) & =  \frac{1}{2\sqrt{2\pi}} \exp\left(-\frac{1}{8}(\theta_1+1)^2 \right), \\
\pi_{\text{post}}(\theta_1) & =  \frac{1}{ \sqrt{2\pi}} \exp\left(-\frac{1}{2}(\theta_2-2)^2 \right). 
\end{split}
\end{equation}

Recall that the posterior PDF given in 
Equation (\ref{eq-example-post}) can be sampled through the expression:
\begin{equation}\label{eq-example-exact-normal}
\boldsymbol{\mu}+\mathbf{C}^{1/2}\mathcal{N}(0,I),
\end{equation}
where $\mathcal{N}(0,I)$ designates a Gaussian joint PDF of zero mean and unit covariance matrix, and
$\mathbf{C}^{1/2}$ is given by:
\begin{equation*}
\mathbf{C}^{1/2} = 
\left[\begin{array}{cc}
2 & 0 \\
0 & 1
\end{array}\right].
\end{equation*}

Thus, in this simple statistical inverse problem, we use QUESO implementation of the Markov chain 
algorithm to sample the posterior \eqref{eq-example-post} via Expression (\ref{eq-example-exact-normal}) and compare the calculated marginal results for $\theta_1$ and $\theta_2$ 
against the analytical formulas given in Equation~(\ref{eq-example-exact-marginal}).

\paragraph*{Note:} Due to the possibility to compare QUESO sampling algorithms to analytical expressions, this example is also used in the verification procedures and regression tests within QUESO, and it is reproduced in the directory \verb+tests/t02_sip_sfp+.

\subsection{Running the Example}\label{sec:sip-run}
 
To run the executable provided (available after QUESO installation), enter the following commands:
\begin{lstlisting}[label={},caption={}]
$ cd $HOME/LIBRARIES/QUESO-0.51.0/
$ cd examples/simpleStatisticalInverseProblem
$ rm outputData/*
$ ./exSimpleStatisticalInverseProblem_gsl example.inp    
$ matlab
   $ simple_ip_plots      # inside matlab
   $ exit                 # inside matlab
$ ls -l outputData/*.png
 simple_ip_autocorrelation_raw_filt.png  simple_ip_hist_filt.png 
 simple_ip_cdf_filt.png                  simple_ip_hist_raw.png
 simple_ip_cdf_raw.png                   simple_ip_kde_filt.png
 simple_ip_chain_pos_filt.png            simple_ip_kde_raw.png
\end{lstlisting}

As a result, the user should have created several of PNG figures containing marginal posterior PDF, chain positions, histograms, cumulative density distributions and autocorrelation of both parameters. The name of the figure files have been chosen to be informative, as shown in the Listing above. 

It is worth noting presence of an argument passed to the executable in the
example, namely `\verb+example.inp+'. The argument is a input file to be provided to QUESO with options for
the solution of the SIP and/or SFP; and it is always required. Each option in
the input file is related to one (or more) of the QUESO classes, and is
presented throughout Chapter~\ref{ch-classes}.

\subsection{Example Code}\label{sec:sip-code}

The source code for the example is composed of 5 files:
\texttt{example\_main.C} (Listing \ref{code:sip-main-c}), \linebreak
\texttt{example\_likelihood.h} and \texttt{example\_likelihood.C} (Listings \ref{fig-like-h} and \ref{fig-like-c}),
\texttt{example\_compute.h} and \texttt{example\_compute.C} (Listings \ref{code:sip-compute-h} and \ref{code:sip-compute-c}).

\begin{lstlisting}[caption=File \texttt{example\_main.C.}, label={code:sip-main-c}, linerange={25-1000}]
#include <example_compute.h>

int main(int argc, char* argv[])
{
  // Initialize environment
#ifdef QUESO_HAS_MPI
  MPI_Init(&argc,&argv);

  UQ_FATAL_TEST_MACRO(argc != 2,
                      QUESO::UQ_UNAVAILABLE_RANK,
                      "main()",
                      "input file must be specified in command line as argv[1], just after executable argv[0]");

  QUESO::FullEnvironment* env =  new QUESO::FullEnvironment(MPI_COMM_WORLD,argv[1],"",NULL);
#else
  QUESO::FullEnvironment* env =  new QUESO::FullEnvironment(argv[1],"",NULL);
#endif

  // Compute
  compute(*env);

  // Finalize environment
  delete env;
#ifdef QUESO_HAS_MPI
  MPI_Finalize();
#endif

  return 0;
}
\end{lstlisting}

\begin{lstlisting}[caption=File \texttt{example\_likelihood.h}., label={fig-like-h}, linerange={25-1000}]
#ifndef EX_LIKELIHOOD_H
#define EX_LIKELIHOOD_H

#include <queso/GslMatrix.h>

struct
likelihoodRoutine_DataType
{
  const QUESO::GslVector* meanVector;
  const QUESO::GslMatrix* covMatrix;
};

double likelihoodRoutine(
  const QUESO::GslVector& paramValues,
  const QUESO::GslVector* paramDirection,
  const void*             functionDataPtr,
  QUESO::GslVector*       gradVector,
  QUESO::GslMatrix*       hessianMatrix,
  QUESO::GslVector*       hessianEffect);

#endif
\end{lstlisting}


\begin{lstlisting}[caption=File \texttt{example\_likelihood.C}., label={fig-like-c}, linerange={25-1000}]
#include <example_likelihood.h>

double likelihoodRoutine(
  const QUESO::GslVector& paramValues,
  const QUESO::GslVector* paramDirection,
  const void*             functionDataPtr,
  QUESO::GslVector*       gradVector,
  QUESO::GslMatrix*       hessianMatrix,
  QUESO::GslVector*       hessianEffect)
{
  // Logic just to avoid warnings from INTEL compiler
  const QUESO::GslVector* aux1 = paramDirection;
  if (aux1) {};
  aux1 = gradVector;
  aux1 = hessianEffect;
  QUESO::GslMatrix* aux2 = hessianMatrix;
  if (aux2) {};

  // Just checking: the user, at the application level, expects
  // vector 'paramValues' to have size 2.
  UQ_FATAL_TEST_MACRO(paramValues.sizeGlobal() != 2,
                      QUESO::UQ_UNAVAILABLE_RANK,
                      "likelihoodRoutine()",
                      "paramValues vector does not have size 2");

  // Actual code
  //
  // This code exemplifies multiple Metropolis-Hastings solvers, each calling this likelihood
  // routine. In this simple example, only node 0 in each subenvironment does the job even
  // though there might be more than one node per sub-environment. In a more realistic
  // situation, if the user is asking for multiple nodes per subenvironment, then the model
  // code in the likelihood routines might really demand more than one node. Here we use
  // 'env.subRank()' only. A realistic application might want to use either 'env.subComm()'
  // or 'env.subComm().Comm()'.

  double result = 0.;
  const QUESO::BaseEnvironment& env = paramValues.env();
  if (env.subRank() == 0) {
    const QUESO::GslVector& meanVector =
      *((likelihoodRoutine_DataType *) functionDataPtr)->meanVector;
    const QUESO::GslMatrix& covMatrix  =
      *((likelihoodRoutine_DataType *) functionDataPtr)->covMatrix;

    QUESO::GslVector diffVec(paramValues - meanVector);

    result= scalarProduct(diffVec,covMatrix.invertMultiply(diffVec));
  }
  else {
    // Do nothing;
  }

  return -.5*result;
}
\end{lstlisting}


\begin{lstlisting}[caption=File \texttt{example\_compute.h.}, label={code:sip-compute-h}, linerange={25-1000}]
#ifndef EX_COMPUTE_H
#define EX_COMPUTE_H

#include <queso/Environment.h>

void compute(const QUESO::FullEnvironment& env);

#endif
\end{lstlisting}

\begin{lstlisting}[caption={File \texttt{example\_compute.C}.}, label={code:sip-compute-c}, linerange={25-1000},numbers=left]
#include <example_compute.h>
#include <example_likelihood.h>
#include <queso/GenericScalarFunction.h>
#include <queso/GslMatrix.h>
#include <queso/UniformVectorRV.h>
#include <queso/StatisticalInverseProblem.h>

void compute(const QUESO::FullEnvironment& env) {
  // Step 1 of 5: Instantiate the parameter space
  QUESO::VectorSpace<QUESO::GslVector,QUESO::GslMatrix>
    paramSpace(env, "param_", 2, NULL);

  // Step 2 of 5: Instantiate the parameter domain
  QUESO::GslVector paramMins(paramSpace.zeroVector());
  paramMins.cwSet(-INFINITY);
  QUESO::GslVector paramMaxs(paramSpace.zeroVector());
  paramMaxs.cwSet( INFINITY);
  QUESO::BoxSubset<QUESO::GslVector,QUESO::GslMatrix>
    paramDomain("param_",paramSpace,paramMins,paramMaxs);

  // Step 3 of 5: Instantiate the likelihood function object
  QUESO::GslVector meanVector(paramSpace.zeroVector());
  meanVector[0] = -1;
  meanVector[1] =  2;

  QUESO::GslMatrix covMatrix(paramSpace.zeroVector());
  covMatrix(0,0) = 4.; covMatrix(0,1) = 0.;
  covMatrix(1,0) = 0.; covMatrix(1,1) = 1.;

  likelihoodRoutine_DataType likelihoodRoutine_Data;
  likelihoodRoutine_Data.meanVector = &meanVector;
  likelihoodRoutine_Data.covMatrix  = &covMatrix;

  QUESO::GenericScalarFunction<QUESO::GslVector,QUESO::GslMatrix>
    likelihoodFunctionObj("like_",
                          paramDomain,
                          likelihoodRoutine,
                          (void *) &likelihoodRoutine_Data,
                          true); // routine computes [ln(function)]

  // Step 4 of 5: Instantiate the inverse problem
  QUESO::UniformVectorRV<QUESO::GslVector,QUESO::GslMatrix>
    priorRv("prior_", paramDomain);
  QUESO::GenericVectorRV<QUESO::GslVector,QUESO::GslMatrix>
    postRv("post_", paramSpace);
  QUESO::StatisticalInverseProblem<QUESO::GslVector,QUESO::GslMatrix>
    ip("", NULL, priorRv, likelihoodFunctionObj, postRv);

  // Step 5 of 5: Solve the inverse problem
  QUESO::GslVector paramInitials(paramSpace.zeroVector());
  paramInitials[0] = 0.1;
  paramInitials[1] = -1.4;

  QUESO::GslMatrix proposalCovMatrix(paramSpace.zeroVector());
  proposalCovMatrix(0,0) = 8.; proposalCovMatrix(0,1) = 4.;
  proposalCovMatrix(1,0) = 4.; proposalCovMatrix(1,1) = 16.;

  ip.solveWithBayesMetropolisHastings(NULL,paramInitials, &proposalCovMatrix);

  return;
}
\end{lstlisting}

\subsection{Input File}\label{sec:sip-input-file}

QUESO reads an input file for solving statistical problems. In the case of a SIP, it expects a list of options for MCMC (or Multilevel),
together with options for QUESO environment; such as the amount of processors to be used and the seed for its random algorithms.
Note that the names of the variables have been designed to be informative:
\begin{description}\vspace{-8pt}
\item[ \texttt{env}:] refers to QUESO environment; \vspace{-8pt}
\item[ \texttt{ip}:] refers to inverse problem;\vspace{-8pt}
\item[ \texttt{mh}:] refers to Metropolis-Hastings;\vspace{-8pt}
\item[ \texttt{dr}:] refers to delayed rejection;\vspace{-8pt}
\item[ \texttt{am}:] refers to adaptive Metropolis;\vspace{-8pt}
\item[ \texttt{rawChain}:] refers to the raw, entire chain; \vspace{-8pt}
\item[ \texttt{filteredChain}:] refers to a filtered chain (related to a specified \texttt{lag});\vspace{-8pt}
\end{description}

The options used for solving this simple SIP are displayed in Listing \ref{code:sip-input-file}.

\begin{lstlisting}[caption={Options for QUESO library used in application code (Listings \ref{code:sip-main-c}-\ref{code:sip-compute-c}})., 
label={code:sip-input-file},]
###############################################
# UQ Environment
###############################################
#env_help                = anything
env_numSubEnvironments   = 1
env_subDisplayFileName   = outputData/display
env_subDisplayAllowAll   = 0
env_subDisplayAllowedSet = 0
env_displayVerbosity     = 2
env_syncVerbosity        = 0
env_seed                 = 0

###############################################
# Statistical inverse problem (ip)
###############################################
#ip_help                 = anything
ip_computeSolution      = 1
ip_dataOutputFileName   = outputData/sipOutput
ip_dataOutputAllowedSet = 0

###############################################
# 'ip_': information for Metropolis-Hastings algorithm
###############################################
#ip_mh_help                 = anything
ip_mh_dataOutputFileName   = outputData/sipOutput
ip_mh_dataOutputAllowedSet = 0 1

ip_mh_rawChain_dataInputFileName    = .
ip_mh_rawChain_size                 = 32768
ip_mh_rawChain_generateExtra        = 0
ip_mh_rawChain_displayPeriod        = 50000
ip_mh_rawChain_measureRunTimes      = 1
ip_mh_rawChain_dataOutputFileName   = outputData/ip_raw_chain
ip_mh_rawChain_dataOutputAllowedSet = 0 1
ip_mh_rawChain_computeStats         = 1

ip_mh_displayCandidates             = 0
ip_mh_putOutOfBoundsInChain         = 1
ip_mh_tk_useLocalHessian            = 0
ip_mh_tk_useNewtonComponent         = 1
ip_mh_dr_maxNumExtraStages          = 1
ip_mh_dr_listOfScalesForExtraStages = 5.
ip_mh_am_initialNonAdaptInterval    = 0
ip_mh_am_adaptInterval              = 100
ip_mh_am_eta                        = 1.92
ip_mh_am_epsilon                    = 1.e-5

ip_mh_filteredChain_generate             = 1
ip_mh_filteredChain_discardedPortion     = 0.
ip_mh_filteredChain_lag                  = 16
ip_mh_filteredChain_dataOutputFileName   = outputData/ip_filt_chain
ip_mh_filteredChain_dataOutputAllowedSet = 0 1

\end{lstlisting}

\subsection{Create your own Makefile}\label{sec:sip-makefile}

Makefiles are special format files that together with the make utility will help one to compile and automatically build and manage projects (programs).  
Listing \ref{code:ip_makefile} presents a Makefile, named `\texttt{Makefile\_sip\_example\_margarida}', that may be used to compile the code and create the executable \verb+simple_sip_example+. Naturally, it must be adapted to the user's settings, i.e., it has to have the correct paths for the user's libraries that have actually been used to compile and install QUESO  (see Sections \ref{sec:Pre_Queso}--\ref{sec:install_Queso_make}).

\begin{lstlisting}[caption={Makefile for the application code in Listings
  \ref{code:sip-main-c}-\ref{code:sip-compute-c}},
  language=bash,
  label={code:ip_makefile}]
  QUESO_DIR = /path/to/queso
  BOOST_DIR = /path/to/boost
  GSL_DIR   = /path/to/gsl

  INC_PATHS = \
     -I. \
     -I$(QUESO_DIR)/include \
     -I$(BOOST_DIR)/include \
     -I$(GSL_DIR)/include

  LIBS = \
     -L$(QUESO_DIR)/lib -lqueso \
     -L$(BOOST_DIR)/lib -lboost_program_options \
     -L$(GSL_DIR)/lib -lgsl

  CXX = mpic++
  CXXFLAGS += -g -Wall -c

  default: all

  .SUFFIXES: .o .C

  all:       example_sip

  clean:
     rm -f *~
     rm -f *.o
     rm -f simple_sip_example

  example_sip: example_main.o example_likelihood.o example_compute.o
     $(CXX) example_main.o \
            example_likelihood.o \
            example_compute.o \
            -o simple_sip_example $(LIBS)

  %.o: %.C
     $(CXX) $(INC_PATHS) $(CXXFLAGS) $<
\end{lstlisting}

Thus, to compile, build and execute the code, the user just needs to run the following commands in the same directory where the files are:
\begin{lstlisting}
$ cd $HOME/LIBRARIES/QUESO-0.51.0/examples/simpleStatisticalInverseProblem/
$ export LD_LIBRARY_PATH=$LD_LIBRARY_PATH:\
  $HOME/LIBRARIES/gsl-1.15/lib/:\
  $HOME/LIBRARIES/boost-1.53.0/lib/:\
  $HOME/LIBRARIES/hdf5-1.8.10/lib:\
  $HOME/LIBRARIES/QUESO-0.51.0/lib 
$ make -f Makefile_example_margarida 
$ ./simple_sip_example example.inp
\end{lstlisting}

The `\verb+export+' instruction above is only necessary if the user has not saved it in his/her \verb+.bashrc+ file.

\subsection{Data Post-Processing and Visualization}\label{sec:sip-results}

There are a few Matlab-ready commands that are very helpful tools for post-processing the data generated by QUESO when solving statistical inverse problems. This section discusses the results computed by QUESO with the code of Section \ref{sec:sip-code}, and shows how to use Matlab for the post-processing of such results. Only the essential Matlab commands are presented; for the complete/detailed codes, please refer to file '\verb+simple_ip_plots.m+'.

According to the specifications of the input file in Listing~\ref{code:sip-input-file}, a folder named `\verb+outputData+' containing the following files should be created: \verb+display_sub0.txt, ip_filt_chain_sub0.m,+ \verb+ip_raw_chain_sub0.m, sipOutput_sub0.m, ip_filt_chain.m, ip_raw_chain.m+

The code bellow shows how to load the data provided by QUESO during the solution process of the SIP described, in the form of 
chains of positions.

\begin{lstlisting}[caption={Matlab code for loading the data in both raw and filtered chains of the SIP, by calling the file \texttt{simple\_ip\_plots.m}.}]
% inside Matlab
>> clear all
>> simple_ip_plots
\end{lstlisting}

\subsubsection{Autocorrelation Plots}

\begin{figure}[htpb]
\centering
\includegraphics[scale=0.35]{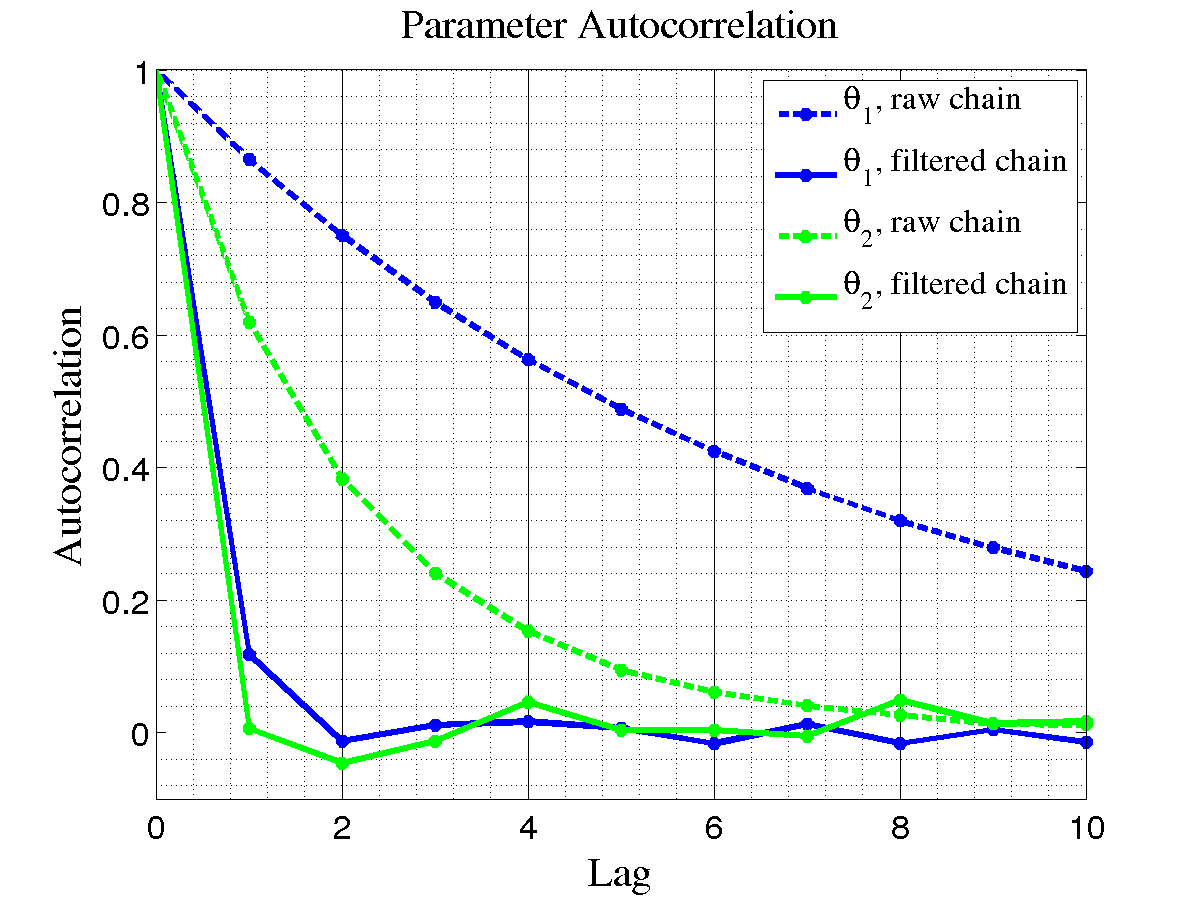}
\vspace{-10pt}
\caption{
Autocorrelation plots obtained with QUESO for the SIP. }
\label{fig:simple_sip_autocorrelation_raw_filt}
\end{figure}

The code presented in Listing \ref{matlab:simple_sip_autocorr} uses Matlab function \verb+autocorr+ to generate Figure \ref{fig:simple_sip_autocorrelation_raw_filt}
which presents the autocorrelation of the parameters $\theta_1$ and $\theta_2$ in both cases: raw and filtered chain. 

\begin{lstlisting}[label=matlab:simple_sip_autocorr,caption={Matlab code for the autocorrelation plots depicted in Figure \ref{fig:simple_sip_autocorrelation_raw_filt}.}]
% inside Matlab
% theta_1
>> nlags=10;
>> [ACF_raw, lags] = autocorr(ip_mh_rawChain_unified(:,1), nlags, 0);
>> [ACF_filt, lags] = autocorr(ip_mh_filtChain_unified(:,1), nlags, 0);
>> [ACF_raw2, lags2] = autocorr(ip_mh_rawChain_unified(:,2), nlags, 0);
>> [ACF_filt2, lags3] = autocorr(ip_mh_filtChain_unified(:,2), nlags, 0);
>> plot(lags,ACF_raw,'b--*',lags,ACF_filt,'b*-',lags2,ACF_raw2,'g--*',lags2,ACF_filt2,'g*-','linewidth',3);
>> h=legend('\theta_1, raw chain','\theta_1, filtered chain','\theta_2, raw chain','\theta_2, filtered chain','location','northeast');
\end{lstlisting}

\subsubsection{KDE Plots}

Matlab function \verb+[f,xi] = ksdensity(x)+ (kernel smoothing density estimate) computes a probability density estimate of the sample in the vector \texttt{x}. \texttt{f} is the vector of density values evaluated at the points in \texttt{xi}. The estimate is based on a normal kernel function, using a window parameter (`width') that is a function of the number of points in \texttt{x}. The density is evaluated at 100 equally spaced points that cover the range of the data in x.  In order to estimate the KDE of the parameters, it is used together with the option `\verb+pdf+'. 

\begin{lstlisting}[label=matlab:ip_kde,caption={Matlab code for the KDE plots displayed in the left of Figure \ref{fig:simple_sip_kde}.}]
% Inside Matlab
% Raw chain
>> [f,x] = ksdensity(ip_mh_rawChain_unified(:,1),'function','pdf');
>> [f2,x2] = ksdensity(ip_mh_rawChain_unified(:,2),'function','pdf');
>> x_p1=sort(ip_mh_rawChain_unified(:,1)); %analytical
>> f_p1=(exp(-(x_p1+1).*(x_p1+1)/8))/2/sqrt(2*pi);
>> x_p2=sort(ip_mh_rawChain_unified(:,1));
>> f_p2=(exp(-(x_p2-2).*(x_p2-2)/2))/sqrt(2*pi);
>> plot(x,f,'b',x2,f2,'g','linewidth',4);
>> hold;
>> plot(x_p1,f_p1,'--k',x_p2,f_p2,'-k','linewidth',2);
>> h=legend('\theta_1', '\theta_2', 'analytical (\theta_1)', 'analytical (\theta_2)', 'location', 'northwest');
\end{lstlisting}

%

\begin{figure}[htpb]
\centering 
\subfloat[Raw chain]{\includegraphics[scale=0.35]{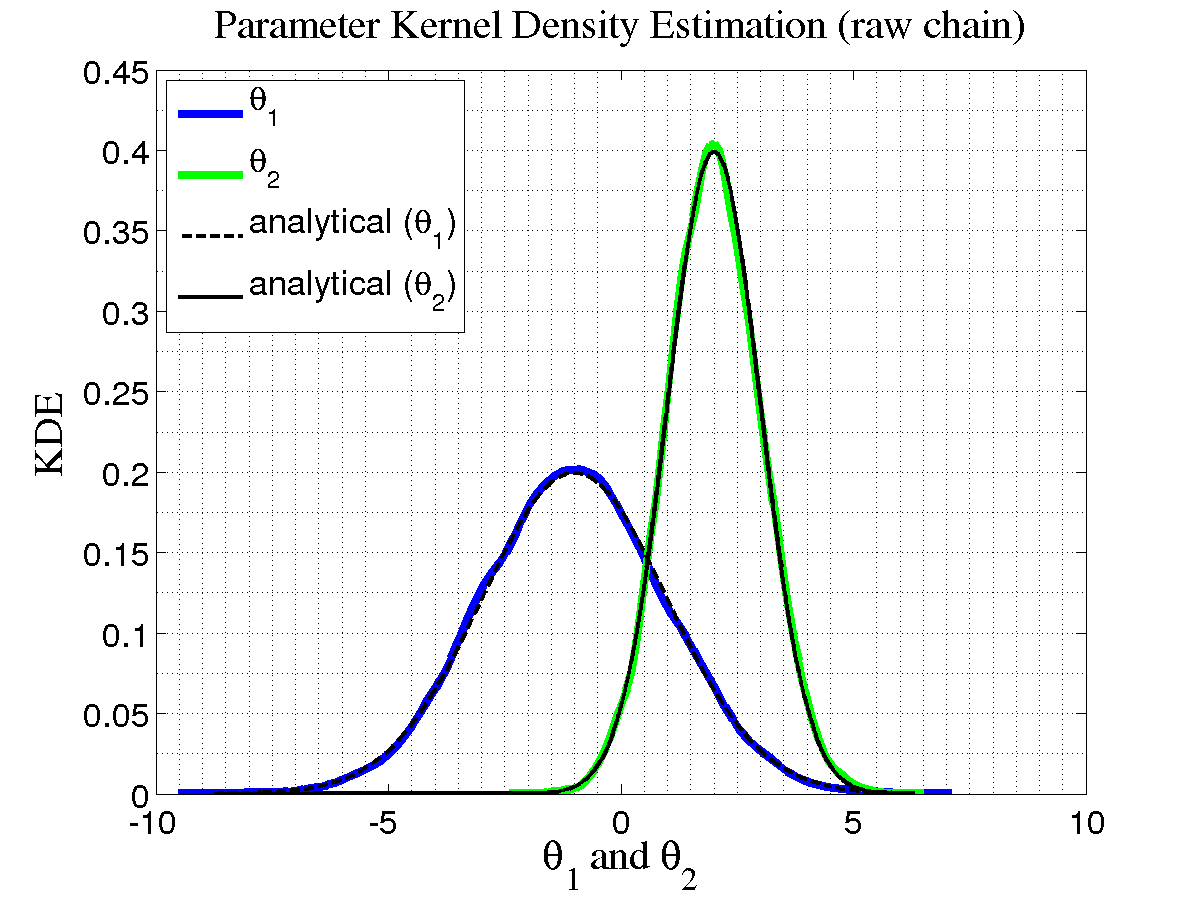}}
\subfloat[Filtered chain]{\includegraphics[scale=0.35]{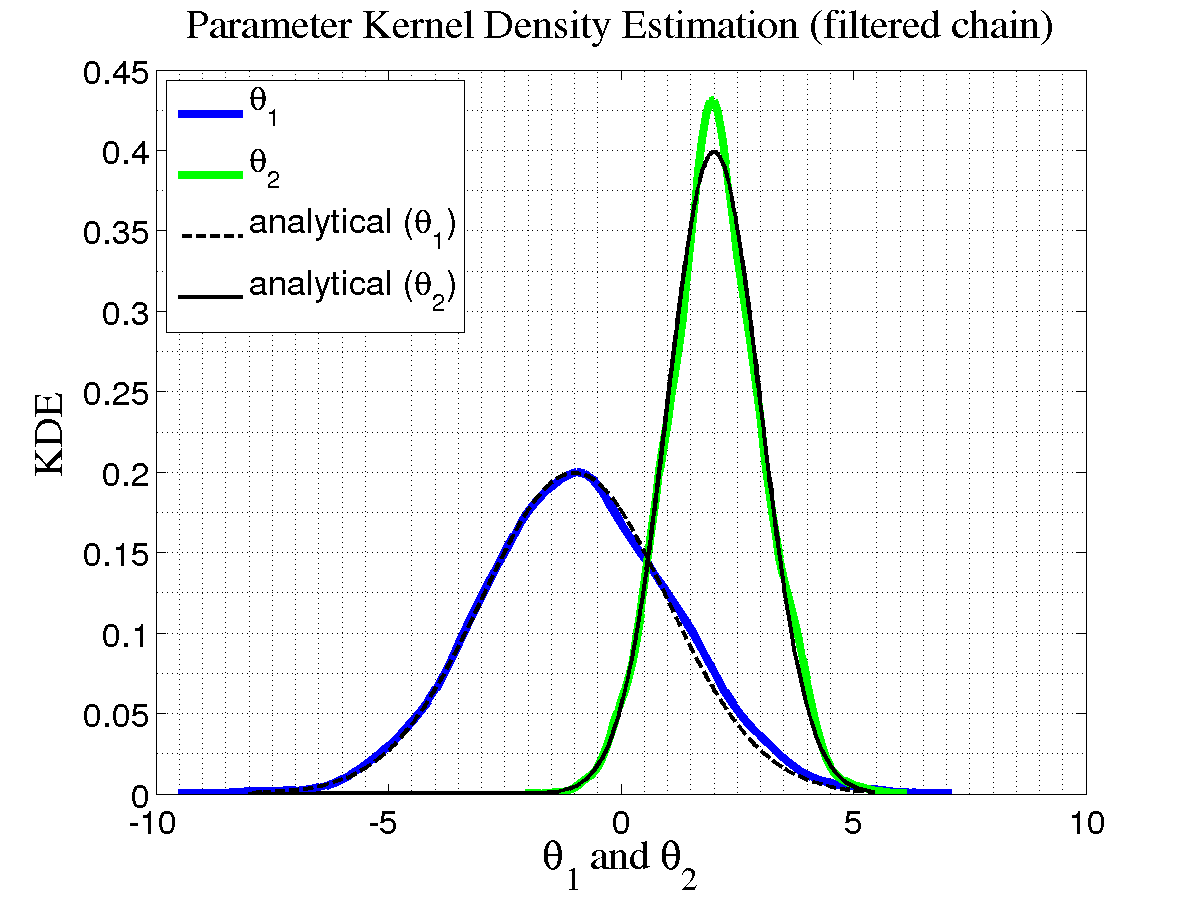}}
\vspace*{-10pt}
\caption{Kernel Density Estimation. QUESO results for estimation of the KDE of $\theta_1$ and $\theta_2$ are plotted against the analytical expressions $\pi_{\text{post}}(\theta_1)  =  \frac{1}{2\sqrt{2\pi}} \exp\left(-\frac{1}{8}(\theta_1+1)^2 \right)$  and $\pi_{\text{post}}(\theta_2)  =  \frac{1}{ \sqrt{2\pi}} \exp\left(-\frac{1}{2}(\theta_2-2)^2 \right)$, respectively.}
\label{fig:simple_sip_kde}
\end{figure}

\subsubsection{Covariance and Correlation Matrices}

Matlab function \verb+cov+ calculates the covariance matrix for a data matrix (where each column represents a separate quantity), 
and \verb+corr+ calculates the correlation matrix.

Listing \ref{matlab:cov_matrix} presents the Matlab steps for calculating the covariance and correlation matrices for the parameters $\theta_1$ and $\theta_2$.


\begin{lstlisting}[label=matlab:ip_cov_matrix,caption={Matlab code for finding covariance and correlation matrices.}]
% inside Matlab
>> cov_matrix_theta1_theta2 = cov(ip_mh_rawChain_unified)

cov_matrix_theta1_theta2 =

    3.8729    0.0259
    0.0259    1.0050
    
>> corr_matrix_theta1_theta2 = corr(ip_mh_rawChain_unified)

corr_matrix_theta1_theta2 =

    1.0000    0.0132
    0.0132    1.0000    
\end{lstlisting}


\section{\texttt{simpleStatisticalForwardProblem}}\label{sec:example_sfp}

In this simple statistical forward problem (SFP), suppose that the quantity of interest $\mathbf{q}$ is a function of a random variable $\bv{\theta}$ of two parameters, namely $\bf{q}:\mathbb{R}^2\rightarrow\mathbb{R}$ such as:
\begin{equation}\label{eq-example-q}
\mathbf{q}(\boldsymbol{\theta}) = \theta_1+\theta_2,\quad\forall\boldsymbol{\theta}=(\theta_1,\theta_2)\in\mathbb{R}^2.
\end{equation}

Suppose also that the parameters in $\theta$ have Gaussian distribution with mean $\bv{\mu}$ and covariance matrix $\bf{C}$ given by:
\begin{equation}\label{eq-example-mu-sfp}
\boldsymbol{\mu} = 
\left(\begin{array}{c}
-1 \\
2
\end{array}\right)
\quad
\text{and}
\quad
\mathbf{C} = 
\left[\begin{array}{cc}
4 & 0 \\
0 & 1
\end{array}\right].
\end{equation}

Notice that since the solution $\mathbf{Q}$ of this SFP is the sum of two random variables $\boldsymbol{\Theta}_1$ and $\boldsymbol{\Theta}_2$, and since these two random variables independent Gaussian by assumption, should have:
\begin{equation}\label{eq-example-E-V}
E[\mathbf{Q}] = E[\boldsymbol{\Theta}_1] + E[\boldsymbol{\Theta}_2] = -1 + 2 = 1 \quad \text{and} \quad
V[\mathbf{Q}] = V[\boldsymbol{\Theta}_1] + V[\boldsymbol{\Theta}_2] = 4 + 1 = 5
\end{equation}
where $E$ and $V$ indicate expectation and variance, respectively. Thus the analytical expression for the solution $\bf{Q}$ is this SFP is the one-dimensional Gaussian distribution of mean 1 and variance 5:
\begin{equation}\label{eq-example-sfp-analytical}
{\bf Q}(x)=   \frac{1}{ \sqrt{10\pi}} \exp\left(-\frac{1}{10}(x-1)^2 \right)
\end{equation}

In this example, we use QUESO Monte Carlo algorithm to sample from the QoI given in Equation (\ref{eq-example-q}) and analyze it. 
Since the parameters have known independent Gaussian distributions, the results obtained by QUESO via sampling the QoI, in Equation (\ref{eq-example-q}), should match the Gaussian distribution given in Equation (\ref{eq-example-sfp-analytical}).

\paragraph*{Note:} Due to the possibility to compare QUESO sampling algorithms to an analytical expression, this example is also used in the verification procedures and regression tests within QUESO. In fact it is the second part of the test \verb+tests/t02_sip_sfp+.

\subsection{Running the Example}\label{sec:sfp-run}
 
To run the executable provided (available after QUESO installation), enter the following commands:
\begin{lstlisting}[label={},caption={}]
$ cd $HOME/LIBRARIES/QUESO-0.51.0/
$ cd examples/simpleStatisticalForwardProblem
$ rm outputData/*
$ ./exSimpleStatisticalForwardProblem_gsl example.inp    
$ matlab
   $ simple_fp_plots      # inside matlab
   $ exit                 # inside matlab
$ ls -l outputData/*.png
 simple_fp_autocorrelation_qoi.png  simple_fp_chain_pos_param.png  
 simple_fp_hist_qoi.png             simple_fp_cdf_qoi.png
 simple_fp_chain_pos_qoi.png        simple_fp_kde_qoi.png
\end{lstlisting}

As a result, the user should have created several of PNG figures containing marginal posterior PDF, chain positions of the parameters and the QoI, histogram, cumulative density distribution and autocorrelation. The name of the figure files have been chosen to be informative, as shown in the Listing above.

\subsection{Example Code}\label{sec:code-sfp}

The source code for the SFP example is composed of 5 files:
\texttt{simple\_sfp\_example\_main.C} (Listing~\ref{code:sfp-main-c}),
\texttt{simple\_sfp\_example\_qoi.h} and \texttt{simple\_sfp\_example\_qoi.C} (Listings \ref{code:sfp-qoi-h} and~\ref{code:sfp-qoi-c}),
\texttt{simple\_sfp\_example\_compute.h}  and \texttt{simple\_sfp\_example\_compute.C} (Listings \ref{code:sfp-compute-h} and \ref{code:sfp-compute-c}).

\begin{lstlisting}[caption=File \texttt{simple\_sfp\_example\_main.C.}, label={code:sfp-main-c}, linerange={25-1000}]
#include <simple_sfp_example_compute.h>

int main(int argc, char* argv[])
{
  // Initialize environment
#ifdef QUESO_HAS_MPI
  MPI_Init(&argc,&argv);

  UQ_FATAL_TEST_MACRO(argc != 2, QUESO::UQ_UNAVAILABLE_RANK, "main()",
                      "input file must be specified in command line as argv[1], just after executable argv[0]");
  QUESO::FullEnvironment* env =
    new QUESO::FullEnvironment(MPI_COMM_WORLD,argv[1],"",NULL);
#else
  QUESO::FullEnvironment* env =
    new QUESO::FullEnvironment(argv[1],"",NULL);
#endif

  // Compute
  compute(*env);

  // Finalize environment
  delete env;
#ifdef QUESO_HAS_MPI
  MPI_Finalize();
#endif

  return 0;
}
\end{lstlisting}

\begin{lstlisting}[caption=File \texttt{simple\_sfp\_example\_qoi.h}., label={code:sfp-qoi-h}, linerange={25-1000}]
#ifndef __EX_QOI_H__
#define __EX_QOI_H__

#include <queso/GslMatrix.h>
#include <queso/DistArray.h>

struct
qoiRoutine_DataType
{
  double coef1;
  double coef2;
};

void
qoiRoutine(
  const QUESO::GslVector&                    paramValues,
  const QUESO::GslVector*                    paramDirection,
  const void*                                functionDataPtr,
        QUESO::GslVector&                    qoiValues,
        QUESO::DistArray<QUESO::GslVector*>* gradVectors,
        QUESO::DistArray<QUESO::GslMatrix*>* hessianMatrices,
        QUESO::DistArray<QUESO::GslVector*>* hessianEffects);

#endif
\end{lstlisting}

\begin{lstlisting}[caption=File \texttt{simple\_sfp\_example\_qoi.C}., label={code:sfp-qoi-c}, linerange={25-1000}]
#include <simple_sfp_example_qoi.h>

void
qoiRoutine(
  const QUESO::GslVector&                    paramValues,
  const QUESO::GslVector*                    paramDirection,
  const void*                                functionDataPtr,
        QUESO::GslVector&                    qoiValues,
        QUESO::DistArray<QUESO::GslVector*>* gradVectors,
        QUESO::DistArray<QUESO::GslMatrix*>* hessianMatrices,
        QUESO::DistArray<QUESO::GslVector*>* hessianEffects)
{
  // Logic just to avoid warnings from INTEL compiler
  const QUESO::GslVector* aux1 = paramDirection;
  if (aux1) {};
  QUESO::DistArray<QUESO::GslVector*>* aux2 = gradVectors;
  if (aux2) {};
  aux2 = hessianEffects;
  QUESO::DistArray<QUESO::GslMatrix*>* aux3 = hessianMatrices;
  if (aux3) {};

  // Just checking: the user, at the application level, expects
  // vector 'paramValues' to have size 2 and
  // vector 'qoiValues' to have size 1.
  UQ_FATAL_TEST_MACRO(paramValues.sizeGlobal() != 2,
                      QUESO::UQ_UNAVAILABLE_RANK,
                      "qoiRoutine()",
                      "paramValues vector does not have size 2");

  UQ_FATAL_TEST_MACRO(qoiValues.sizeGlobal() != 1,
                      QUESO::UQ_UNAVAILABLE_RANK,
                      "qoiRoutine()",
                      "qoiValues vector does not have size 1");

  // Actual code
  //
  // This code exemplifies multiple Monte Carlo solvers, each calling this qoi routine.
  // In this simple example, only node 0 in each sub-environment does the job even though
  // there might be more than one node per sub-environment.
  // In a more realistic situation, if the user is asking for multiple nodes per sub-
  // environment, then the model code in the qoi routine might really demand more than one
  // node. Here we use 'env.subRank()' only. A realistic application might want to use
  // either 'env.subComm()' or 'env.subComm().Comm()'.

  const QUESO::BaseEnvironment& env = paramValues.env();
  if (env.subRank() == 0) {
    double coef1 = ((qoiRoutine_DataType *) functionDataPtr)->coef1;
    double coef2 = ((qoiRoutine_DataType *) functionDataPtr)->coef2;
    qoiValues[0] = (coef1*paramValues[0] + coef2*paramValues[1]);
  }
  else {
    qoiValues[0] = 0.;
  }

  return;
}
\end{lstlisting}

\begin{lstlisting}[caption=File \texttt{simple\_sfp\_example\_compute.h.}, label={code:sfp-compute-h}, linerange={25-1000}]
#ifndef __EX_COMPUTE_H__
#define __EX_COMPUTE_H__

#include <queso/Environment.h>

void compute(const QUESO::FullEnvironment& env);

#endif
\end{lstlisting}

\begin{lstlisting}[caption={File \texttt{simple\_sfp\_example\_compute.C}.}, label={code:sfp-compute-c}, linerange={25-1000},numbers=left]
#include <simple_sfp_example_compute.h>
#include <simple_sfp_example_qoi.h>
#include <queso/GslMatrix.h>
#include <queso/StatisticalForwardProblem.h>
#include <queso/GenericVectorFunction.h>
#include <queso/GaussianVectorRV.h>

void compute(const QUESO::FullEnvironment& env) {

  // Step 1 of 6: Instantiate the parameter space
  QUESO::VectorSpace<QUESO::GslVector,QUESO::GslMatrix>
    paramSpace(env, "param_", 2, NULL);

  // Step 2 of 6: Instantiate the parameter domain
  QUESO::GslVector paramMins(paramSpace.zeroVector());
  paramMins.cwSet(-INFINITY);
  QUESO::GslVector paramMaxs(paramSpace.zeroVector());
  paramMaxs.cwSet( INFINITY);
  QUESO::BoxSubset<QUESO::GslVector,QUESO::GslMatrix>
    paramDomain("param_",paramSpace,paramMins,paramMaxs);

  // Step 3 of 6: Instantiate the qoi space
  QUESO::VectorSpace<QUESO::GslVector,QUESO::GslMatrix>
    qoiSpace(env, "qoi_", 1, NULL);

  // Step 4 of 6: Instantiate the qoi function object
  qoiRoutine_DataType qoiRoutine_Data;
  qoiRoutine_Data.coef1 = 1.;
  qoiRoutine_Data.coef2 = 1.;
  QUESO::GenericVectorFunction<QUESO::GslVector,QUESO::GslMatrix,
                               QUESO::GslVector,QUESO::GslMatrix>
    qoiFunctionObj("qoi_",
                   paramDomain,
                   qoiSpace,
                   qoiRoutine,
                   (void *) &qoiRoutine_Data);

  // Step 5 of 6: Instantiate the forward problem
  // Parameters are Gaussian RV
  QUESO::GslVector meanVector( paramSpace.zeroVector() );
  meanVector[0] = -1;
  meanVector[1] = 2;

  QUESO::GslMatrix covMatrix = QUESO::GslMatrix(paramSpace.zeroVector());
  covMatrix(0,0) = 4.;
  covMatrix(0,1) = 0.;
  covMatrix(1,0) = 0.;
  covMatrix(1,1) = 1.;

  QUESO::GaussianVectorRV<QUESO::GslVector,QUESO::GslMatrix>
    paramRv("param_",paramDomain,meanVector,covMatrix);

  QUESO::GenericVectorRV<QUESO::GslVector,QUESO::GslMatrix>
    qoiRv("qoi_", qoiSpace);

  QUESO::StatisticalForwardProblem<QUESO::GslVector,QUESO::GslMatrix,
                                   QUESO::GslVector,QUESO::GslMatrix>
    fp("", NULL, paramRv, qoiFunctionObj, qoiRv);

  // Step 6 of 6: Solve the forward problem
  fp.solveWithMonteCarlo(NULL);

  return;
}
\end{lstlisting}
\subsection{Input File}\label{sec:sfp-input-file}

In the case of a SFP, QUESO expects a list of options for Monte Carlo algorithm,
together with options for QUESO environment; such as the name of the output files and which sub-environments will write to to them. 
Note that the names of the variables have been designed to be informative:
\begin{description}\vspace{-8pt}
\item[ \texttt{env}:] refers to QUESO environment; \vspace{-8pt}
\item[ \texttt{fp}:] refers to forward problem;\vspace{-8pt}
\item[ \texttt{mc}:] refers to Monte Carlo;\vspace{-8pt}
\item[ \texttt{pseq}:] refers to the parameter sequence; and\vspace{-8pt}
\item[ \texttt{qseq}:] refers to the quantity of interest sequence.
\end{description}

The options used for solving this simple SFP are displayed in Listing \ref{code:sfp-input-file}.

\begin{lstlisting}[caption={File name \texttt{simple\_sfp\_example.inp} with options for QUESO library used in application code (Listings \ref{code:sfp-main-c}--\ref{code:sfp-compute-c}})., 
label={code:sfp-input-file},]
###############################################
# UQ Environment
###############################################
env_numSubEnvironments   = 1
env_subDisplayFileName   = outputData/display
env_subDisplayAllowAll   = 0
env_subDisplayAllowedSet = 0
env_displayVerbosity     = 2
env_syncVerbosity        = 0
env_seed                 = 0

###############################################
# Statistical forward problem (fp)
###############################################
fp_computeSolution      = 1
fp_computeCovariances   = 1
fp_computeCorrelations  = 1
fp_dataOutputFileName   = outputData/sfpOutput
fp_dataOutputAllowedSet = 0 1

###############################################
# 'fp_': information for Monte Carlo algorithm
###############################################
fp_mc_dataOutputFileName   = outputData/sfpOutput
fp_mc_dataOutputAllowedSet = 0 1

fp_mc_pseq_dataOutputFileName   = outputData/fp_p_seq
fp_mc_pseq_dataOutputAllowedSet = 0 1

fp_mc_qseq_size                 = 20000
fp_mc_qseq_displayPeriod        = 2000
fp_mc_qseq_measureRunTimes      = 1
fp_mc_qseq_dataOutputFileName   = outputData/fp_q_seq
fp_mc_qseq_dataOutputAllowedSet = 0 1

\end{lstlisting}


%
%
\subsection{Create your own Makefile}\label{sec:sfp-makefile}

Listing \ref{code:makefile} presents a Makefile, named `\texttt{Makefile\_sfp\_example\_margarida}', that may be used to compile the code and create the executable \verb+simple_sfp_example+. Naturally, it must be adapted to the user's settings, i.e., it has to have the correct paths for the user's libraries that have actually been used to compile and install QUESO.

\begin{lstlisting}[caption={Makefile for the application code in Listings
  \ref{code:sfp-main-c}--\ref{code:sfp-compute-c}},
  label={code:sfp-makefile},
  language={bash}]
  QUESO_DIR = /path/to/queso
  BOOST_DIR = /path/to/boost
  GSL_DIR   = /path/to/gsl

  INC_PATHS = \
     -I. \
     -I$(QUESO_DIR)/include \
     -I$(BOOST_DIR)/include \
     -I$(GSL_DIR)/include

  LIBS = \
     -L$(QUESO_DIR)/lib -lqueso \
     -L$(BOOST_DIR)/lib -lboost_program_options \
     -L$(GSL_DIR)/lib -lgsl

  CXX = mpic++
  CXXFLAGS += -g -Wall -c

  default: all

  .SUFFIXES: .o .C

  all:       example_sfp

  clean:
     rm -f *~
     rm -f *.o
     rm -f simple_sfp_example

  example_sfp: simple_sfp_example_main.o simple_sfp_example_qoi.o simple_sfp_example_compute.o
     $(CXX) simple_sfp_example_main.o \
            simple_sfp_example_qoi.o \
            simple_sfp_example_compute.o \
            -o simple_sfp_example $(LIBS)

  %.o: %.C
     $(CXX) $(INC_PATHS) $(CXXFLAGS) $<
\end{lstlisting}

Thus, to compile, build and execute the code, the user just needs to run the following commands in the same directory where the files are:
\begin{lstlisting}
$ cd HOME/LIBRARIES/QUESO-0.51.0/examples/simpleStatisticalForwardProblem 
$ export LD_LIBRARY_PATH=$LD_LIBRARY_PATH:\
  $HOME/LIBRARIES/gsl-1.15/lib/:\
  $HOME/LIBRARIES/boost-1.53.0/lib/:\
  $HOME/LIBRARIES/hdf5-1.8.10/lib:\
  $HOME/LIBRARIES/QUESO-0.51.0/lib 
$ make -f Makefile_sfp_example_margarida 
$ ./simple_sfp_example simple_sfp_example.inp
\end{lstlisting}

The `\verb+export+' instruction above is only necessary if the user has not saved it in his/her \verb+.bashrc+ file.

\subsection{Data Post-Processing and Visualization}\label{sec:sfp-results}

This section discusses the results computed by QUESO with the code of Section \ref{sec:code-sfp}, and shows how to use Matlab for the post-processing of the data generated by QUESO when solving SFPs. Only the essential Matlab commands are presented; for the complete/detailed codes, please refer to file '\verb+simple_fp_plots.m+'.

According to the specifications of the input file in Listing~\ref{code:sfp-input-file}, a folder named `\verb+outputData+' containing the following files should be created: \verb+display_sub0.txt, fp_p_seq.m,+ \linebreak \verb+fp_p_seq_sub0.m, fp_q_seq.m, fp_q_seq_sub0.m,+ and \verb+sfpOutput_sub0.m+.

The code below shows how to load the data provided by QUESO during the solution
process of the SFP described, in the form of chains of positions.
\begin{lstlisting}[caption={Matlab code for loading the data in both parameter and QoI chains of the SFP.}]
% inside Matlab
>> clear all
>> fp_p_seq.m
>> fp_q_seq.m
\end{lstlisting}

Alternatively, the user may call the file \texttt{simple\_fp\_plots.m}, which
contains the above commands, together with a variety of others, for data
visualization:
\begin{lstlisting}[caption={Matlab code for loading the data in both parameter and QoI chains of the SFP, by calling the file \texttt{simple\_fp\_plots.m}.}]
% inside Matlab
>> clear all
>> simple_fp_plots
\end{lstlisting}

\subsubsection{Histogram Plots}

In order to plot a histogram of the QoI, you may use the pre-defined Matlab function \verb+hist+.
The Matlab code presented in Listing \ref{matlab:fp_hist_qoi} below shows how to create the Figure~\ref{fig:fp_qoi_hist}.

\begin{lstlisting}[label=matlab:fp_hist_qoi,caption={Matlab code for the QoI histogram plot.}]
% inside Matlab
>> fp_q_seq  %if commands of Listings 3.19/3.20 have not been called
>> nbins=20;
>> hist(fp_mc_QoiSeq_unified,nbins);
>> title('QoI Histogram','fontsize',20);
>> xlabel('QoI=\theta_1+\theta_2','fontname', 'Times', 'fontsize',20)
>> ylabel('Frequency','fontsize',20);
\end{lstlisting}

\begin{figure}[p]
\centering 
\includegraphics[scale=0.35]{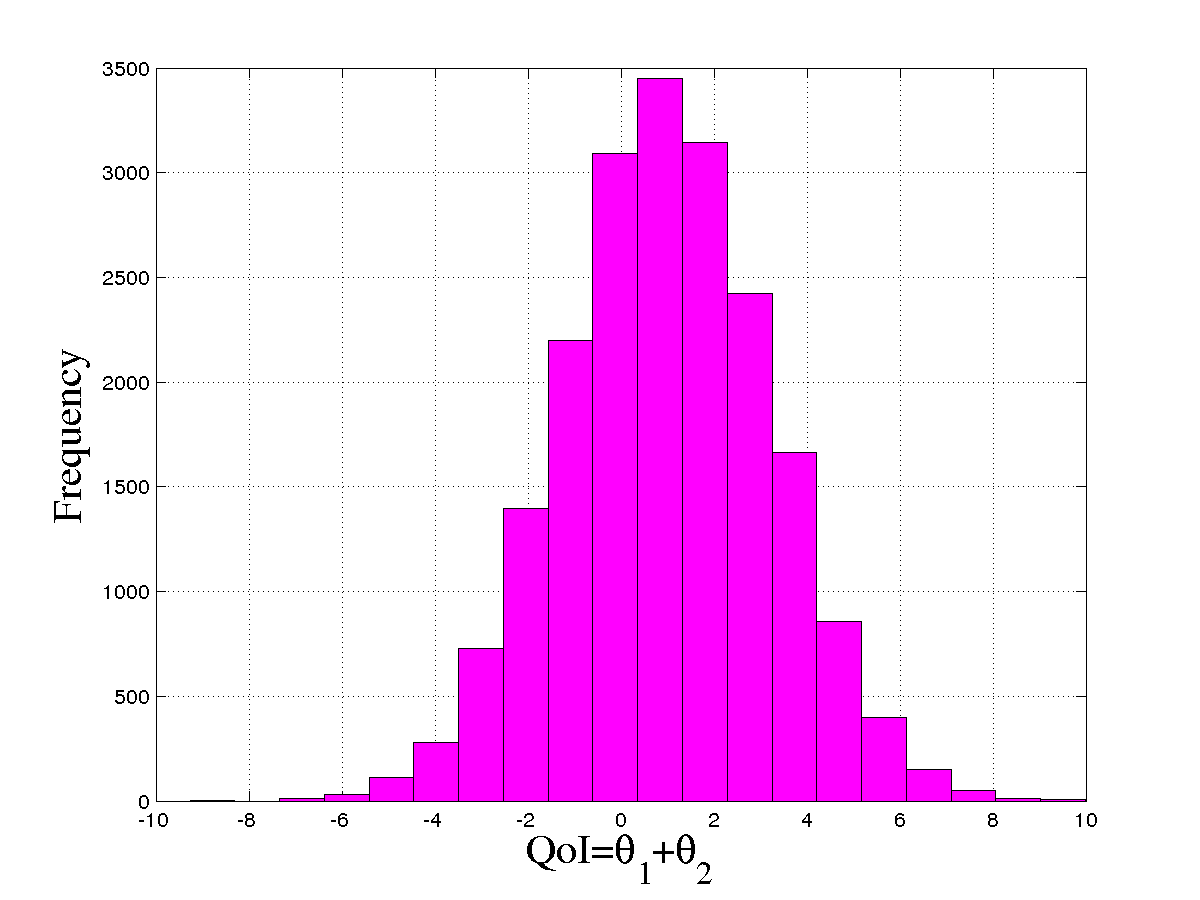}
\vspace{-10pt}
\caption{QoI histogram.}
\label{fig:fp_qoi_hist}
\end{figure}

\subsubsection{KDE Plot}

Matlab function \verb+ksdensity+ (Kernel smoothing density estimate) together with the option `\verb+pdf+' may be used to estimate the KDE of the QoI. 

\begin{lstlisting}[label=matlab:fp_kde_qoi,caption={Matlab code for the KDE displayed in Figure \ref{fig:simple_sfp_kde}}]
% inside Matlab
>> fp_q_seq  %if commands of Listing 5.19 have not been called
>> [fi,xi] = ksdensity(fp_mc_QoiSeq_unified,'function','pdf');
>> x=sort(fp_mc_QoiSeq_unified);
>> mu=1;
>> sigma2=5;
>> f=(exp(-(x-mu).*(x-mu)/sigma2/2))/sqrt(2*pi*sigma2);
>> plot(xi,fi,'-m','linewidth',4);
>> hold;
>> plot(x,f,'--k','linewidth',2);
>> h=legend('QoI = \theta_1+\theta_2','analytical','location','northwest');
\end{lstlisting}

\begin{figure}[p]
\centering 
\includegraphics[scale=0.35]{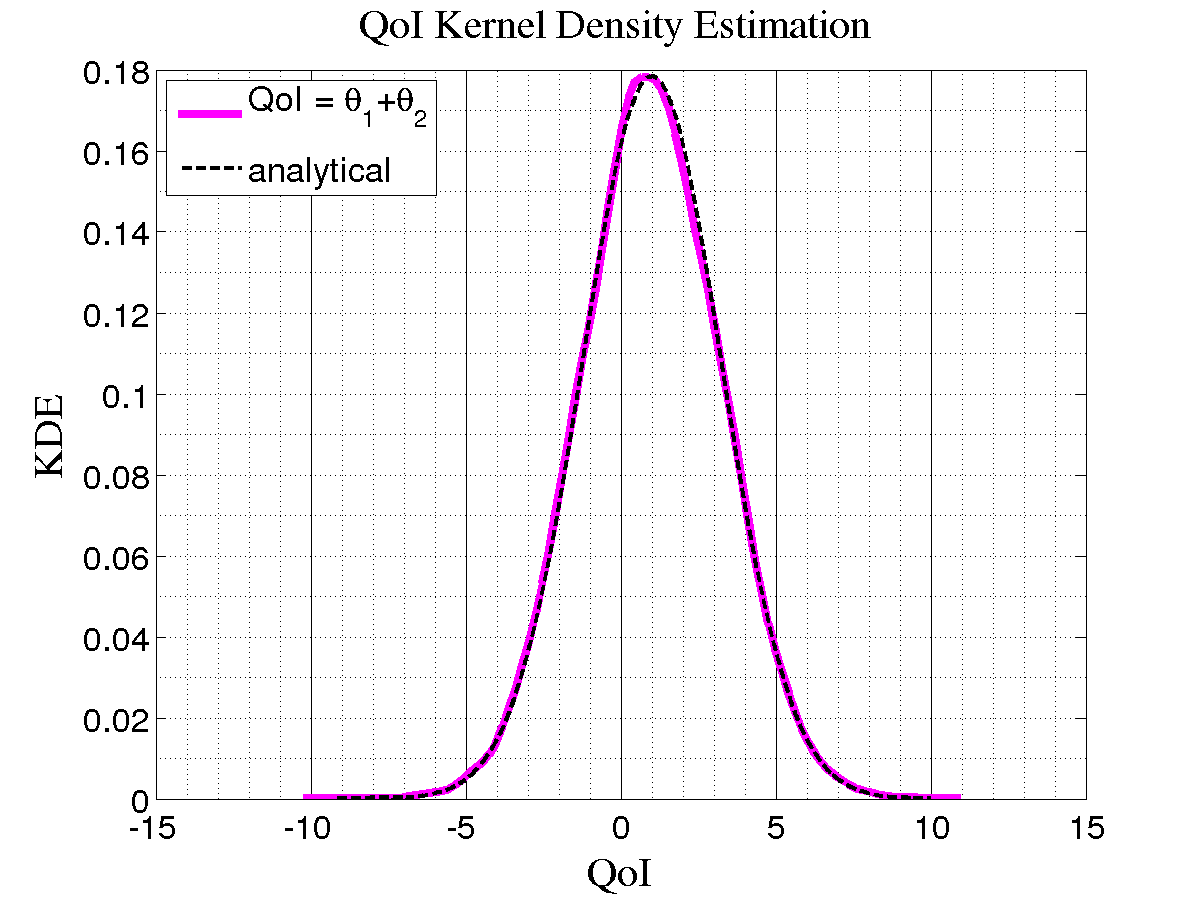}
\vspace{-10pt}
\caption{Kernel Density Estimation. QUESO results are plotted against the PDF of a Gaussian distribution $Q(x)=   \frac{1}{ \sqrt{10\pi}} \exp\left(-\frac{1}{10}(x-1)^2 \right)$, where $\mu=1$ and $\sigma^2=5$.}
\label{fig:simple_sfp_kde}
\end{figure}

\subsubsection{CDF Plot}

Matlab function \verb+ksdensity+ with \verb+'cdf'+ option may also be used for plotting the Cumulative Distribution Function of the QoI.

\begin{lstlisting}[label=matlab:fp_cdf_qoi,caption={Matlab code for the QoI CDF plot displayed in Figure \ref{fig:simple_sfp_cdf}.}]
% inside Matlab
>> fp_q_seq  %if commands of Listing 5.19 have not been called
>> [f,xi] = ksdensity(fp_mc_QoiSeq_unified,'function','cdf');
>> plot(xi,f,'-m','linewidth',3)
\end{lstlisting}

\begin{figure}[p]
\centering 
\includegraphics[scale=0.35]{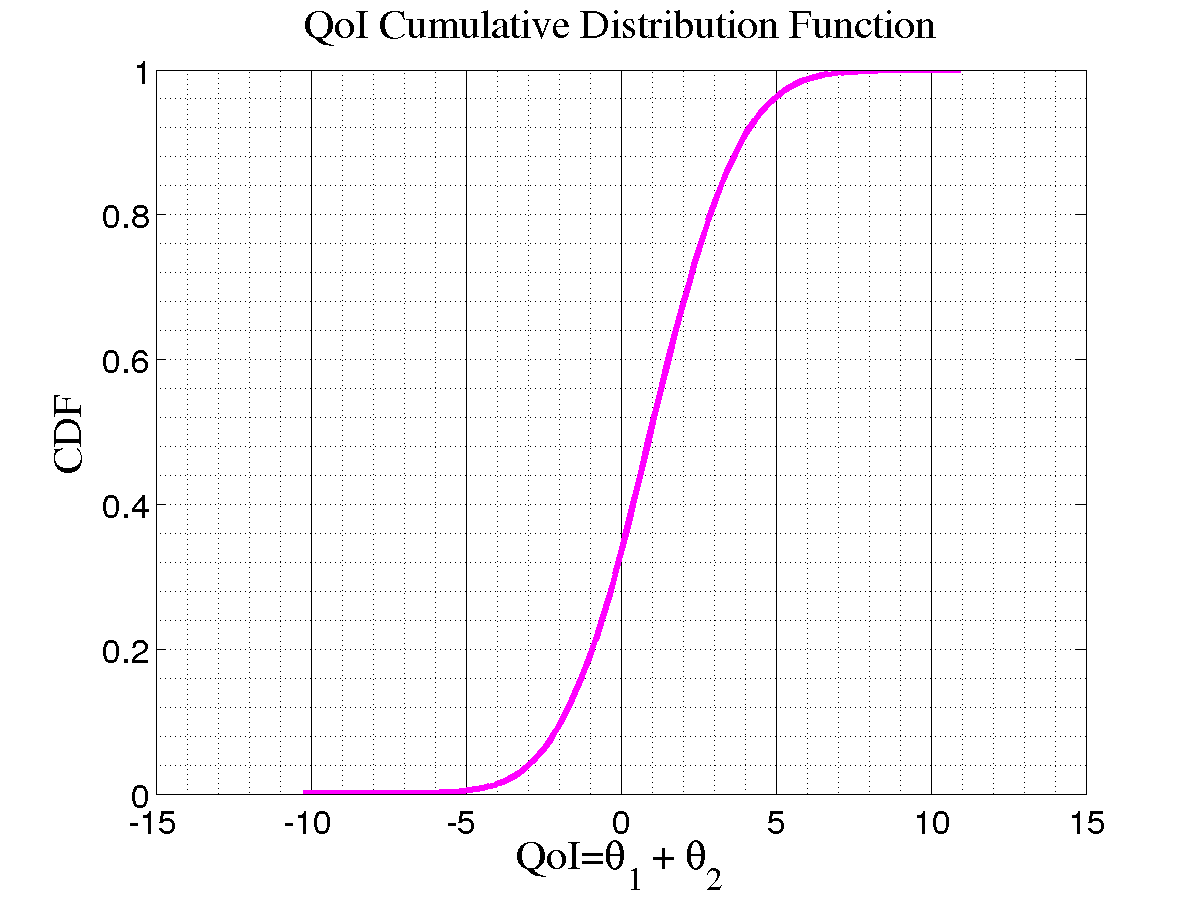}
\vspace*{-10pt}
\caption{Cumulative Distribution Function.}
\label{fig:simple_sfp_cdf}
\end{figure}


\section{\texttt{gravity}}\label{sec:example_gravity}

This section presents an example of how to use QUESO in order to develop an application that solves a 
statistical inverse problem (SIP) and a statistical forward problem (SFP), where the solution of the former 
serves as input to the later. During the SIP, the acceleration due to gravity for an object in free 
fall near the surface of the Earth is inferred. During the SFP, the distance traveled by a projectile 
launched at a given angle and altitude is calculated using the calibrated magnitude of the acceleration of gravity.

In this section we describe a statistical forward problem  of predicting the described in Section 
\ref{sec:gravity-ip}.

\subsection{Statistical Inverse Problem}\label{sec:gravity-ip}



A possible deterministic mathematical model for the vertical motion of an object in free fall near the 
surface of the Earth is given by
\begin{equation}\label{eq:gravity01}
h(t)=-\frac{1}{2} g t^2 + v_0 t + h_0.
\end{equation}
where
$v_0$ [$m/s$] is the initial velocity,
$h_0$ [$m$] is the initial altitude,
$h(t)$ [$m$] is the altitude with respect to time,
$t$ [$s$] is the elapsed time, and
$g$ [$m/s^2$] is the magnitude of the acceleration due to gravity 
(the parameter which cannot be directly measured and will be statistically inferred).

\subsubsection{Experimental Data}
\begin{figure}[!ht]
\centering
\setlength{\unitlength}{4144sp}%
\begingroup\makeatletter\ifx\SetFigFont\undefined%
\gdef\SetFigFont#1#2#3#4#5{%
  \reset@font\fontsize{#1}{#2pt}%
  \fontfamily{#3}\fontseries{#4}\fontshape{#5}%
  \selectfont}%
\fi\endgroup%
\begin{picture}(1754,1477)(249,-803)
{\color[rgb]{0,0,1}\thinlines
\put(1171,-61){\circle*{90}}
}%
{\color[rgb]{0,0,0}\multiput(1171,569)(0.00000,-106.36364){6}{\line( 0,-1){ 53.182}}
\put(1171,-16){\vector( 0,-1){0}}
}%
\thicklines
{\color[rgb]{0,0,0}\put(271,-781){\line( 1, 0){1710}}
}%
\thinlines
{\color[rgb]{0,0,0}\put(271,569){\line( 1, 0){1215}}
}%
{\color[rgb]{0,0,0}\put(721,569){\vector( 0, 1){  0}}
\put(721,569){\vector( 0,-1){1350}}
}%
\put(465,-264){\makebox(0,0)[lb]{\smash{{\SetFigFont{12}{14.4}{\rmdefault}{\mddefault}{\updefault}{\color[rgb]{0,0,0}$h_0$}%
}}}}
\put(1711,524){\makebox(0,0)[lb]{\smash{{\SetFigFont{12}{14.4}{\rmdefault}{\mddefault}{\updefault}{\color[rgb]{0,0,0}$v_0=0$}%
}}}}
\put(1711,-61){\makebox(0,0)[lb]{\smash{{\SetFigFont{12}{14.4}{\rmdefault}{\mddefault}{\updefault}{\color[rgb]{0,0,0}$h(t)=-\frac{1}{2} g\,t^2+h_0$}%
}}}}
\end{picture}%
\vspace*{-8pt}
\caption{An object falls from altitude $h_0$ with zero initial velocity ($v_0=0$).}
\label{fig:free_fall}
\end{figure}
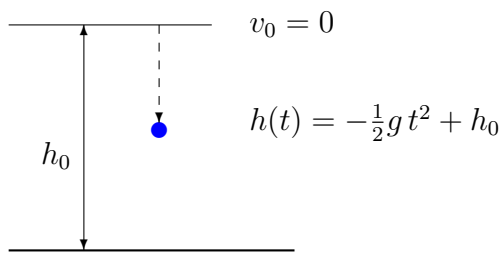

We assume that the experiment of allowing an object to fall from different altitudes with zero initial 
velocity has been repeatedly conducted (See Figure \ref{fig:free_fall}). The data collected, 
e.g.  $\mathbf{d}$, is displayed in Table \ref{table:data}; the standard deviations, $\sigma$'s, 
refer to the uncertainties in the measured times during the experiment execution~\cite{interactagram}. 

\begin{table}[htp]
\caption{Measurement data $\mathbf{d}$ of size $n_d=14$.
The object falls from altitude $h_0$ in $t$ seconds, with standard deviation of $\sigma$ 
seconds in the time measurement~\cite{interactagram}.
}
\vspace{-8pt}
\begin{center}
\begin{tabular}{ccc}
\toprule
altitude [$m$] & time [$s$]  & Std. Dev. $\sigma$ [$s$]\\
\midrule
\midrule
$~$10	&	1.41	&	0.02	\\
$~$20	&	2.14	&	0.12	\\
$~$30	&	2.49	&	0.02	\\
$~$40	&	2.87	&	0.01	\\
$~$50	&	3.22	&	0.03	\\
$~$60	&	3.49	&	0.01	\\
$~$70	&	3.81	&	0.03	\\
$~$80	&	4.07	&	0.03	\\
$~$90	&	4.32	&	0.03	\\
100	&	4.47	&	0.05	\\
110	&	4.75	&	0.01	\\
120	&	4.99	&	0.04	\\
130	&	5.16	&	0.01	\\
140	&	5.26	&	0.09	\\
\bottomrule
\end{tabular}
\end{center}
\label{table:data}
\end{table}

\subsubsection{The Prior RV, Likelihood and Posterior RV}

In a straightforward classical interpretation of Bayesian inference, the prior signifies the 
modeler's honest opinion about the unknown.
For the gravity inference problem, let's assume that gravity varies uniformly in the 
interval [8,11], or, in other words, we chose uniform prior distribution in that interval:

\begin{equation}\label{eq-g-prior}
\pi_{\text{prior}}=\mathcal{U}(8,11).
\end{equation}

We choose the usual likelihood function:
\begin{equation}\label{eq:like02}
\pi_{\text{like}}(\mathbf{d} | \boldsymbol{\theta})
\varpropto
\exp
\left\{
-\frac{1}{2}
[\mathbf{y}(\boldsymbol{\theta})-\mathbf{d}]^T
\left[\mathbf{C}(\boldsymbol{\theta})\right]^{-1}
[\mathbf{y}(\boldsymbol{\theta})-\mathbf{d}]
\right\},
\end{equation}
where $\mathbf{C}(\boldsymbol{\theta})$ is a given covariance matrix, $\mathbf{d}$ denotes 
experimental data, $\mathbf{y}(\boldsymbol{\theta})$ is the model output data.

Recalling the deterministic model for the acceleration of gravity (\ref{eq:gravity01}) 
with zero initial velocity,  the information provided in Table \ref{table:data}, and 
Equation (\ref{eq:like02}); and, additionally, invoking the nomenclature used in Section 
\ref{sec:statistical_concepts}, we have:
\begin{equation}\label{eq:like03}
\boldsymbol{\theta} \stackrel{\text{\small{def.}}}{=} g,
\quad
\mathbf{y}(\boldsymbol{\theta})= 
\left[
\begin{array}{c}
\sqrt{\dfrac{2 h_1}{g}}\\	
\sqrt{\dfrac{2 h_2}{g}}\\	
\vdots\\	
\sqrt{\dfrac{2 h_{n_d}}{g}}
\end{array}
\right],
\quad 
\mathbf{d} = 
\left[
\begin{array}{c}
t_1    \\
t_2    \\ 
\vdots \\	
t_{n_d}
\end{array}
\right],
\quad
\mathbf{C}(\boldsymbol{\theta})=
\left[
\begin{array}{cccc}
\sigma^2_1 & 0	        & \cdots & 0 \\
0          & \sigma^2_2 & \cdots & 0 \\
\vdots     & \vdots     & \ddots & 0 \\
0          & 0          & \cdots & \sigma^2_{n_d}
\end{array}
\right],
\end{equation}
where $n_d=14$ is the number of data points in Table \ref{table:data}.

Now we are ready to evoke Bayes' formula in order to obtain the posterior PDF 
$\pi_{\text{post}}(\boldsymbol{\theta})$:
\begin{equation}\label{eq-Bayes-g}
\pi_{\text{post}}(\boldsymbol{\theta}|\mathbf{d})\varpropto  \pi_{\text{like}}(\mathbf{d}|\boldsymbol{\theta}) \, \pi_{\text{prior}}(\boldsymbol{\theta}).
\end{equation}

\subsection{Statistical Forward Problem}

Projectile motion refers to the motion of an object projected into the air at an angle, e.g. a soccer ball 
being kicked, a baseball being thrown, or an athlete long jumping. Supposing the object does not have a 
propulsion system and neglecting air resistance, then the only force acting on the object is a 
constant gravitational acceleration $g$.

\begin{figure}[!ht]
\centering
\includegraphics[scale=1]{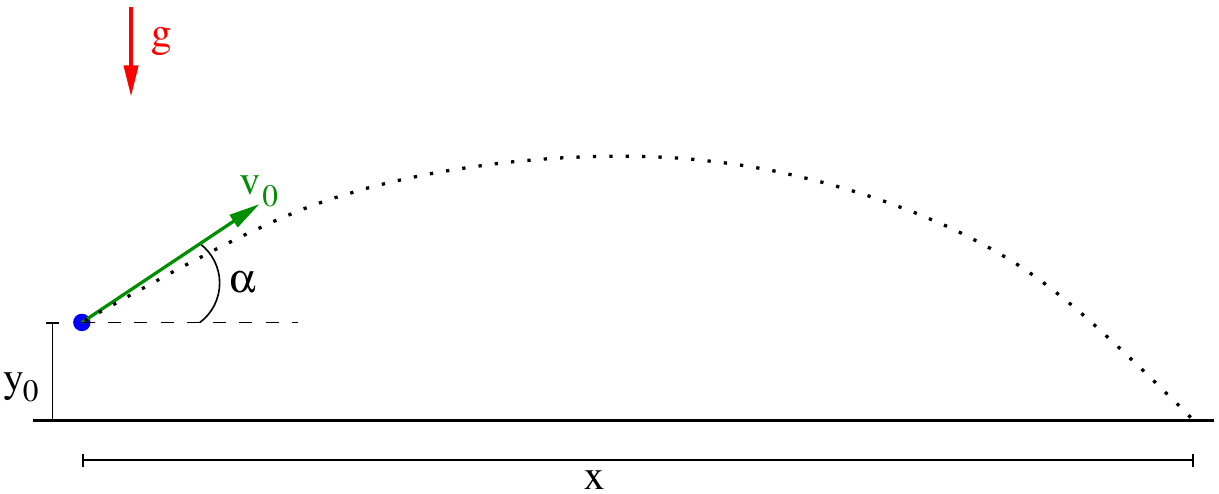}
\vspace*{-8pt}
\caption{Object traveling with projectile motion. }
\label{fig:projectile}
\end{figure}

A possible deterministic two-dimensional mathematical model for the vertical motion of an object projected 
from near the surface of the Earth is given by
\begin{align}\label{eq:fwd01}
v_x &= v_{0x} \\ 
v_y &= v_{0y} - gt \\ 
  x &= v_{0x}t \\ 
  h &= h_0 + v_{0y}t - \frac{1}{2} g t^2  
\end{align}
where
$h_0$ is the initial height, $x=x(t)$ is the distance traveled by the object, $\bv{v_0}=(v_{0x},v_{0y})$ 
is the initial velocity, $v_{0x} = v_{0} \cos(\alpha)$, $v_{0y} = v_{0} \sin(\alpha)$, and $v_0=\|\bv{v_0}\|^2$.
Figure \ref{fig:projectile} displays the projectile motion of an object in these conditions.

For this example, we assume that $h_0 =0$ m, $\alpha = \pi/4$ radians, $v_0 = 5$ m/s, all 
deterministic variables; and $g$ is the solution of the SIP described in Section \ref{sec:gravity-ip}.

Since a PDF is assigned to parameter $g$; thus, the output of the mathematical model (\ref{eq:fwd01}) 
becomes a random variable, thus we have a statistical forward problem. 

\subsubsection{The Input RV, QoI Function and Output RV}
 
The input random variable for the statistical forward problem is the acceleration of gravity $g$, 
which is also the solution (posterior PDF) of the inverse problem described in Section \ref{sec:gravity-ip}. 
The output random variable for this example is the distance $x$ traveled by an object in projectile motion. 
Note that, since there is uncertainty in the parameter $g$ ($g$ is given as a PDF), one can 
expect that this uncertainty will be propagated to $x$, which will also be given as a PDF.

Combining the expressions in Equation \ref{eq:fwd01} and rearranging them, we have that QoI 
function for $x$ 
is: 
\begin{equation}\label{eq:fp_deterministic}
x=\dfrac{ v_0 \cos \alpha }{g} \left( v_0  \sin \alpha  + \sqrt{ ( v_0  \sin \alpha)^2 + 2g\, y_0 }\right).                                                                                        
\end{equation}
where $y$ is the distance traveled and our quantity of interest (QoI). 

\subsection{Running the Example}\label{sec:gravity-run}
 
To run the executable provided (available after QUESO installation), enter the following commands:
\begin{lstlisting}[label={},caption={}]
$ cd $HOME/LIBRARIES/QUESO-0.51.0/examples/gravity
$ rm outputData/*
$ ./gravity_gsl gravity_inv_fwd.inp
\end{lstlisting}

The console output of the program is:
\begin{lstlisting}[caption={Console output of program \texttt{gravity\_gsl}}, label={code:console_output},language={bash}]
kemelli@violeta:~/LIBRARIES/QUESO-0.51.0/examples/gravity$ ./gravity_gsl gravity_inv_fwd.inp 
---------------------------------------------------------------------
QUESO Library: Version = 0.47.1 (47.1)

Development Build

Build Date   = 2013-04-29 17:05
Build Host   = violeta
Build User   = kemelli
Build Arch   = x86_64-unknown-linux-gnu
Build Rev    = 38998M

C++ Config   = mpic++ -g -O2 -Wall

Trilinos DIR = 
GSL Libs     = -L/home/kemelli/LIBRARIES/gsl-1.15/lib -lgsl -lgslcblas -lm
GRVY DIR     = 
GLPK DIR     = 
HDF5 DIR     = /home/kemelli/LIBRARIES/hdf5-1.8.10
--------------------------------------------------------------------------------------------------------------
Beginning run at Mon Apr 29 17:27:32 2013

MPI node of worldRank 0 has fullRank 0, belongs to subEnvironment of id 0, and has subRank 0
MPI node of worldRank 0 belongs to sub communicator with full ranks 0
MPI node of worldRank 0 also belongs to inter0 communicator with full ranks 0, and has inter0Rank 0


Beginning run of 'Gravity + Projectile motion' example at Mon Apr 29 17:27:32 2013

 my fullRank = 0
 my subEnvironmentId = 0
 my subRank = 0
 my interRank = 0

Beginning 'SIP -> Gravity estimation' at Mon Apr 29 17:27:32 2013

Solving the SIP with Metropolis Hastings

Beginning 'SFP -> Projectile motion' at Mon Apr 29 17:27:33 2013

Solving the SFP with Monte Carlo

Ending run of 'Gravity + Projectile motion' example at Mon Apr 29 17:27:33 2013

Ending run at Mon Apr 29 17:27:33 2013
Total run time = 1 seconds
kemelli@violeta:~/LIBRARIES/QUESO-0.51.0/examples/gravity$ 
\end{lstlisting}

In order to generate chain plots, histograms, KDEs, etc., the user may use Matlab/GNU Octave and call 
the following command lines:
\begin{lstlisting}
$ matlab
   $ gravity_plots_ip      # inside matlab
   $ gravity_plots_fp      # inside matlab
   $ exit                  # inside matlab
$ ls -l outputData/*.png
  sfp_gravity_autocorrelation.png  sfp_gravity_cdf.png
  sfp_gravity_chain_pos.png        sfp_gravity_hist.png
  sfp_gravity_kde.png              sip_gravity_autocorrelation_raw_filt.png
  sip_gravity_cdf_filt.png         sip_gravity_cdf_raw.png
  sip_gravity_chain_pos_filt.png   sip_gravity_chain_pos_raw.png
  sip_gravity_hist_filt.png        sip_gravity_hist_raw.png
  sip_gravity_kde_filt.png         sip_gravity_kde_raw.png
\end{lstlisting}

As a result, the user should have created several of PNG figures containing marginal posterior PDF, 
chain positions of the parameters and the QoI, histogram, cumulative density distribution and 
autocorrelation. The name of the figure files have been chosen to be informative, as shown in the listing above.

\subsection{Example Code}\label{sec:gravity_code}

The source code for the SIP and the SFP is composed of 7 files.
Three of them are common for both problems: \texttt{gravity\_main.C, gravity\_compute.h} and 
\texttt{gravity\_compute.C}; they combine both problems and use the solution of the SIP 
(the posterior PDF for the gravity) as an input for the SFP and are presented, respectively, 
in Listings \ref{code:gravity_main}, \ref{code:gravity_compute_h} and \ref{code:gravity_compute_C}.
Two of files specifically  handle the SIP: \texttt{gravity\_likelihood.h}, and \texttt{gravity\_likelihood.C}, 
and are displayed in Listings \ref{code:gravity_like_h} and \ref{code:gravity_like_C}. Finally, the files 
specific for the SFP are \texttt{gravity\_qoi.h} and \texttt{gravity\_qoi.C}, and they are presented in 
Listings \ref{code:gravity_qoi_h} and \ref{code:gravity_qoi_C}.

\begin{lstlisting}[caption=File \texttt{gravity\_main.C.}, label=code:gravity_main, linerange={25-1000}]
 /*------------------------------------------------------------------
 * Brief description of this file:
 *
 * This is an example of how to use QUESO classes and algorithms in order to define and solve
 * a statistical inverse problem (SIP) and/or a statistical forward problem (SFP).
 * The SIP consists on calibrating the magnitude 'g' of acceleration gravity using
 * measurements of the time that it takes for an object in free fall to reach the ground from
 * a given height and zero initial velocity. The solution of the SIP is the posterior
 * probability density function (PDF) of 'g'.
 * The SFP consists of calculating the maximum distance traveled by an object in projectile
 * motion. The posterior PDF of 'g' from the SIP might be used as input to the SFP.
 *
 * The code consists of 7 files:
 * - 'gravity_main.C' (this file)
 * - 'gravity_compute.C' (the driving application code)
 * - 'gravity_compute.h'
 * - 'gravity_likelihood.C' (necessary for the SIP)
 * - 'gravity_likelihood.h'
 * - 'gravity_qoi.C' (necessary for the SFP)
 * - 'gravity_qoi.h'
 *-----------------------------------------------------------------*/

#include <gravity_compute.h>

int main(int argc, char* argv[])
{
  // Initialize QUESO environment
#ifdef QUESO_HAS_MPI
  MPI_Init(&argc,&argv);
  QUESO::FullEnvironment* env =
    new QUESO::FullEnvironment(MPI_COMM_WORLD,argv[1],"",NULL);
#else
  QUESO::FullEnvironment* env =
    new QUESO::FullEnvironment(argv[1],"",NULL);
#endif

  // Call application
  computeGravityAndTraveledDistance(*env);

  // Finalize QUESO environment
  delete env;
#ifdef QUESO_HAS_MPI
  MPI_Finalize();
#endif

  return 0;
}
\end{lstlisting}
 
\begin{lstlisting}[caption=File \texttt{gravity\_compute.h.}, label=code:gravity_compute_h, linerange={32-1000}]
#define EX_COMPUTE_H

#include <queso/Environment.h>

void computeGravityAndTraveledDistance(const QUESO::FullEnvironment& env);

#endif
\end{lstlisting}

\begin{lstlisting}[caption={File \texttt{gravity\_compute.C}. The first part of the code (lines 60--150) handles the statistical forward problem, whereas the second part of the code (lines 151--216) handles the statistical forward problem.\\}, label=code:gravity_compute_C, linerange={26-1000},numbers=left]
 * This file is divided in two parts:
 * - the first one handles the statistical inverse problem (SIP) for estimating
 *   the magnitude 'g' of gravity acceleration; and
 * - the second part handles the statistical forward problem (SFP) for
 *   predicting the maximum distance traveled by a projectile.
 *
 * The SIP definition requires a user defined likelihood function; refer to
 * files 'gravity_likelihood.h' and 'gravity_likelihood.C'. The SFP definition
 * requires a user defined qoi function; refer to files 'gravity_qoi.h' and
 * 'gravity_qoi.C'.
 */

#include <cmath>
#include <sys/time.h>

#include <queso/GslMatrix.h>
#include <queso/GenericScalarFunction.h>
#include <queso/GenericVectorFunction.h>
#include <queso/GaussianVectorRV.h>
#include <queso/UniformVectorRV.h>
#include <queso/GenericVectorRV.h>
#include <queso/StatisticalInverseProblem.h>
#include <queso/StatisticalForwardProblem.h>

#include <gravity_compute.h>
#include <gravity_likelihood.h>
#include <gravity_qoi.h>

void computeGravityAndTraveledDistance(const QUESO::FullEnvironment& env) {
  struct timeval timevalNow;

  gettimeofday(&timevalNow, NULL);
  if (env.fullRank() == 0) {
    std::cout << "\nBeginning run of 'Gravity + Projectile motion' example at "
              << ctime(&timevalNow.tv_sec)
              << "\n my fullRank = "         << env.fullRank()
              << "\n my subEnvironmentId = " << env.subId()
              << "\n my subRank = "          << env.subRank()
              << "\n my interRank = "        << env.inter0Rank()
               << std::endl << std::endl;
  }

  // Just examples of possible calls
  if ((env.subDisplayFile()       ) &&
      (env.displayVerbosity() >= 2)) {
    *env.subDisplayFile() << "Beginning run of 'Gravity + Projectile motion' example at "
                          << ctime(&timevalNow.tv_sec)
                          << std::endl;
  }
  env.fullComm().Barrier();
  env.subComm().Barrier();  // Just an example of a possible call

  //================================================================
  // Statistical inverse problem (SIP): find posterior PDF for 'g'
  //================================================================
  gettimeofday(&timevalNow, NULL);
  if (env.fullRank() == 0) {
    std::cout << "Beginning 'SIP -> Gravity estimation' at "
              << ctime(&timevalNow.tv_sec)
              << std::endl;
  }

  //------------------------------------------------------
  // SIP Step 1 of 6: Instantiate the parameter space
  //------------------------------------------------------
  QUESO::VectorSpace<> paramSpace(env, "param_", 1, NULL);

  //------------------------------------------------------
  // SIP Step 2 of 6: Instantiate the parameter domain
  //------------------------------------------------------
  QUESO::GslVector paramMinValues(paramSpace.zeroVector());
  QUESO::GslVector paramMaxValues(paramSpace.zeroVector());

  paramMinValues[0] = 8.;
  paramMaxValues[0] = 11.;

  QUESO::BoxSubset<> paramDomain("param_", paramSpace, paramMinValues,
      paramMaxValues);

  //------------------------------------------------------
  // SIP Step 3 of 6: Instantiate the likelihood function
  // object to be used by QUESO.
  //------------------------------------------------------
  Likelihood<> lhood("like_", paramDomain);

  //------------------------------------------------------
  // SIP Step 4 of 6: Define the prior RV
  //------------------------------------------------------
  QUESO::UniformVectorRV<> priorRv("prior_", paramDomain);

  //------------------------------------------------------
  // SIP Step 5 of 6: Instantiate the inverse problem
  //------------------------------------------------------
  // Extra prefix before the default "rv_" prefix
  QUESO::GenericVectorRV<> postRv("post_", paramSpace);

  // No extra prefix before the default "ip_" prefix
  QUESO::StatisticalInverseProblem<> ip("", NULL, priorRv, lhood, postRv);

  //------------------------------------------------------
  // SIP Step 6 of 6: Solve the inverse problem, that is,
  // set the 'pdf' and the 'realizer' of the posterior RV
  //------------------------------------------------------

  QUESO::GslVector paramInitials(paramSpace.zeroVector());
  priorRv.realizer().realization(paramInitials);

  QUESO::GslMatrix proposalCovMatrix(paramSpace.zeroVector());
  proposalCovMatrix(0,0) = std::pow(std::abs(paramInitials[0]) / 20.0, 2.0);

  ip.solveWithBayesMetropolisHastings(NULL, paramInitials, &proposalCovMatrix);

  //================================================================
  // Statistical forward problem (SFP): find the max distance
  // traveled by an object in projectile motion; input pdf for 'g'
  // is the solution of the SIP above.
  //================================================================
  gettimeofday(&timevalNow, NULL);
  std::cout << "Beginning 'SFP -> Projectile motion' at "
            << ctime(&timevalNow.tv_sec)
            << std::endl;

  //------------------------------------------------------
  // SFP Step 1 of 6: Instantiate the parameter *and* qoi spaces.
  // SFP input RV = FIP posterior RV, so SFP parameter space
  // has been already defined.
  //------------------------------------------------------
  QUESO::VectorSpace<> qoiSpace(env, "qoi_", 1, NULL);

  //------------------------------------------------------
  // SFP Step 2 of 6: Instantiate the parameter domain
  //------------------------------------------------------

  // Not necessary because input RV of the SFP = output RV of SIP.
  // Thus, the parameter domain has been already defined.

  //------------------------------------------------------
  // SFP Step 3 of 6: Instantiate the qoi object
  // to be used by QUESO.
  //------------------------------------------------------
  Qoi<> qoi("qoi_", paramDomain, qoiSpace);

  //------------------------------------------------------
  // SFP Step 4 of 6: Define the input RV
  //------------------------------------------------------

  // Not necessary because input RV of SFP = output RV of SIP
  // (postRv).

  //------------------------------------------------------
  // SFP Step 5 of 6: Instantiate the forward problem
  //------------------------------------------------------
  QUESO::GenericVectorRV<> qoiRv("qoi_", qoiSpace);

  QUESO::StatisticalForwardProblem<> fp("", NULL, postRv, qoi, qoiRv);

  //------------------------------------------------------
  // SFP Step 6 of 6: Solve the forward problem
  //------------------------------------------------------
  std::cout << "Solving the SFP with Monte Carlo"
            << std::endl << std::endl;
  fp.solveWithMonteCarlo(NULL);

  gettimeofday(&timevalNow, NULL);
  if ((env.subDisplayFile()       ) &&
      (env.displayVerbosity() >= 2)) {
    *env.subDisplayFile() << "Ending run of 'Gravity + Projectile motion' example at "
                          << ctime(&timevalNow.tv_sec)
                          << std::endl;
  }
  if (env.fullRank() == 0) {
    std::cout << "Ending run of 'Gravity + Projectile motion' example at "
              << ctime(&timevalNow.tv_sec)
              << std::endl;
  }
}
\end{lstlisting}

\begin{lstlisting}[caption=File \texttt{gravity\_likelihood.h}., label=code:gravity_like_h, linerange={32-1000}]
class Likelihood : public QUESO::BaseScalarFunction<V, M>
{
public:
  Likelihood(const char * prefix, const QUESO::VectorSet<V, M> & domain);
  virtual ~Likelihood();
  virtual double lnValue(const V & domainVector, const V * domainDirection,
      V * gradVector, M * hessianMatrix, V * hessianEffect) const;
  virtual double actualValue(const V & domainVector, const V * domainDirection,
      V * gradVector, M * hessianMatrix, V * hessianEffect) const;

private:
  std::vector<double> m_heights; // heights
  std::vector<double> m_times;   // times
  std::vector<double> m_stdDevs; // uncertainties in time measurements
};

#endif
\end{lstlisting}

\begin{lstlisting}[caption=File \texttt{gravity\_likelihood.C}., label=code:gravity_like_C, linerange={31-1000}]
#include <queso/StatisticalInverseProblem.h>
#include <queso/ScalarFunction.h>
#include <queso/VectorSet.h>

#include <gravity_likelihood.h>

template<class V, class M>
Likelihood<V, M>::Likelihood(const char * prefix,
    const QUESO::VectorSet<V, M> & domain)
  : QUESO::BaseScalarFunction<V, M>(prefix, domain),
    m_heights(0),
    m_times(0),
    m_stdDevs(0)
{
  // Data available in /inputData/data02.dat
  double const heights[] = {10, 20, 30, 40, 50, 60, 70, 80, 90, 100, 110,
                            120, 130, 140};

  double const times  [] = {1.41, 2.14, 2.49, 2.87, 3.22, 3.49, 3.81, 4.07,
                            4.32, 4.47, 4.75, 4.99, 5.16, 5.26};

  double const stdDevs[] = {0.020, 0.120, 0.020, 0.010, 0.030, 0.010, 0.030,
                            0.030, 0.030, 0.050, 0.010, 0.040, 0.010, 0.09};

  std::size_t const n = sizeof(heights) / sizeof(*heights);
  m_heights.assign(heights, heights + n);
  m_times.assign(times, times + n);
  m_stdDevs.assign(stdDevs, stdDevs + n);
}

template<class V, class M>
Likelihood<V, M>::~Likelihood()
{
  // Deconstruct here
}

template<class V, class M>
double
Likelihood<V, M>::lnValue(const V & domainVector, const V * domainDirection,
    V * gradVector, M * hessianMatrix, V * hessianEffect) const
{
  double g = domainVector[0];

  double misfitValue = 0.0;
  for (unsigned int i = 0; i < m_heights.size(); ++i) {
    double modelTime = std::sqrt(2.0 * m_heights[i] / g);
    double ratio = (modelTime - m_times[i]) / m_stdDevs[i];
    misfitValue += ratio * ratio;
  }

  return -0.5 * misfitValue;
}

template<class V, class M>
double
Likelihood<V, M>::actualValue(const V & domainVector,
    const V * domainDirection, V * gradVector, M * hessianMatrix,
    V * hessianEffect) const
{
  return std::exp(this->lnValue(domainVector, domainDirection, gradVector,
        hessianMatrix, hessianEffect));
}

template class Likelihood<QUESO::GslVector, QUESO::GslMatrix>;
\end{lstlisting}

\begin{lstlisting}[caption=File \texttt{gravity\_qoi.h}., label=code:gravity_qoi_h, linerange={32-1000}]
#define QUESO_EXAMPLE_GRAVITY_QOI_H

#include <queso/VectorFunction.h>
#include <queso/DistArray.h>

template<class P_V = QUESO::GslVector, class P_M = QUESO::GslMatrix,
         class Q_V = QUESO::GslVector, class Q_M = QUESO::GslMatrix>
class Qoi : public QUESO::BaseVectorFunction<P_V, P_M, Q_V, Q_M>
{
public:
  Qoi(const char * prefix, const QUESO::VectorSet<P_V, P_M> & domainSet,
      const QUESO::VectorSet<Q_V, Q_M> & imageSet);
  virtual ~Qoi();
  virtual void compute(const P_V & domainVector, const P_V * domainDirection,
      Q_V & imageVector, QUESO::DistArray<P_V *> * gradVectors,
      QUESO::DistArray<P_M *> * hessianMatrices,
      QUESO::DistArray<P_V *> * hessianEffects) const;

  void setAngle(double angle);
  void setInitialVelocity(double velocity);
  void setInitialHeight(double height);

private:
  double m_angle;
  double m_initialVelocity;
  double m_initialHeight;
};

#endif
\end{lstlisting}

\begin{lstlisting}[caption=File \texttt{gravity\_qoi.C}., label=code:gravity_qoi_C, linerange={32-96}]

#include <queso/GslVector.h>
#include <queso/GslMatrix.h>
#include <gravity_qoi.h>

template<class P_V, class P_M, class Q_V, class Q_M>
Qoi<P_V, P_M, Q_V, Q_M>::Qoi(const char * prefix,
    const QUESO::VectorSet<P_V, P_M> & domainSet,
    const QUESO::VectorSet<Q_V, Q_M> & imageSet)
  : QUESO::BaseVectorFunction<P_V, P_M, Q_V, Q_M>(prefix, domainSet, imageSet),
    m_angle(M_PI / 4.0),
    m_initialVelocity(5.0),
    m_initialHeight(0.0)
{
}

template<class P_V, class P_M, class Q_V, class Q_M>
Qoi<P_V, P_M, Q_V, Q_M>::~Qoi()
{
  // Deconstruct here
}

template<class P_V, class P_M, class Q_V, class Q_M>
void
Qoi<P_V, P_M, Q_V, Q_M>::compute(const P_V & domainVector,
    const P_V * domainDirection,
    Q_V & imageVector, QUESO::DistArray<P_V *> * gradVectors,
    QUESO::DistArray<P_M *> * hessianMatrices,
    QUESO::DistArray<P_V *> * hessianEffects) const
{
  if (domainVector.sizeLocal() != 1) {
    queso_error_msg("domainVector does not have size 1");
  }
  if (imageVector.sizeLocal() != 1) {
    queso_error_msg("imageVector does not have size 1");
  }

  double g = domainVector[0];  // Sample of the RV 'gravity acceleration'
  double distanceTraveled = 0.0;
  double aux = m_initialVelocity * std::sin(m_angle);
  distanceTraveled = (m_initialVelocity * std::cos(m_angle) / g) *
    (aux + std::sqrt(std::pow(aux, 2) + 2.0 * g * m_initialHeight));

  imageVector[0] = distanceTraveled;
}

template class Qoi<QUESO::GslVector, QUESO::GslMatrix, QUESO::GslVector,
                   QUESO::GslMatrix>;
\end{lstlisting}

\subsection{Input File}\label{sec:gravity-input-file}

QUESO reads an input file for solving statistical problems.
In the case of a SIP, it expects a list of options for MCMC, while in case of SFP it expects a list of 
options for Monte Carlo. The  input file `\texttt{gravity\_inv\_fwd.inp} used in this example is 
presented in Listing \ref{code:gravity_inv_fwd}.

\begin{lstlisting}[caption=Some options for QUESO library used in application code (Listings \ref{code:gravity_main}-\ref{code:gravity_like_C})., label={code:gravity_inv_fwd}]
###############################################
# UQ Environment
###############################################
env_numSubEnvironments   = 1 
env_subDisplayFileName   = outputData/display_env
env_subDisplayAllowAll   = 0
env_subDisplayAllowedSet = 0 1 2 3 4 5 6 7
env_displayVerbosity     = 2 
env_seed                 = -1 

###############################################
# Statistical inverse problem (ip)
###############################################
ip_computeSolution      = 1
ip_dataOutputFileName   = outputData/sip_gravity
ip_dataOutputAllowedSet = 0 1

###############################################
# Information for Metropolis-Hastings algorithm
###############################################
ip_mh_dataOutputFileName   = outputData/sip_gravity
ip_mh_dataOutputAllowedSet = 0 1

ip_mh_rawChain_dataInputFileName    = . 
ip_mh_rawChain_size                 = 20000
ip_mh_rawChain_generateExtra        = 0
ip_mh_rawChain_displayPeriod        = 2000
ip_mh_rawChain_measureRunTimes      = 1
ip_mh_rawChain_dataOutputFileName   = outputData/sip_gravity_raw_chain
ip_mh_rawChain_dataOutputAllowedSet = 0 1 2 3 4 5 6 7

ip_mh_displayCandidates             = 0
ip_mh_putOutOfBoundsInChain         = 0 
ip_mh_dr_maxNumExtraStages          = 3
ip_mh_dr_listOfScalesForExtraStages = 5. 10. 20.
ip_mh_am_initialNonAdaptInterval    = 0 
ip_mh_am_adaptInterval              = 100
ip_mh_am_eta                        = 1.98  	#(2.4^2)/d, d is the dimension of the problem
ip_mh_am_epsilon                    = 1.e-5

ip_mh_filteredChain_generate             = 1
ip_mh_filteredChain_discardedPortion     = 0.
ip_mh_filteredChain_lag                  = 20
ip_mh_filteredChain_dataOutputFileName   = outputData/sip_gravity_filtered_chain
ip_mh_filteredChain_dataOutputAllowedSet = 0 1

###############################################
# Statistical forward problem (fp)
###############################################
fp_help                 = anything
fp_computeSolution      = 1
fp_computeCovariances   = 1
fp_computeCorrelations  = 1
fp_dataOutputFileName   = outputData/sfp_gravity
fp_dataOutputAllowedSet = 0 1

###############################################
# 'fp_': information for Monte Carlo algorithm
###############################################
fp_mc_help                 = anything
fp_mc_dataOutputFileName   = outputData/sfp_gravity
fp_mc_dataOutputAllowedSet = 0 1

fp_mc_pseq_dataOutputFileName   = outputData/sfp_gravity_p_seq
fp_mc_pseq_dataOutputAllowedSet = 0 1

fp_mc_qseq_dataInputFileName    = . 
fp_mc_qseq_size                 = 16384
fp_mc_qseq_displayPeriod        = 20000
fp_mc_qseq_measureRunTimes      = 1
fp_mc_qseq_dataOutputFileName   = outputData/sfp_gravity_qoi_seq
fp_mc_qseq_dataOutputAllowedSet = 0 1

\end{lstlisting}

Moreover, for the gravity inverse problem, one may notice that QUESO will use the Metropolis-Hastings algorithm 
to sample the posterior PDF
(indicated by the prefix \texttt{mh\_}in the variable names) without adaptive steps
(indicated by the zero value assigned to the variable \linebreak \texttt{ip\_mh\_am\_initialNonAdaptInterval}, 
which can also be achieved by setting zero to \linebreak
\verb+ip_mh_am_adaptInterval+) and with delayed rejection (indicated by the one-value assigned to the 
variable \texttt{ip\_mh\_dr\_maxNumExtraStages}).

\subsection{Create your own Makefile}\label{sec:gravity-makefile}

Listing \ref{code:makefile} presents a Makefile, named \texttt{Makefile\_example\_violeta}, that may be 
used to compile the code and create the executable \verb+gravity_gsl+. Naturally, it must be adapted 
to the user's settings, i.e., it has to have the correct paths for the user's libraries that were actually 
used to compile and install QUESO (see Sections \ref{sec:Pre_Queso}--\ref{sec:install_Queso_make}).

\begin{lstlisting}[caption={Makefile for the application code in Listings
  \ref{code:gravity_main}-\ref{code:gravity_like_C}},
  label={code:makefile},
  language={bash}]
  QUESO_DIR = /path/to/queso
  BOOST_DIR = /path/to/boost
  GSL_DIR   = /path/to/gsl

  INC_PATHS = \
     -I. \
     -I$(QUESO_DIR)/include \
     -I$(BOOST_DIR)/include \
     -I$(GSL_DIR)/include

  LIBS = \
     -L$(QUESO_DIR)/lib -lqueso \
     -L$(BOOST_DIR)/lib -lboost_program_options \
     -L$(GSL_DIR)/lib -lgsl

  CXX = mpic++
  CXXFLAGS += -g -Wall -c

  default: all

  .SUFFIXES: .o .C

  all:       example_gravity_gsl

  clean:
     rm -f *~
     rm -f *.o
     rm -f gravity_gsl

  example_gravity_gsl: gravity_main.o gravity_likelihood.o gravity_compute.o gravity_qoi.o
     $(CXX) gravity_main.o \
            gravity_likelihood.o \
            gravity_compute.o \
            gravity_qoi.o \
            -o gravity_gsl $(LIBS)

  %.o: %.C
     $(CXX) $(INC_PATHS) $(CXXFLAGS) $<
\end{lstlisting}

\subsection{Running the Gravity Example with Several Processors}

Even though the application described in Section \ref{sec:gravity_code} is a serial code, it is 
possible to run it using more than one processor, i.e., in parallel mode. 
Supposing the user's workstation has $N_p=8$ processors, then, the user my choose to have $N_s =$ 
8, 4 or 2 subenvironments. This complies with the requirement that the total number of processors 
in the environment must be a multiple of the specified number of subenvironments.

Thus, to build and run the application code with $N_p = 8$, and $N_s=8$ subenvironments, the must 
set the variable \texttt{env\_numSubEnvironments = 8} in the input file 
(Listing~\ref{code:gravity_inv_fwd}) and enter the following command lines:

\begin{lstlisting}[caption={}, label={},language={bash}]
cd $HOME/LIBRARIES/QUESO-0.51.0/examples/gravity/
mpirun -np 8 ./gravity_gsl gravity_inv_fwd.inp
\end{lstlisting}

The steps above will create a total number of 8 raw chains, of size defined by the variable \texttt{ip\_mh\_rawChain\_size}. QUESO internally combines these 8 chains into a single chain of size $8\; \times\,$\texttt{ip\_mh\_rawChain\_size} and saves it in a file named according to the variable \texttt{ip\_mh\_rawChain\_dataOutputFileName}. 
QUESO also provides the user with the option of writing each chain -- handled by its corresponding processor -- in a separate file, which is accomplished by setting the variable \texttt{ip\_mh\_rawChain\_dataOutputAllowedSet = 0 1 ... Ns-1}.\\

\noindent
{\bf Note:} Although the discussion in the previous paragraph refers to the raw chain of a SIP, the analogous is true for the filtered chains (SIP), and for the samples employed in the SFP (\texttt{ip\_mh\_filteredChain\_size},    \texttt{fp\_mc\_qseq\_size} and \texttt{fp\_mc\_qseq\_size}, respectively).

\subsection{Data Post-Processing and Visualization}\label{sec:gravity-results}

According to the specifications of the input file in Listing~\ref{code:gravity_inv_fwd}, 
both a folder named \verb+outputData+ and a the following files should be generated:
\begin{verbatim}
sfp_gravity_sub0.m,         sip_gravity_sub0.m, 
sfp_gravity_p_seq.m,        sip_gravity_filtered_chain.m,,
sfp_gravity_p_seq_sub0.m    sip_gravity_filtered_chain_sub0.m,
sfp_gravity_qoi_seq.m,      sip_gravity_raw_chain.m,       
sfp_gravity_qoi_seq_sub0.m  sip_gravity_raw_chain_sub0.m,
display_env_sub0.txt 
\end{verbatim}

In this section, a convenient capability of QUESO of internally handling possible conflicts in 
chain size is presented. Recalling the input file \verb+gravity_inv_fwd.inp+ presented in 
Listing~\ref{code:gravity_inv_fwd}, one may notice that  the raw chain size for the SIP is 
chosen to have 20000 positions (\verb+ip_mh_rawChain_size = 20000+); the lag of the filtered chain 
is chosen to be 20 (\verb+ip_mh_filteredChain_lag = 20+) and the chain size for the SFP has 16384 
positions (\verb+fp_mc_qseq_size = 16384+). Because the solution of the SIP, ie, the posterior PDF, 
is used as input PDF for the SFP, QUESO internally sets \verb+fp_mc_qseq_size = 20000+, as can be 
seen in the file \verb+display_env_sub0.txt+.  The file \verb+display_env_sub0.txt+ contains information 
from the subenvironment `0' that was generated during the run of the application code.

\subsubsection{Statistical Inverse Problem}

There are a few Matlab-ready commands that are very helpful tools for post-processing the data 
generated by QUESO when solving statistical inverse problems.
This section discusses the results computed by QUESO with the code of Section 
\ref{sec:gravity_code}, and shows how to use Matlab for the post-processing of such results.

\paragraph{Chain Plots}\

\begin{figure}[p]
\centering 
\subfloat[Raw chain]{\includegraphics[scale=0.35]{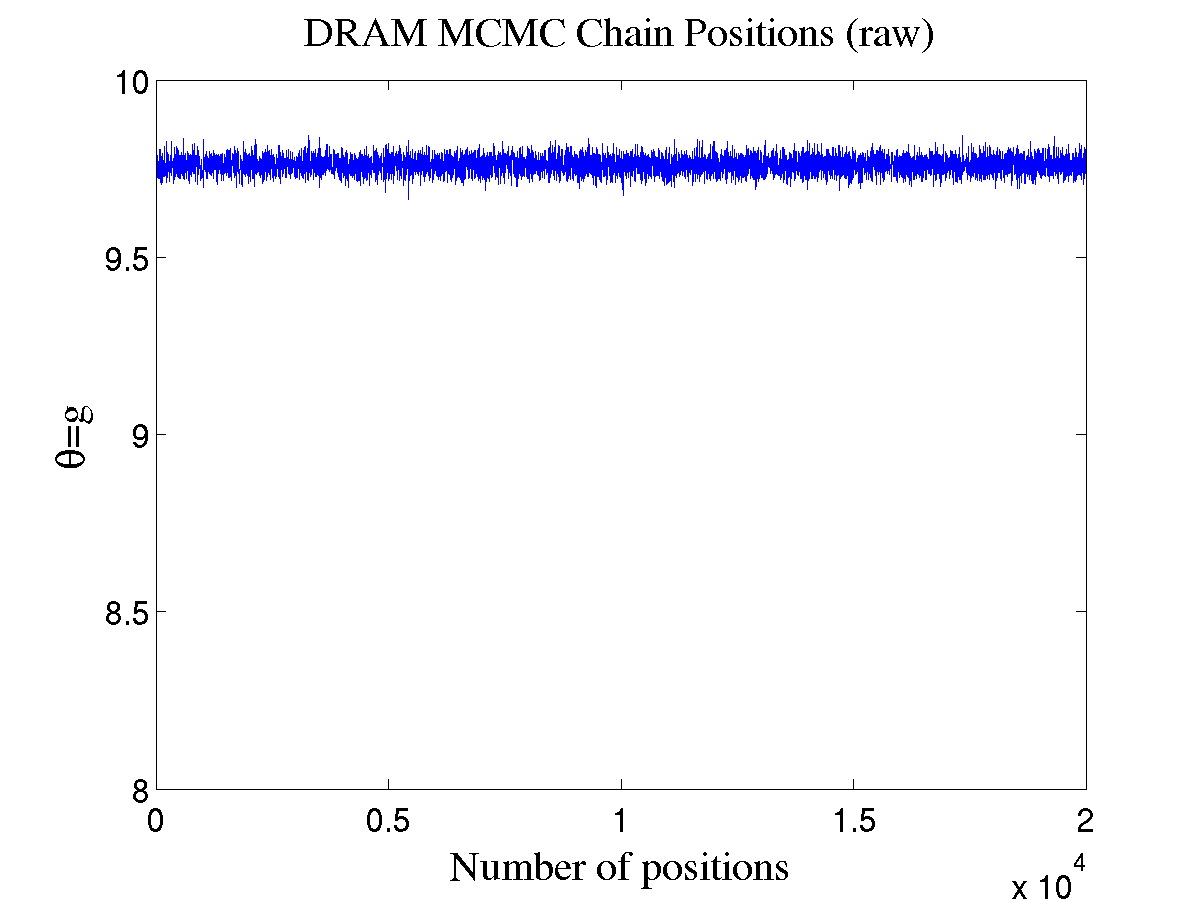}\label{fig:sip_gravity_chain_pos_raw}}
\subfloat[Filtered chain]{\includegraphics[scale=0.35]{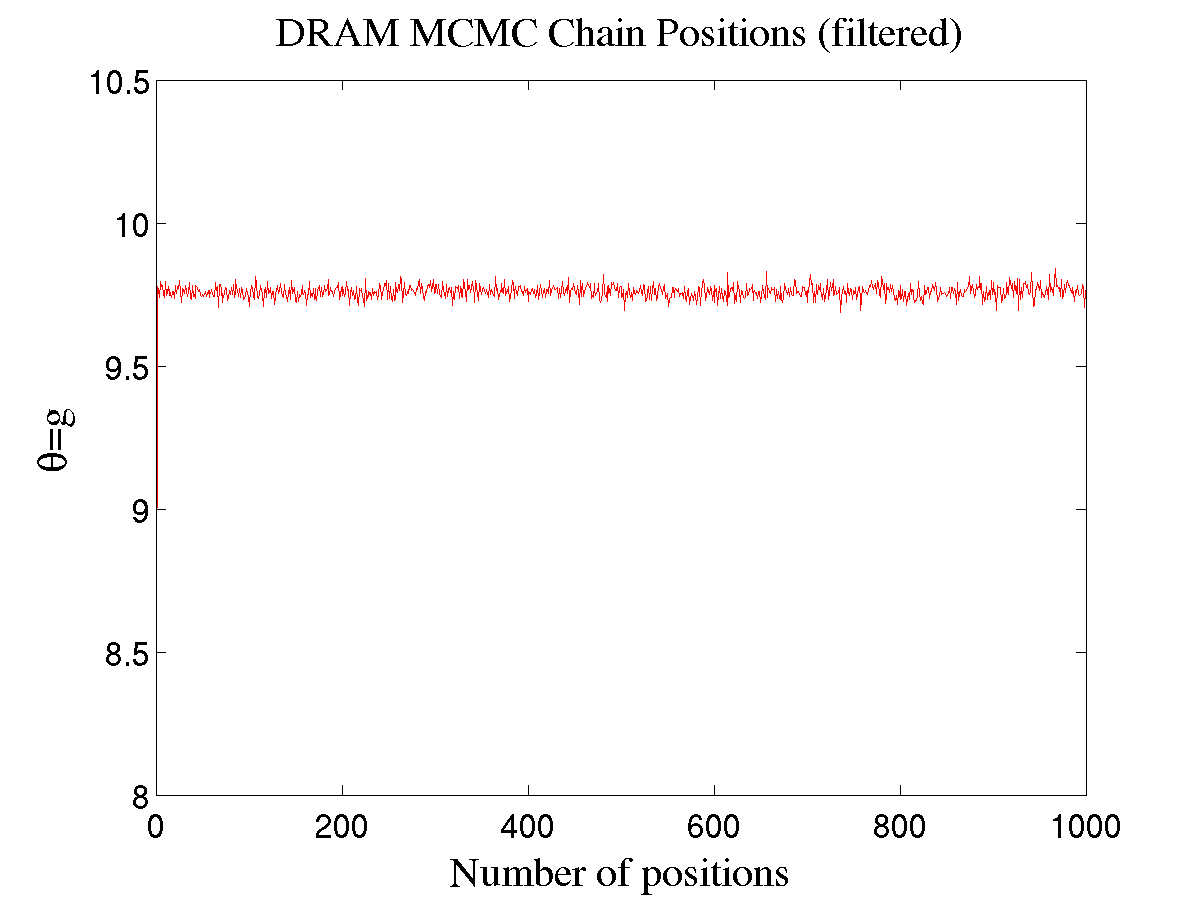}\label{fig:sip_gravity_chain_pos_filtered}}
\vspace*{-10pt}
\caption{MCMC raw chain with \chainsizeresults{} positions and a filtered chain with lag of 20 positions.}
\end{figure}

It is quite simple to plot, using Matlab, the chain of positions used in the DRAM algorithm 
implemented within QUESO. 
The sequence of Matlab commands presented in Listing \ref{matlab:chain} generates the 
graphic depicted in Figure \ref{fig:sip_gravity_chain_pos_raw}.
Figure~\ref{fig:sip_gravity_chain_pos_filtered} is obtained analogously. 

\begin{lstlisting}[label=matlab:chain,caption={Matlab code for the chain plot.}]
% inside Matlab
>> sip_gravity_raw_chain
>> plot(ip_mh_rawChain_unified)
>> ylabel('\theta=g','fontsize',20);
>> xlabel('Number of positions','fontsize',20);
>> title('DRAM Chain Positions (raw)','fontsize',20);
\end{lstlisting}

\paragraph{Histogram Plots}\

In order to plot histograms of the parameter using either the raw chain or the filtered chain, 
you simply have to use the pre-defined Matlab function \verb+hist+.

\begin{lstlisting}[label=matlab:hist,caption={Matlab code for the histogram plot.}]
% inside Matlab
>> sip_gravity_raw_chain
>> nbins=100;
>> hist(ip_mh_rawChain_unified,nbins)
>> title('Parameter Histogram (raw chain)','fontsize',20);
>> xlabel('Gravity (m/s^2)','fontsize',20);
>> ylabel('Frequency','fontsize',20);
>> grid on;
\end{lstlisting}

\begin{figure}[p]
\centering 
\subfloat[Raw chain]{\includegraphics[scale=0.35]{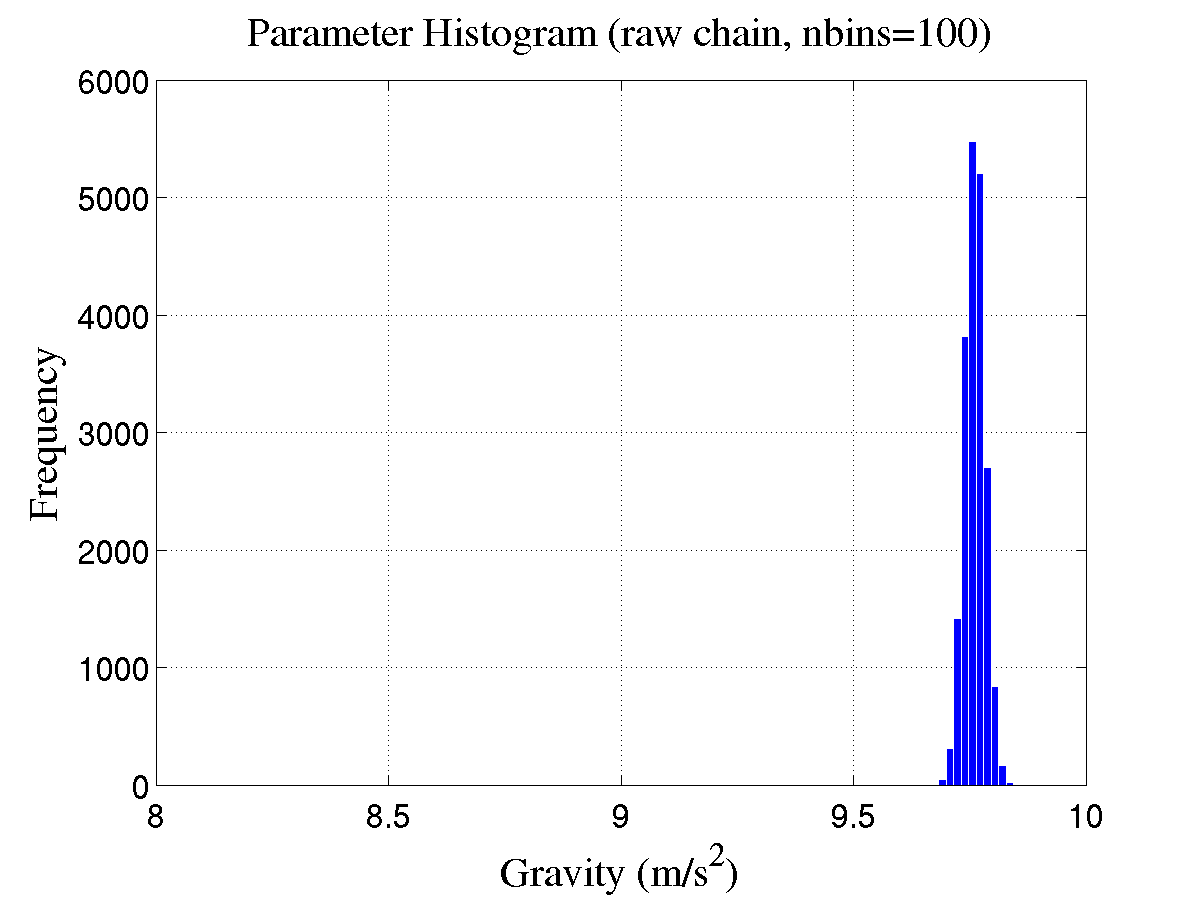}\label{fig:sip_gravity_hist_raw}}
\subfloat[Filtered chain]{\includegraphics[scale=0.35]{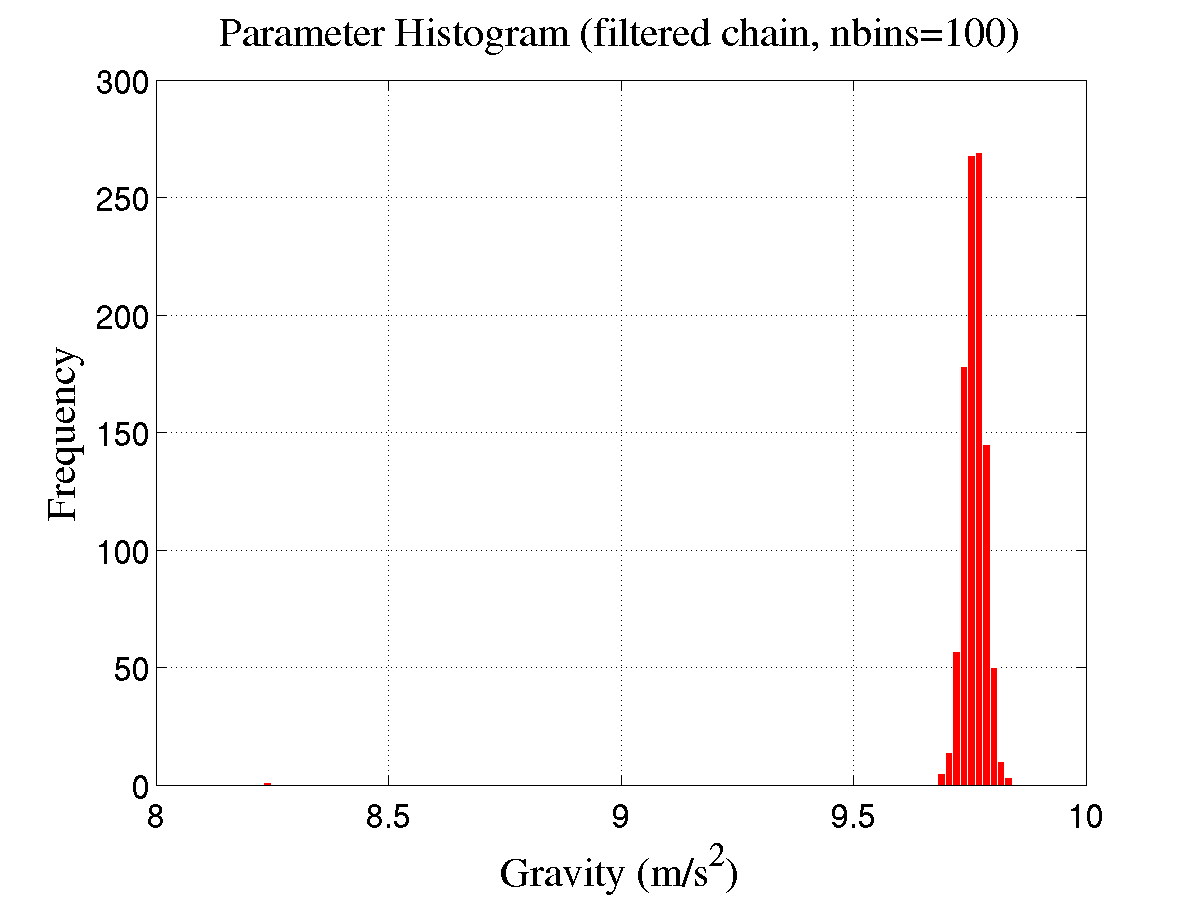}\label{fig:sip_gravity_hist_filtered}}
\vspace*{-10pt}
\caption{Histograms of parameter $\theta=g$. }
\end{figure}

\paragraph{KDE Plots} \

Matlab function \verb+ksdensity+ (Kernel smoothing density estimate) together with the 
option \verb+'pdf'+ may be used for plotting the KDE of the parameter.
\begin{lstlisting}[label=matlab:kde,caption={Matlab code for the KDE plot.}]
% inside Matlab
>> sip_gravity_raw_chain
>> [f,xi] = ksdensity(ip_mh_rawChain_unified,'function','pdf');
>> plot(xi,f,'-b','linewidth',3)
>> title('Parameter Kernel Density Estimation','fontsize',20);
>> xlabel('Gravity (m/s^2)','fontsize',20);
>> ylabel('KDE','fontsize',20);
>> grid on;
\end{lstlisting}

\begin{figure}[p]
\centering 
\subfloat[Raw chain]{\includegraphics[scale=0.35]{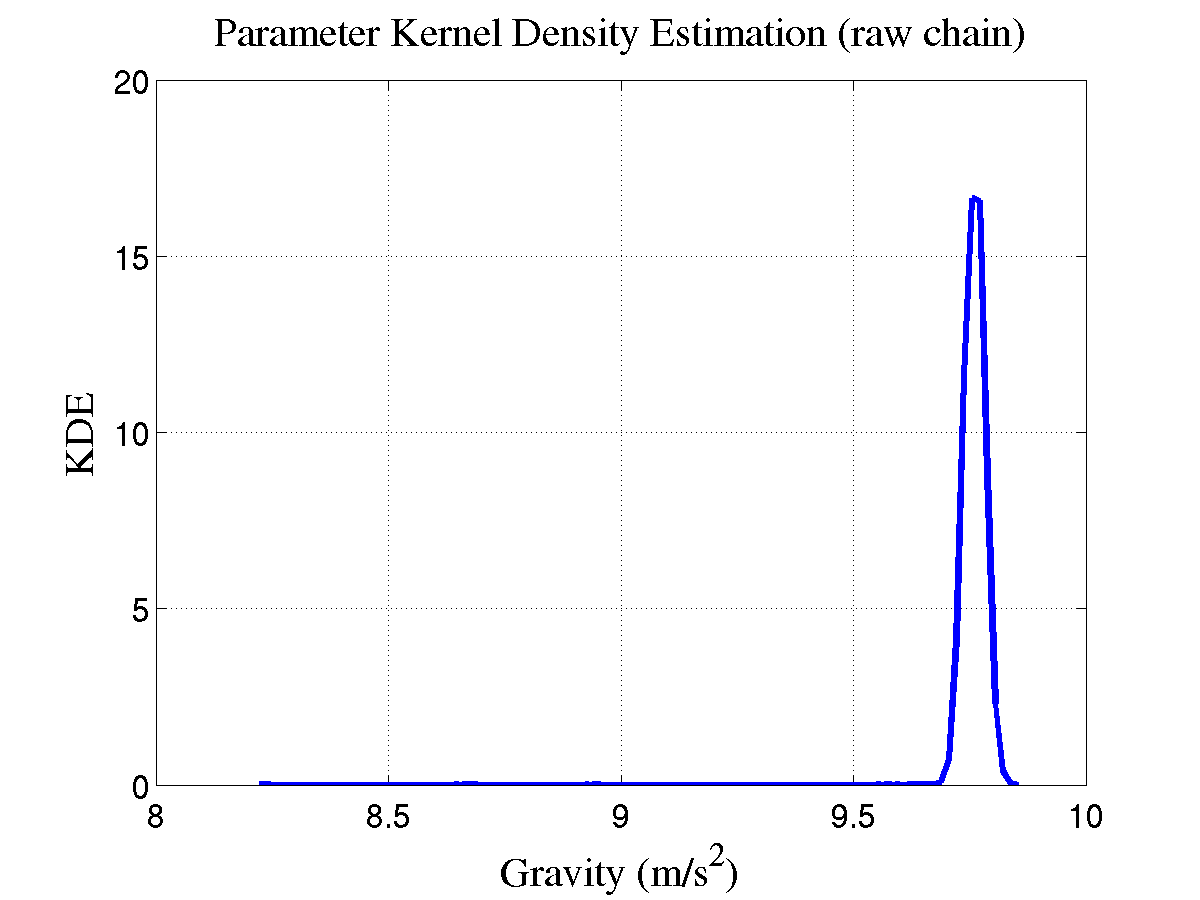}\label{fig:sip_gravity_kde_raw}}
\subfloat[Filtered chain]{\includegraphics[scale=0.35]{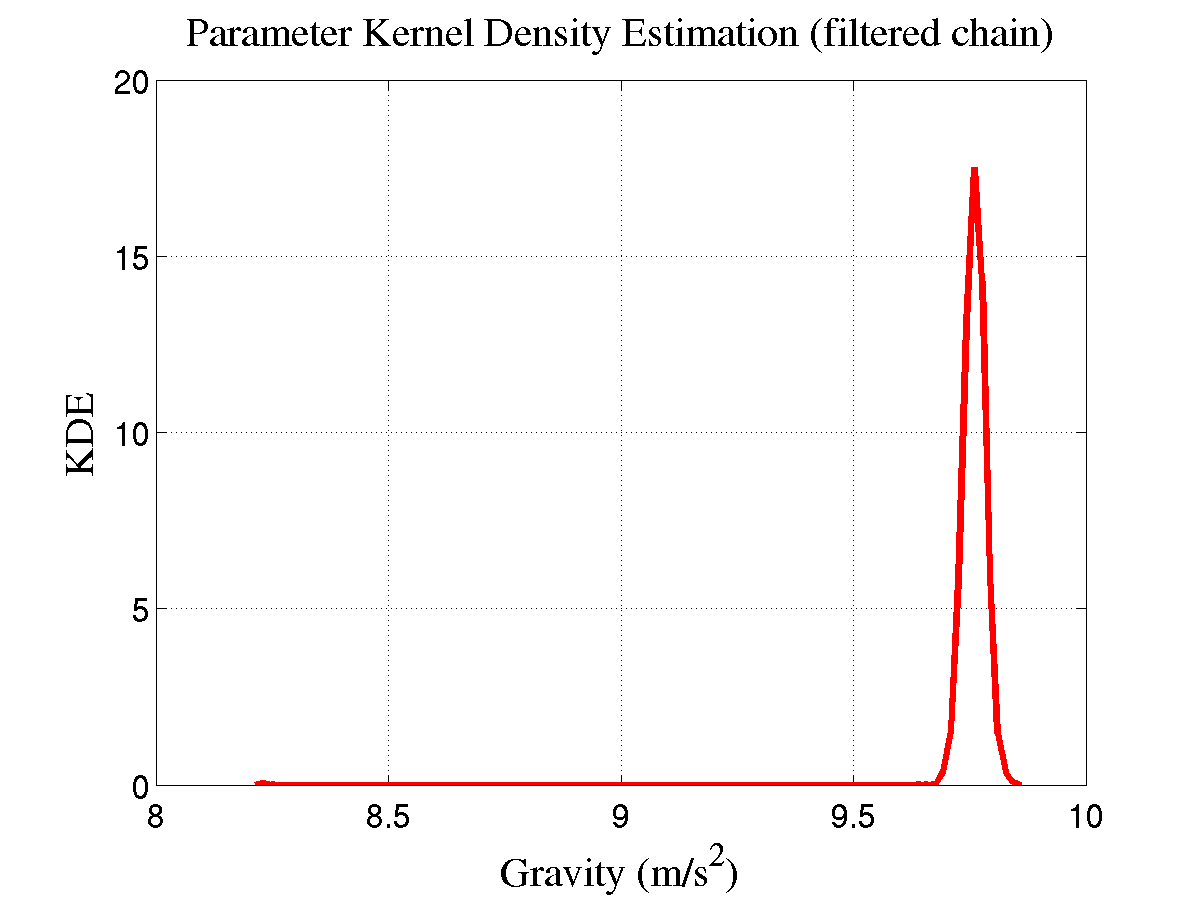}\label{fig:sip_gravity_kde_filtered}}
\vspace*{-10pt}
\caption{Kernel Density Estimation. }
\end{figure}

\paragraph{CDF Plots} \

Matlab function \verb+ksdensity+ (Kernel smoothing density estimate) with \verb+'cdf'+ option 
may also be used for plotting the Cumulative Distribution Function of the parameter.

\begin{lstlisting}[label=matlab:cdf,caption={Matlab code for the CDF plot.}]
% inside Matlab
>> sip_gravity_raw_chain
>> [f,xi] = ksdensity(ip_mh_rawChain_unified,'function','cdf');
>> plot(xi,f,'-b','linewidth',3)
>> title('Parameter Cumulative Distribution Function','fontsize',20);
>> xlabel('Gravity (m/s^2)','fontsize',20);
>> ylabel('CDF','fontsize',20);
>> grid on;
\end{lstlisting}

\begin{figure}[p]
\centering 
\subfloat[Raw chain]{\includegraphics[scale=0.35]{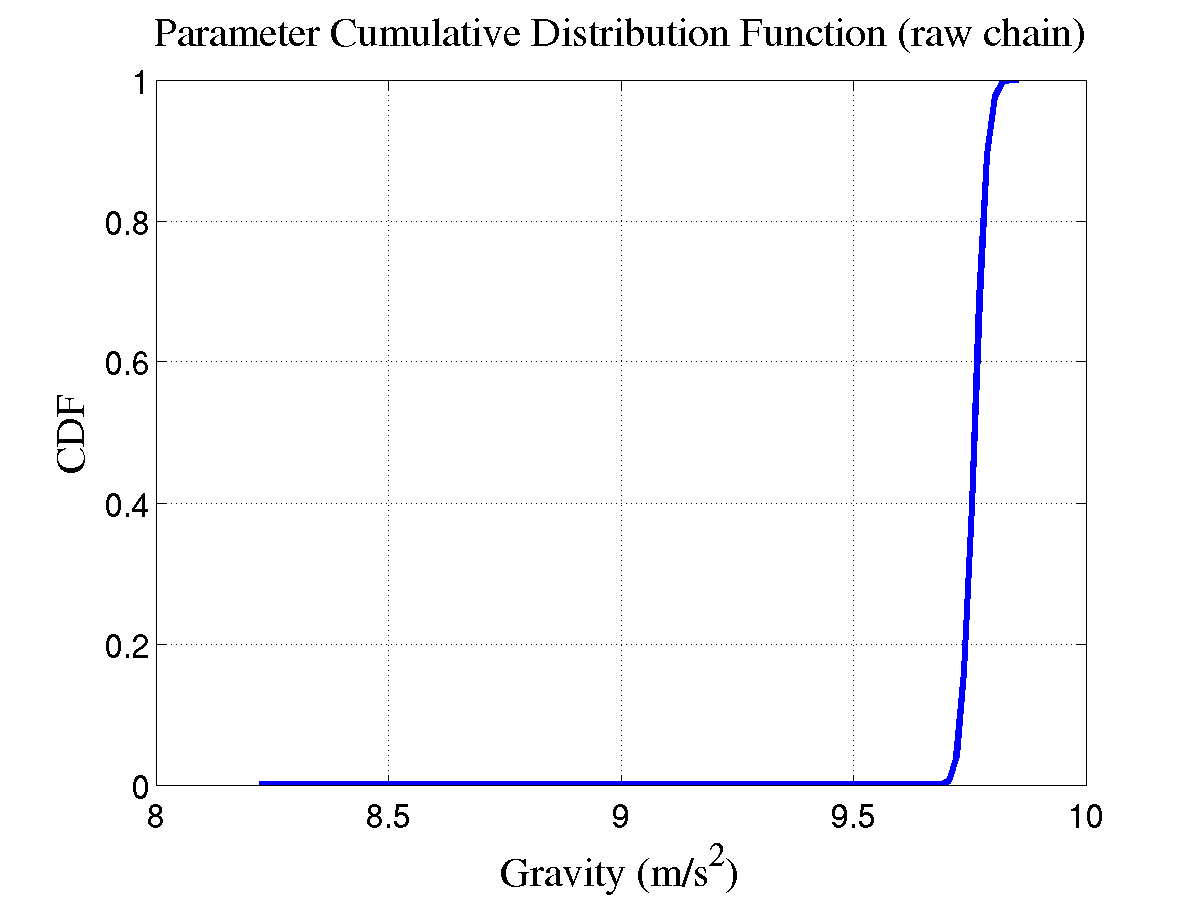}\label{fig:sip_gravity_cdf_raw}}
\subfloat[Filtered chain]{\includegraphics[scale=0.35]{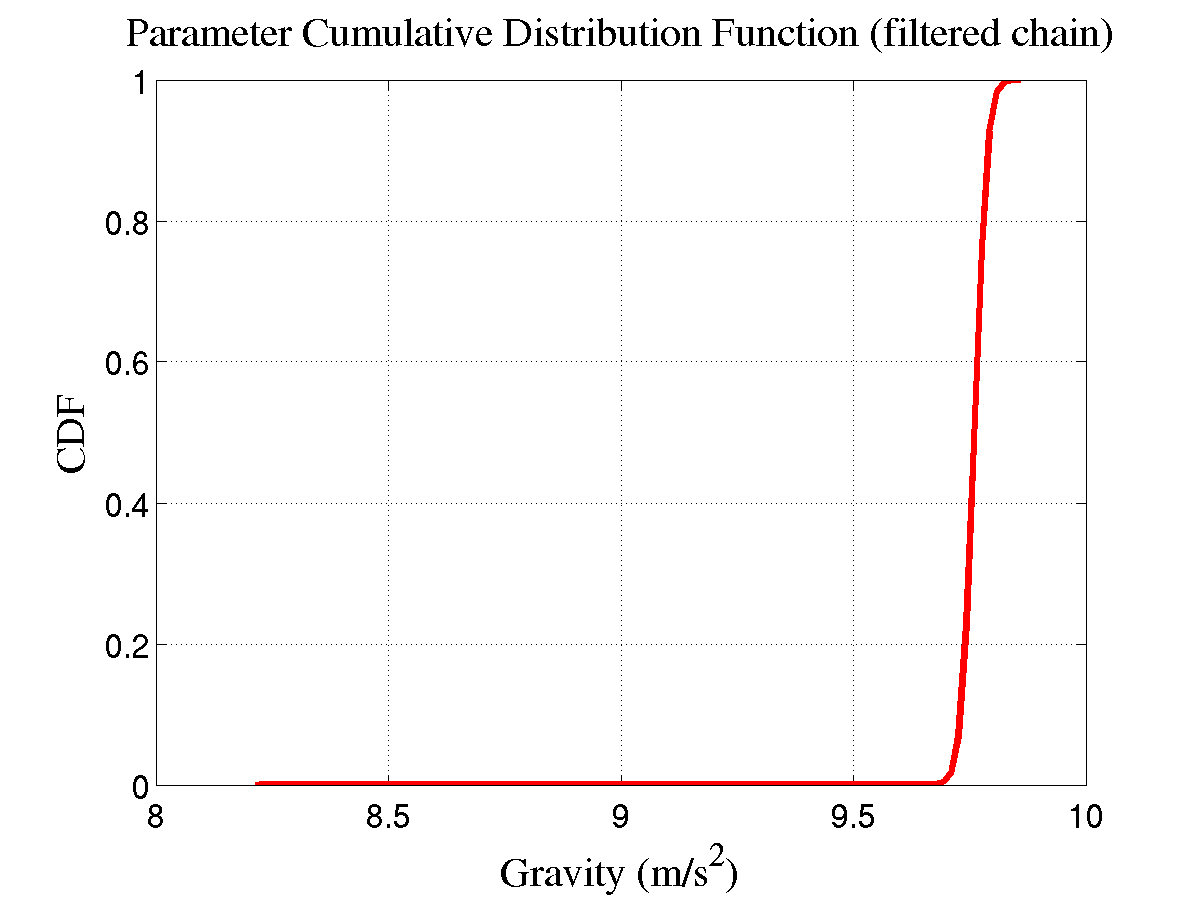}\label{fig:sip_gravity_cdf_filtered}}
\vspace*{-10pt}
\caption{Cumulative Distribution Function. }
\end{figure}

\paragraph{Autocorrelation Plots}\

\begin{figure}[p]
\centering
\includegraphics[scale=0.35]{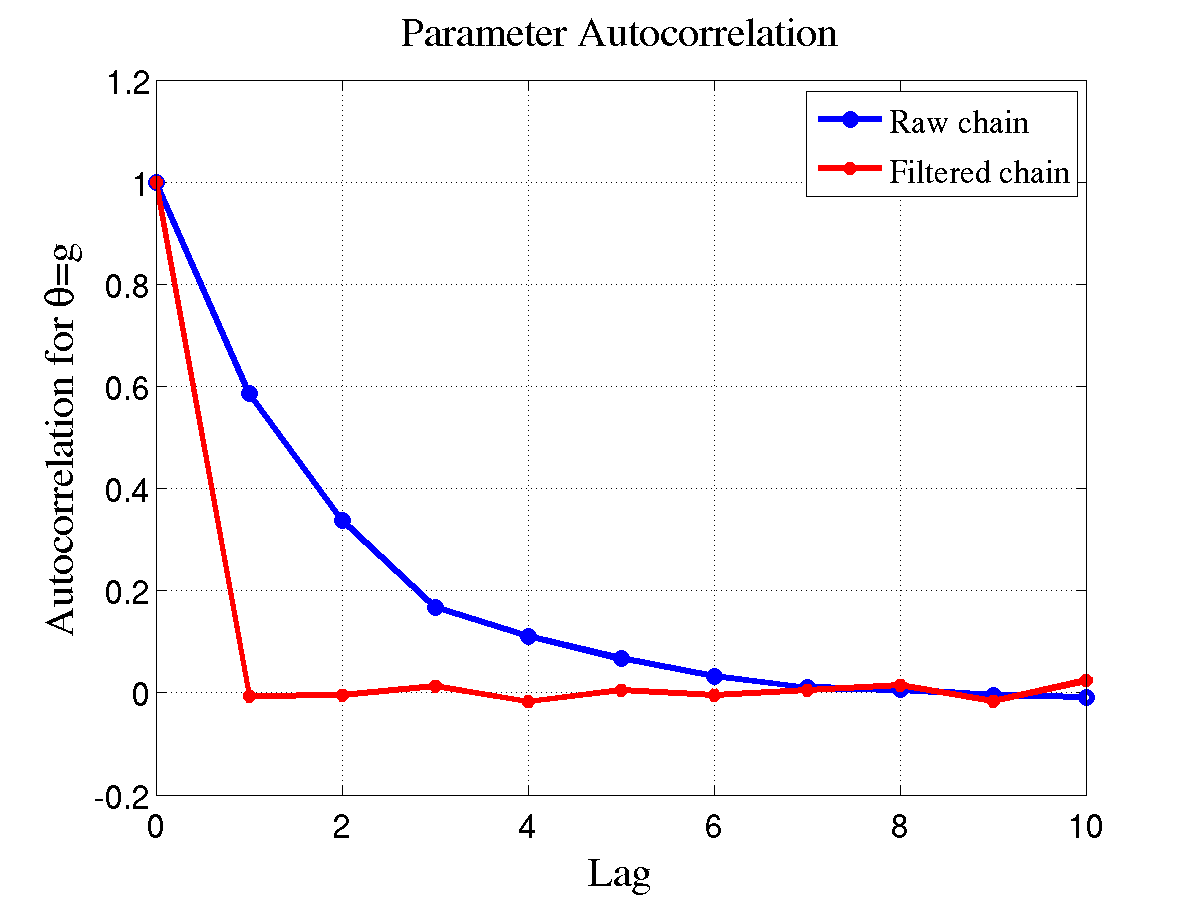}
\vspace{-8pt}
\caption{
Autocorrelation plots. }
\label{fig:sip_gravity_autocorrelation_raw_filt}
\end{figure}

The code presented in Listing \ref{matlab:autocorr} uses matlab function \verb+autocorr+ to generate 
Figure \ref{fig:sip_gravity_autocorrelation_raw_filt}
which presents the autocorrelation of the parameter $g$ in both cases: raw and filtered chain.

\begin{lstlisting}[label=matlab:autocorr,caption={Matlab code for the autocorrelation plots.}]
% inside Matlab
>> sip_gravity_raw_chain
>> sip_gravity_filtered_chain
>> nlags=10;
>> [ACF_raw,lags,bounds]= autocorr(ip_mh_rawChain_unified, nlags, 0);
>> [ACF_filt,lags,bounds]=autocorr(ip_mh_filtChain_unified,nlags, 0);
>> plot(lags,ACF_raw,'bo-',lags,ACF_filt,'r*-','linewidth',3);
>> ylabel('Autocorrelation for \theta=g','fontsize',20);
>> xlabel('Lag','fontsize',20);
>> title('Parameter Autocorrelation','fontsize',20);
>> grid on;
>> h=legend('raw chain','filtered chain','location','northeast');
>> set(h,'fontsize',16);
\end{lstlisting}

\paragraph{Covariance and Correlation Matrices}\

Matlab function \verb+cov+ calculates the covariance matrix for a data matrix 
(where each column represents a separate quantity), and \verb+corr+ calculates the correlation matrix.
Since our statistical inverse problem has only one parameter (the acceleration $g$ due to gravity), 
both covariance and correlation matrices have dimension $1 \times 1$, i.e., they are scalars.

\begin{lstlisting}[label=matlab:cov_matrix,caption={Matlab code for finding the covariance matrix.}]
% inside Matlab
>> sip_gravity_raw_chain;
>> cov_matrix_g = cov(ip_mh_rawChain_unified)
   
cov_matrix_g =

   6.8709e-04
>> corr_matrix_g = corr(ip_mh_rawChain_unified)

corr_matrix_g =

     1
>>
\end{lstlisting}

\newpage

\subsubsection{Statistical Forward Problem}

\paragraph{Chain Plots} \

\begin{figure}[h]
\centering 
\subfloat[Chain positions]{\includegraphics[scale=0.35]{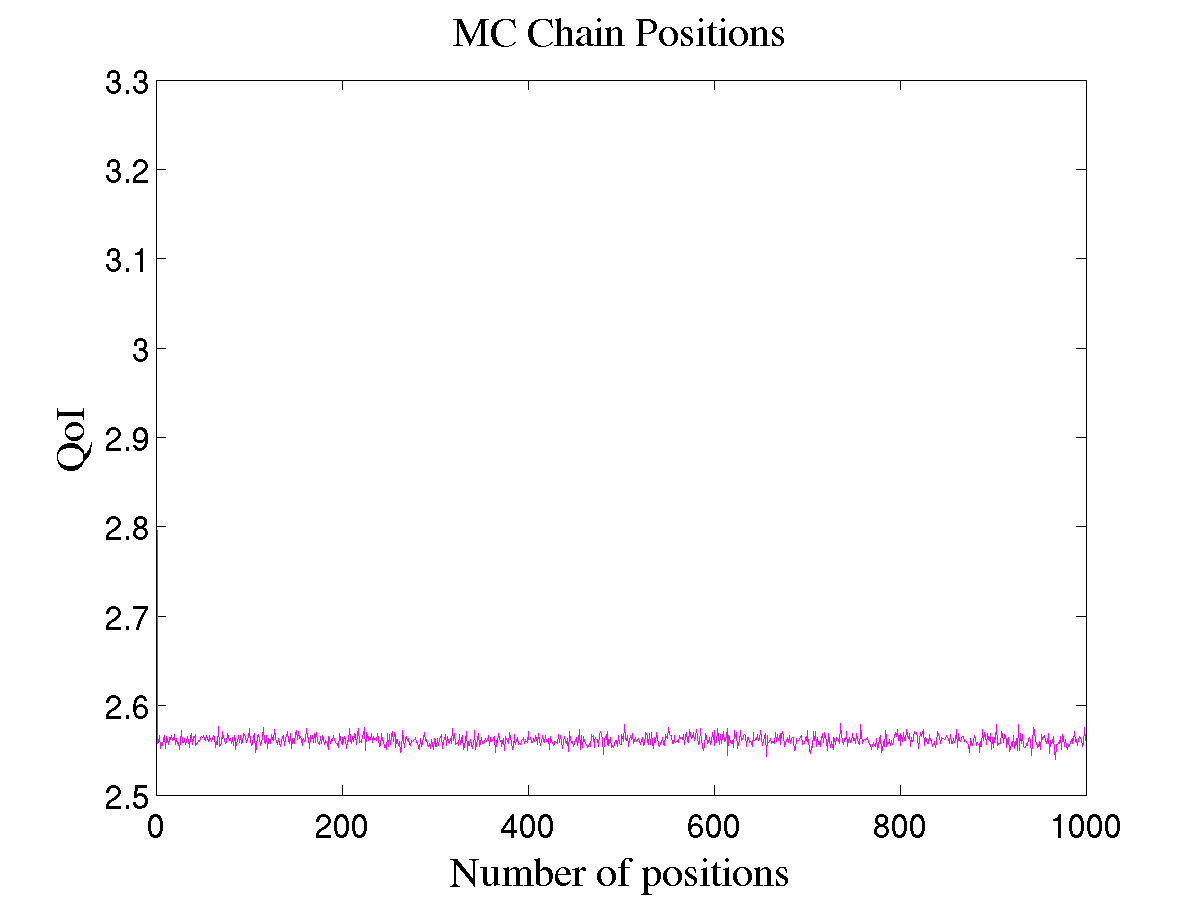}\label{fig:sfp_gravity_chain}}
\subfloat[Hystogram]{\includegraphics[scale=0.35]{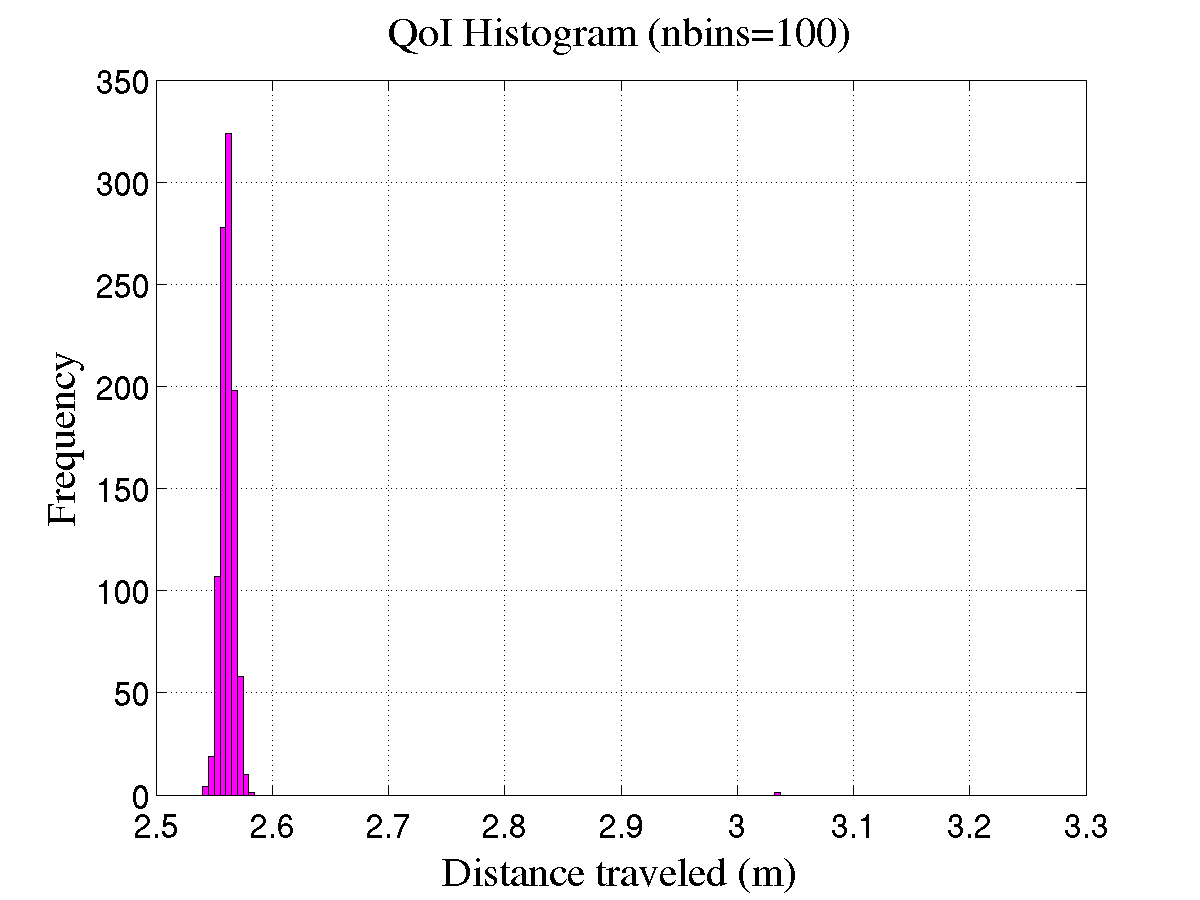}\label{fig:sfp_gravity_hist}}
\vspace{-6pt}
\caption{MC chain positions and histogram of QoI $=d$.}
\end{figure}

It is quite simple to plot, using Matlab, the chain of positions generated by the Monte Carlo algorithm 
implemented within QUESO and called during the solution of the statistical forward problem. 
The sequence of Matlab commands presented bellow generates the graphic depicted in 
Figure~\ref{fig:sfp_gravity_chain}. 

\begin{lstlisting}[label=matlab:chain_qoi,caption={Matlab code for the chain plot.}]
% inside Matlab
>> sfp_gravity_qoi_seq.m
>> plot(fp_mc_QoiSeq_unified);
>> ylabel('QoI','fontsize',20);
>> xlabel('Number of positions','fontsize',20);
>> title('MC Chain Positions','fontsize',20);
\end{lstlisting}


\paragraph{Histogram Plots} \

In order to plot a histogram of the QoI, you may use the pre-defined Matlab function \verb+hist+.
The Matlab code presented in below shows how to create the Figure~\ref{fig:sfp_gravity_hist}.

\begin{lstlisting}[label=matlab:hist_qoi,caption={Matlab code for the QoI histogram plot.}]
>> sfp_gravity_qoi_seq.m
>> nbins=100;
>> hist(fp_mc_QoiSeq_unified);
>> title('QoI Histogram','fontsize',20);
>> xlabel('Distance traveled (m)','fontsize',20);
>> ylabel('Frequency','fontsize',20);
>> grid on;
\end{lstlisting}

\paragraph{KDE Plots} \

\begin{figure}[htp]
\centering 
\subfloat[KDE]{\includegraphics[scale=0.35]{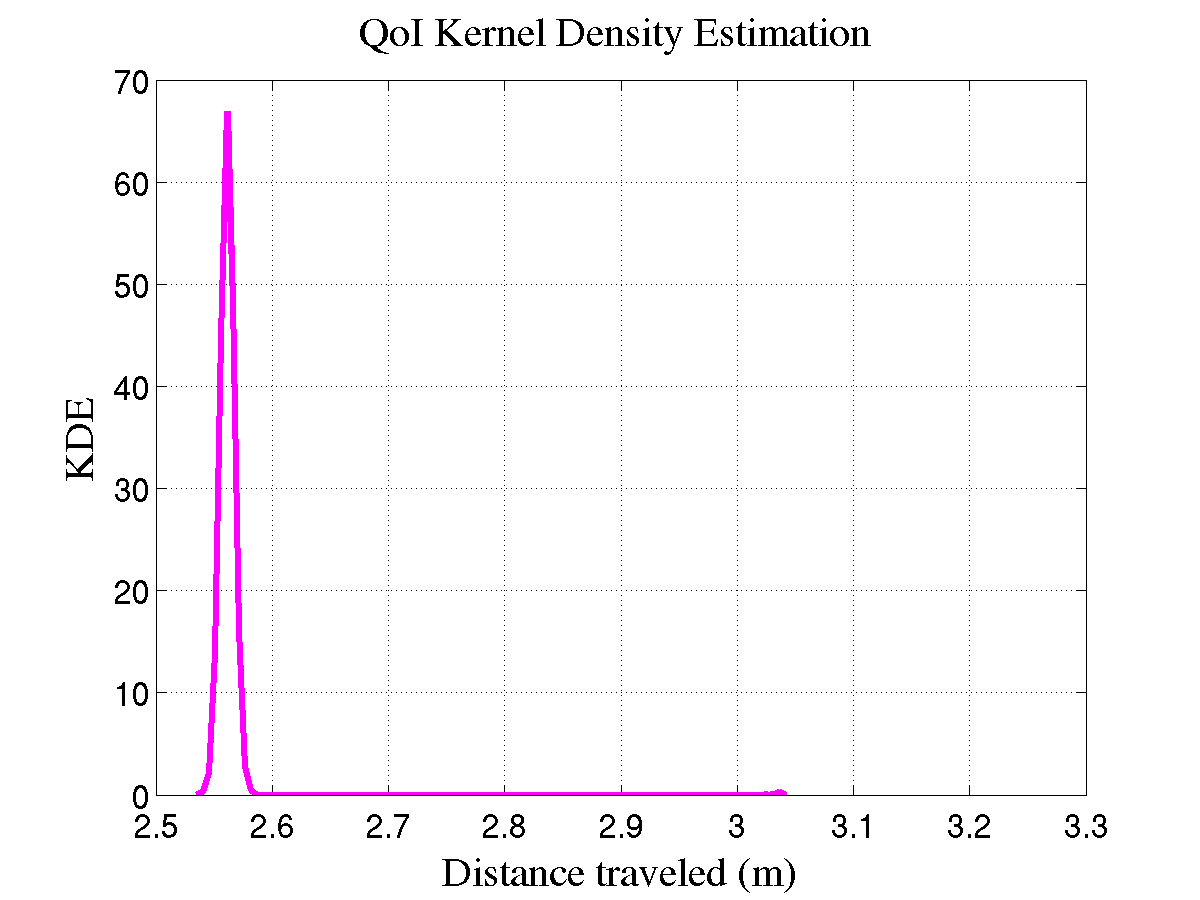}\label{fig:sfp_gravity_kde}}
\subfloat[CDF]{\includegraphics[scale=0.35]{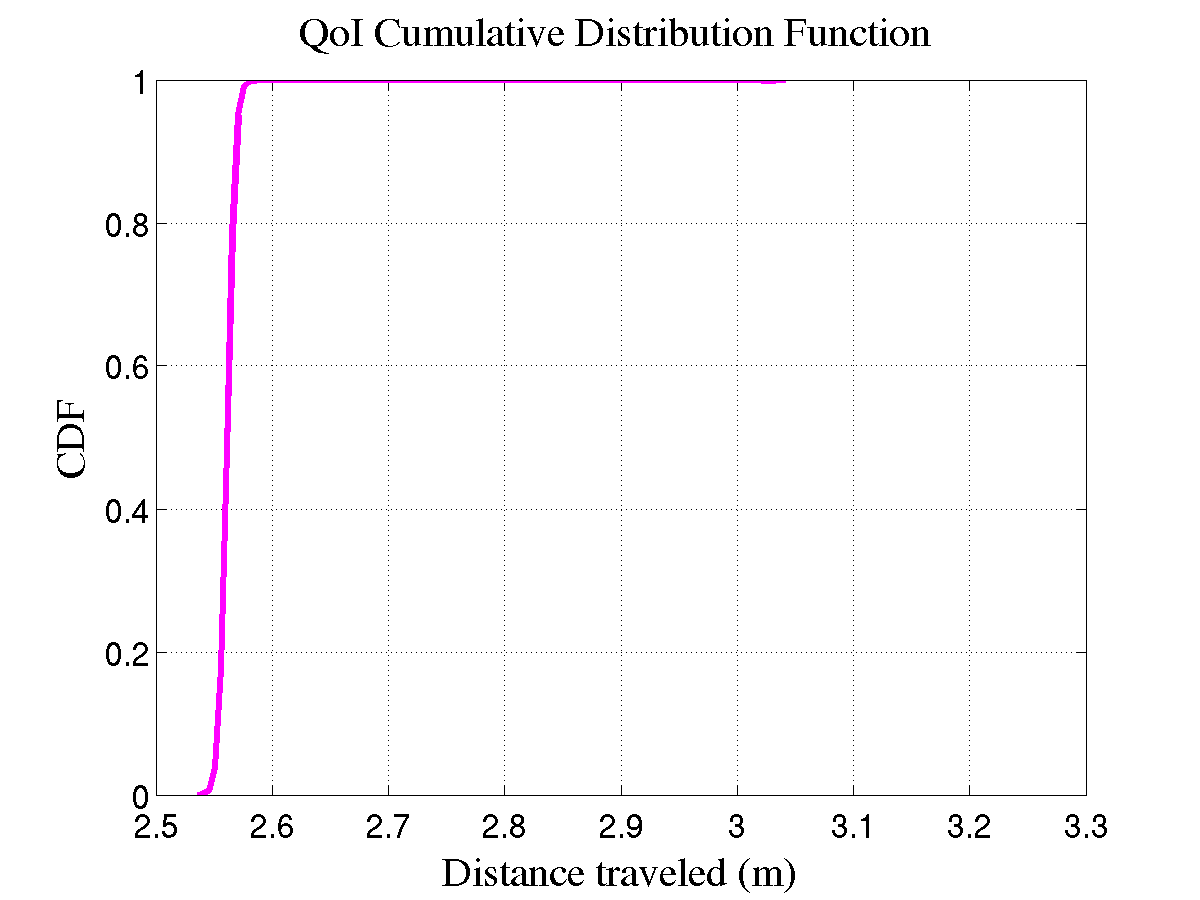}\label{fig:sfp_gravity_cdf}}
\vspace*{-6pt}
\caption{Kernel Density Estimation and Cumulative Distribution Function of QoI.}
\end{figure}

Matlab function \verb+ksdensity+ (Kernel smoothing density estimate) together with the 
option \verb+'pdf'+ may be used for plotting the KDE of the he QoI, displayed in Figure \ref{fig:sfp_gravity_kde}.

\begin{lstlisting}[label=matlab:kde_qoi,caption={Matlab code for the QoI KDE plot.}]
% inside Matlab
>> sfp_gravity_qoi_seq.m
>> [f,xi] = ksdensity(fp_mc_QoiSeq_unified,'function','pdf');
>> plot(xi,f,'-b','linewidth',3)
>> title('QoI Kernel Density Estimation ','fontsize',20);
>> xlabel('Distance traveled (m)','fontsize',20);
>> ylabel('KDE','fontsize',20);
>> grid on;
\end{lstlisting}



\paragraph{CDF Plots} \

Matlab function \verb+ksdensity+ (Kernel smoothing density estimate) with \verb+'cdf'+ option may also be 
used for plotting the Cumulative Distribution Function of the QoI, displayed in Figure~\ref{fig:sfp_gravity_cdf}.

\begin{lstlisting}[label=matlab:cdf_qoi,caption={Matlab code for the QoI CDF plot.}]
% inside Matlab
>> sfp_gravity_qoi_seq.m
>> [f,xi] = ksdensity(fp_mc_QoiSeq_unified,'function','cdf');
>> plot(xi,f,'-b','linewidth',3)
>> title('QoI Cumulative Distribution Function ','fontsize',20);
>> xlabel('Distance traveled (m)','fontsize',20);
>> ylabel('CDF','fontsize',20);
>> grid on;
\end{lstlisting}
%

\paragraph{Autocorrelation Plots} \

\begin{figure}[htp]
\centering
\includegraphics[scale=0.35]{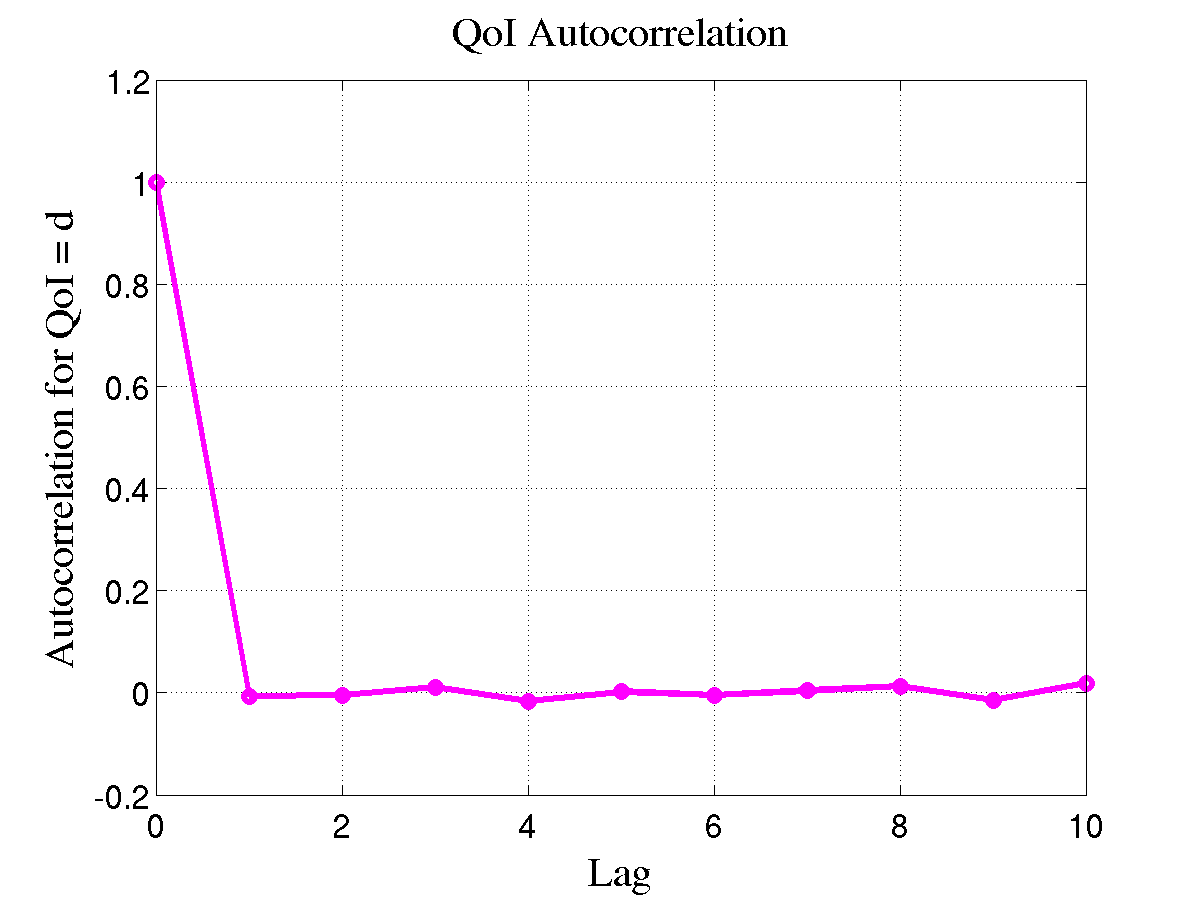}
\vspace*{-10pt}
\caption{Autocorrelation plot.}
\label{fig:sfp_gravity_autocorrelation}
\end{figure}

The code presented in Listing \ref{matlab:autocorr_qoi} uses Matlab function \verb+autocorr+ 
to generate Figure~\ref{fig:sfp_gravity_autocorrelation}, which presents the autocorrelation of the QoI $d$.

\begin{lstlisting}[label=matlab:autocorr_qoi,caption={Matlab code for the QoI autocorrelation plot.}]
% inside Matlab
>> sfp_gravity_qoi_seq.m
>> nlags=10;
>> [ACF, lags, bounds] = autocorr(fp_mc_QoiSeq_unified, nlags, 0);
>> plot(lags,ACF,'bo-','linewidth',3);
>> ylabel('Autocorrelation for QoI = d','fontsize',20);
>> xlabel('Lag','fontsize',20);
>> title('QoI Autocorrelation','fontsize',20);
>> grid on;
\end{lstlisting}

\paragraph{Covariance and Correlation Matrices} \

For a matrix input \verb+X+, where each row is an observation, and each column is a 
variable, the Matlab function \verb+cov(X)+ may be used to calculate the covariance matrix.

Thus,  in order to calculated the covariance matrix between the parameter and the quantity of interest 
sequences generated by Monte Carlo sampler with QUESO,
one may simply define \verb+X=[fp_mc_ParamSeq_unified fp_mc_QoiSeq_unified]+.
The code presented in Listing \ref{matlab:cov_pqoi} shows the usage of Matlab commands for finding such the matrix.

\begin{lstlisting}[label=matlab:cov_pqoi,caption={Matlab code for the matrix of covariance between parameter $g$ and QoI $d$.}]
% inside Matlab
>> sfp_gravity_qoi_seq;
>> sfp_gravity_p_seq;
>> X=[fp_mc_ParamSeq_unified fp_mc_QoiSeq_unified];
>> cov_p_QoI = cov(X)

cov_p_QoI =
	  [ 2.826e-03 	-8.555e-04 ] 
	  [-8.555e-04 	 2.599e-04 ]

\end{lstlisting}

Analogously, the Matlab function \verb+corrcoef(X)+ returns a matrix of correlation coefficients 
calculated from an input matrix \verb+X+ whose rows are observations and whose columns are variables.
In order to calculated the correlation matrix between the parameter and the QoI sequences, 
one may simply define \verb+X=[fp_mc_ParamSeq_unified fp_mc_QoiSeq_unified]+.

\begin{lstlisting}[label=matlab:corr_param_qoi,caption={Matlab code for the matrix of correlation between parameter $g$ and quantity of interest $d$.}]
% inside Matlab
>> sfp_gravity_qoi_seq;
>> sfp_gravity_p_seq;
>> X=[fp_mc_ParamSeq_unified fp_mc_QoiSeq_unified];
>> corr_p_QoI = corrcoef(X)

corr_p_QoI =
	  [ 1.000e+00 	-9.981e-01 ] 
	  [-9.981e-01 	 1.000e+00 ]
>>
\end{lstlisting}


\section{\texttt{validationCycle}}\label{sec:example_tga}

This is the last and more complex of all \Queso{} examples. 
In this example, we numerically solve a statistical inverse problem related to a thermogravimetric experiment, where a material sample has its mass measured while loosing it through a controlled heating process. 

Given a simple mass evolution model that has a temperature profile and material
properties as input parameters, and given thermogravimetric measurements with
prescribed variances, the statistical inverse problems ask for the
specification of the random variables that represent the unknown material
properties. We compute probability density functions with the Bayesian approach
and compute sets of realizations through the Metropolis-Hastings algorithm with
delayed rejection.

We qualitatively analyze the sensitivity of the solutions with respect to
problem characteristics, namely ``amount'' of data and ``quality'' of data, and
also with respect to algorithm options, namely chain initial position, number
of delayed rejections
and chain size.

\subsection{Thermogravimetric Experiments and a Simple Model}

Suppose a given material sample of initial mass $m_0$ and at initial temperature $T_0$ is heated with constant heating rate $\beta$ (K/min). Heating is maintained until the sample fully ablates (decomposes). The sample mass $m(T)$ is measured at temperatures $T>T_0$.
Let $w(T)~=~m(T)/m_0$ denote the mass fraction.

%
\begin{figure}[!ht]
\centering
\includegraphics[scale=0.4]{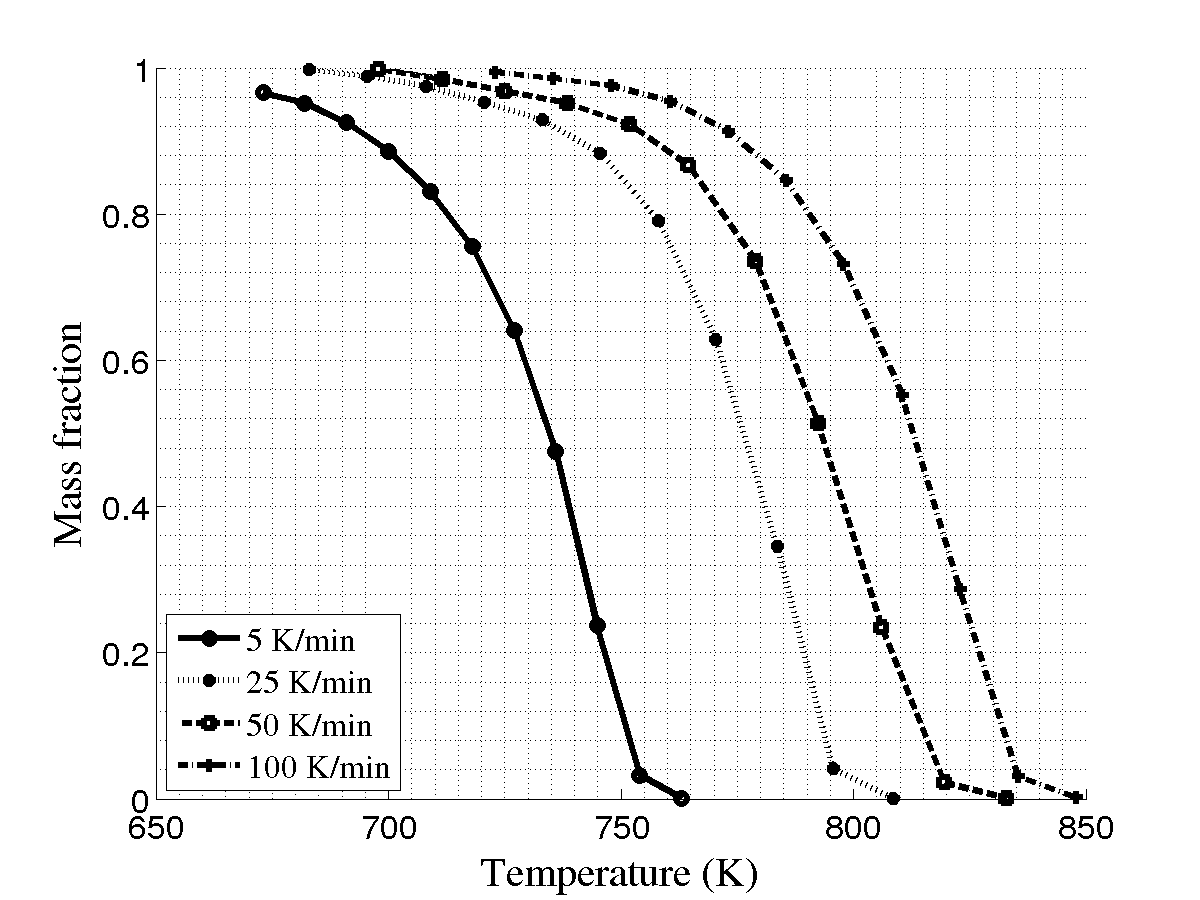}
\vspace*{-8pt}
\caption{Mass fraction decay over temperature given different heating rates. Data from J. A. Conesa, R. F. Marcilla and J. A. Caballero, ``Thermogravimetric studies on the thermal decomposition of polyethylene'', {\it J. Anal. Appl. Pyrolysis}, 36(1):1-15, 1996}
\label{fig:tga-exp-data}
\end{figure}

It is convenient to transform the kinetic equation to a \emph{per unit temperature}
form by dividing through by $\beta$. Thus, a simple approach to
the simulation of a thermogravimetric phenomenon consists on modeling the
sample as a homogeneous material of scalar properties $A>0$ and $E>0$ whose
relative mass $w$ obeys the following initial value ordinal differential
equation problem:
\begin{equation}\label{eq:tga-model}
\left\{
\begin{array}{rcll}
{\displaystyle \frac{dw}{dt} } & = & -\dfrac{A w}{\beta} \exp\left(-\frac{E}{RT}\right), & t \geqslant 0,\\
& & \\
w(0) & = & 1, &
\end{array}
\right.
\end{equation}
where the kinetic parameters $A$ and $E$ are referred to, respectively, as pre-exponential factor (min$^{-1}$) and activation energy (J/mol).

In this combined SIP--SFP, we  calibrate both model parameters $A$ and $E$ given the mathematical model in Equation \eqref{eq:tga-model} and experimental data (Section \ref{sec:tga-data}). Then the inferred values for $A$ and $E$ are then used for uncertainty propagation on the remaining mass fraction at time $t=3.9$ s when $\beta=250$ K/min, i.e., our quantity of interest is $w(t=3.9)$. 

\subsection{Statistical Inverse Problem}\label{sec:tga-sip}

Let $\mathbf{m}=(A,E)$ be the vector of model parameters and $M=\mathbb{R}_{+}^{2}$ be the space of model parameters.
Let
$V_T$ denote the space of functions $f:\mathbb{R}_{+}\rightarrow\mathbb{R}_{+}$ that are weakly differentiable.
$V_T$ will be the space of temperature profiles.
Finally, let
$V_w$ denote the space of functions $f:\mathbb{R}_{+}\rightarrow[0,1]$ that are weakly differentiable.
$V_w$ will be the space of relative mass evolutions.
We will denote by
\begin{equation*}
w(\mathbf{m},T)\in V_w
\end{equation*}
the solution of Equation \eqref{eq:tga-model} for given $\mathbf{m}\in M$ and $T\in V_T$.

%

\subsubsection{Misfit Functions $\mathcal{F}(\mathbf{m})$}

Let
$V_S$ denote the space of all functions $f:\mathbb{R}_{+}\rightarrow\mathbb{R}_{+}$ that are square-Lebesgue-integrable
over any finite interval.
$V_S$ will be the space of misfit weight functions.
Let
$V_{\sigma}$ denote the space of all functions $f:\mathbb{R}_{+}\rightarrow\mathbb{R}_+^{*}$ such that $1/f$ is square-Lebesgue-integrable
over any finite interval.
$V_{\sigma}$ will be the space of variance functions.

Given a reference relative mass evolution function $\text{$d$}\in V_w$,
a temperature profile $T\in V_T$,
and some $t_{_{\text{F}}}>0$,
let $\mathcal{F}:M\rightarrow\mathbb{R}$ be the functional defined by
\begin{equation*}
\mathcal{F}(\mathbf{m}) = \int_{0}^{t_{_{\text{F}}}}~\left\{[w(\mathbf{m},T)](t)-\text{$d$}(t)\right\}^2\cdot S(t)~dt,
\end{equation*}
or simply
\begin{equation}\label{eq-F}
\mathcal{F}(\mathbf{m}) = \int_{0}^{t_{_{\text{F}}}}~(w-d)^2\cdot S~dt.
\end{equation}

The functional \eqref{eq-F} is general enough for our studies, since it can properly describe
not only the case where one has continuous measurements $\text{$d$}$,
but also the case of a finite set of $N_{\text{meas}}$ discrete measurements $0\leqslant d_j\leqslant 1$,
$1\leqslant i\leqslant N_{\text{meas}}$ at instants $0\leqslant t_1 < t_2 < \ldots < t_{N_{\text{meas}}}$.

In the case of continuous measurements, for instance, one can set
\begin{equation*}
\mathcal{F}_1(\mathbf{m}) = \int_{0}^{t_{_{\text{F}}}}~\left\{[w(\mathbf{m},T)](t)-\text{$d$}(t)\right\}^2\cdot\frac{1}{\sigma^2(t)}~dt,
\end{equation*}
for some given variance function $\sigma^2\in V_S$ satisfying $\sigma(t)>0$ for all $t\geqslant 0$.

On the other hand, when measurements are discrete and a corresponding finite set of variances $\sigma_j^2>0,~j=1,2,\ldots,N_{\text{meas}}$ is given, one can set
\begin{equation*}
\mathcal{F}_2(\mathbf{m}) = \int_0^{t_{_F}}~\{[w(\mathbf{m},T)](t)-\hat{d}(t)\}^2\cdot\left[\sum_{j=1}^{N_{\text{meas}}}\frac{\delta(t-t_j)}{{\hat{\sigma}}^2(t)}\right]~dt,
\end{equation*}
where
$\hat{d}\in V_w$ and $\hat{\sigma}\in V_{\sigma}$ are any functions satisfying
$\hat{d}(t_j)=d_j$ and $\hat{\sigma}(t_j)=\sigma_j$, $j=1,2,\ldots,N_{\text{meas}}$,
in which case the functional simply becomes
\begin{equation*}
\mathcal{F}_2(\mathbf{m}) = \sum_{j=1}^{N_{\text{meas}}}~\frac{\{[w(\mathbf{m},T)](t_j)-d_j\}^2}{\sigma_j^2},
\end{equation*}
assuming, without loss of generality, that $t_{_F}\geqslant t_{N_{\text{meas}}}$.

\subsubsection{Bayesian Approach: Prior RV, Likelihood and Posterior RV}

In \underline{deterministic} inverse problems treat the unknown parameters as scalars or vectors and the goal is
the calculation of their best values according to a given criteria, usually least squares, e.g. solving the unconstrained optimization problem
\begin{equation}\label{eq-unconstrained-min}
\underset{\mathbf{m}\in M}{\text{min}}~\mathcal{F}(\mathbf{m}).
\end{equation}

In \underline{statistical} inverse problems, the unknown parameters are treated
as random variables (RVs) and the goal is the specification of their
probability density functions (PDFs)~\cite{KaSo05}.

Applying the Bayesian approach
\begin{equation*}
\pi_{\text{posterior}}(\mathbf{m})\propto \pi_{\text{prior}}(\mathbf{m})\cdot\pi_{\text{likelihood}}(\mathbf{m})
\end{equation*}
we have that for the TGA SIP, the prior distribution and the likelihood are, respectively:
\begin{equation*}
\pi_{\text{prior}}(\mathbf{m}) \propto e^{-\frac{1}{2}V(\mathbf{m})}\quad\text{and}\quad
\pi_{\text{likelihood}}(\mathbf{m}) \propto e^{-\frac{1}{2}\mathcal{F}(\mathbf{m})}.
\end{equation*}

Thus, we chose parameters $(A,E)$ to have joint uniform prior PDF over the open
square domain, i.e.:
$$\pi_{\text{prior}}=\mathcal{U}((1.0\times 10^{10},5.0\times 10^{11})\times (4.0\times 10^{5},6.0\times 10^{5})).$$

\subsubsection{Data from experiments}\label{sec:tga-data}

Table \ref{table:data-tga} presents the data collected in the TGA experiment. 

\begin{table}[htb]
\begin{center}
\begin{tabular}{cllc}
\toprule
Observation   & Temperature     & Relative mass             & Variance \\
index ``$i$'' & $T_i$ (K)       & $m^*_{\text{obs},i}$ (\%) & $V_i$    \\
\midrule
\midrule
 1            & $\quad$ 673.034 & $\quad$ 96.5855    & 0.1      \\
 2            & $\quad$ 682.003 & $\quad$ 95.1549    & 0.1      \\
 3            & $\quad$ 690.985 & $\quad$ 92.5048    & 0.1      \\
 4            & $\quad$ 699.979 & $\quad$ 88.6353    & 0.1      \\
 5            & $\quad$ 708.989 & $\quad$ 83.0585    & 0.1      \\
 6            & $\quad$ 718.02  & $\quad$ 75.5306    & 0.1      \\
 7            & $\quad$ 727.089 & $\quad$ 64.1003    & 0.1      \\
 8            & $\quad$ 735.96  & $\quad$ 47.5474    & 0.1      \\
 9            & $\quad$ 744.904 & $\quad$ 23.6777    & 0.1      \\
 10           & $\quad$ 754.062 & $\quad$ 03.2234    & 0.1      \\
 11           & $\quad$ 763.049 & $\quad$ 00.0855448 & 0.1      \\
\bottomrule
\end{tabular}
\vspace{-.2cm}
\caption{Experimental data.}\label{table:data-tga}
\end{center}
\end{table}

\subsection{Statistical Forward Problem}\label{sec:tga-sfp}

In spacecraft design, ablation is used to both cool and protect mechanical parts and/or payloads that would otherwise be damaged by extremely high temperatures. Two principal applications are heat shields for spacecraft entering a planetary atmosphere from space and cooling of rocket engine nozzles~\cite{wiki:ablation}. 

Suppose that an object about to re-enter the Earth atmosphere has a thermal
protection layer (shield) of composition of the same sample material described
in Section \ref{sec:tga-sip}. Also, as the object re-enters the atmosphere, its
shield loses mass according to Equation \eqref{eq:tga-model}.  The initial
sample temperature is $T_0=0.1$ K and it is then heated with constant rate
$\beta=5$ K/m.

We are interested in answering the following question: at scenario $\beta=250$
K/min, what is the remaining mass fraction at time $t=3.9$ s? In other words,
the quantity of interest is $w(t=3.9s)$.


\subsubsection{The Input RV, QoI Function and Output RV}

The input random variables for this SFP are the inferred parameters $A$ and $E$ which are the solution (posterior PDF) of the inverse problem described in Section \ref{sec:tga-sip}. The output random variable for this example is the remaining mass fraction at 3.9 s, i.e. $w(t=3.9)$. Note that, since there is uncertainty in the parameters $A$ and $E$ (both given as PDFs), one can expect that this uncertainty will be propagated to $w(t=3.9)$, which will also be given as a PDF. Finally, the QoI function for $w$ is the solution of the Equation \eqref{eq:tga-model} evaluated when $t=3.9$ s, which is calculated using numerical integration with adjustable and acceptable time-stepping using GSL function \verb+gsl_odeiv_evolve_apply+\footnote{\url{http://www.gnu.org/software/gsl/manual/html_node/Evolution.html\#index-gsl_005fodeiv2_005fevolve_005fapply}}. 

\subsection{Running the Example}\label{sec:tga-run}

To run the executable provided (available after QUESO installation), and generate figures for the chains, PDFs, CDFs, etc., enter the following commands:
\begin{lstlisting}[label={},caption={}]
$ cd $HOME/LIBRARIES/QUESO-0.50.0/examples/validationCycle
$ rm outputData/*
$ ./exTgaValidationCycle_gsl tagCycle.inp    
$ matlab
   $ tga_cycle_plot.m     # inside matlab
   $ exit                 # inside matlab
$ ls -l outputData/*.png
  cal_parameter1_prior.png                cal_parameter2_prior.png                
  cal_val_parameter1_PDF.png              cal_val_parameter2_PDF.png              
  cal_val_parameter1_CDF.png              cal_val_parameter2_CDF.png              
  cal_val_parameter1_autocorr.png         cal_val_parameter2_autocorr.png
  cal_val_QoI_CDF.png                     cal_val_QoI_PDF.png
  cal_val_QoI_autocorrelation.png            
\end{lstlisting}

As a result, the user should have created several of PNG figures containing marginal posterior PDF, chain positions of the parameters and the QoI, histogram, cumulative density distribution and autocorrelation. The name of the figure files have been chosen to be informative, as shown in the Listing above.

\subsection{TGA Example Code}\label{sec:code-tga}

The program example given in this paper is compatible with version 0.47.1 of QUESO.
The source code for the example is composed of 5 files:
 \texttt{exTgaValidationCycle\_gsl.C} (Listing \ref{code:tga-main-c}),
 \texttt{exTgaValidationCycle\_appl.h} (Listing \ref{code:tga-appl-h}), 
 \texttt{exTgaValidationCycle\_likelihood.h}  (Listing \ref{code:tga-like-h}) and 
\texttt{exTgaValidationCycle\_qoi.C} (Listing \ref{code:tga-qoi-h}).

\begin{lstlisting}[caption=File \texttt{exTgaValidationCycle\_gsl.C}., label={code:tga-main-c}, linerange={28-1000}]
/* This is an example of how to define and run a PECOS validation cycle for a
 * thermogravimetric analysis (TGA) model
 * using QUESO classes and algorithms. The code itself is in the templated
 * routine 'uqAppl(*env)'. This routine is called right after the initialization
 * of the MPI environment and of the QUESO environment and is available in
 * file 'exTgaValidationCycle_gsl.C'.
 *
 * It is easier to understand this example after undertanding three other simpler
 * examples:
 * (1) examples/simpleStatisticalInverseProblem/src/example_main.C
 * (2) examples/simpleStatisticalForwardProblem/src/simple_sfp_example_main.C
 * (3) examples/gravity/src/gravity_main.C
 *
 * Example (1) focuses on the templated class StatisticalInverseProblem<P_V,P_M>.
 * Example (2) focuses on the templated class StatisticalForwardProblem<P_V,P_M,Q_V,Q_M>.
 * Example (3) focuses on both classes, since it uses the solution of the inverse problem
 * as input to the forward problem.
 *
 * This example with TGA uses both statistical problem templated classes, twice each in
 * fact, but it also:
 * - uses the templated class ValidationCycle<P_V,P_M,Q_V,Q_M>,
 * - reads experimental data from files in order to compute the likelihood function
 *   at candidate parameter vectors generated by the Metropolis-Hastings algorithm, and
 * - compares the qoi cdfs computed from the two statistical forward problems at
 *   the prediction scenario.
 *--------------------------------------------------------------------------
 *-------------------------------------------------------------------------- */

#include <exTgaValidationCycle_appl.h>
#include <queso/GslMatrix.h>
#include <queso/EnvironmentOptions.h>

int main(int argc, char* argv[])
{
  //************************************************
  // Initialize environment
  //************************************************
#ifdef QUESO_HAS_MPI
  MPI_Init(&argc,&argv);
#endif

  QUESO::EnvOptionsValues* envOptionsValues = NULL;
#ifdef UQ_EXAMPLES_USES_QUESO_INPUT_FILE
  UQ_FATAL_TEST_MACRO(argc != 2,
                      QUESO::UQ_UNAVAILABLE_RANK,
                      "main()",
                      "input file must be specified in command line as argv[1], just after executable argv[0]");
#ifdef QUESO_HAS_MPI
  QUESO::FullEnvironment* env = new QUESO::FullEnvironment(MPI_COMM_WORLD,argv[1],"",envOptionsValues);
#else
  QUESO::FullEnvironment* env = new QUESO::FullEnvironment(argv[1],"",envOptionsValues);
#endif

#else
  envOptionsValues = new QUESO::EnvOptionsValues();
  envOptionsValues->m_subDisplayFileName   = "outputData/display";
  envOptionsValues->m_subDisplayAllowedSet.insert(0);
  envOptionsValues->m_subDisplayAllowedSet.insert(1);
  envOptionsValues->m_displayVerbosity     = 2;
  envOptionsValues->m_seed                 = 0;

#ifdef QUESO_HAS_MPI
  QUESO::FullEnvironment* env = new QUESO::FullEnvironment(MPI_COMM_WORLD,"","",envOptionsValues);
#else
  QUESO::FullEnvironment* env = new QUESO::FullEnvironment("","",envOptionsValues);
#endif
#endif

  //************************************************
  // Run application
  //************************************************
  uqAppl<QUESO::GslVector, // type for parameter vectors
         QUESO::GslMatrix, // type for parameter matrices
         QUESO::GslVector, // type for qoi vectors
         QUESO::GslMatrix  // type for qoi matrices
        >(*env);

  //************************************************
  // Finalize environment
  //************************************************
  delete env;
  delete envOptionsValues;
#ifdef QUESO_HAS_MPI
  MPI_Finalize();
#endif
  return 0;
}
\end{lstlisting}

\lstinputlisting[caption=File \texttt{exTgaValidationCycle\_appl.h}., label={code:tga-appl-h}, linerange={27-1000}]{exTgaValidationCycle_appl.h}

\lstinputlisting[caption=File \texttt{exTgaValidationCycle\_likelihood.h.}, label={code:tga-like-h}, linerange={27-1000}]{exTgaValidationCycle_likelihood.h}

\begin{lstlisting}[caption=File \texttt{exTgaValidationCycle\_qoi.h}., label={code:tga-qoi-h}, linerange={27-1000}]
#ifndef EX_TGA_VALIDATION_CYCLE_QOI_H
#define EX_TGA_VALIDATION_CYCLE_QOI_H

#include <queso/Defines.h>
#include <queso/DistArray.h>
#include <gsl/gsl_odeiv.h>

//********************************************************
// The (user defined) data class that carries the data
// needed by the (user defined) qoi routine
//********************************************************
template<class P_V,class P_M,class Q_V, class Q_M>
struct
qoiRoutine_Data
{
  double m_beta;
  double m_criticalMass;
  double m_criticalTime;
};

// The actual (user defined) qoi routine
template<class P_V,class P_M,class Q_V,class Q_M>
void qoiRoutine(const P_V&                    paramValues,
                const P_V*                    paramDirection,
                const void*                   functionDataPtr,
                      Q_V&                    qoiValues,
                      QUESO::DistArray<P_V*>* gradVectors,
                      QUESO::DistArray<P_M*>* hessianMatrices,
                      QUESO::DistArray<P_V*>* hessianEffects)
{
  double A             = paramValues[0];
  double E             = paramValues[1];
  double beta          = ((qoiRoutine_Data<P_V,P_M,Q_V,Q_M> *) functionDataPtr)->m_beta;
  double criticalMass  = ((qoiRoutine_Data<P_V,P_M,Q_V,Q_M> *) functionDataPtr)->m_criticalMass;
  double criticalTime  = ((qoiRoutine_Data<P_V,P_M,Q_V,Q_M> *) functionDataPtr)->m_criticalTime;

  double params[]={A,E,beta};
      	
  // integration
  const gsl_odeiv_step_type *T   = gsl_odeiv_step_rkf45; //rkf45; //gear1;
        gsl_odeiv_step      *s   = gsl_odeiv_step_alloc(T,1);
        gsl_odeiv_control   *c   = gsl_odeiv_control_y_new(1e-6,0.0);
        gsl_odeiv_evolve    *e   = gsl_odeiv_evolve_alloc(1);
        gsl_odeiv_system     sys = {func, NULL, 1, (void *)params}; 
	
  double temperature = 0.1;
  double h = 1e-3;
  double Mass[1];
  Mass[0]=1.;
  
  double temperature_old = 0.;
  double M_old[1];
  M_old[0]=1.;
	
  double crossingTemperature = 0.;
  //unsigned int loopSize = 0;
  while ((temperature < criticalTime*beta) &&
         (Mass[0]     > criticalMass     )) {
    int status = gsl_odeiv_evolve_apply(e, c, s, &sys, &temperature, criticalTime*beta, &h, Mass);
    UQ_FATAL_TEST_MACRO((status != GSL_SUCCESS),
                        paramValues.env().fullRank(),
                        "qoiRoutine()",
                        "gsl_odeiv_evolve_apply() failed");
    //printf("t = %6.1lf, mass = %10.4lf\n",t,Mass[0]);
    //loopSize++;

    if (Mass[0] <= criticalMass) {
      crossingTemperature = temperature_old + (temperature - temperature_old) * (M_old[0]-criticalMass)/(M_old[0]-Mass[0]);
    }
		
    temperature_old=temperature;
    M_old[0]=Mass[0];
  }

  if (criticalMass > 0.) qoiValues[0] = crossingTemperature/beta; // QoI = time to achieve critical mass
  if (criticalTime > 0.) qoiValues[0] = Mass[0];                  // QoI = mass fraction remaining at critical time
	
  //printf("loopSize = %d\n",loopSize);
  if ((paramValues.env().displayVerbosity() >= 3) && (paramValues.env().fullRank() == 0)) {
    printf("In qoiRoutine(), A = %g, E = %g, beta = %.3lf, criticalTime = %.3lf, criticalMass = %.3lf: qoi = %lf.\n",A,E,beta,criticalTime,criticalMass,qoiValues[0]);
  }

  gsl_odeiv_evolve_free (e);
  gsl_odeiv_control_free(c);
  gsl_odeiv_step_free   (s);

  return;
}

#endif // EX_TGA_VALIDATION_CYCLE_QOI_H
\end{lstlisting}

\subsection{Input File}\label{sec:tga-input-file}

The input file used with this TGA SIP--SFP QUESO provides QUESO with options
for its environments, and for both  MCMC and Monte-Carlo algorithms. It is
displayed in Listing~\ref{code:tga-input-file}.

\begin{lstlisting}[caption={File name \texttt{tgaCycle.inp} with options for QUESO library used in application code (Listings \ref{code:tga-main-c}-\ref{code:tga-like-h}})., 
label={code:tga-input-file},]
###############################################
# UQ Environment
###############################################
#env_help                 = anything
env_numSubEnvironments   = 1
env_subDisplayFileName   = outputData/display
env_subDisplayAllowAll   = 0
env_subDisplayAllowedSet = 0 1
env_displayVerbosity     = 2
env_syncVerbosity        = 0
env_seed                 = 0

###############################################
# Calibration (cal) stage: statistical inverse problem (ip)
###############################################
cycle_cal_ip_help                 = anything
cycle_cal_ip_computeSolution      = 1
cycle_cal_ip_dataOutputFileName   = outputData/tgaCalOutput
cycle_cal_ip_dataOutputAllowedSet = 0 1

###############################################
# 'cal_ip_': information for Metropolis-Hastings algorithm
#
# '_size' examples 16K   128K   1M      2M      16M
#                  16384 131072 1048576 2097152 16777216
###############################################
cycle_cal_ip_mh_help                 = anything
cycle_cal_ip_mh_dataOutputFileName   = outputData/tgaCalOutput
cycle_cal_ip_mh_dataOutputAllowedSet = 0 1

cycle_cal_ip_mh_initialPosition_dataInputFileName          = . # inputData/initPos
cycle_cal_ip_mh_initialPosition_dataInputFileType          = m
cycle_cal_ip_mh_initialProposalCovMatrix_dataInputFileName = . # inputData/adaptedMatrix_am910000
cycle_cal_ip_mh_initialProposalCovMatrix_dataInputFileType = m
cycle_cal_ip_mh_rawChain_dataInputFileName    = . # outputData/file_cal_ip_raw_input
cycle_cal_ip_mh_rawChain_size                 = 1048576
cycle_cal_ip_mh_rawChain_generateExtra        = 0
cycle_cal_ip_mh_rawChain_displayPeriod        = 20000
cycle_cal_ip_mh_rawChain_measureRunTimes      = 1
cycle_cal_ip_mh_rawChain_dataOutputFileName   = outputData/file_cal_ip_raw
cycle_cal_ip_mh_rawChain_dataOutputAllowedSet = 0 1

cycle_cal_ip_mh_displayCandidates             = 0
cycle_cal_ip_mh_putOutOfBoundsInChain         = 1
cycle_cal_ip_mh_tk_useLocalHessian            = 0
cycle_cal_ip_mh_tk_useNewtonComponent         = 1
cycle_cal_ip_mh_dr_maxNumExtraStages          = 1
cycle_cal_ip_mh_dr_listOfScalesForExtraStages = 5. 4. 3.
cycle_cal_ip_mh_am_initialNonAdaptInterval    = 0 # 10000
cycle_cal_ip_mh_am_adaptInterval              = 100 # 10000
cycle_cal_ip_mh_am_adaptedMatrices_dataOutputPeriod   = 60000
cycle_cal_ip_mh_am_adaptedMatrices_dataOutputFileName = . # outputData/adaptedMatrix
cycle_cal_ip_mh_am_adaptedMatrices_dataOutputFileType = m
cycle_cal_ip_mh_am_eta                        = 1.92
cycle_cal_ip_mh_am_epsilon                    = 1.e-5

cycle_cal_ip_mh_filteredChain_generate             = 1
cycle_cal_ip_mh_filteredChain_discardedPortion     = 0.
cycle_cal_ip_mh_filteredChain_lag                  = 20
cycle_cal_ip_mh_filteredChain_dataOutputFileName   = .
cycle_cal_ip_mh_filteredChain_dataOutputAllowedSet = 0 1

###############################################
# Calibration (cal) stage: statistical forward problem (fp)
###############################################
cycle_cal_fp_help                 = anything
cycle_cal_fp_computeSolution      = 1
cycle_cal_fp_computeCovariances   = 1
cycle_cal_fp_computeCorrelations  = 1
cycle_cal_fp_dataOutputFileName   = outputData/tgaCalOutput
cycle_cal_fp_dataOutputAllowedSet = 0 1

###############################################
# 'cal_fp_': information for Monte Carlo algorithm
#
# '_size' examples 16K   128K   1M      2M      16M
#                  16384 131072 1048576 2097152 16777216
###############################################
cycle_cal_fp_mc_help                 = anything
cycle_cal_fp_mc_dataOutputFileName   = outputData/tgaCalOutput
cycle_cal_fp_mc_dataOutputAllowedSet = 0 1

cycle_cal_fp_mc_pseq_dataOutputFileName   = .
cycle_cal_fp_mc_pseq_dataOutputAllowedSet = 0 1

cycle_cal_fp_mc_qseq_dataInputFileName    = . # outputData/file_cal_fp_qoi1
cycle_cal_fp_mc_qseq_size                 = 1048576
cycle_cal_fp_mc_qseq_displayPeriod        = 20000
cycle_cal_fp_mc_qseq_measureRunTimes      = 1
cycle_cal_fp_mc_qseq_dataOutputFileName   = outputData/file_cal_fp_qoi2
cycle_cal_fp_mc_qseq_dataOutputAllowedSet = 0 1

###############################################
# Validation (val) stage: statistical inverse problem (ip)
###############################################
cycle_val_ip_help                 = anything
cycle_val_ip_computeSolution      = 1
cycle_val_ip_dataOutputFileName   = outputData/tgaValOutput
cycle_val_ip_dataOutputAllowedSet = 0 1

###############################################
# 'val_ip_': information for Metropolis-Hastings algorithm
#
# '_size' examples 16K   128K   1M      2M      16M
#                  16384 131072 1048576 2097152 16777216
###############################################
cycle_val_ip_mh_help                 = anything
cycle_val_ip_mh_dataOutputFileName   = outputData/tgaValOutput
cycle_val_ip_mh_dataOutputAllowedSet = 0 1

cycle_val_ip_mh_rawChain_dataInputFileName    = . # outputData/file_val_ip_raw_input
cycle_val_ip_mh_rawChain_size                 = 1048576
cycle_val_ip_mh_rawChain_generateExtra        = 0
cycle_val_ip_mh_rawChain_displayPeriod        = 20000
cycle_val_ip_mh_rawChain_measureRunTimes      = 1
cycle_val_ip_mh_rawChain_dataOutputFileName   = outputData/file_val_ip_raw
cycle_val_ip_mh_rawChain_dataOutputAllowedSet = 0 1

cycle_val_ip_mh_displayCandidates             = 0
cycle_val_ip_mh_putOutOfBoundsInChain         = 1
cycle_val_ip_mh_tk_useLocalHessian            = 0
cycle_val_ip_mh_tk_useNewtonComponent         = 1
cycle_val_ip_mh_dr_maxNumExtraStages          = 1
cycle_val_ip_mh_dr_listOfScalesForExtraStages = 5. 4. 3.
cycle_val_ip_mh_am_initialNonAdaptInterval    = 0
cycle_val_ip_mh_am_adaptInterval              = 100
cycle_val_ip_mh_am_eta                        = 1.92
cycle_val_ip_mh_am_epsilon                    = 1.e-5

cycle_val_ip_mh_filteredChain_generate             = 1
cycle_val_ip_mh_filteredChain_discardedPortion     = 0.
cycle_val_ip_mh_filteredChain_lag                  = 20
cycle_val_ip_mh_filteredChain_dataOutputFileName   = .
cycle_val_ip_mh_filteredChain_dataOutputAllowedSet = 0 1

###############################################
# Validation (val) stage: statistical forward problem (fp)
###############################################
cycle_val_fp_help                 = anything
cycle_val_fp_computeSolution      = 1
cycle_val_fp_computeCovariances   = 1
cycle_val_fp_computeCorrelations  = 1
cycle_val_fp_dataOutputFileName   = outputData/tgaValOutput
cycle_val_fp_dataOutputAllowedSet = 0 1

###############################################
# 'val_fp_': information for Monte Carlo algorithm
#
# '_size' examples 16K   128K   1M      2M      16M
#                  16384 131072 1048576 2097152 16777216
###############################################
cycle_val_fp_mc_help                 = anything
cycle_val_fp_mc_dataOutputFileName   = outputData/tgaValOutput
cycle_val_fp_mc_dataOutputAllowedSet = 0 1

cycle_val_fp_mc_pseq_dataOutputFileName   = .
cycle_val_fp_mc_pseq_dataOutputAllowedSet = 0 1


cycle_val_fp_mc_qseq_dataInputFileName    = . # outputData/file_val_fp_qoi1
cycle_val_fp_mc_qseq_size                 = 1048576
cycle_val_fp_mc_qseq_displayPeriod        = 20000
cycle_val_fp_mc_qseq_measureRunTimes      = 1
cycle_val_fp_mc_qseq_dataOutputFileName   = outputData/file_val_fp_qoi2
cycle_val_fp_mc_qseq_dataOutputAllowedSet = 0 1

\end{lstlisting}

\subsection{Data Post-Processing and Visualization}\label{sec:tga-results}

According to the specifications of the input file in Listing~\ref{code:tga-input-file}, both a folder named \verb+outputData+ and a the following files should be generated:
\begin{verbatim}
file_cal_ip_raw.m        file_val_ip_raw.m        
file_cal_ip_raw_sub0.m   file_val_ip_raw_sub0.m
file_cal_fp_qoi2.m       file_val_fp_qoi2.m      
file_cal_fp_qoi2_sub0.m  file_val_fp_qoi2_sub0.m     
tgaCalOutput_sub0.m      tgaValOutput_sub0.m
display_sub0.txt    
\end{verbatim}

The sequence of Matlab commands is identical to the ones presented in Sections
\ref{sec:sip-results}, \ref{sec:sfp-results} and \ref{sec:gravity-results};
therefore, are omitted here. The reader is invited to explore the Matlab file
\texttt{tga\_cycle\_plot.m}  
for details of how the figures have been generated.

\subsubsection{KDE Plots of Parameters}
\begin{figure}[htpb]
\centering 
\subfloat{\includegraphics[scale=0.3]{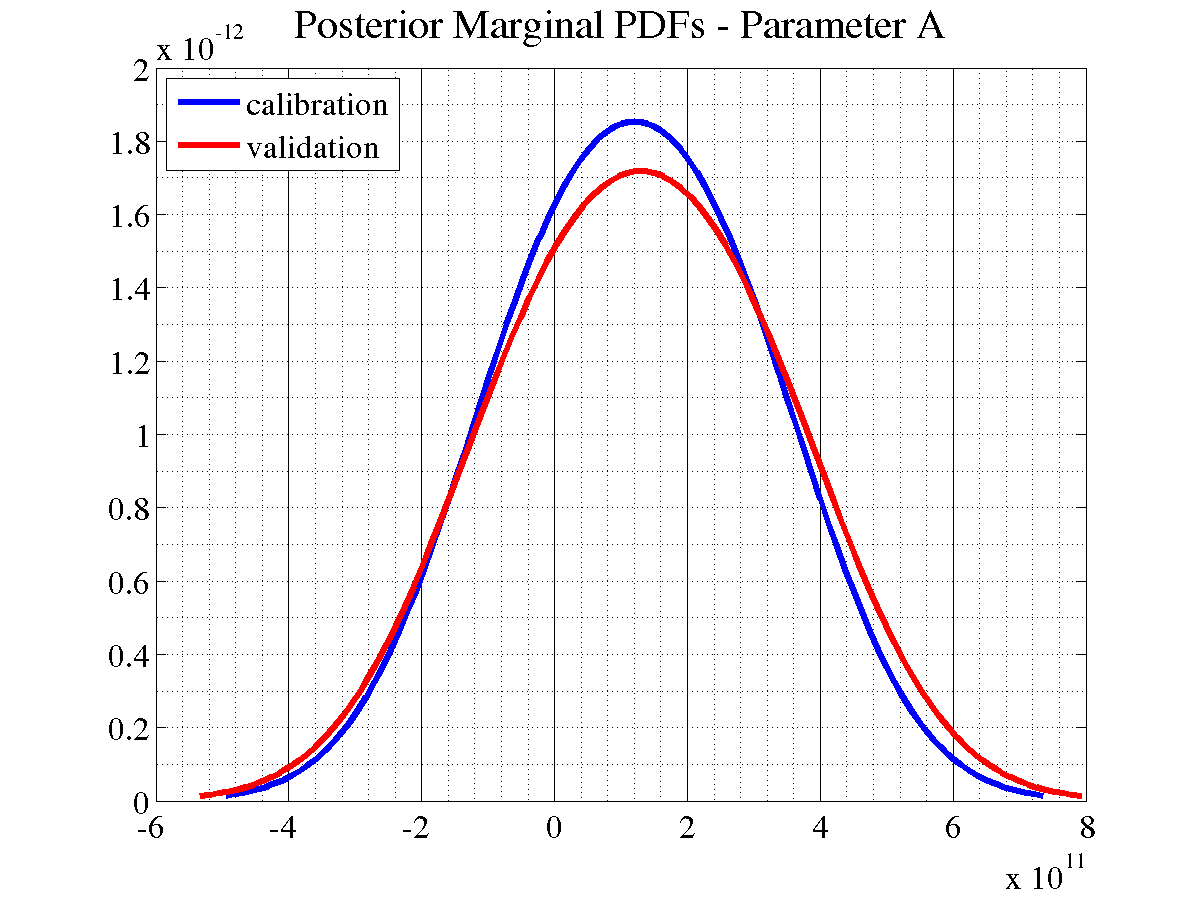}}
\subfloat{\includegraphics[scale=0.3]{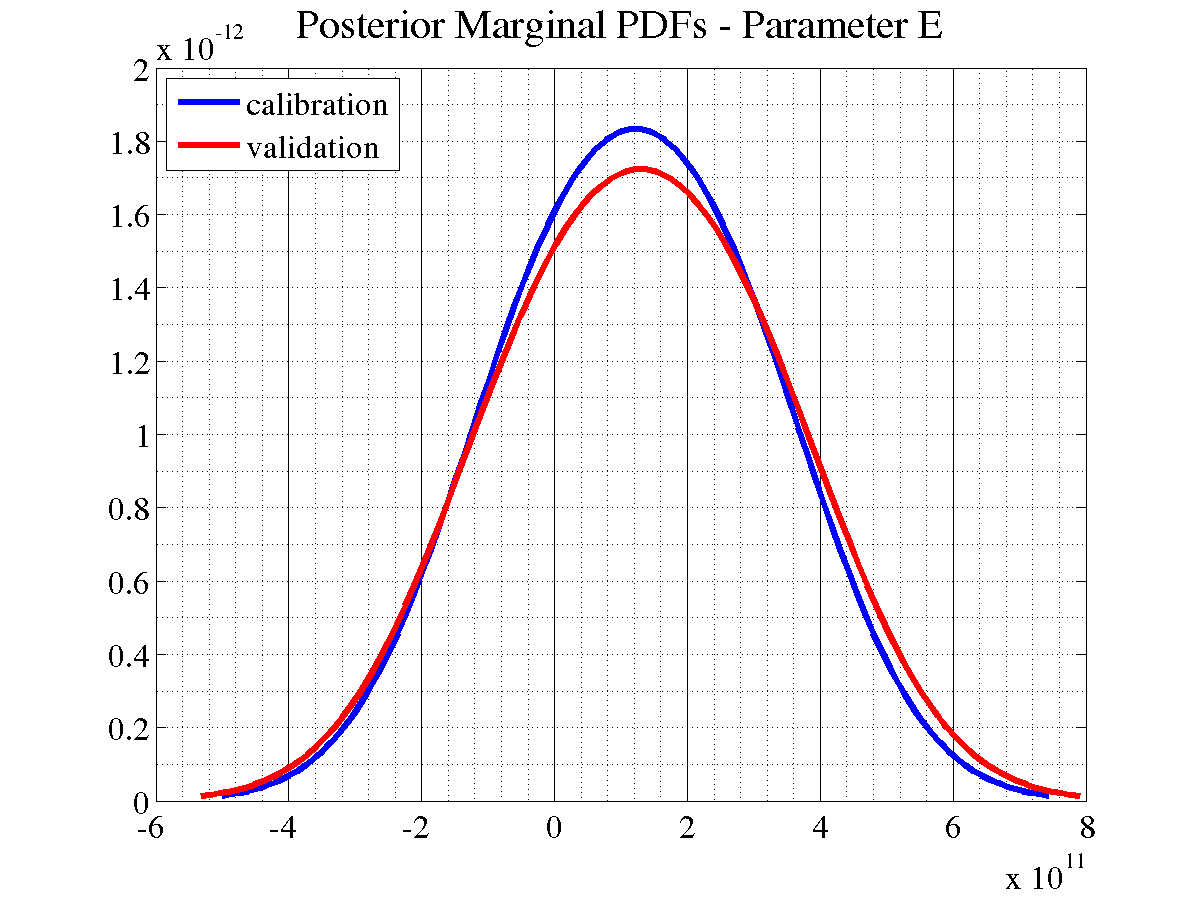}}
\vspace*{-10pt}
\caption{Posterior distributions of parameters $A$ and $E$.}
\label{fig:tga_ip_pdf}
\end{figure}

Matlab function \verb+ksdensity+ (Kernel smoothing density estimate) together
with the option `\verb+pdf+' may be used to estimate the KDE of the parameters,
as illustrated in Figure \ref{fig:tga_ip_pdf}.
%

\subsubsection{CDF Plots of Parameters}

\begin{figure}[htpb]
\centering 
\subfloat{\includegraphics[scale=0.3]{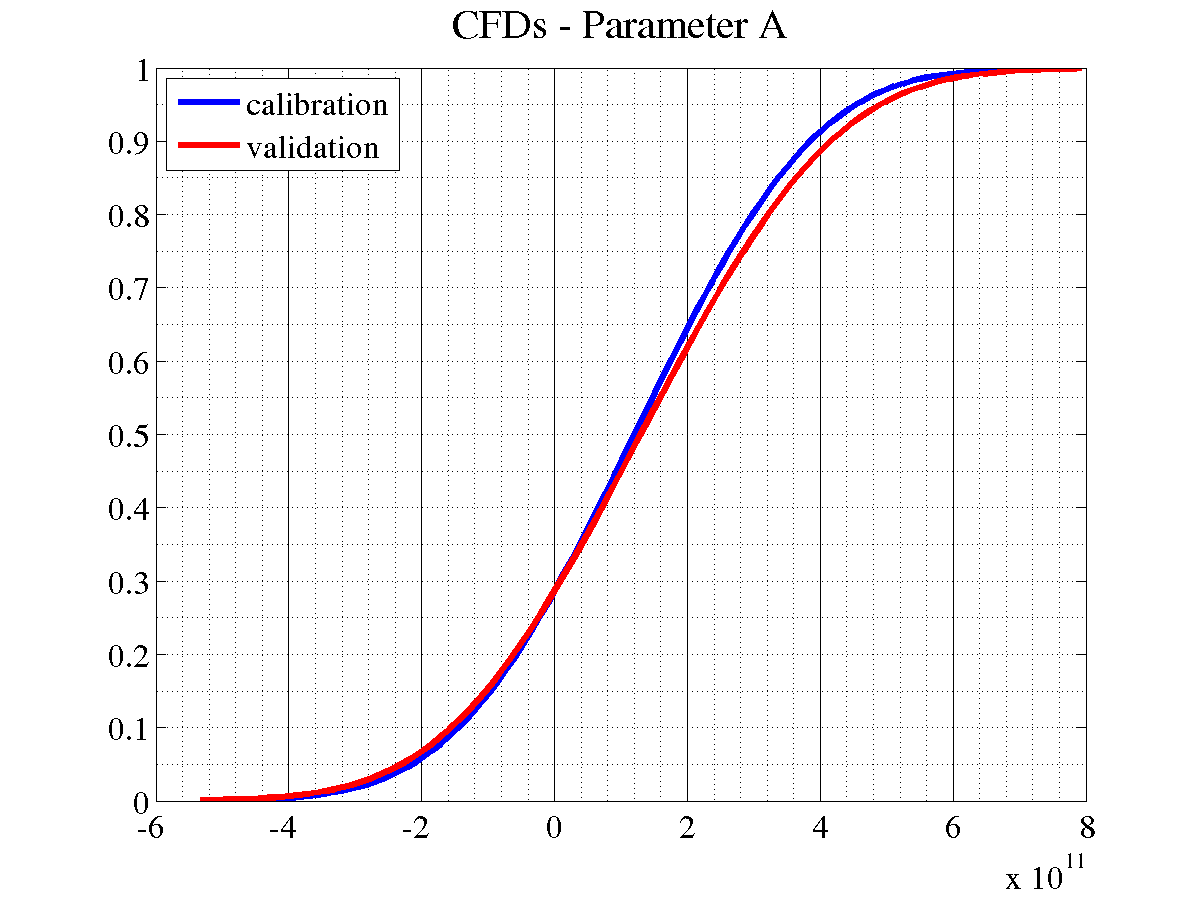}}
\subfloat{\includegraphics[scale=0.3]{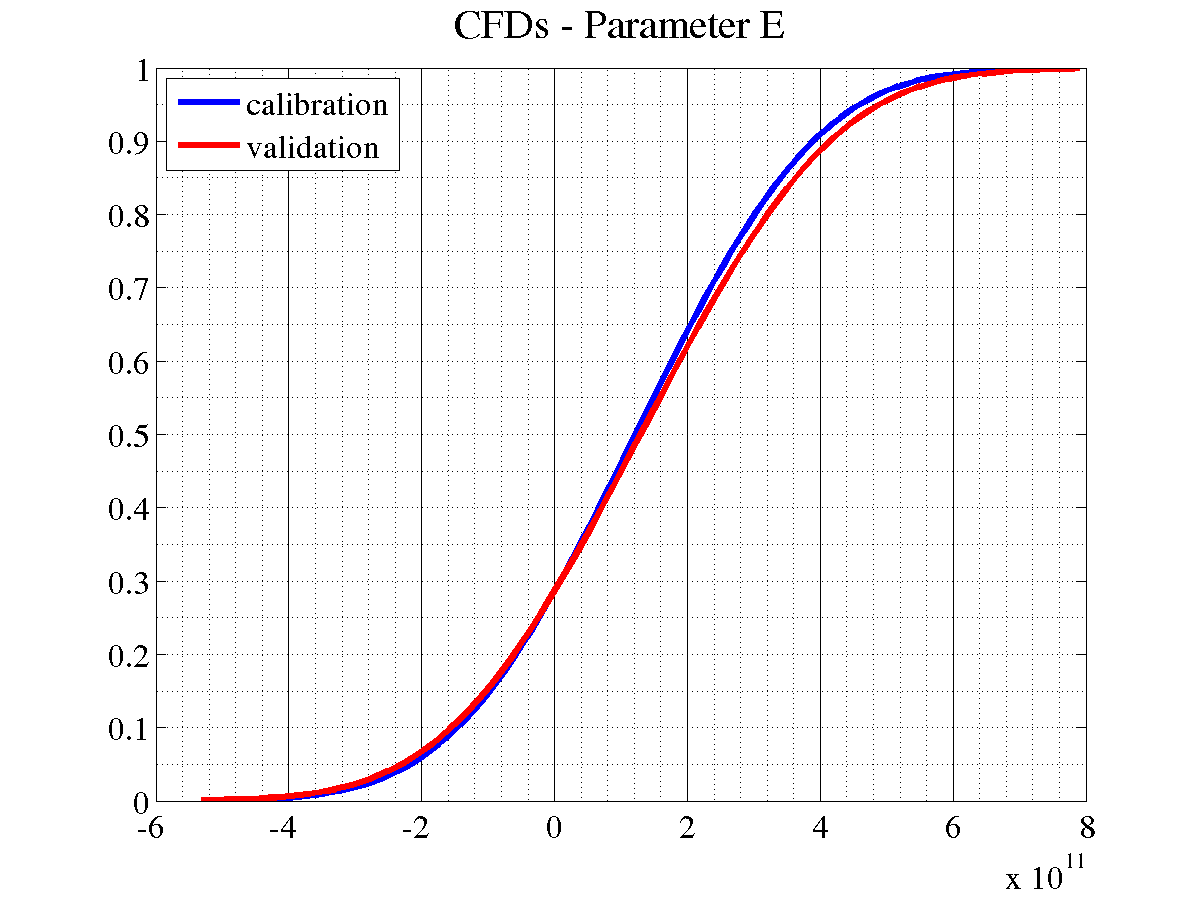}}
\vspace*{-10pt}
\caption{Cumulative density functions of parameters $A$ and $E$.}
\label{fig:tga_ip_cdf}
\end{figure}

Matlab function \verb+ksdensity+ with \verb+'cdf'+ option may also be used for plotting the Cumulative Distribution Function of each one of the parameters, as illustrated in Figure \ref{fig:tga_ip_cdf}.

\subsubsection{Autocorrelation Plots of Parameters}

\begin{figure}[p]
\centering 
\subfloat{\includegraphics[scale=0.3]{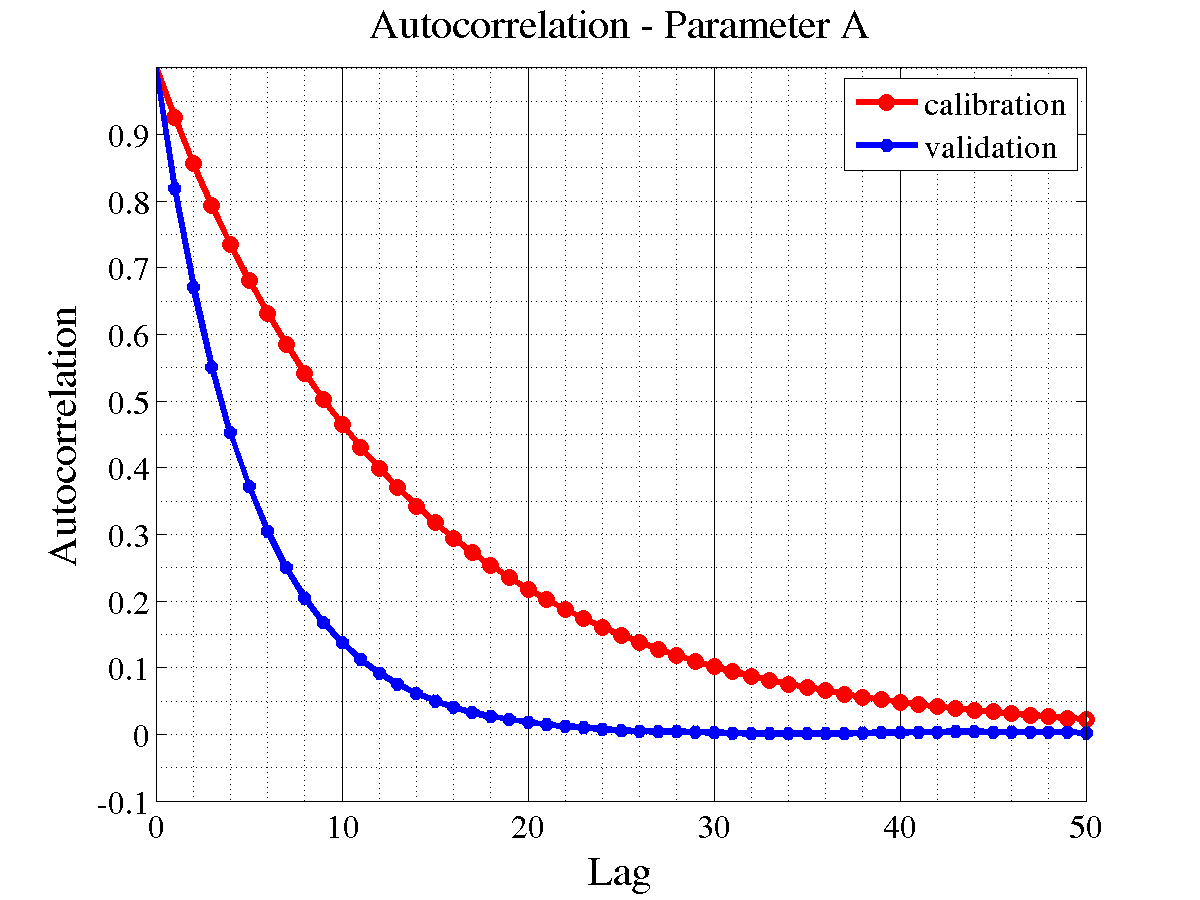}}
\subfloat{\includegraphics[scale=0.3]{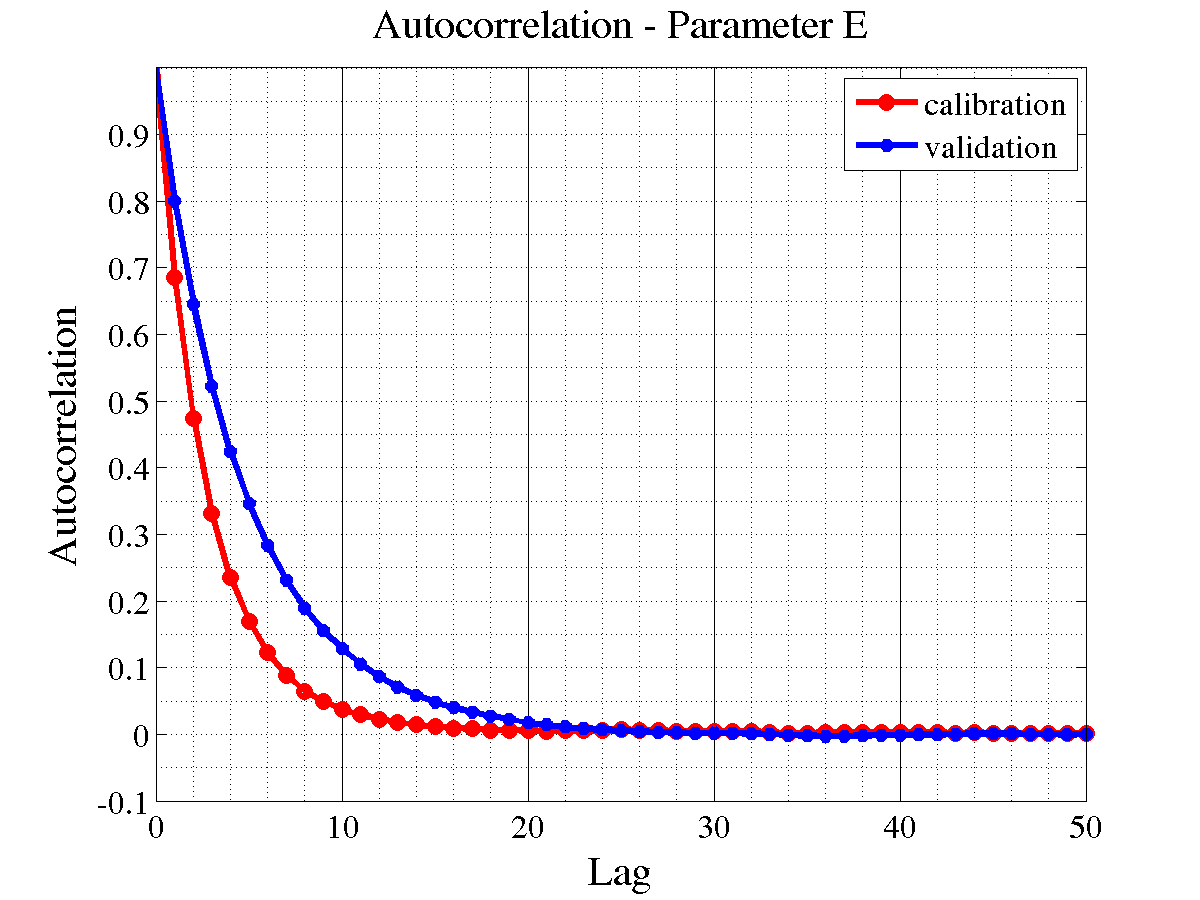}}
\vspace*{-10pt}
\caption{Autocorrelation of parameters $A$ and $E$ (filtered chain).}
\label{fig:tga_ip_autocorrelation_param}
\end{figure}

Figure \ref{fig:tga_ip_autocorrelation_param} presents the autocorrelation of the parameters $A$ and $E$ in both cases: calibration and validation stages.

\subsubsection{KDE, CDF and Autocorrelation Plots of QoI}
\begin{figure}[p]
\centering 
\subfloat[QoI PDF]{\includegraphics[scale=0.3]{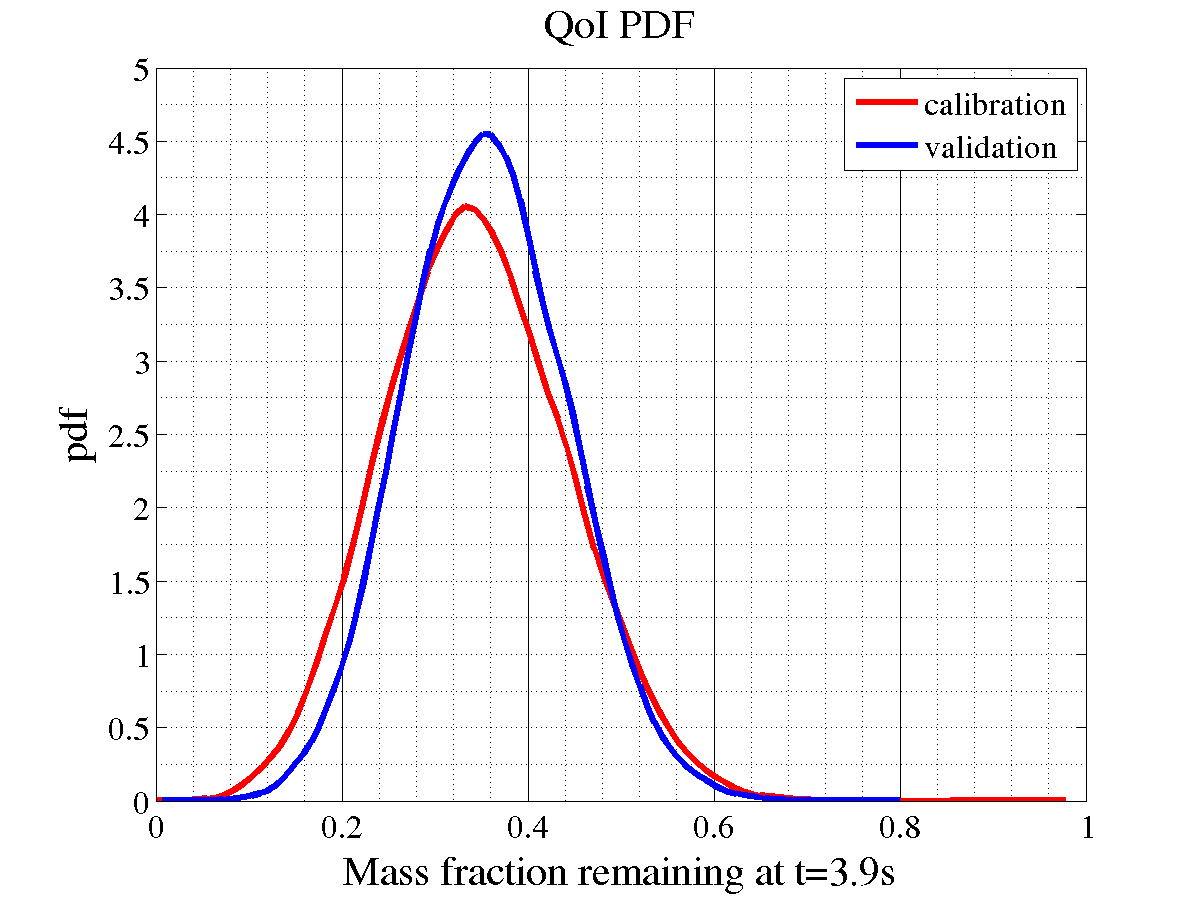}\label{fig:tga_pdf_qoi}}
\subfloat[QoI CDF]{\includegraphics[scale=0.3]{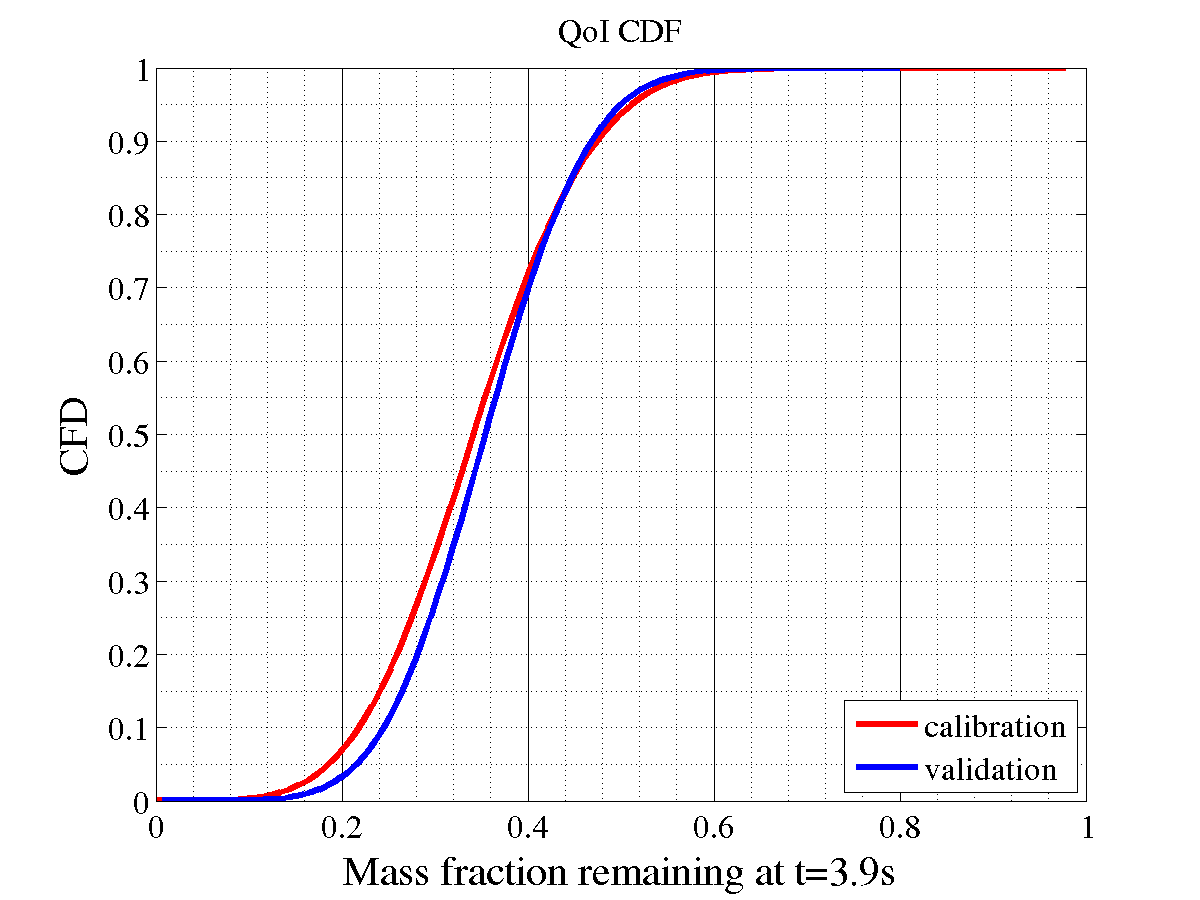}\label{fig:tga_cdf_qoi}}
\vspace*{-10pt}
\caption{QoI PDF and CDF, during calibration and validation stages.}
\end{figure}

\begin{figure}[p]
\centering 
\includegraphics[scale=0.3]{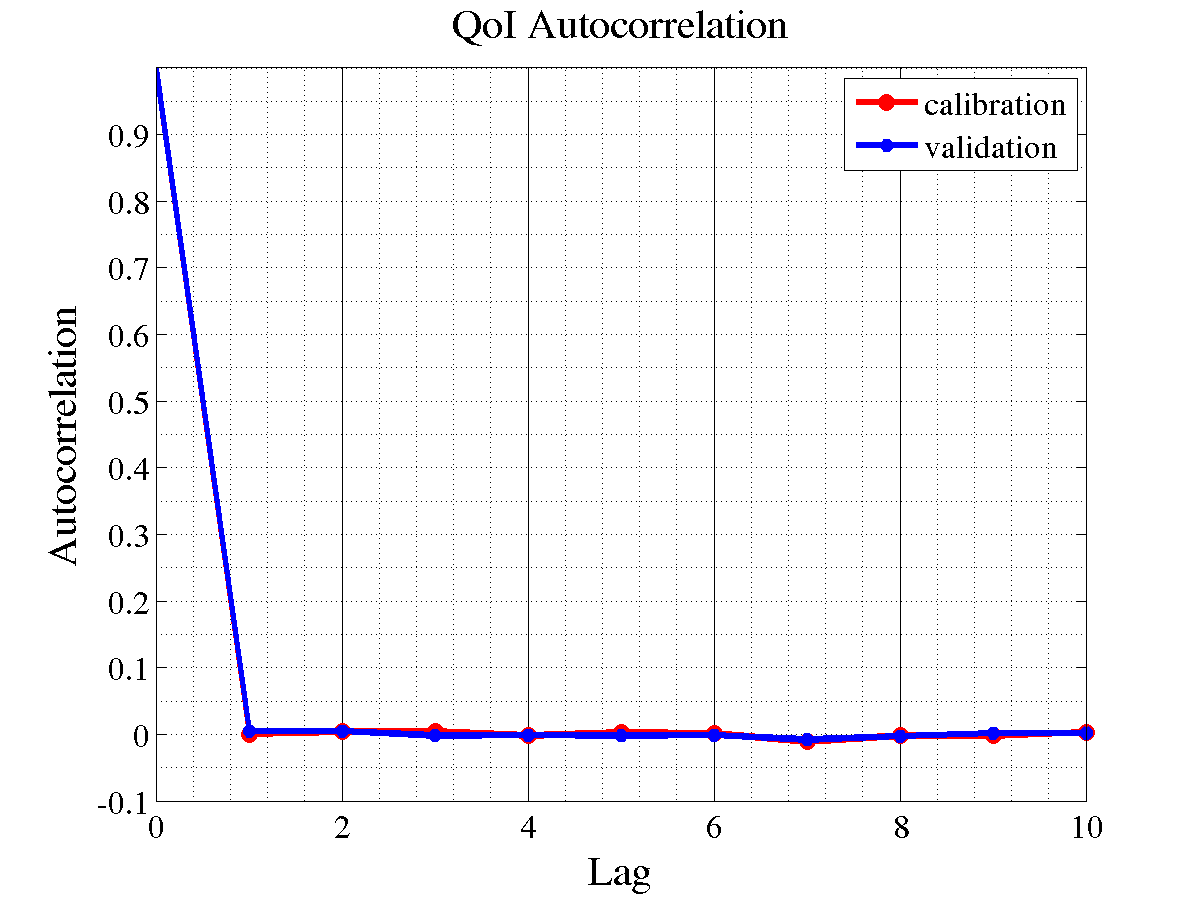}
\vspace*{-10pt}
\caption{QoI autocorrelation.}
\label{fig:tga_autocorrelation_qoi}
\end{figure}

Figures \ref{fig:tga_pdf_qoi}  and \ref{fig:tga_cdf_qoi} present PDF and CDF of QoI, respectively and Figure \ref{fig:tga_autocorrelation_qoi} presents its autocorrelation.

%

\section{\texttt{modal}}\label{sec:example_modal}

This example presents a combination of two statistical inverse problems in one. 
It presents the capability of the Multilevel method in sampling from a target distribution that has either one or two modes (distinct peaks). The random variable of interest has three parameters, i.e., $\bv{\theta}=(\theta_1,\theta_2, \sigma^2) \in \mathbb{R}^3$, where the third parameter may be seen as variation.

The example also it gives the user the opportunity to chose either one single type of prior distribution, uniform, for the three components of the random variable, or two different priors: a uniform and a beta distribution.

Choosing between a one-mode or a two-mode target distribution is done at execution level, as presented in the following code line:

\begin{lstlisting}[label={},caption={}]
$ cd $HOME/LIBRARIES/QUESO-0.51.0/
$ cd examples/modal
$ rm outputData/*
$ ./modal_gsl example.inp <num_of_nodes>
\end{lstlisting}
where \verb+<num_of_nodes>+ is either 1 or 2.
%
%
%
%
%
%
%
%
\subsection{One-mode distribution}

In this case, the target distribution is assumed to have only one mode.
Suppose also that the random variable $\bv{\theta}$  can either have a uniform prior distribution for all its components, i.e.:
$$
\pi_{\text{prior}}=\mathcal{U}([0,3]) \times \mathcal{U}([0,3]) \times \mathcal{U}([0,0.3]).
$$
or, the prior distribution is defined as a combination of uniform prior for $\theta_1$ and $\theta_2$, with a beta prior for $\sigma^2$:
$$
\pi_{\text{prior}}=\mathcal{U}([0,3]) \times \mathcal{U}([0,3]) \times \mathcal{B}(\alpha,\beta), \quad \text{with} \quad \alpha=3 \quad\text{and}\quad \beta=0.09709133373799.
$$

The likelihood function is defined as follows:
\begin{equation}
\begin{split} \small
\quad f(\D|\bv{\theta})= -\dfrac{5}{2} \log\left(2 \pi \sigma^2\right)-\dfrac{1}{2\sigma^2} &\Bigg[
 \left(10 \sqrt{10 \theta_1+20 \theta_2+10 \sqrt{\theta_1^2+4 \theta_2^2}}-72.0470\right)^2 +\\
&+\left(10 \sqrt{10 \theta_1+20 \theta_2+10 \sqrt{\theta_1^2+4 \theta_2^2}}-71.8995\right)^2 +\\
&+\left(10 \sqrt{10 \theta_1+20 \theta_2+10 \sqrt{\theta_1^2+4 \theta_2^2}}-72.2801\right)^2 +\\
&+\left(10 \sqrt{10 \theta_1+20 \theta_2+10 \sqrt{\theta_1^2+4 \theta_2^2}}-71.9421\right)^2 +\\
&+\left(10 \sqrt{10 \theta_1+20 \theta_2+10 \sqrt{\theta_1^2+4 \theta_2^2}}-72.3578\right)^2 \Bigg].
\end{split}
\end{equation}

\subsubsection{Running the One-Mode Example}
 
To run the executable provided considering a \underline{one-mode} distribution, enter the following commands:
\begin{lstlisting}[label={},caption={Running the example with a one-mode distribution.}]
$ cd $HOME/LIBRARIES/QUESO-0.51.0/
$ cd examples/modal
$ rm outputData/*
$ ./modal_gsl example.inp 1      #one mode!
$ matlab
   $  plot_modal_all_levels_1mode  # inside matlab
   $ exit                          # inside matlab
$ ls -l outputData/*.png
modal_1_mode_kde_target.png  modal_1_mode_level_1.png  modal_1_mode_level_5.png
modal_1_mode_kde_theta1.png  modal_1_mode_level_2.png  modal_1_mode_level_6.png
modal_1_mode_kde_theta2.png  modal_1_mode_level_3.png  modal_1_mode_level_7.png
modal_1_mode_kde_theta3.png  modal_1_mode_level_4.png
\end{lstlisting}

As a result, the user should have created several of PNG figures scatter plots of each one of the levels and the kernel density estimation of the parameters, for each level in the Multilevel method. The name of the figure files have been chosen to be informative, as shown in the Listing above.

\subsection{Two-mode distribution}

In this case, the target distribution is assumed to have two modes.
Suppose that $\bv{\theta}$ has a either uniform distribution for all its components, i.e.:
$$
\pi_{\text{prior}}=\mathcal{U}([0,3]) \times \mathcal{U}([0,3]) \times \mathcal{U}([0,0.3]).
$$
or, the prior distribution is defined as a combination of uniform prior for the $\theta_1$, with a beta prior for $\theta_2$:
$$
\pi_{\text{prior}}=\mathcal{U}([0,3]) \times \mathcal{U}([0,3]) \times \mathcal{B}(\alpha,\beta), \quad \text{with} \quad \alpha=3 \quad\text{and}\quad \beta=0.08335837191688.
$$

The likelihood function is defined as follows:
\begin{equation}
\begin{split}\small
f(\D|\bv{\theta})=  -5 \log\left(2 \pi \sigma^2\right)- \dfrac{1}{2\sigma^2} &\Bigg[ 
  \left(10 \sqrt{10 \theta_1+20 \theta_2+10 \sqrt{\theta_1^2+4 \theta_2^2}}-72.0470\right)^2+\\
&+\left(10 \sqrt{10 \theta_1+20 \theta_2+10 \sqrt{\theta_1^2+4 \theta_2^2}}-71.8995\right)^2+\\
&+\left(10 \sqrt{10 \theta_1+20 \theta_2+10 \sqrt{\theta_1^2+4 \theta_2^2}}-72.2801\right)^2+\\
&+\left(10 \sqrt{10 \theta_1+20 \theta_2+10 \sqrt{\theta_1^2+4 \theta_2^2}}-71.9421\right)^2+\\
&+\left(10 \sqrt{10 \theta_1+20 \theta_2+10 \sqrt{\theta_1^2+4 \theta_2^2}}-72.3578\right)^2+\\
&+\left(10 \sqrt{10 \theta_1+20 \theta_2-10 \sqrt{\theta_1^2+4 \theta_2^2}}-28.0292\right)^2+\\
&+\left(10 \sqrt{10 \theta_1+20 \theta_2-10 \sqrt{\theta_1^2+4 \theta_2^2}}-27.3726\right)^2+\\
&+\left(10 \sqrt{10 \theta_1+20 \theta_2-10 \sqrt{\theta_1^2+4 \theta_2^2}}-27.5388\right)^2+\\
&+\left(10 \sqrt{10 \theta_1+20 \theta_2-10 \sqrt{\theta_1^2+4 \theta_2^2}}-27.0357\right)^2+\\
&+\left(10 \sqrt{10 \theta_1+20 \theta_2-10 \sqrt{\theta_1^2+4 \theta_2^2}}-27.1588\right)^2 \Bigg].
\end{split} 
\end{equation}

\subsubsection{Running the Two-Mode Example}
 
To run the executable provided considering a \underline{two-modes} distribution, enter the following commands:
\begin{lstlisting}[label={},caption={Running the example with a two-mode distribution.}]
$ cd $HOME/LIBRARIES/QUESO-0.51.0/
$ cd examples/modal
$ rm outputData/*
$ ./modal_gsl example.inp 2         # two modes!
$ matlab
   $  plot_modal_all_levels_2modes  # inside matlab
   $ exit                           # inside matlab
$ ls -l outputData/*.png
modal_2_modes_kde_target.png  modal_2_modes_level_1.png  modal_2_modes_level_5.png
modal_2_modes_kde_theta1.png  modal_2_modes_level_2.png  modal_2_modes_level_6.png
modal_2_modes_kde_theta2.png  modal_2_modes_level_3.png  modal_2_modes_level_7.png
modal_2_modes_kde_theta3.png  modal_2_modes_level_4.png  modal_2_modes_level_8.png
\end{lstlisting}

As a result, the user should have created several of PNG figures scatter plots of each one of the levels and the kernel density estimation of the parameters, for each level in the Multilevel method. The name of the figure files have been chosen to be informative, as shown in the Listing above.

\subsection{Example Code}\label{sec:modal-code}

The source code for the example is composed of 5 files:
\texttt{example\_main.C} (Listing \ref{code:modal-main-c}), \linebreak
\texttt{example\_likelihood.h} and \texttt{example\_likelihood.C} (Listings \ref{fig-like-modal-h} and \ref{fig-like-modal-c}),
\texttt{example\_compute.h} and \texttt{example\_compute.C} (Listings \ref{code:modal-compute-h} and \ref{code:modal-compute-c}).

\begin{lstlisting}[caption=File \texttt{example\_main.C.}, label={code:modal-main-c}, linerange={33-1000}]
#include <example_compute.h>

int main(int argc, char* argv[])
{
  // Initialize environment
  MPI_Init(&argc,&argv);

  UQ_FATAL_TEST_MACRO(argc != 3,
                      QUESO::UQ_UNAVAILABLE_RANK,
                      "main()",
                      "after executable argv[0], input file must be specified in command line as argv[1], then numModes (1 or 2) must be specified as argv[2]");

  QUESO::FullEnvironment* env =
    new QUESO::FullEnvironment(MPI_COMM_WORLD,argv[1],"",NULL);

  // Compute
  unsigned int numModes = (unsigned int) atoi(argv[2]);

  compute(*env,numModes);

  // Finalize environment
  delete env;
  MPI_Finalize();

  std::cout << std::endl << "FIM!" << std::endl << std::endl;

  return 0;
}
\end{lstlisting}

\begin{lstlisting}[caption=File \texttt{example\_likelihood.h}., label={fig-like-modal-h}, linerange={32-1000}]
#ifndef __EX_LIKELIHOOD_H__
#define __EX_LIKELIHOOD_H__

#include <queso/GslMatrix.h>

struct
likelihoodRoutine_DataType
{
  unsigned int numModes;
};

double likelihoodRoutine(
  const QUESO::GslVector& paramValues,
  const QUESO::GslVector* paramDirection,
  const void*             functionDataPtr,
  QUESO::GslVector*       gradVector,
  QUESO::GslMatrix*       hessianMatrix,
  QUESO::GslVector*       hessianEffect);

#endif
\end{lstlisting}

\lstinputlisting[caption=File \texttt{example\_likelihood.C}., label={fig-like-modal-c}, linerange={33-1000}]{modal_example_likelihood.C}

\begin{lstlisting}[caption=File \texttt{example\_compute.h.}, label={code:modal-compute-h}, linerange={32-1000}]
#ifndef __EX_COMPUTE_H__
#define __EX_COMPUTE_H__

#include <queso/Environment.h>

void compute(const QUESO::FullEnvironment& env, unsigned int numModes);

#endif
\end{lstlisting}

Note that in line 12 of Listings \ref{code:modal-compute-c} the \verb+#define+ directive creates the macro
 \linebreak
\verb+APPLS_MODAL_USES_CONCATENATION+. Such macro, together with the directives \verb+#ifdef+, \verb+#else+, and \verb+#endif+, tells the compiler that the application will use concatenated priors, by controlling compilation of portions of file \texttt{example\_compute.C}. Commenting line 12 of Listings \ref{code:modal-compute-c} will make the application to use uniform priors only:

\lstinputlisting[caption={File \texttt{example\_compute.C}.}, label={code:modal-compute-c}, linerange={33-1000},numbers=left]{modal_example_compute.C}

\subsection{Input File}\label{sec:modal-input-file}

QUESO reads an input file for solving statistical problems, which provides options for the Multilevel or MCMC method. In this example, the Multilevel method is chosen to sample from the distribution. Many variables are common to both MCMC and Multilevel method, especially because the Multilevel method also has the option of delaying the rejection of a candidate. The names of the variables have been designed to be informative in this case as well:
\begin{description}\vspace{-8pt}
\item[ \texttt{env}:] refers to QUESO environment; \vspace{-8pt}
\item[ \texttt{ip}:] refers to inverse problem;\vspace{-8pt}
\item[ \texttt{ml}:] refers to Multilevel;\vspace{-8pt}
\item[ \texttt{dr}:] refers to delayed rejection;\vspace{-8pt}
\item[ \texttt{rawChain}:] refers to the raw, entire chain; \vspace{-8pt}
\item[ \texttt{filteredChain}:] refers to a filtered chain (related to a specified \texttt{lag});\vspace{-8pt}
\item[ \texttt{last}:] refers to instructions specific for the last level of the Multilevel algorithm.
\end{description}

The user may select options for a specific level by naming its number, i.e., in case the user wants to write the raw chain of the level 3 in a separate file, say \verb+'rawChain_ml_level3.m'+, he/she may include the line: 
\begin{lstlisting}
ip_ml_3_rawChain_dataOutputFileName = outputData/rawChain_ml_level3 
\end{lstlisting}
in the input file.

The options used for solving this example are displayed in Listing \ref{code:modal-input-file}. 

\lstinputlisting[caption={Options for QUESO library used in application code (Listings \ref{code:modal-main-c}-\ref{code:modal-compute-c}})., 
label={code:modal-input-file},]{modal_example.inp}

\subsection{Create your own Makefile}\label{sec:modal-makefile}

Makefiles are special format files that together with the make utility will help one to compile and automatically build and manage projects (programs).  
Listing \ref{code:modal_makefile} presents a Makefile, named `\texttt{Makefile\_modal\_example\_violeta}', that may be used to compile the code and create the executable \verb+modal_gsl+. Naturally, it must be adapted to the user's settings, i.e., it has to have the correct paths for the user's libraries that have actually been used to compile and install QUESO  (see Sections \ref{sec:Pre_Queso}--\ref{sec:install_Queso_make}).

\begin{lstlisting}[caption={Makefile for the application code in Listings
  \ref{code:modal-main-c}-\ref{code:modal-compute-c}},
  label={code:modal_makefile},
  language={bash}]
  QUESO_DIR = /path/to/queso
  BOOST_DIR = /path/to/boost
  GSL_DIR   = /path/to/gsl

  INC_PATHS = \
     -I. \
     -I$(QUESO_DIR)/include \
     -I$(BOOST_DIR)/include \
     -I$(GSL_DIR)/include

  LIBS = \
     -L$(QUESO_DIR)/lib -lqueso \
     -L$(BOOST_DIR)/lib -lboost_program_options \
     -L$(GSL_DIR)/lib -lgsl

  CXX = mpic++
  CXXFLAGS += -g -Wall -c

  default: all

  .SUFFIXES: .o .C

  all:       modal_example_gsl

  clean:
     rm -f *~
     rm -f *.o
     rm -f modal_gsl

  modal_example_gsl: example_main.o example_likelihood.o example_compute.o
     $(CXX) example_main.o \
            example_likelihood.o \
            example_compute.o \
            -o modal_gsl $(LIBS)

  %.o: %.C
     $(CXX) $(INC_PATHS) $(CXXFLAGS) $<
\end{lstlisting}

Thus, to compile, build and execute the code, the user just needs to run the following commands in the same directory where the files are:
\begin{lstlisting}
$ cd $HOME/LIBRARIES/QUESO-0.51.0/examples/modal/
$ export LD_LIBRARY_PATH=$LD_LIBRARY_PATH:\
  $HOME/LIBRARIES/gsl-1.15/lib/:\
  $HOME/LIBRARIES/boost-1.53.0/lib/:\
  $HOME/LIBRARIES/hdf5-1.8.10/lib:\
  $HOME/LIBRARIES/QUESO-0.51.0/lib 
$ make -f Makefile_modal_violeta 
$ ./modal_gsl example.inp <num_modes>
\end{lstlisting}

The `\verb+export+' instruction above is only necessary if the user has not saved it in his/her \verb+.bashrc+ file.

\subsection{Data Post-Processing and Visualization}\label{sec:modal-results}

According to the specifications of the input file in Listing~\ref{code:modal-input-file}, both a folder named \verb+outputData+ and a the following files should be generated:
\begin{verbatim}
rawChain_ml.m 
display_sub0.txt    
\end{verbatim}

The sequence of Matlab commands is identical to the ones presented in Sections
\ref{sec:sip-results}, \ref{sec:sfp-results}, \ref{sec:gravity-results} and \ref{sec:tga-results};
therefore, are omitted here. The reader is invited to explore the Matlab files
\texttt{plot\_modal\_all\_levels\_1mode.m}  and/or \texttt{plot\_modal\_all\_levels\_2modes.m}  
for details of how the figures have been generated.

\subsubsection{Scatter Plots}

\begin{figure}[htpb]
\centering
\subfloat{\includegraphics[scale=0.3]{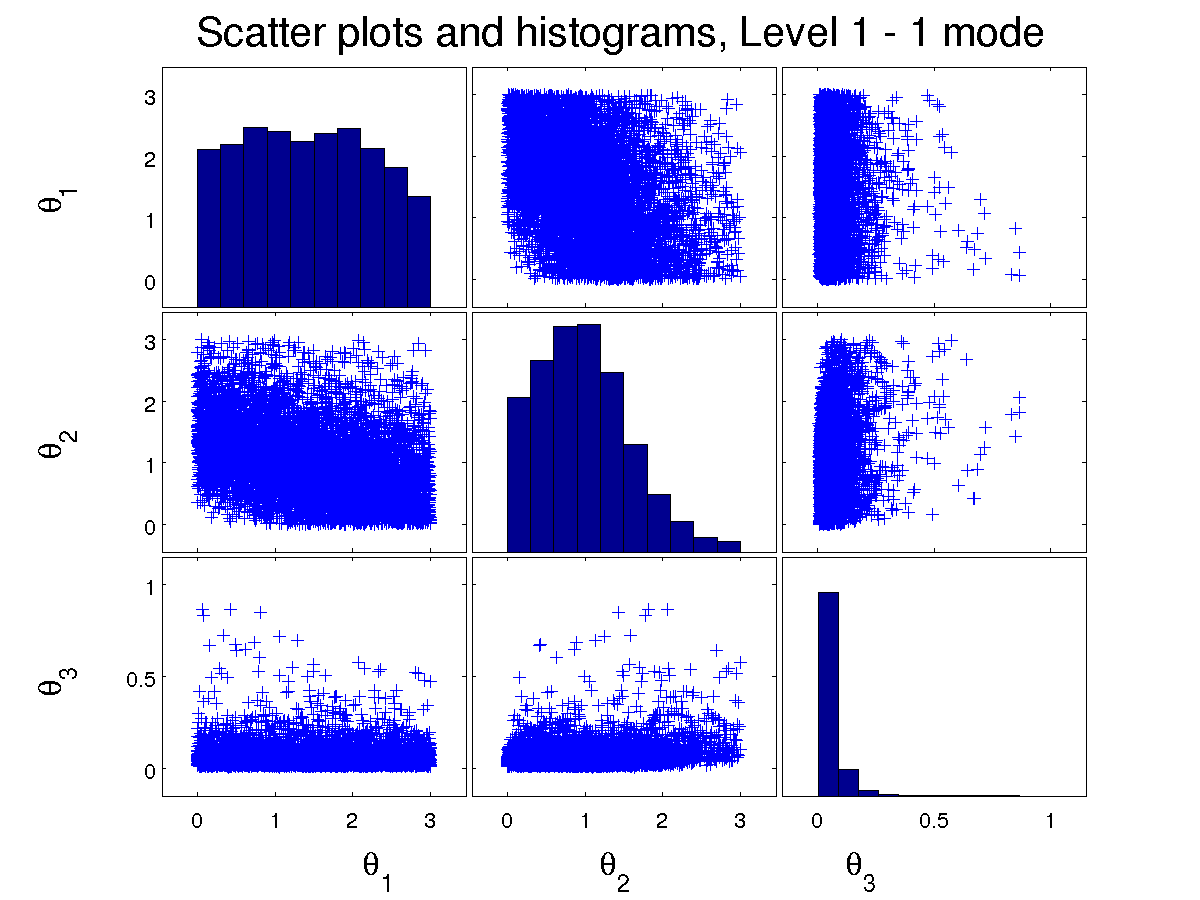}}
\subfloat{\includegraphics[scale=0.3]{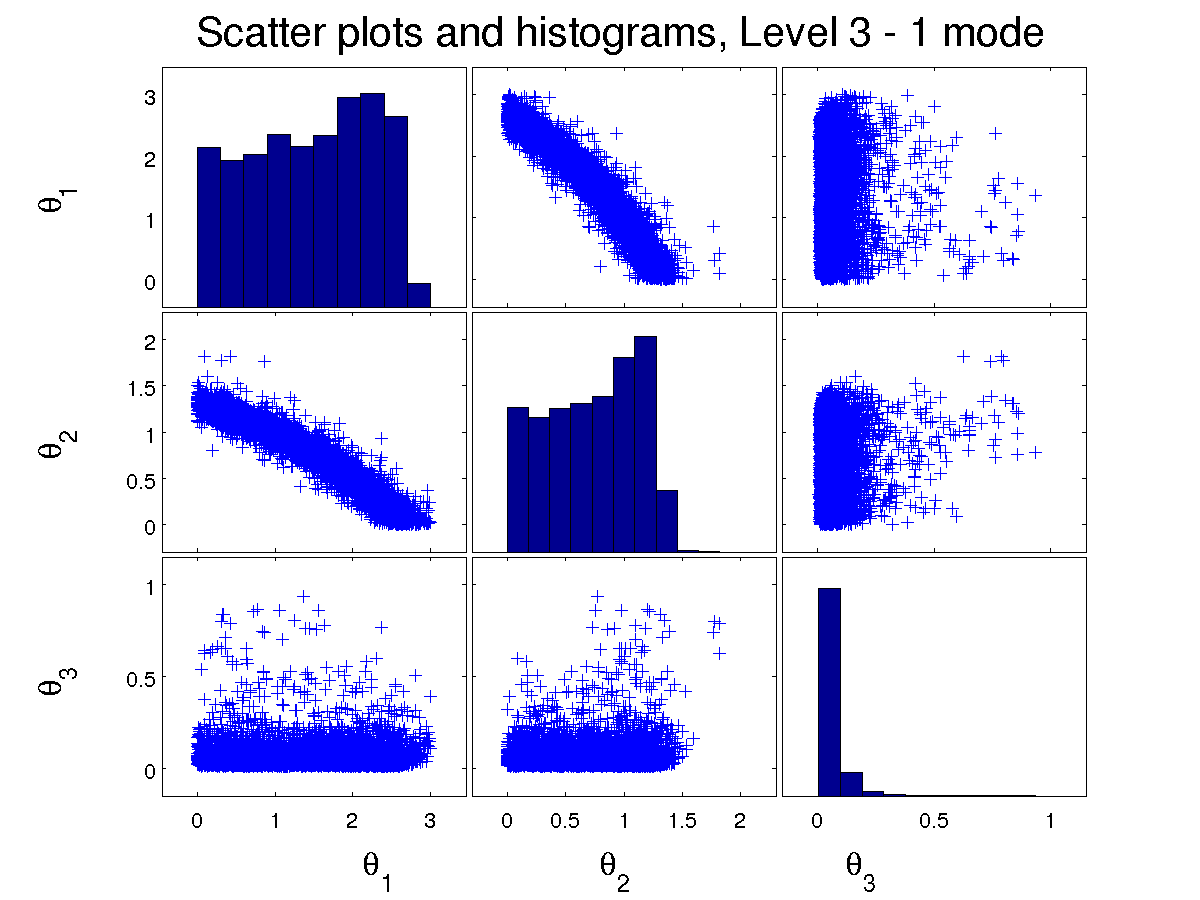}}\\
\subfloat{\includegraphics[scale=0.3]{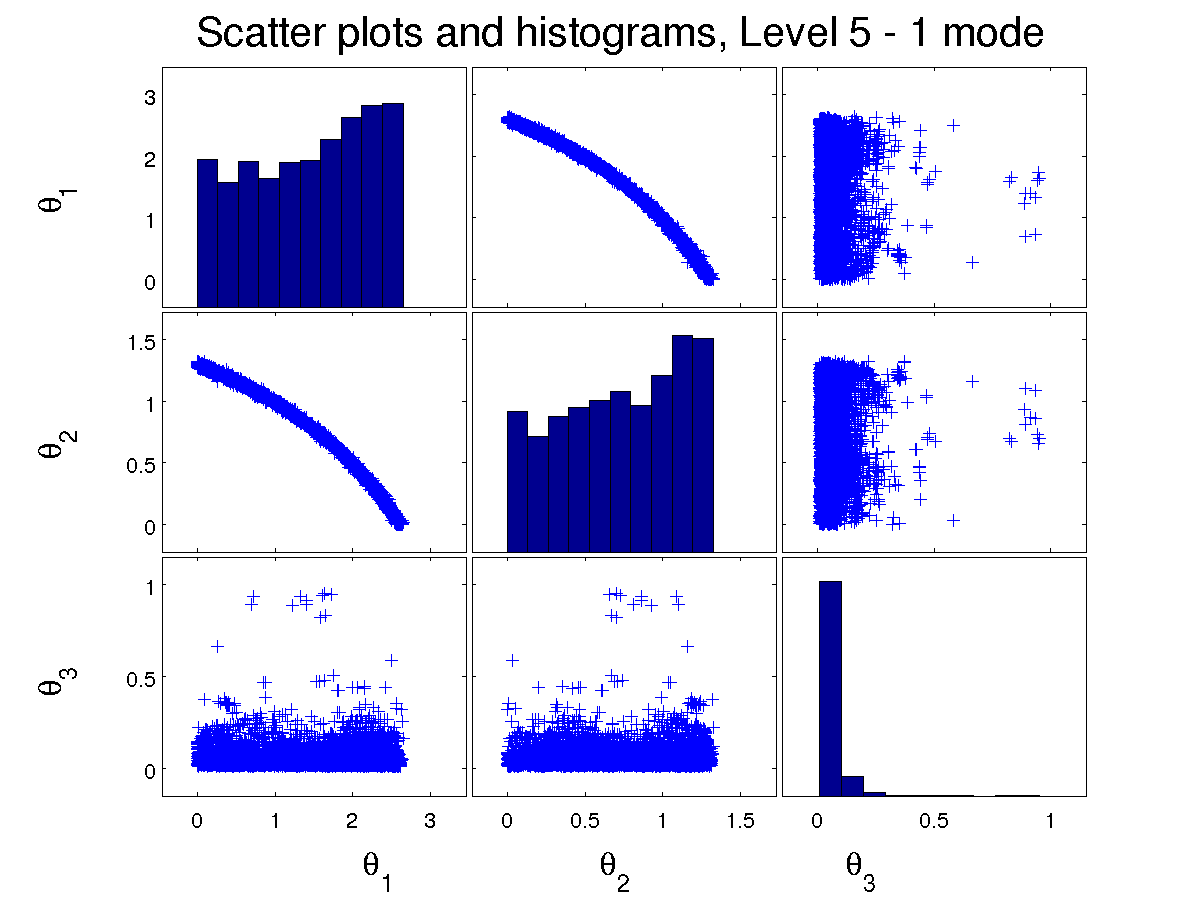}}
\subfloat{\includegraphics[scale=0.3]{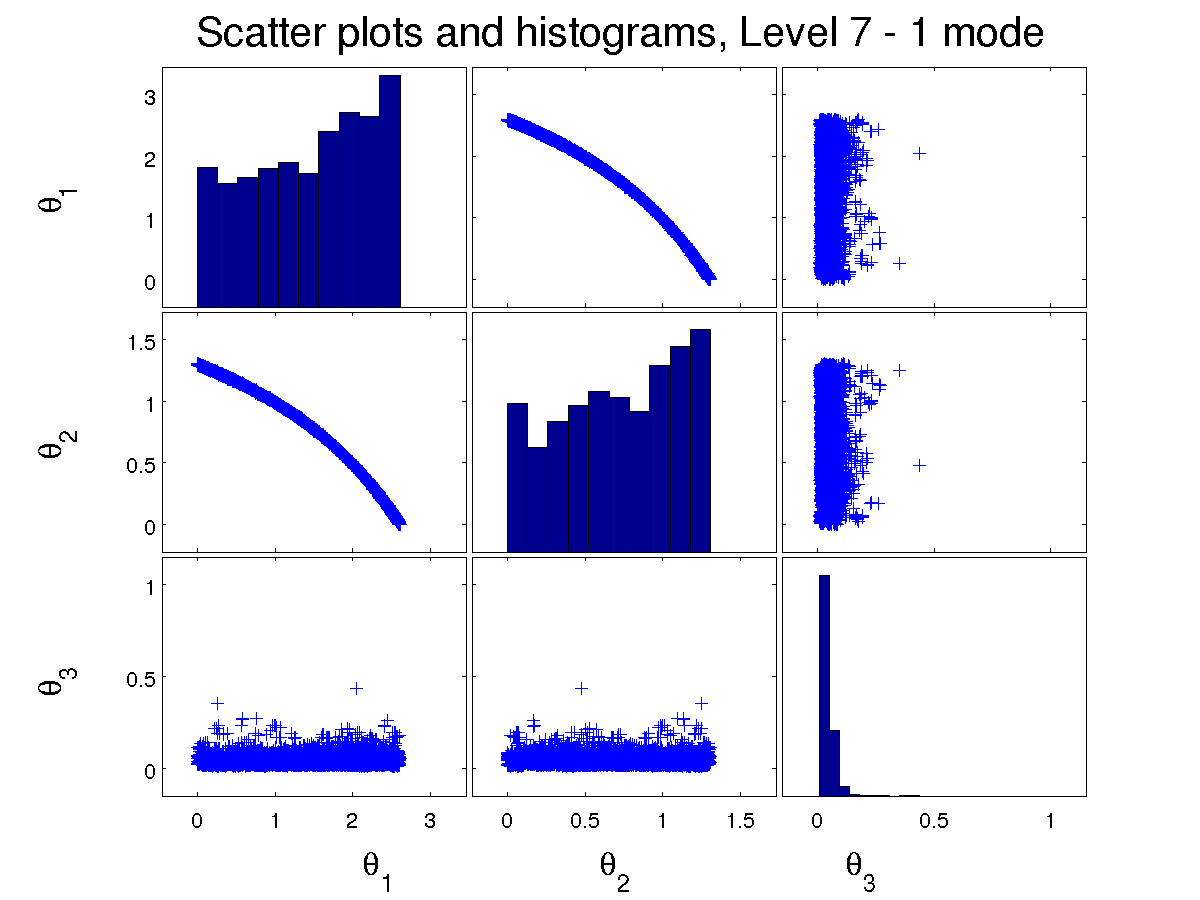}}
\vspace{-10pt}
\caption{Scatter plots for $\theta_1$, $\theta_2$ and $\theta_3=\sigma^2$, levels 1, 3, 5 and 7 (last). One mode distribution.}
\label{fig:modal_scatter_1mode}
\end{figure}

\begin{figure}[htpb]
\centering
\subfloat{\includegraphics[scale=0.3]{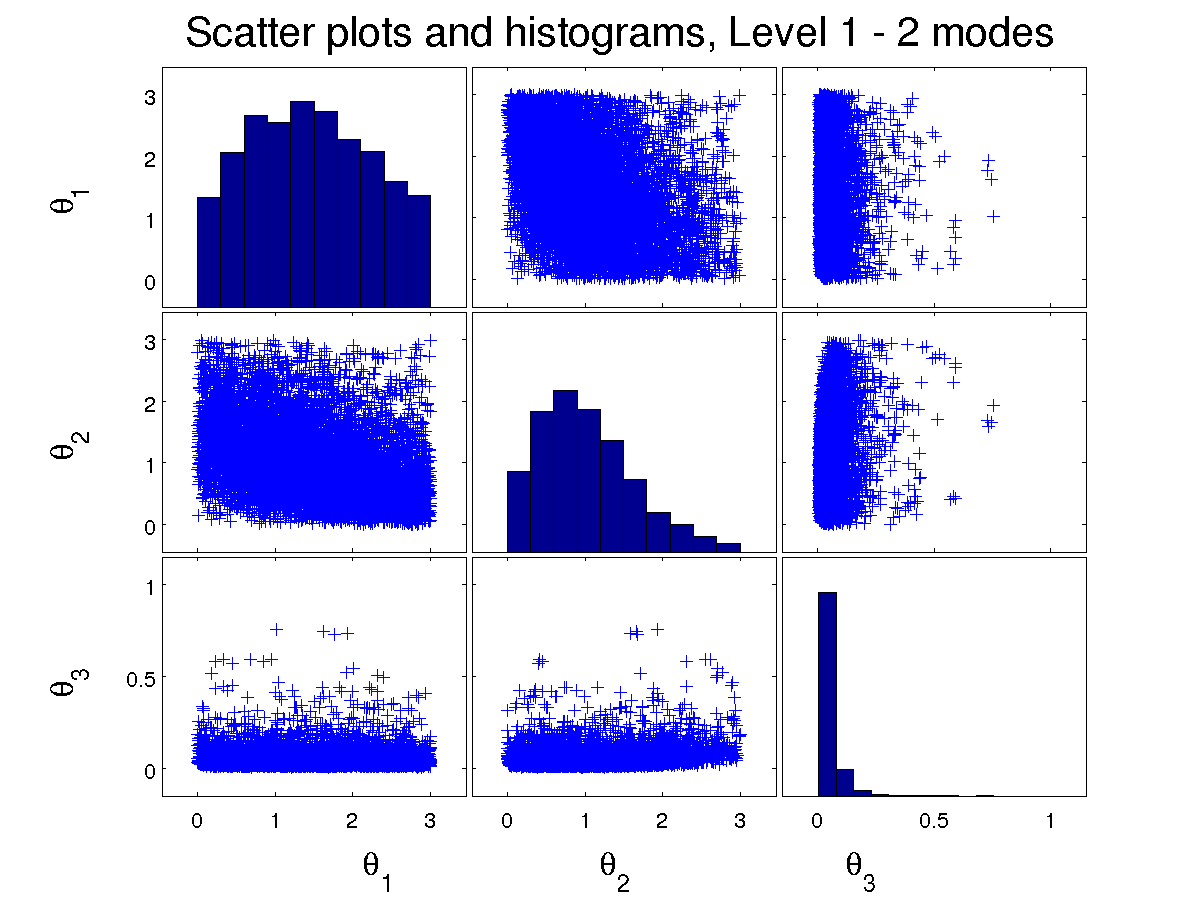}}
\subfloat{\includegraphics[scale=0.3]{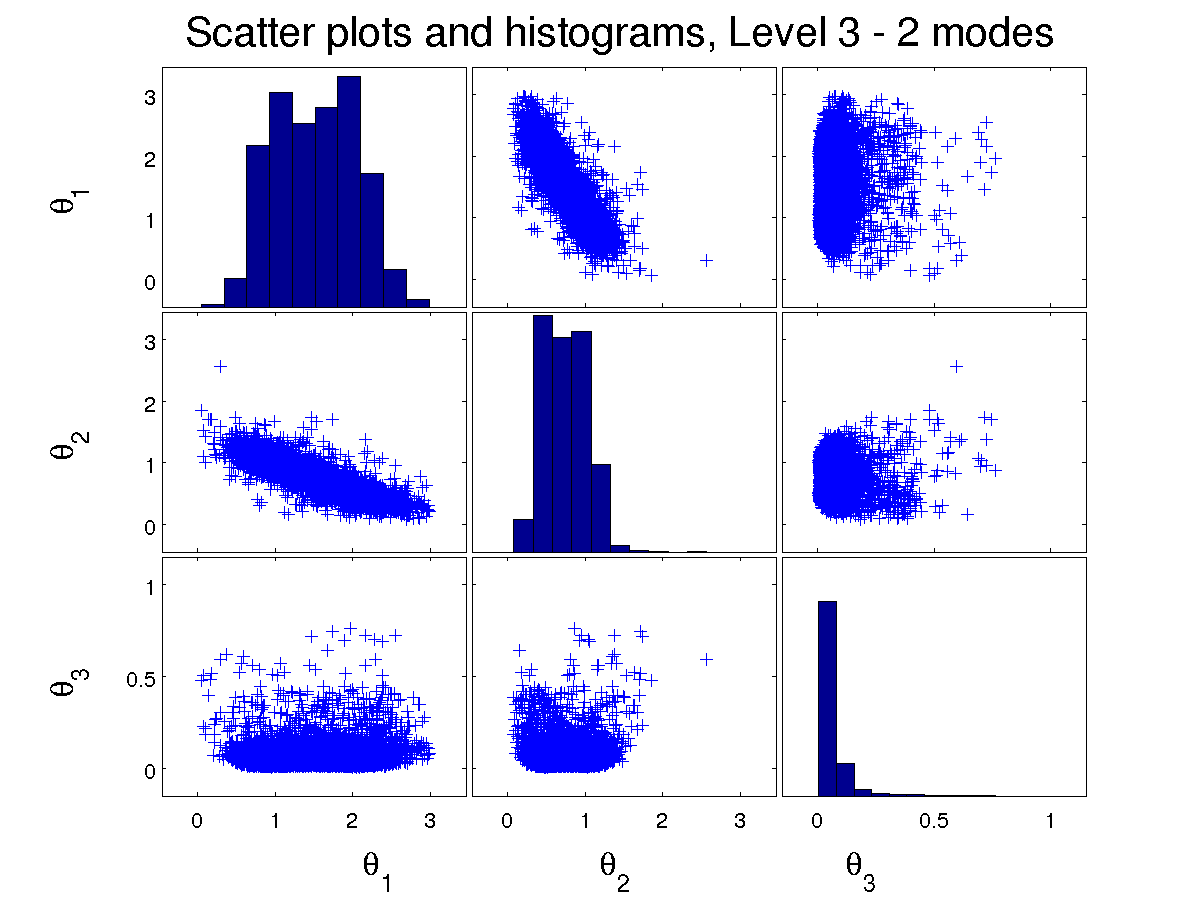}}\\
\subfloat{\includegraphics[scale=0.3]{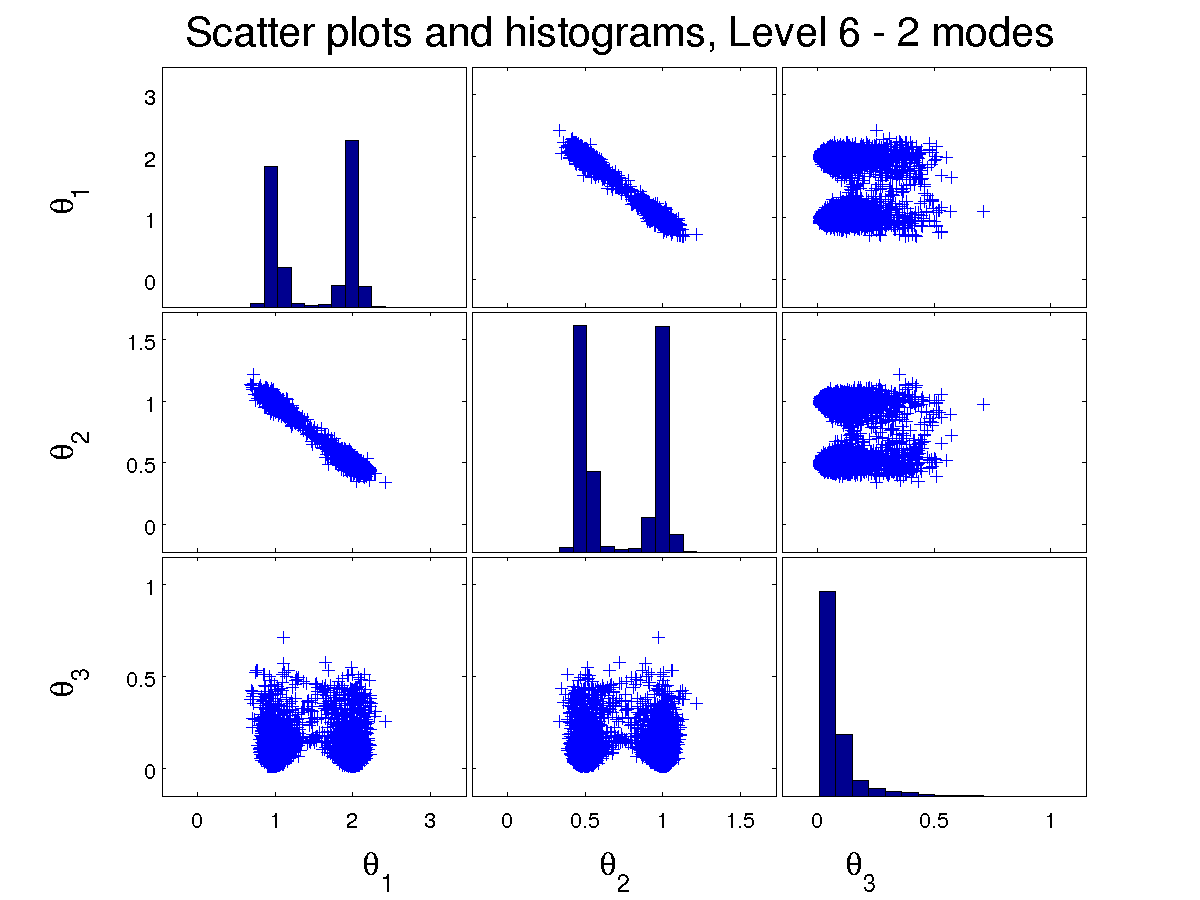}}
\subfloat{\includegraphics[scale=0.3]{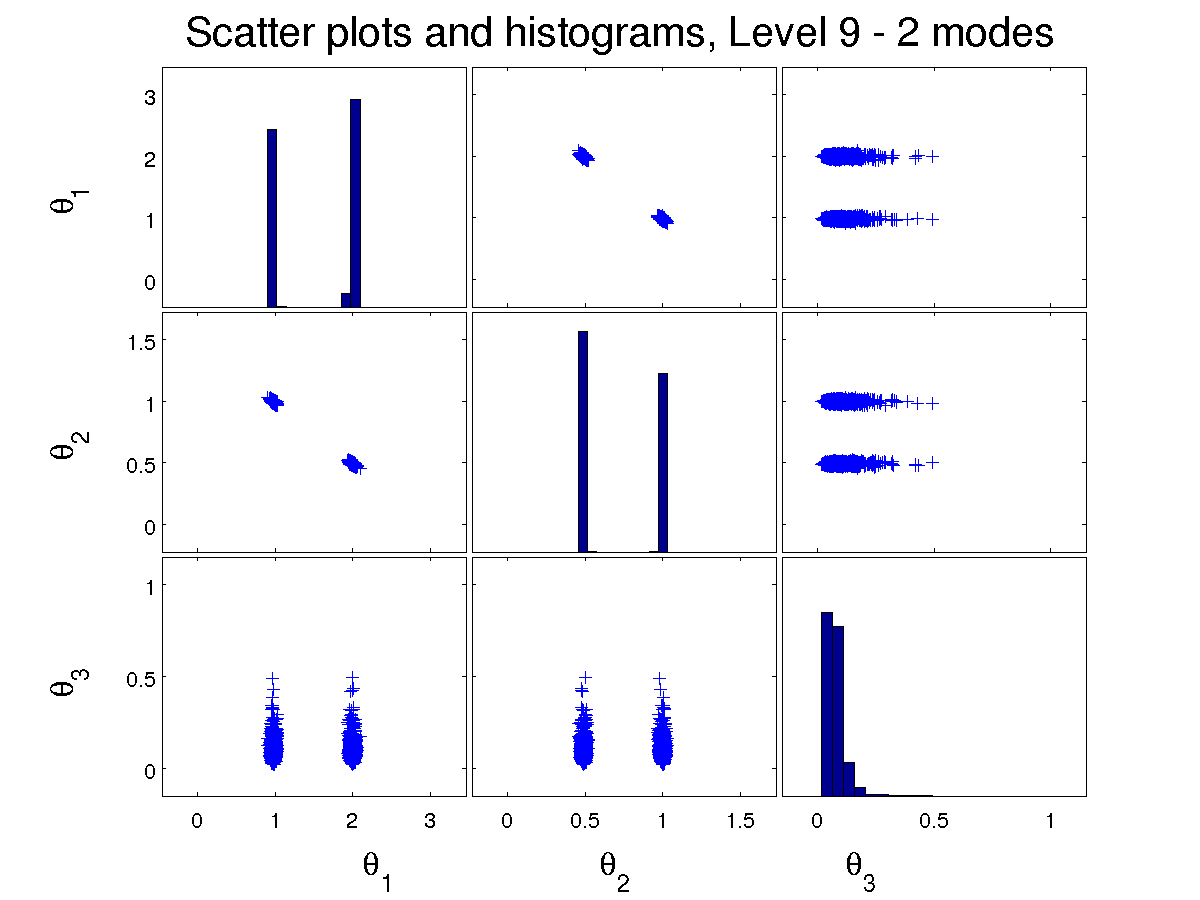}}
\vspace{-10pt}
\caption{Scatter plots for $\theta_1$, $\theta_2$ and $\theta_3=\sigma^2$, levels 1, 3, 6 and 9 (last). Two-mode distribution.}
\label{fig:modal_scatter_2modes}
\end{figure}

The code presented in Listing \ref{matlab:modal_scatter} uses Matlab function \verb+plotmatrix+ to generate Figures \ref{fig:modal_scatter_1mode} and \ref{fig:modal_scatter_2modes}
which presents the scatter plots and histograms of the parameters $\theta_1$ and $\theta_2$, based on the generated raw chains.

\begin{lstlisting}[label=matlab:modal_scatter,caption={Matlab code for the scatter plots depicted in Figures \ref{fig:modal_scatter_1mode} and \ref{fig:modal_scatter_2modes}.}]
fprintf(1,'Scatter plots and histograms of raw chains - Level 1 <press any key>\n');
plotmatrix(ip_ml_1_rawChain_unified, '+b')
set(gca,'fontsize',20); 
xlabel('\theta_1                  \theta_2                   \theta_3','fontsize',16);
ylabel('\theta_3                  \theta_2                   \theta_1','fontsize',16);
title('Scatter plots and histograms, Level 1 - 1 mode')
\end{lstlisting}

\subsubsection{KDE Plots}

\begin{figure}[hptb]
\centering
\subfloat{\includegraphics[scale=0.3]{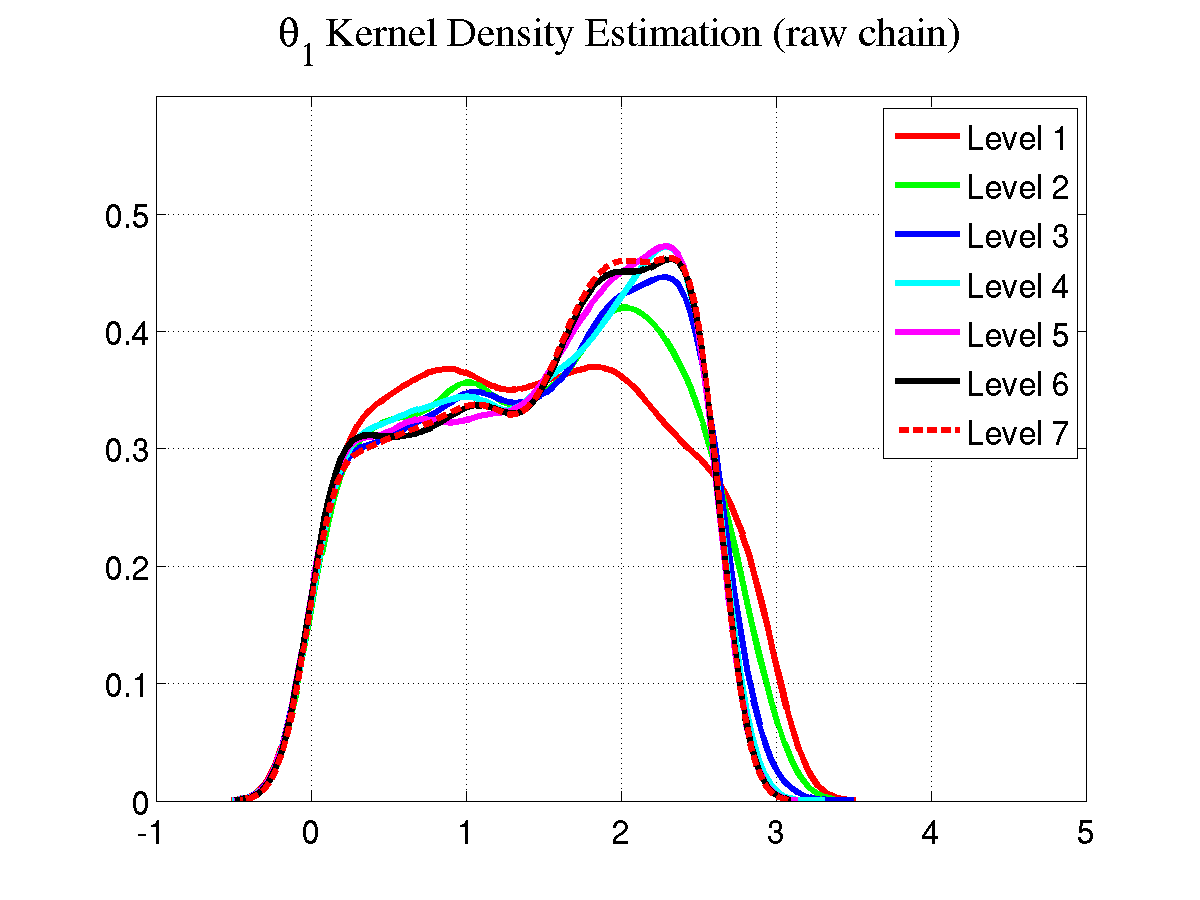}}
\subfloat{\includegraphics[scale=0.3]{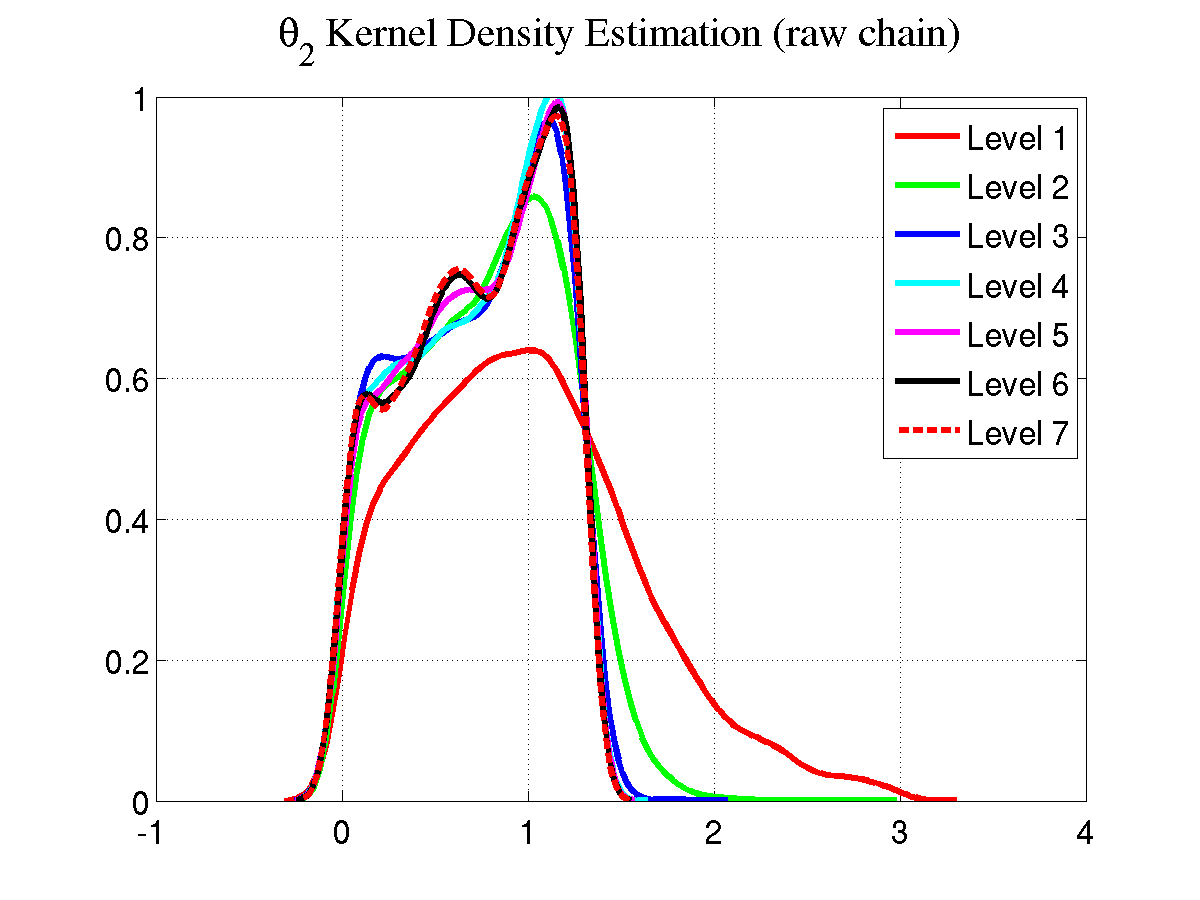}}\\
\subfloat{\includegraphics[scale=0.3]{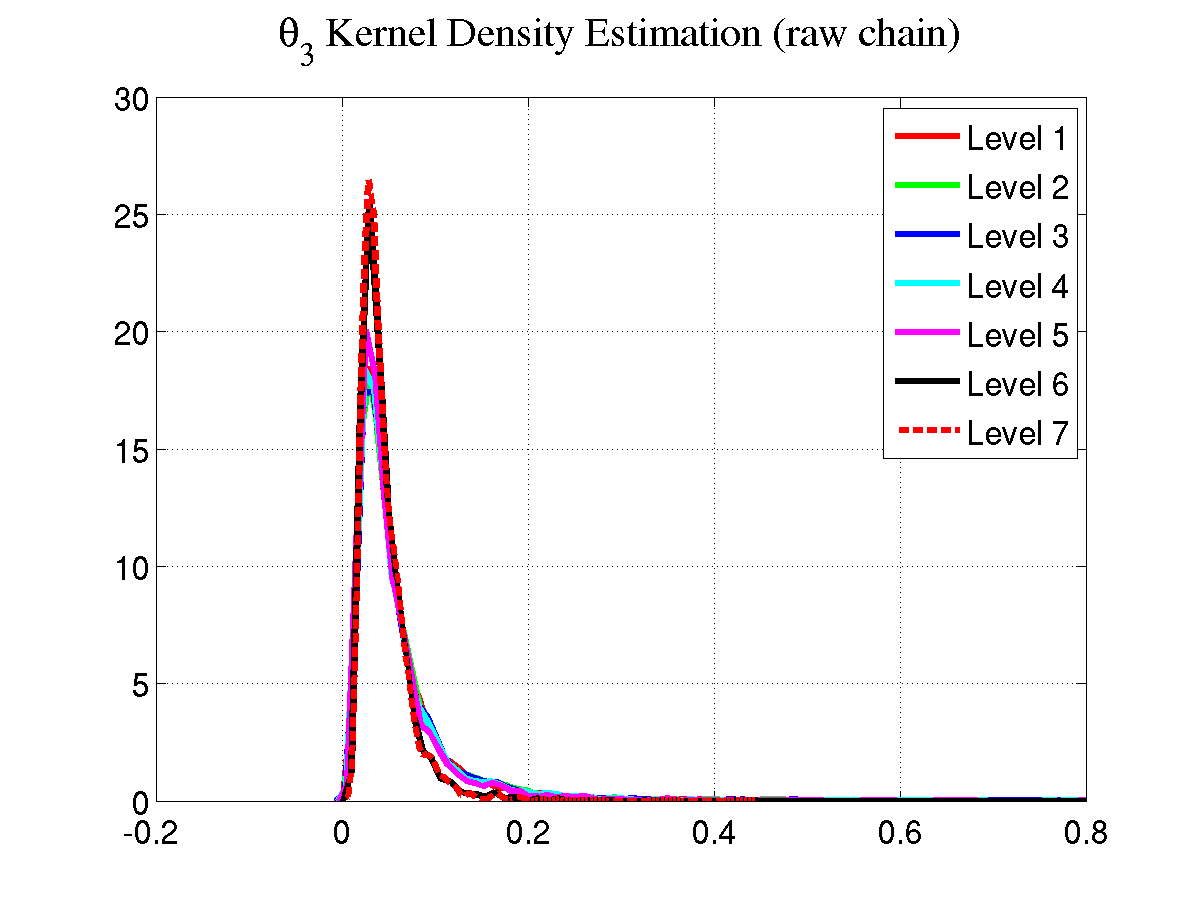}}
\subfloat{\includegraphics[scale=0.3]{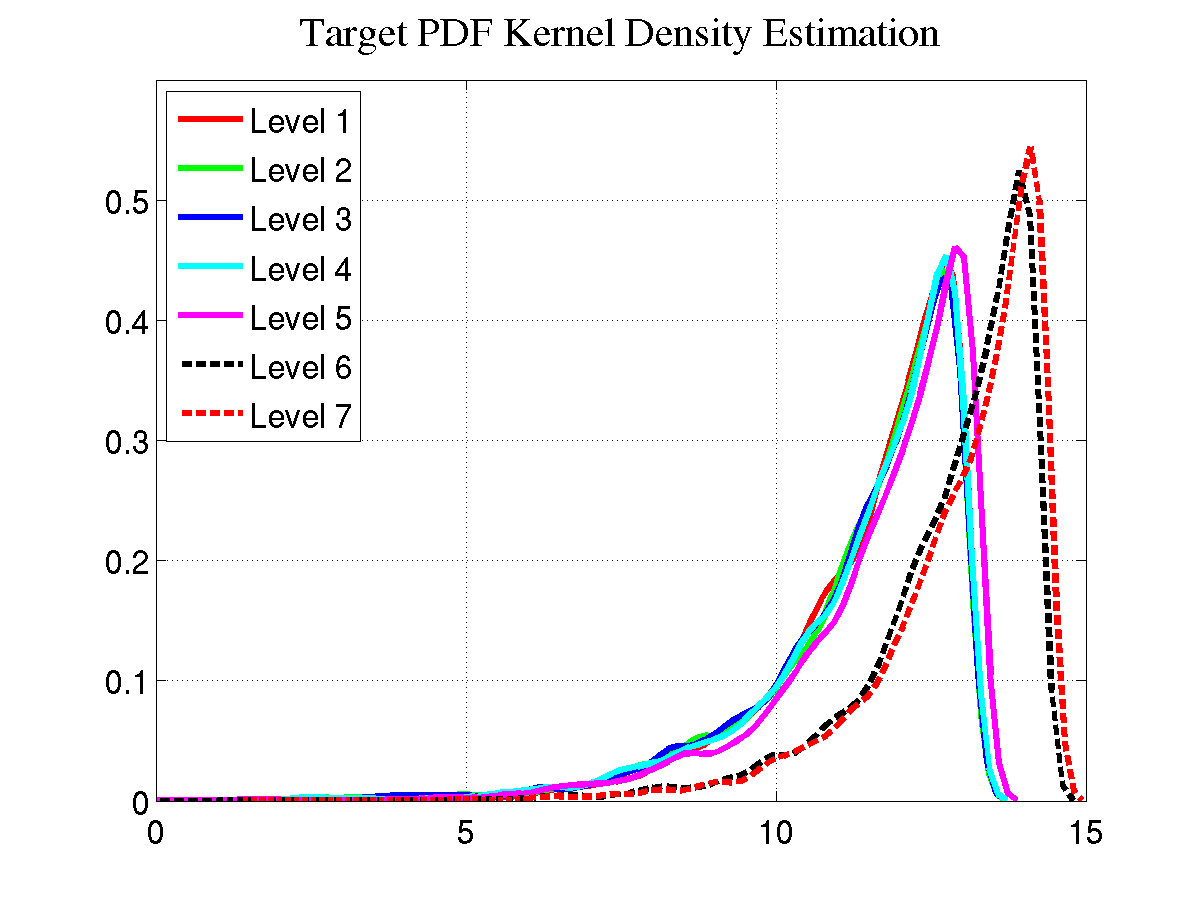}}
\vspace{-10pt}
\caption{KDE plots for $\theta_1$, $\theta_2$, $\theta_3=\sigma^2$, and the target PDF. One mode distribution.}
\label{fig:modal_kde_1mode}
\end{figure}

\begin{figure}[hptb]
\centering
\subfloat{\includegraphics[scale=0.3]{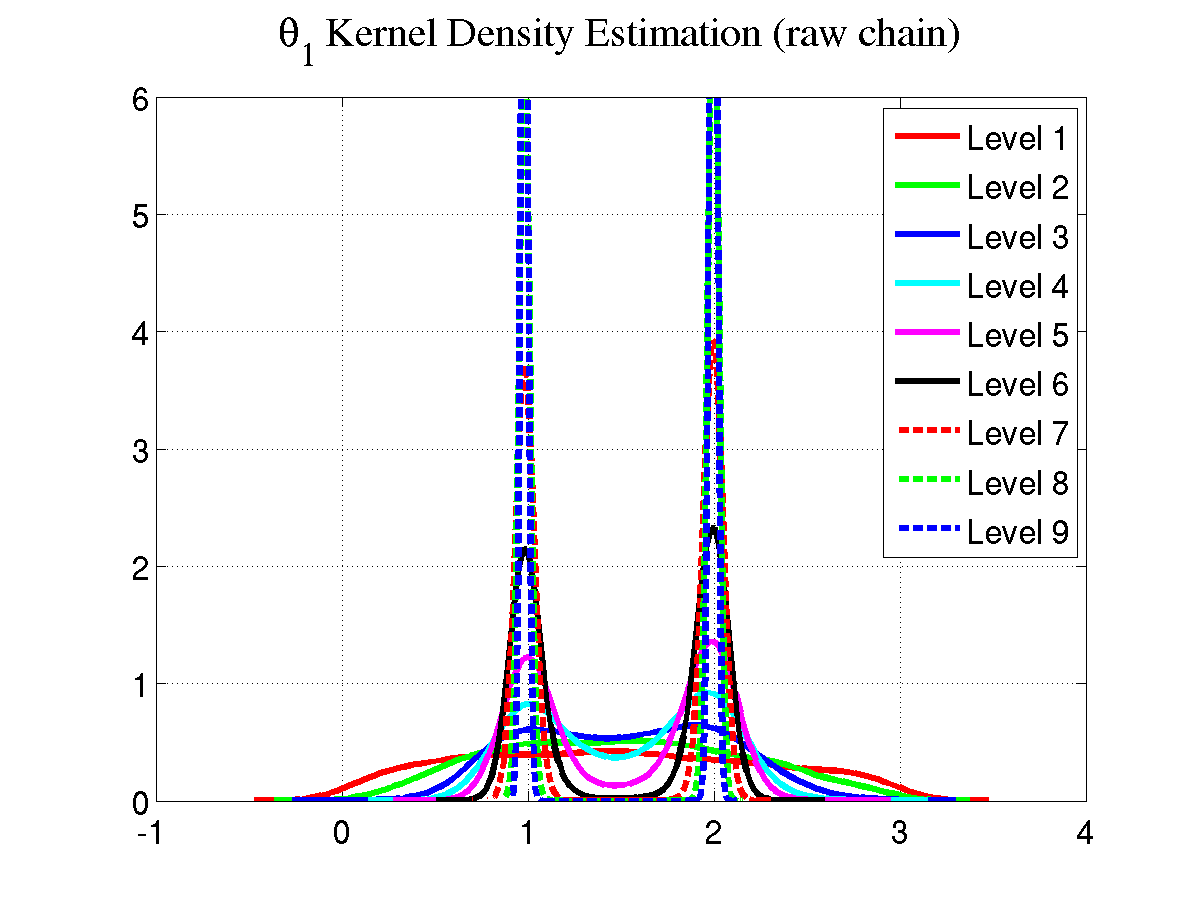}}
\subfloat{\includegraphics[scale=0.3]{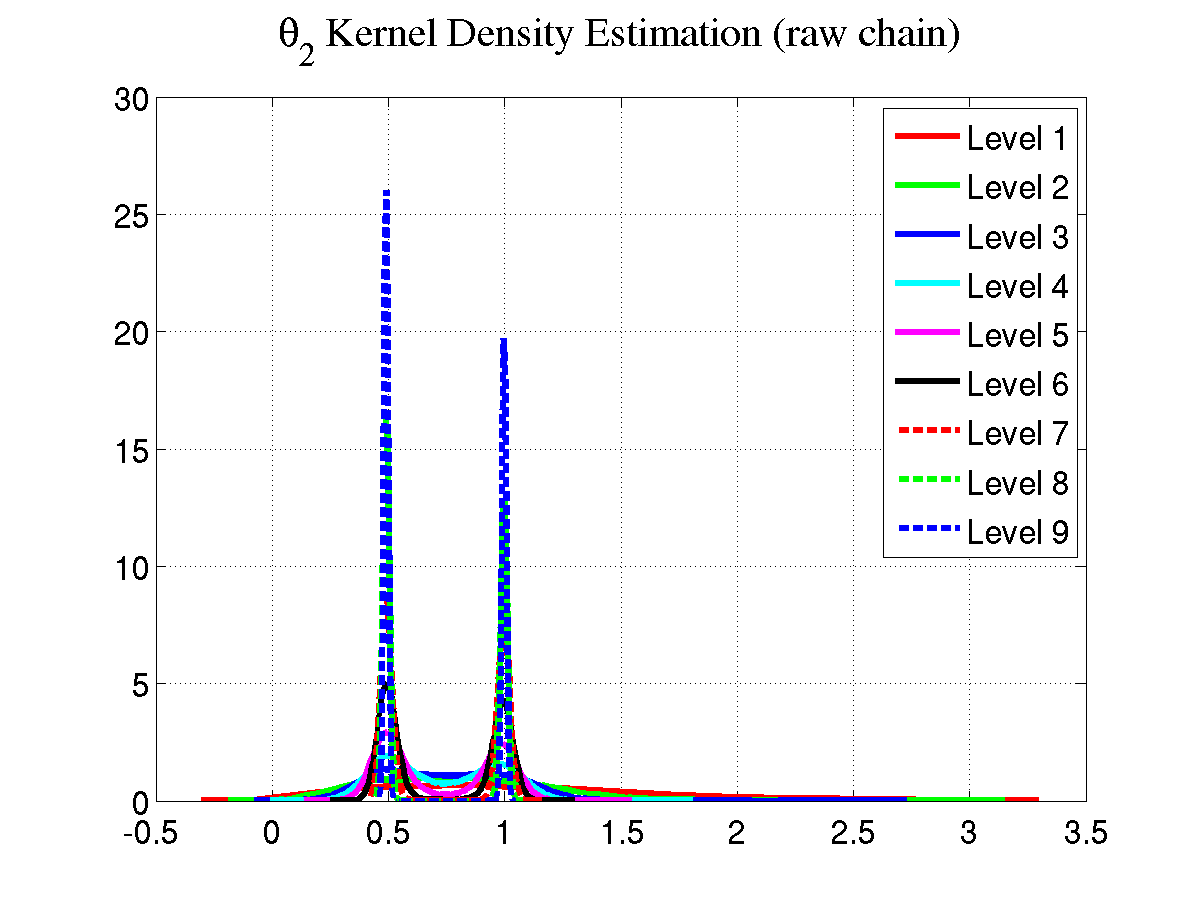}}\\
\subfloat{\includegraphics[scale=0.3]{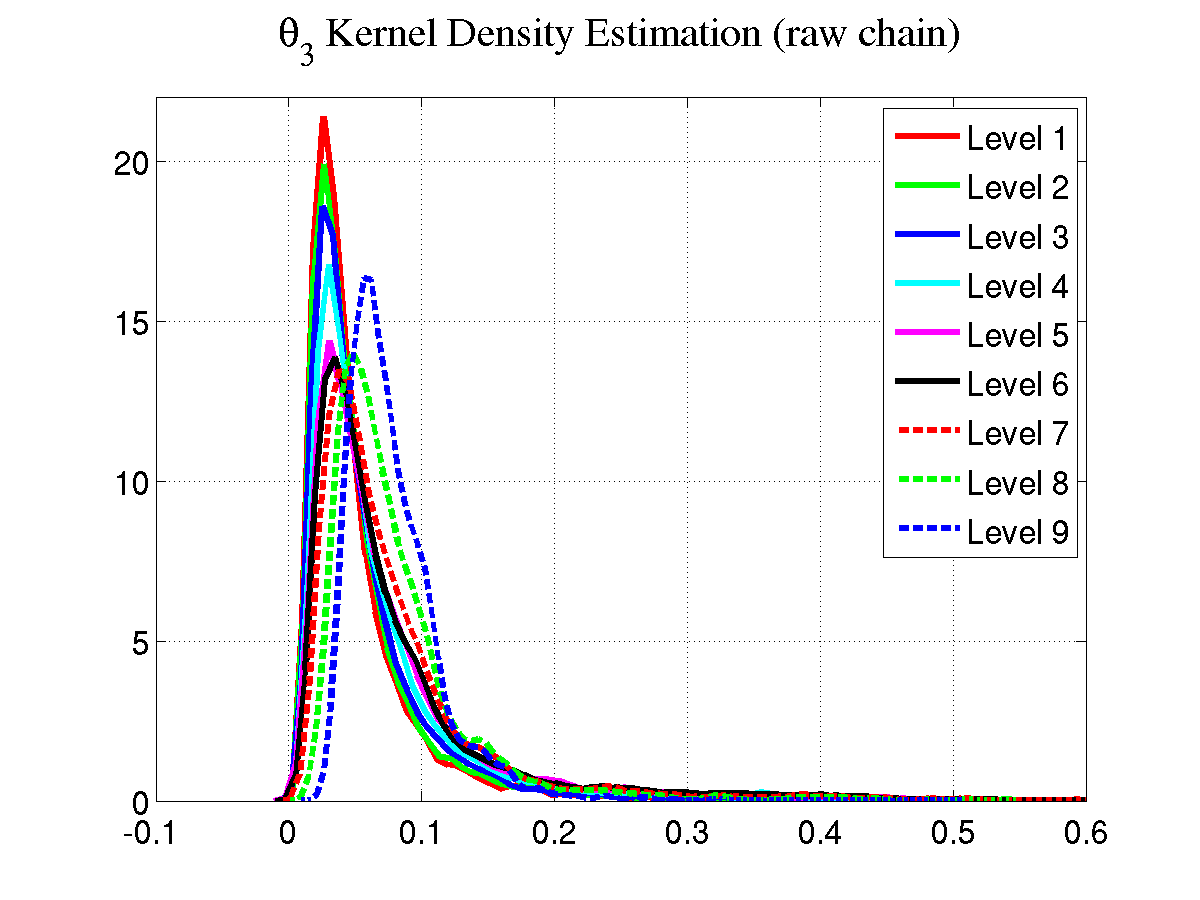}}
\subfloat{\includegraphics[scale=0.3]{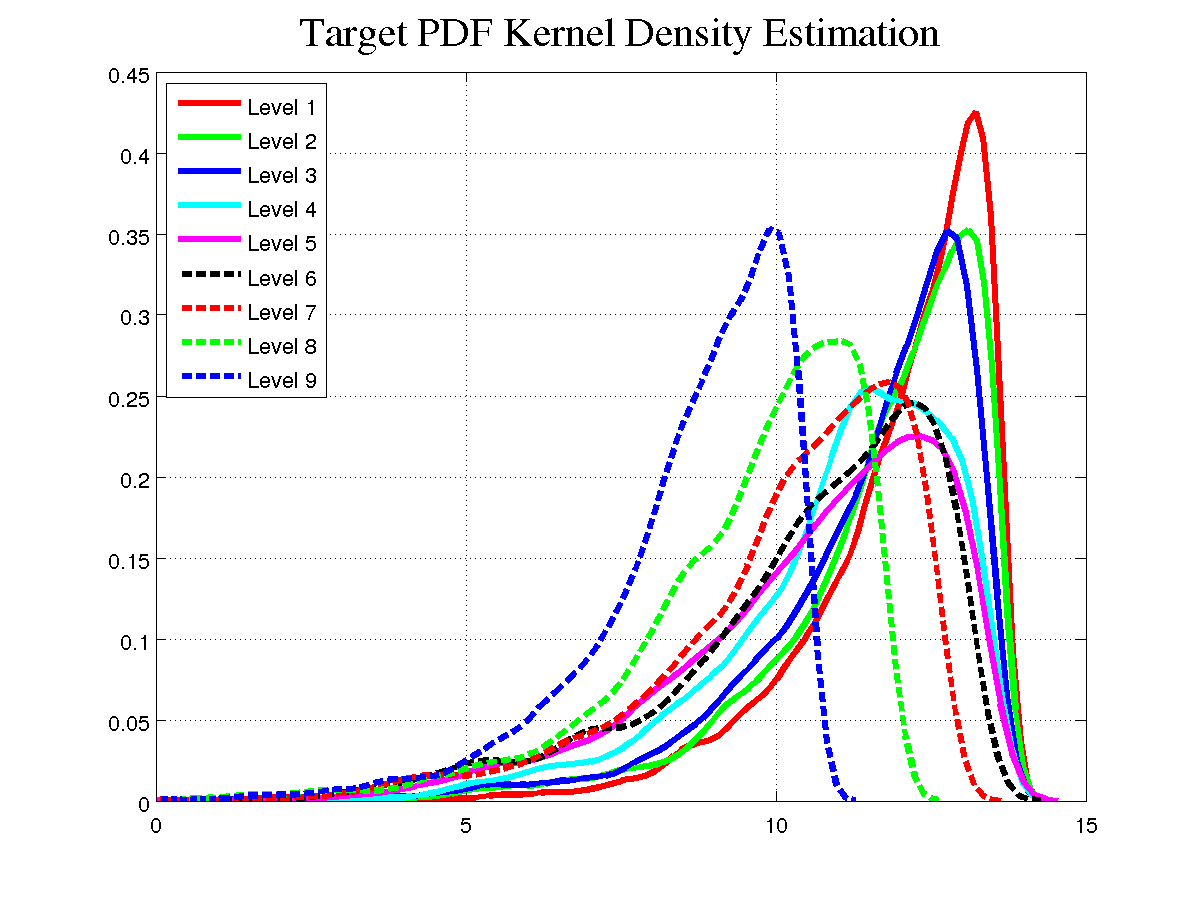}}
\vspace{-10pt}
\caption{KDE plots for $\theta_1$, $\theta_2$, $\theta_3=\sigma^2$, and the target PDF. Two-mode distribution.}
\label{fig:modal_kde_2modes}
\end{figure}

Figures \ref{fig:modal_kde_1mode} and \ref{fig:modal_kde_2modes} present the KDE plots of the parameters $\theta_1$, $\theta_2$, $\theta_3$ and target PDF in both cases: one-mode and two-modes distribution.

\subsubsection{Autocorrelation Plots}

\begin{figure}[htpb]
\centering
\subfloat{\includegraphics[scale=0.25]{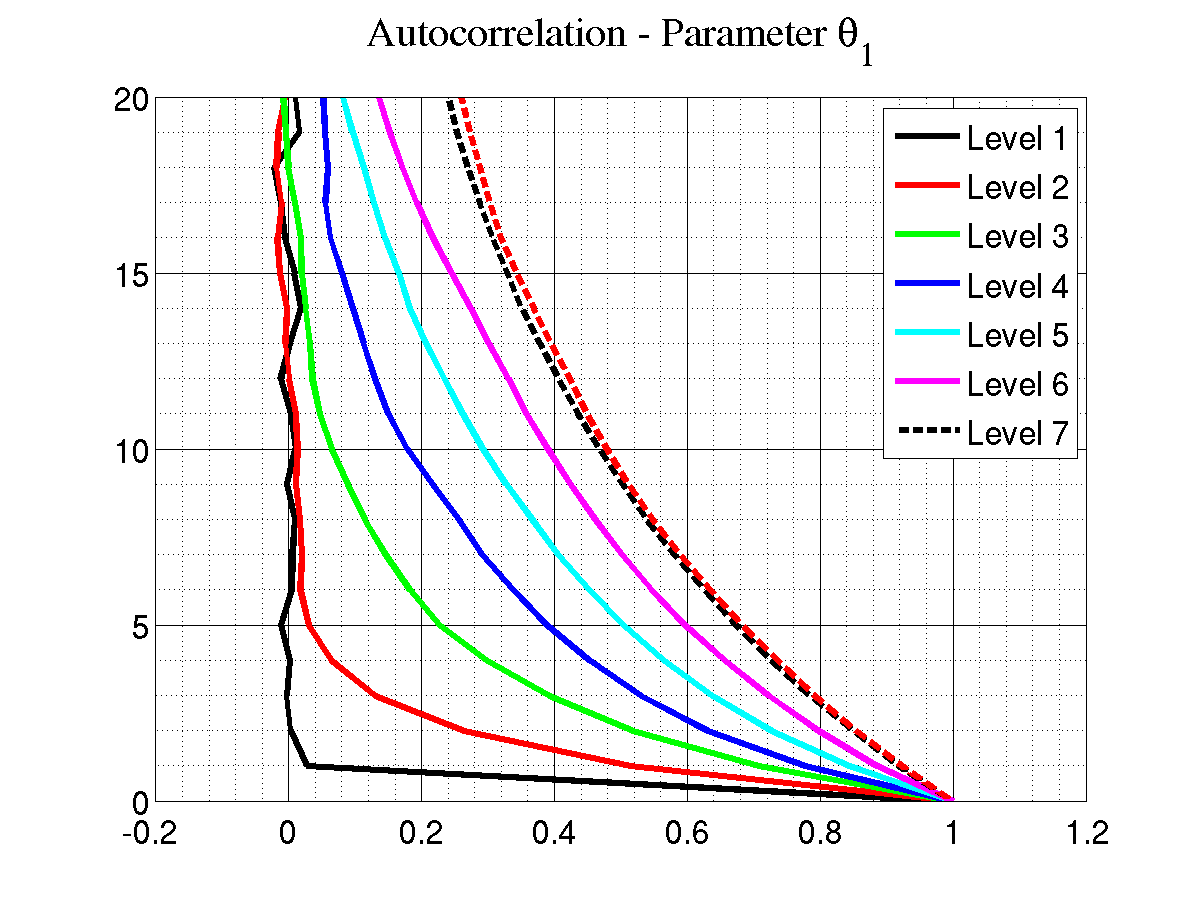}}
\subfloat{\includegraphics[scale=0.25]{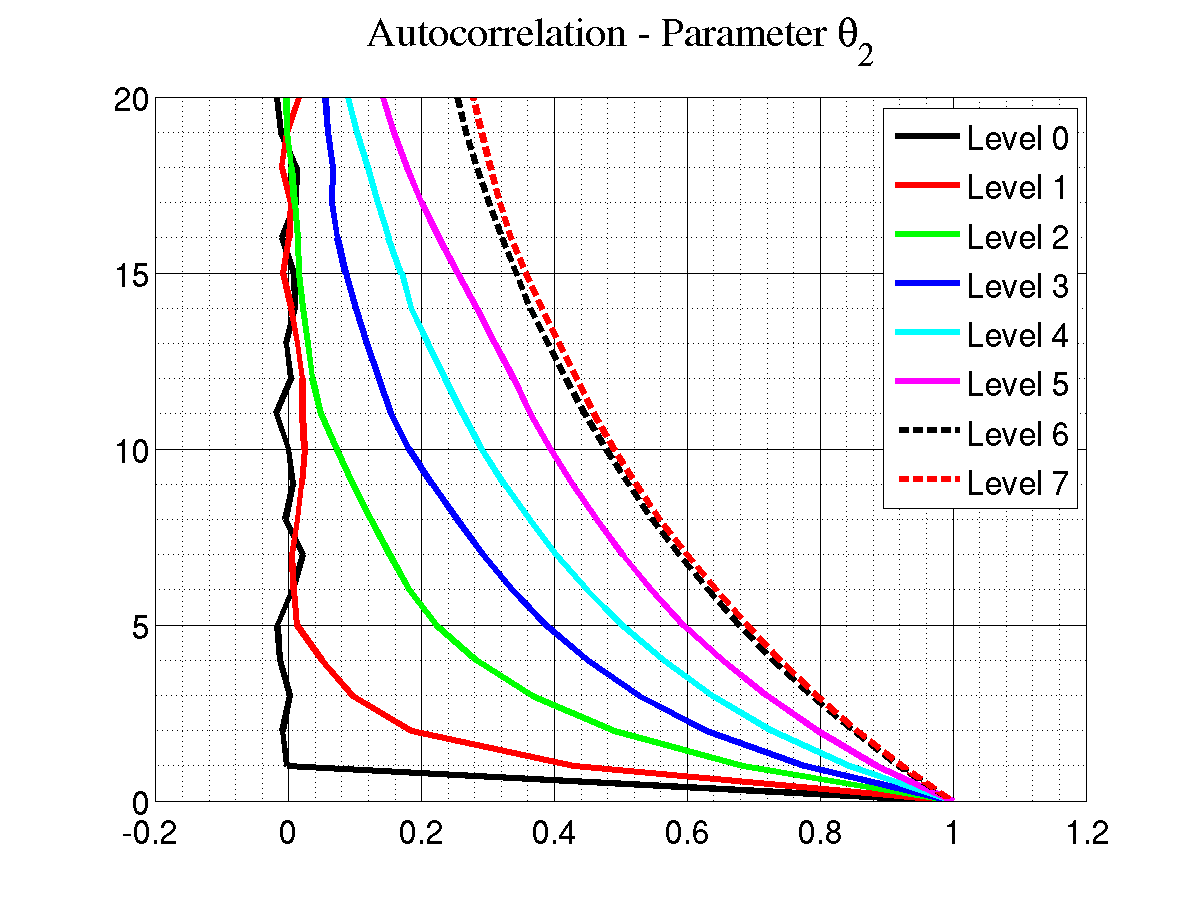}}
\subfloat{\includegraphics[scale=0.25]{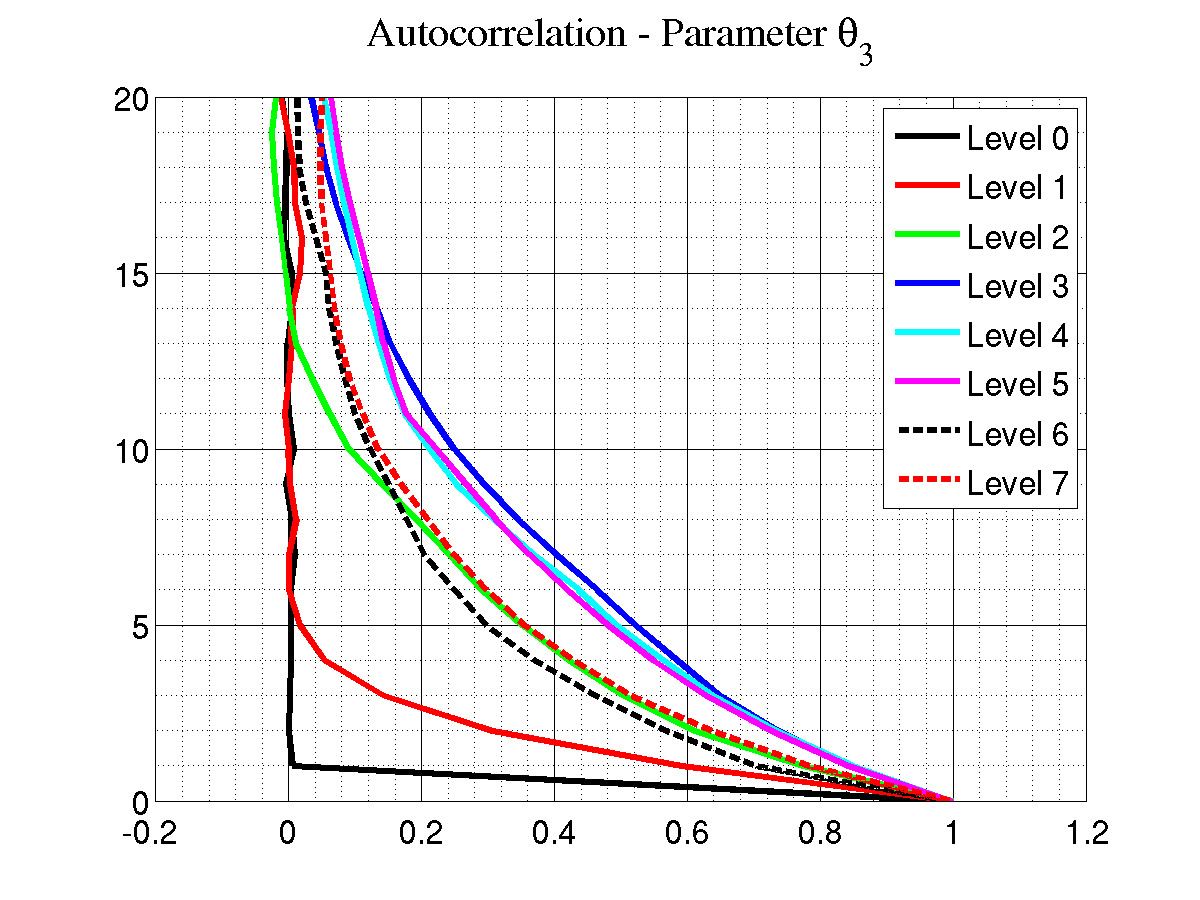}}
\vspace{-10pt}
\caption{Autocorrelation plots for $\theta_1$, $\theta_2$ and $\theta_3=\sigma^2$. One-mode distribution.}
\label{fig:modal_autocorr_1mode}
\end{figure}

\begin{figure}[htpb]
\centering
\subfloat{\includegraphics[scale=0.25]{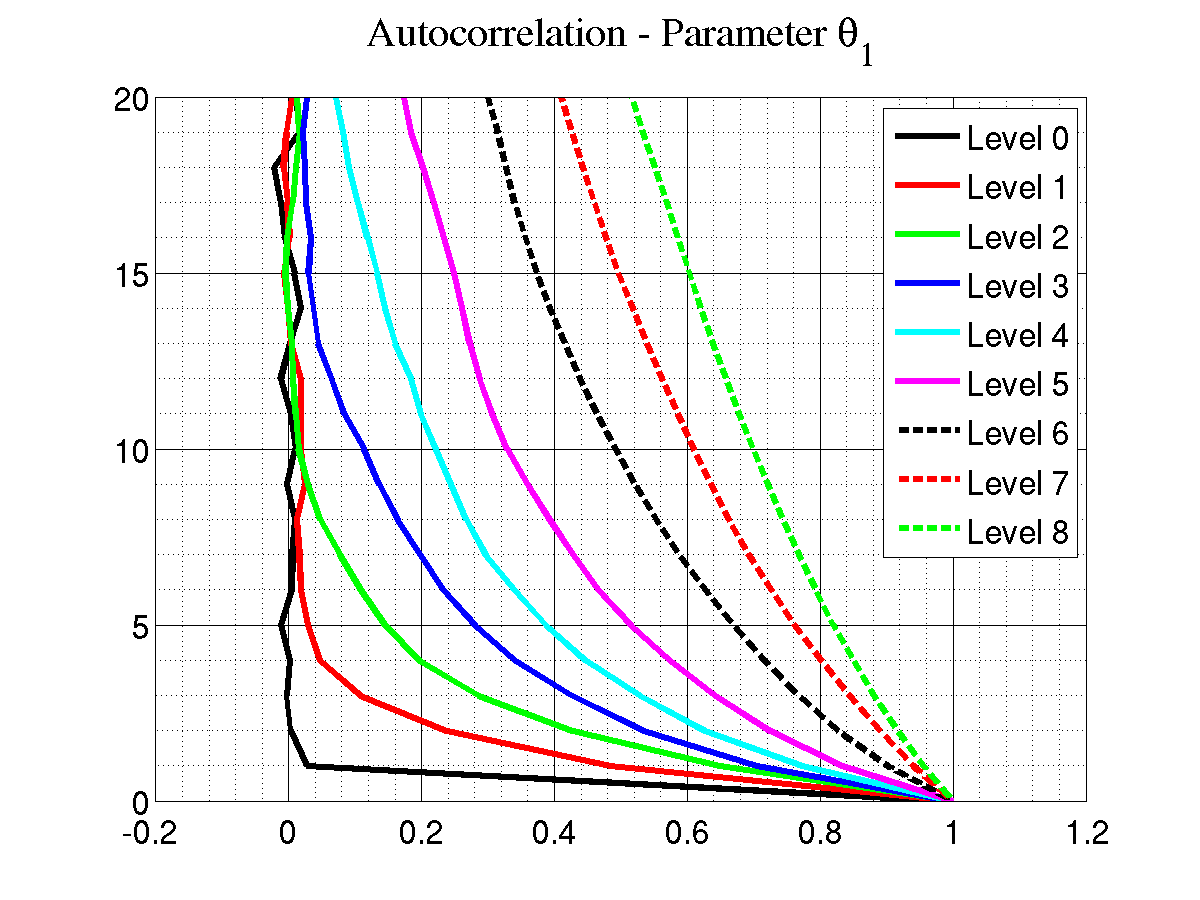}}
\subfloat{\includegraphics[scale=0.25]{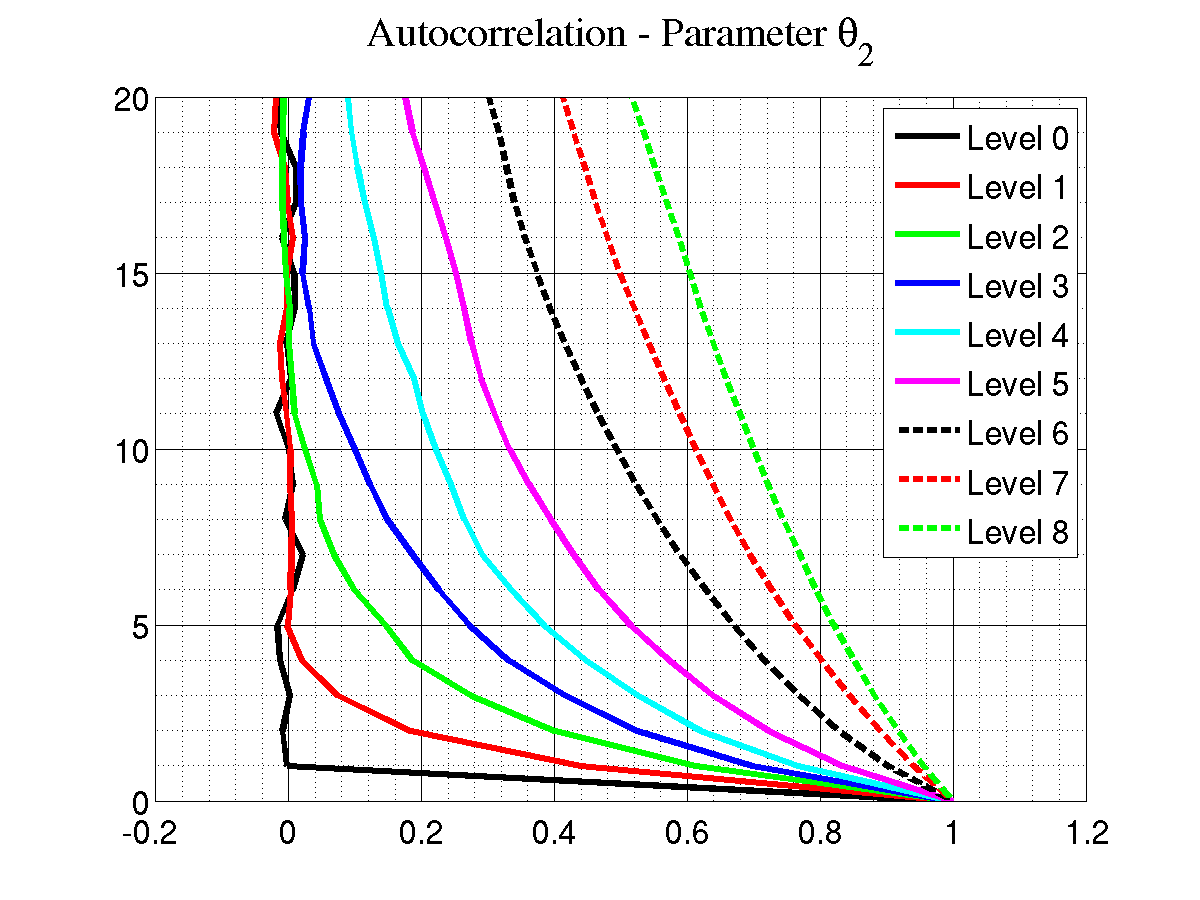}}
\subfloat{\includegraphics[scale=0.25]{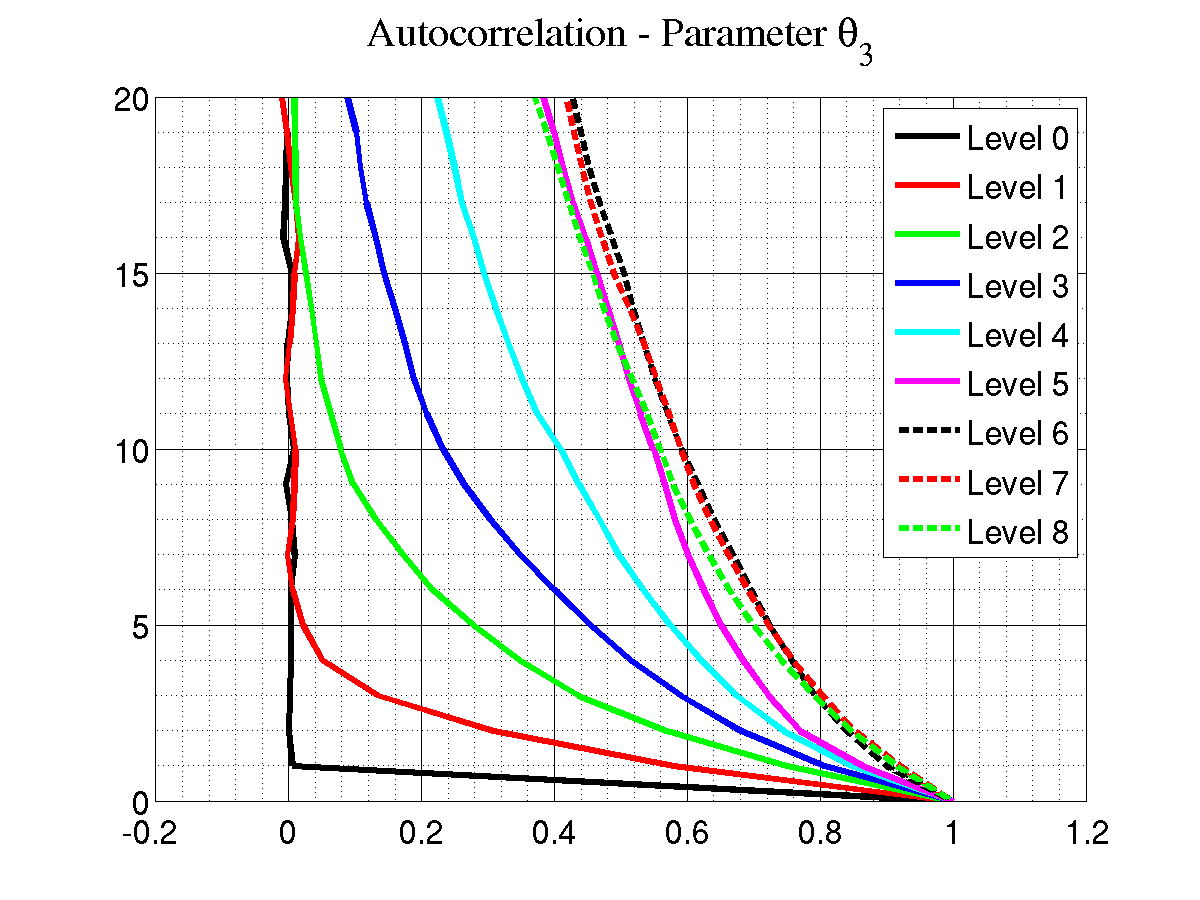}}
\vspace{-10pt}
\caption{Autocorrelation plots for $\theta_1$, $\theta_2$ and $\theta_3=\sigma^2$. Two-mode distribution.}
\label{fig:modal_autocorr_2modes}
\end{figure}

Figures \ref{fig:modal_autocorr_1mode} and \ref{fig:modal_autocorr_2modes} present the autocorrelation of the parameters $\theta_1$, $\theta_2$ and $\theta_3$ in both cases: one-mode and two-modes distribution.

\section{\texttt{bimodal}}\label{sec:example_bimodal}

This example replicates the problem in ``Section 4.1 A 1D Problem'' of~\cite{CheungPrudencio2012}: it presents how to use QUESO and the Multilevel method for sampling from a posterior PDF composed of the sum of two Gaussian distributions. 

Let's define $D=[-250,250]$ and the three distributions $\pi_\prior: D \rightarrow \mathbb{R}_+ $, $f_1: \mathbb{R} \rightarrow \mathbb{R}_+ $ and $f_2: \mathbb{R} \rightarrow \mathbb{R}_+$ by:
\begin{equation}
\begin{split}\label{eq:bimodal_functions}
\pi_\prior &=  \dfrac{1}{|D|} = \dfrac{1}{500}, \quad \forall \,  \theta \in D \\
f_1(\theta) &= \dfrac{1}{(2\pi)^{1/2} \sqrt{|V_1|}} \exp \left(-\dfrac{1}{2}(\theta - \mu_1)^T \, V_1^{-1} \, (\theta - \mu_1) \right), \quad \forall \,  \theta \in \mathbb{R} \\
f_2(\theta) &= \dfrac{1}{(2\pi)^{1/2} \sqrt{|V_2|}} \exp \left(-\dfrac{1}{2}(\theta - \mu_2)^T \, V_2^{-1} \, (\theta - \mu_2) \right), \quad \forall \, \theta \in \mathbb{R},
\end{split}
\end{equation}
where
\begin{equation*}
\mu_1 = 10, \quad  V_1 = 1^2, \quad \mu_2 = 100, \quad V_2= 5^2.
\end{equation*}

\begin{figure}[htpb]
\centering
\includegraphics[scale=0.35]{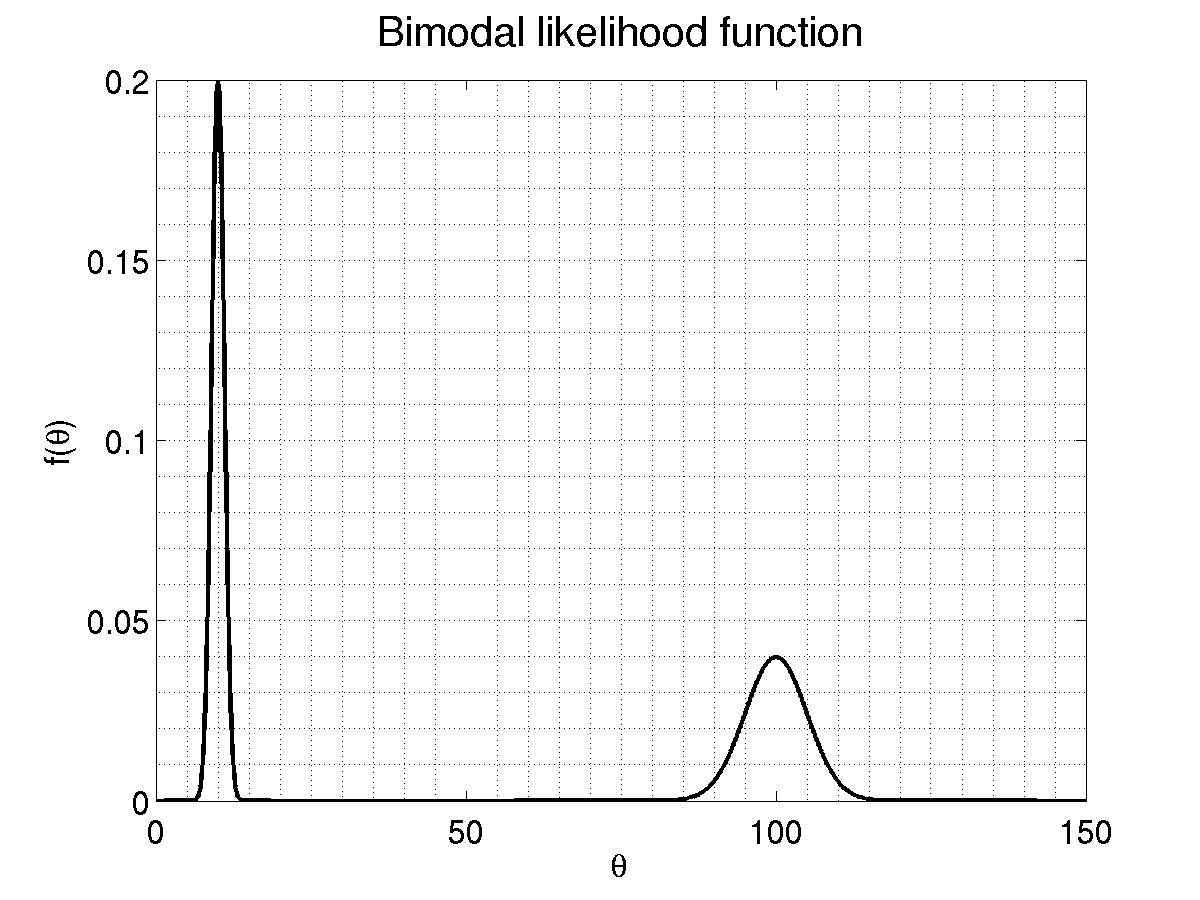}
\vspace{-10pt}
\caption{Likelihood function given by $f=f_1/2+ f_2/2$, where $f_1$ and $f_2$ are defined in Equation \eqref{eq:bimodal_functions}.}
\label{fig:bimodal:likelihood}
\end{figure}

In this example, we want to sample the posterior PDF given by:
\begin{equation}
\pi_\post(\theta)  \propto \left[\dfrac{1}{2} f_1(\theta) + \dfrac{1}{2} f_2(\theta) \right] \cdot \pi_\prior = f(\theta) \cdot \pi_\prior
\end{equation}
where $f(\theta)= \dfrac{1}{2} f_1(\theta) + \dfrac{1}{2} f_2(\theta)$ is the likelihood function, which is depicted in Figure \ref{fig:bimodal:likelihood}.

\subsection{Running the Example}\label{sec:bimodal-run}

To run the executable provided (available after QUESO installation), and generate figures for the chains, PDFs, CDFs, etc., enter the following commands:
\begin{lstlisting}[label={},caption={}]
$ cd $HOME/LIBRARIES/QUESO-0.51.0/examples/bimodal
$ rm outputData/*
$ ./bimodal_gsl bimodal_1chain.inp    
$ matlab
   $ plot_all.m	                          # inside matlab
   $ plot_likelihood_normalized_taus.m    # inside matlab
   $ plot_likelihood_unnormalized_taus.m  # inside matlab
   $ exit                                 # inside matlab
$ ls -l outputData/*.png
bimodal_autocorrelation_rawchain.png  bimodal_likelihood.png
bimodal_cdf_rawchain.png	          bimodal_likelihood_taus_normalized.png
bimodal_kde_rawchain.png	          bimodal_likelihood_taus.png
\end{lstlisting}

As a result, the user should have created several of PNG figures containing marginal posterior PDF, cumulative density distribution and autocorrelation. The name of the figure files have been chosen to be informative, as shown in the Listing above.

\subsection{Example Code}\label{sec:bimodal-code}

The source code for the example is composed of 5 files:
\texttt{bimodal\_main.C} (Listing \ref{code:bimodal-main-c}), \linebreak
\texttt{bimodal\_likelihood.h} and \texttt{bimodal\_likelihood.C} (Listings \ref{fig-like-bimodal-h} and \ref{fig-like-bimodal-c}),
\texttt{bimodal\_compute.h} and \texttt{bimodal\_compute.C} (Listings \ref{code:bimodal-compute-h} and \ref{code:bimodal-compute-c}).

\lstinputlisting[caption=File \texttt{bimodal\_main.C.}, label={code:bimodal-main-c}, linerange={25-1000}]{bimodal_main.C}

\lstinputlisting[caption=File \texttt{bimodal\_likelihood.h}., label={fig-like-bimodal-h}, linerange={25-1000}]{bimodal_likelihood.h}

\lstinputlisting[caption=File \texttt{bimodal\_likelihood.C}., label={fig-like-bimodal-c}, linerange={25-1000}]{bimodal_likelihood.C}

\lstinputlisting[caption=File \texttt{bimodal\_compute.h.}, label={code:bimodal-compute-h}, linerange={25-1000}]{bimodal_compute.h}

Note that in line 57 of Listings \ref{code:bimodal-compute-c} the `\verb+#if 0+' directive tells the compiler that the application will not use DRAM algorithm, but rather the Multilevel solver (line 65). Naturally, the user may chose to use the DRAM algorithm by changing the directive in line 57 to `\verb+#if 1+'.

\lstinputlisting[caption={File \texttt{bimodal\_compute.C}.}, label={code:bimodal-compute-c}, linerange={25-1000},numbers=left]{bimodal_compute.C}

\subsection{Input File}\label{sec:bimodal-input-file}

QUESO reads an input file for solving statistical problems, which provides options for the Multilevel or MCMC method. In this example, the Multilevel method is chosen to sample from the distribution. Many variables are common to both MCMC and Multilevel method, especially because the Multilevel method also has the option of delaying the rejection of a candidate. The names of the variables have been designed to be informative in this case as well:
\begin{description}\vspace{-8pt}
\item[ \texttt{env}:] refers to QUESO environment; \vspace{-8pt}
\item[ \texttt{ip}:] refers to inverse problem;\vspace{-8pt}
\item[ \texttt{ml}:] refers to Multilevel;\vspace{-8pt}
\item[ \texttt{dr}:] refers to delayed rejection;\vspace{-8pt}
\item[ \texttt{rawChain}:] refers to the raw, entire chain; \vspace{-8pt}
\item[ \texttt{filteredChain}:] refers to a filtered chain (related to a specified \texttt{lag});\vspace{-8pt}
\item[ \texttt{last}:] refers to instructions specific for the last level of the Multilevel algorithm.
\end{description}

The user may select options for a specific level by naming its number, i.e., in case the user wants to define a different number of extra stages together with the scales for each stage (in the DRAM part of the ML algorithm) for the level 3, he/she may include the following instructions:
\begin{lstlisting}
ip_ml_3_dr_maxNumExtraStages          = 1
ip_ml_3_dr_listOfScalesForExtraStages = 3.333
\end{lstlisting}
in the input file.

The options used for solving this example are displayed in Listing \ref{code:bimodal-input-file}. 

\lstinputlisting[caption={Options for QUESO library used in application code (Listings \ref{code:bimodal-main-c}-\ref{code:bimodal-compute-c}})., 
label={code:bimodal-input-file},]{bimodal_1chain.inp}

\subsection{Create your own Makefile}\label{sec:bimodal-makefile}

Similarly to the other examples presented in this user's manual and also available with QUESO distribution, a user-created makefile is available: `\texttt{Makefile\_bimodal\_violeta}'. When adapted to the user's settings, namely paths for  QUESO required libraries, it may be used to compile the code and create the executable \verb+bimodal_gsl+. 

Thus, to compile, build and execute the code, the user just needs to run the following commands in the same directory where the files are:
\begin{lstlisting}
$ cd $HOME/LIBRARIES/QUESO-0.51.0/examples/bimodal/
$ export LD_LIBRARY_PATH=$LD_LIBRARY_PATH:\
  $HOME/LIBRARIES/gsl-1.15/lib/:\
  $HOME/LIBRARIES/boost-1.53.0/lib/:\
  $HOME/LIBRARIES/hdf5-1.8.10/lib:\
  $HOME/LIBRARIES/QUESO-0.51.0/lib 
$ make -f Makefile_bimodal_violeta 
$ ./bimodal_gsl example.inp
\end{lstlisting}

Again, the `\verb+export+' instruction above is only necessary if the user has not saved it in his/her \verb+.bashrc+ file.

\subsection{Data Post-Processing and Visualization}\label{sec:bimodal-results}

According to the specifications of the input file in Listing~\ref{code:bimodal-input-file}, both a folder named \verb+outputData+ and a the following files should be generated:
\begin{verbatim}
rawChain_ml.m 
display_sub0.txt    
\end{verbatim}

The sequence of Matlab commands is identical to the ones presented in Sections
\ref{sec:sip-results}, \ref{sec:sfp-results}, \ref{sec:gravity-results} and \ref{sec:tga-results};
therefore, are omitted here. The reader is invited to explore the Matlab files
\texttt{plot\_likelihood\_normalized\_taus.m},
\texttt{plot\_likelihood\_unnormalized\_taus.m} and/or \texttt{plot\_all.m},  for details of how the figures have been generated.

\subsubsection{KDE and CDF Plots}

\begin{figure}[hptb]
\centering
\subfloat[KDE]{\includegraphics[scale=0.3]{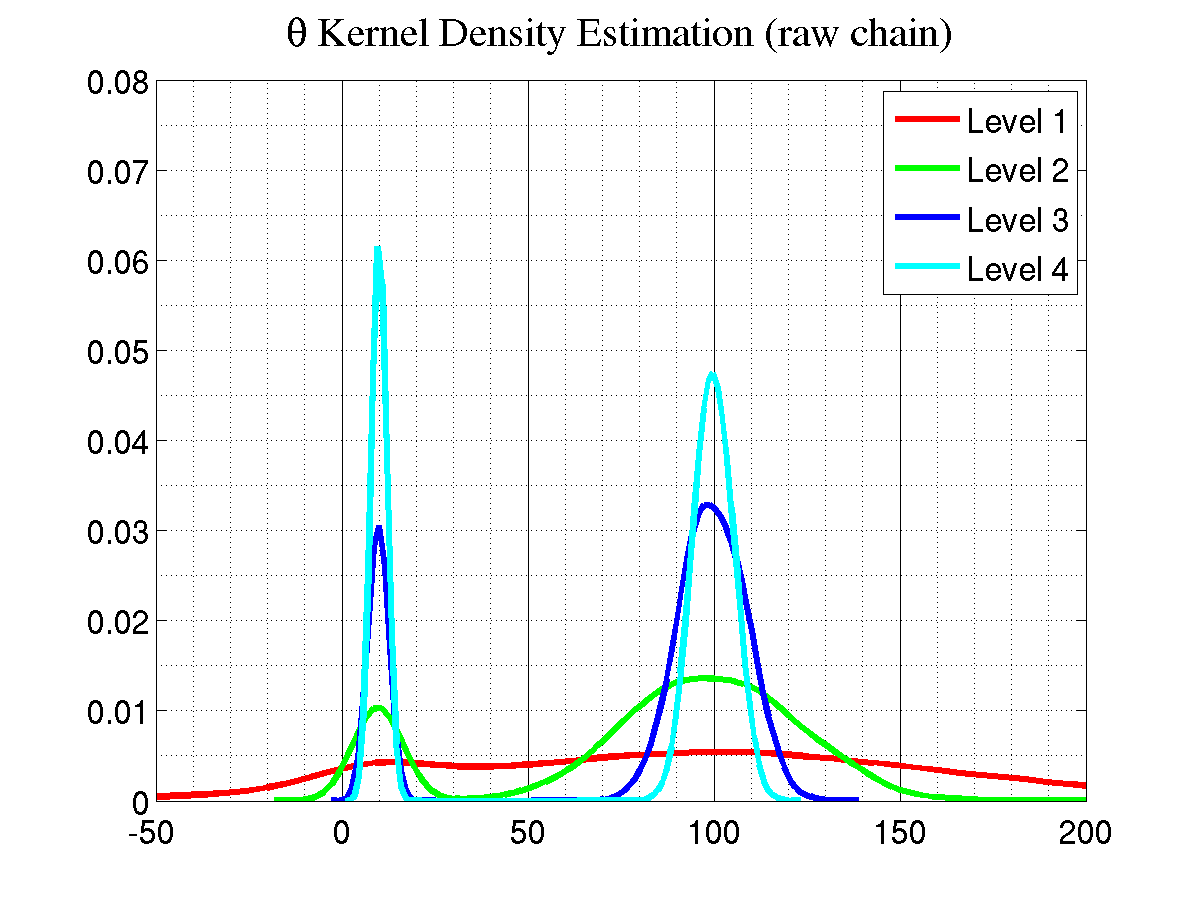}}
\subfloat[CDF]{\includegraphics[scale=0.3]{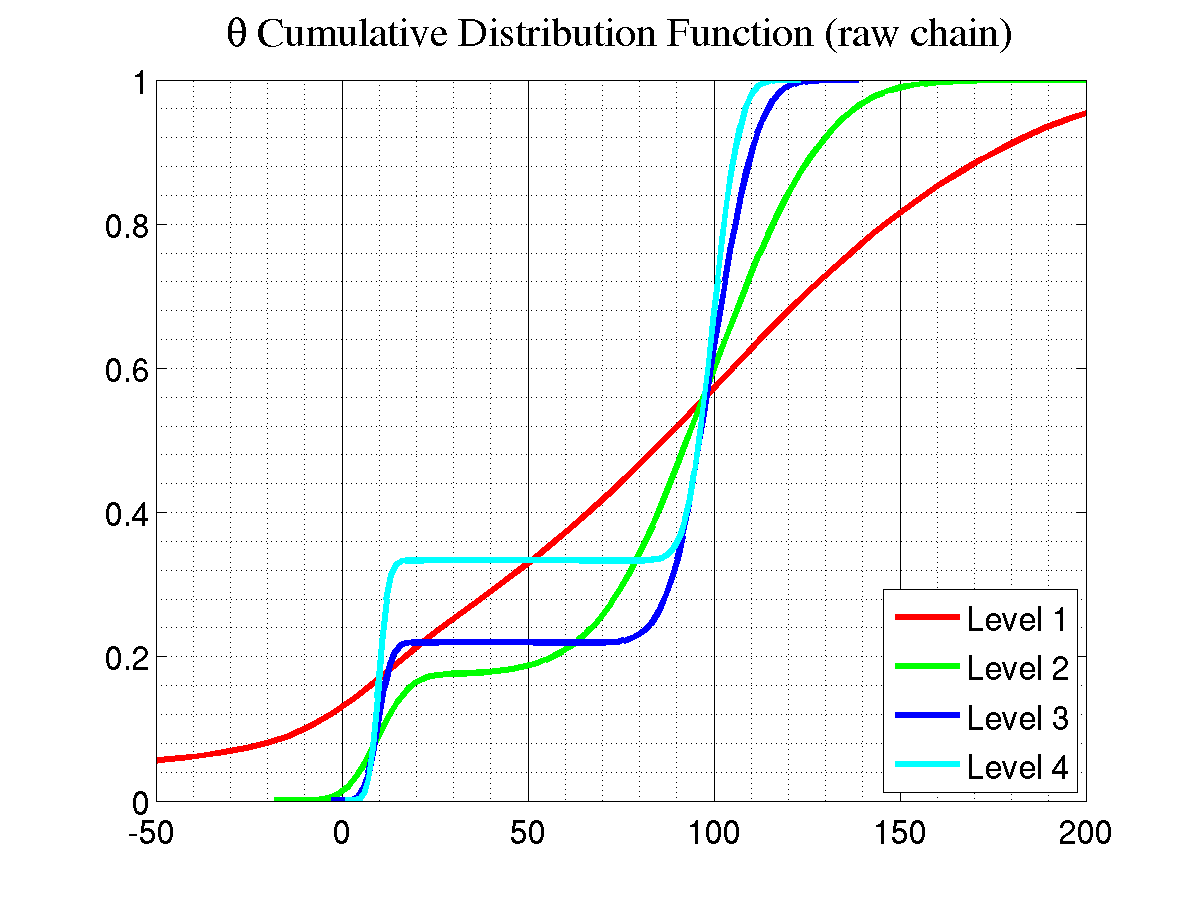}}
\vspace{-8pt}
\caption{KDE and CDF plots of parameter $\theta$, for all fours levels.}
\label{fig:bimodal_kde}
\end{figure}

Figure \ref{fig:bimodal_kde} presents the KDE and CDF plots of the parameter $\theta$.

\subsubsection{Autocorrelation Plots}

\begin{figure}[htpb]
\centering
\includegraphics[scale=0.3]{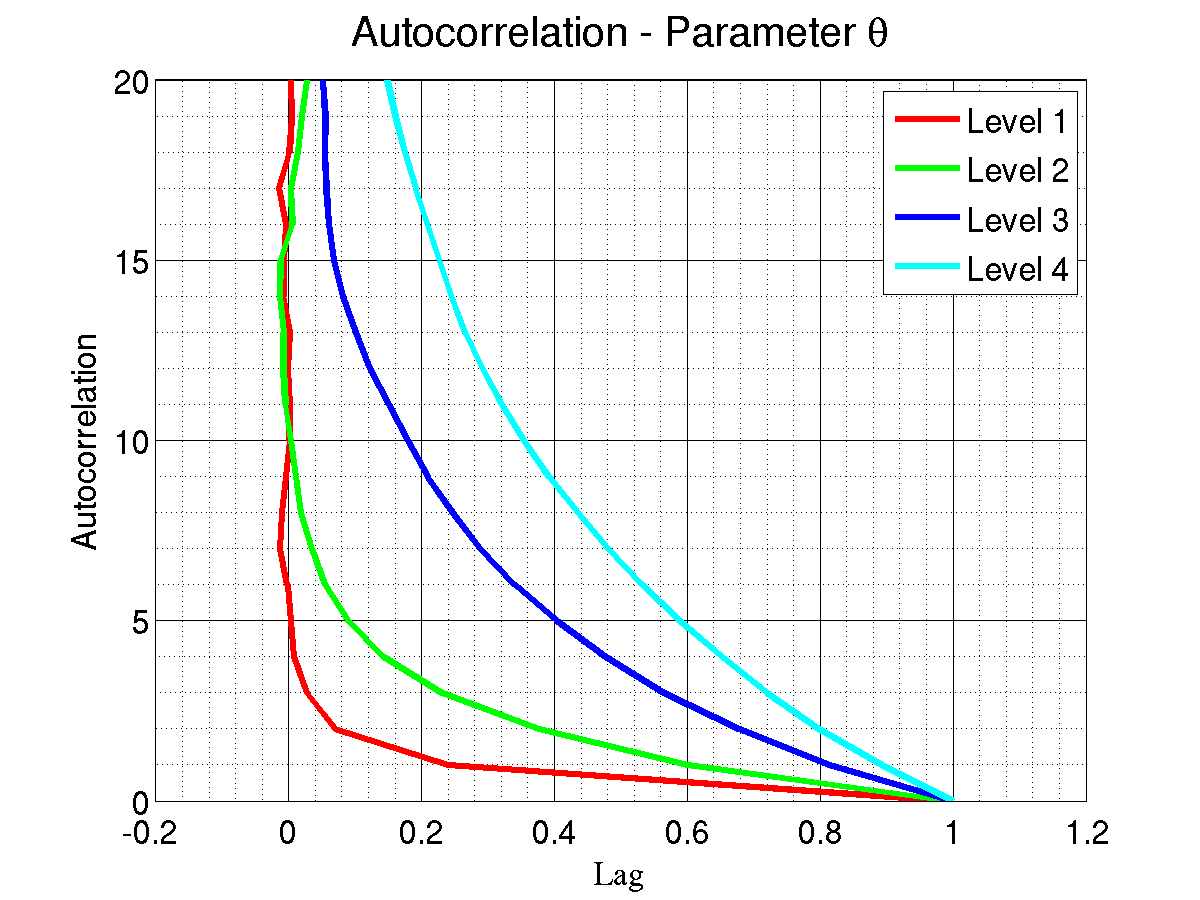}
\vspace{-10pt}
\caption{Autocorrelation plots for $\theta$, all four levels.}
\label{fig:bimodal_autocorr}
\end{figure}

Figure \ref{fig:bimodal_autocorr} presents the autocorrelation of the parameter $\theta$, in each one of the intermediate levels.

\subsubsection{Intermediary Likelihood Plots}
\begin{figure}[htpb]
\centering
\subfloat[normalized]{\includegraphics[scale=0.3]{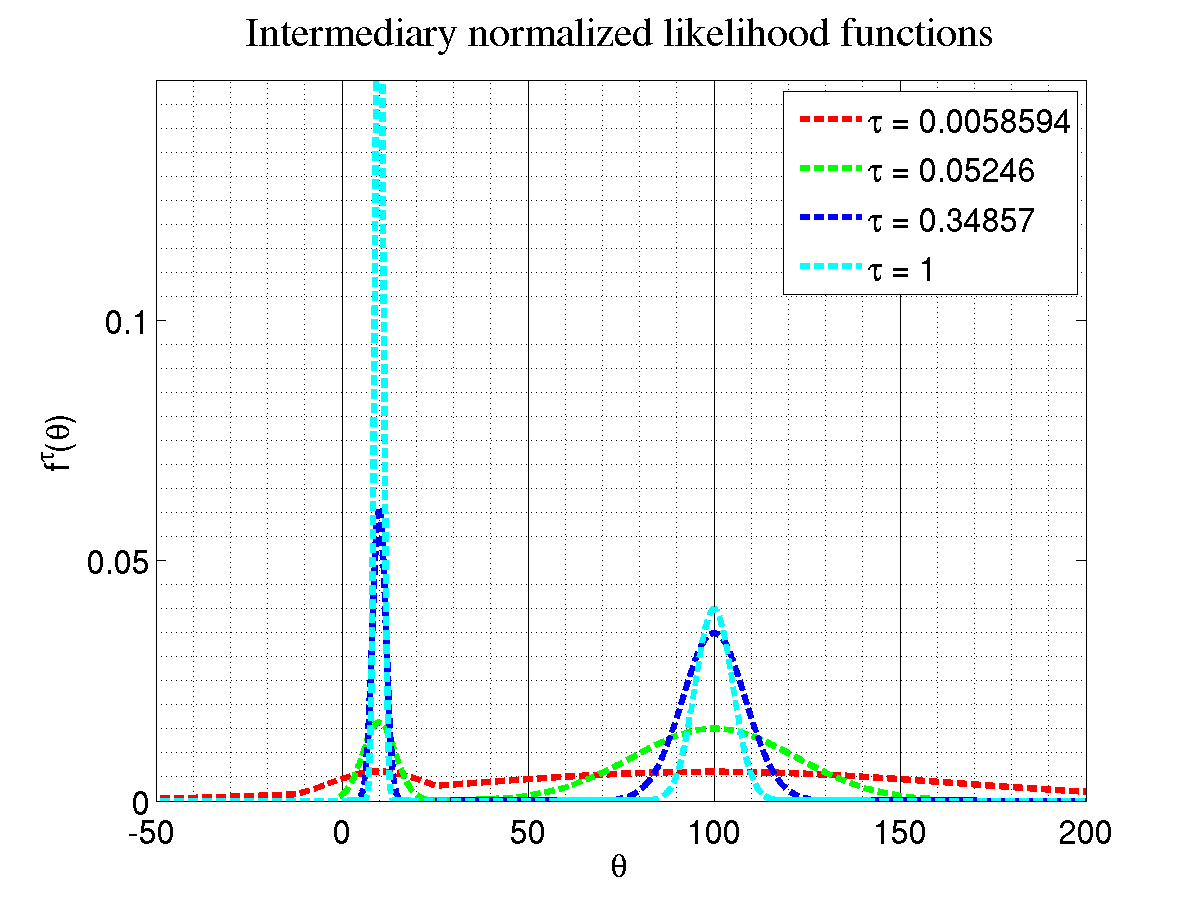}}
\subfloat[unnormalized]{\includegraphics[scale=0.3]{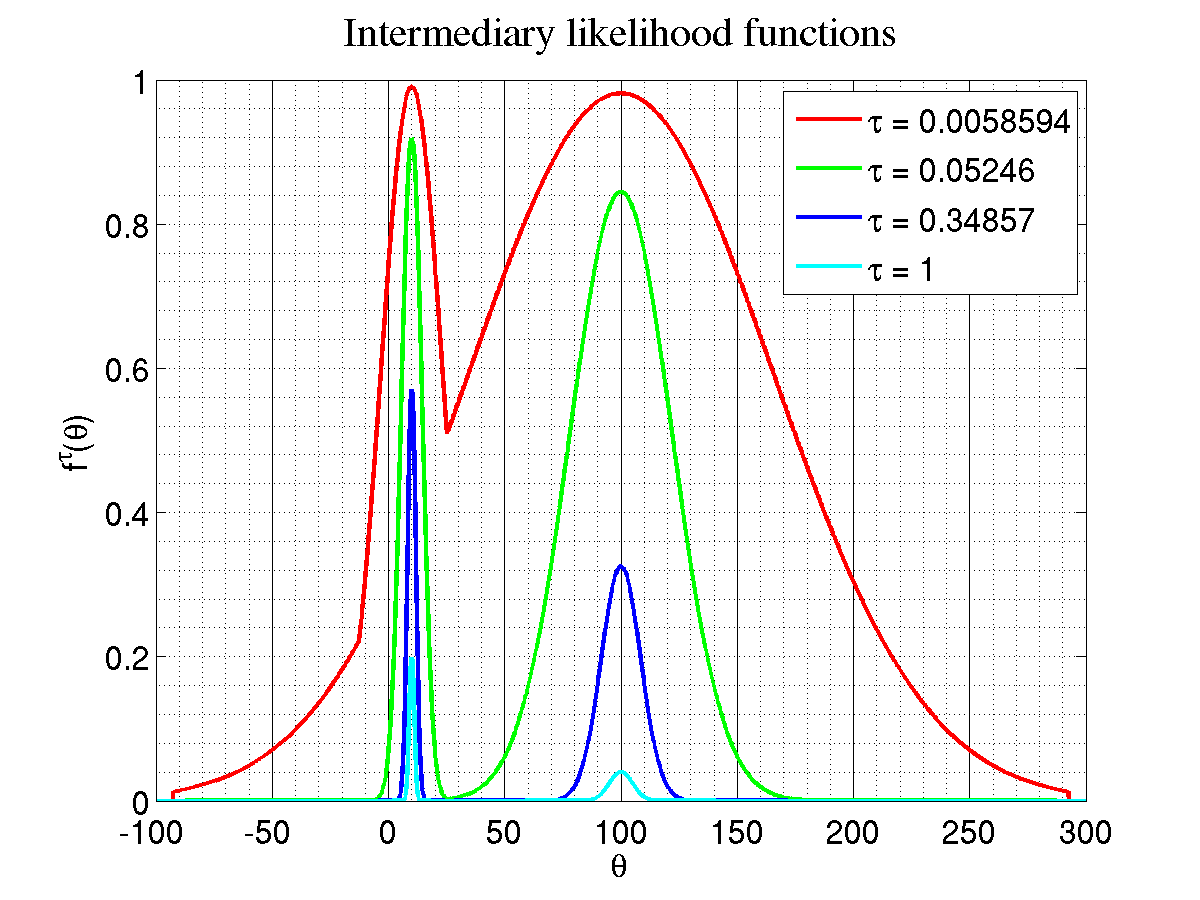}}
\vspace{-10pt}
\caption{Intermediary likelihood functions $f(\theta)^\tau$, where $\tau_i$ is the exponent computed at the $i$-th level of Multilevel algorithm. In this simple problem, only four levels are needed, i.e. $i=1\ldots 4$. The cyan-colored curve (exponent $\tau = 1$) is the same curve as in Figure \ref{fig:bimodal:likelihood}.}
\label{fig:bimodal:likelihood_taus}
\end{figure}


\section{\texttt{hysteretic}}\label{sec:example_hysteretic}

This example replicates the problem in ``Section 4.3 A Hysteretic Model Class
'' of~\cite{CheungPrudencio2012}, and which is also discussed in \cite{Cheung_2009A}.
In this example we consider the nonlinear seismic response of a four-story building. 
This response is modeled with
an inelastic shear building model with some linear viscous damping and hysteretic bilinear interstory restoring forces~\cite{Cheung_2009A}. More
specifically, let $t\geq0 $ denote time, let $a_g(t)$ be a given total acceleration at the base (Fig. 13), and for the i-th floor [degree of freedom (dof)], $1 \leq i \leq N_o \equiv 4$, let us denote:
\begin{equation}
\begin{split}
m_i &= \text{lumped mass},\\
q_i(t) &= \text{horizontal displacement},\\
F_i(t) &= \text{hysteretic restoring force}
\end{split}
\end{equation}

The hysteretic restoring force is illustrated in Figure \ref{fig:hyst_rest_force} and the horizontal base (ground) acceleration (input data) used in~\cite{CheungPrudencio2012} is illustrated in \ref{hig:hist_base_acceleration}. 
\begin{figure}[h!]
\centerline{
\setlength{\unitlength}{4144sp}%
\begingroup\makeatletter\ifx\SetFigFont\undefined%
\gdef\SetFigFont#1#2#3#4#5{%
  \reset@font\fontsize{#1}{#2pt}%
  \fontfamily{#3}\fontseries{#4}\fontshape{#5}%
  \selectfont}%
\fi\endgroup%
\begin{picture}(3419,1922)(789,-1140)
\thicklines
{\color[rgb]{0,0,0}\put(811,-331){\vector( 1, 0){3375}}
}%
{\color[rgb]{0,0,0}\put(1981,-331){\vector( 3, 4){615.600}}
}%
{\color[rgb]{0,0,0}\put(2600,490){\vector( 3, 1){891}}
}%
{\color[rgb]{0,0,0}\put(1120,-940){\vector( 2, 3){400}}
}%
{\color[rgb]{0,0,0}\multiput(2611,479)(0.00000,-108.00000){8}{\line( 0,-1){ 54.000}}
}%
\thinlines
{\color[rgb]{0,0,0}\multiput(2611,479)(102.85714,0.00000){4}{\line( 1, 0){ 51.429}}
\multiput(2971,479)(0.00000,90.00000){2}{\line( 0, 1){ 45.000}}
}%
\thicklines
{\color[rgb]{0,0,0}\put(1981,-1096){\vector( 0, 1){1755}}
}%
\thinlines
{\color[rgb]{0,0,0}\multiput(2071,-196)(102.85714,0.00000){4}{\line( 1, 0){ 51.429}}
\multiput(2431,-196)(0.00000,128.57143){4}{\line( 0, 1){ 64.286}}
}%
\thicklines
{\color[rgb]{0,0,0}\put(2341,-940){\vector(-1, 0){1215}}
}%
{\color[rgb]{0,0,0}\put(3490,780){\vector(-2,-3){1149.231}}
}%
\put(3916,-241){\makebox(0,0)[lb]{\smash{{\SetFigFont{10}{12.0}{\rmdefault}{\mddefault}{\updefault}{\color[rgb]{0,0,0}i-th floor}%
}}}}
\put(3916,-466){\makebox(0,0)[lb]{\smash{{\SetFigFont{10}{12.0}{\rmdefault}{\mddefault}{\updefault}{\color[rgb]{0,0,0}interstory}%
}}}}
\put(2251,-310){\makebox(0,0)[lb]{\smash{{\SetFigFont{9}{12.0}{\rmdefault}{\mddefault}{\updefault}{\color[rgb]{0,0,0}1}%
}}}}
\put(2791,360){\makebox(0,0)[lb]{\smash{{\SetFigFont{9}{12.0}{\rmdefault}{\mddefault}{\updefault}{\color[rgb]{0,0,0}1}%
}}}}
\put(2440,-16){\makebox(0,0)[lb]{\smash{{\SetFigFont{10}{12.0}{\rmdefault}{\mddefault}{\updefault}{\color[rgb]{0,0,0}$k_i$}%
}}}}
\put(2515,-450){\makebox(0,0)[lb]{\smash{{\SetFigFont{10}{12.0}{\rmdefault}{\mddefault}{\updefault}{\color[rgb]{0,0,0}$u_i$}%
}}}}
\put(3000,480){\makebox(0,0)[lb]{\smash{{\SetFigFont{10}{12.0}{\rmdefault}{\mddefault}{\updefault}{\color[rgb]{0,0,0}$r_i k_i$}%
}}}}
\end{picture}%
\\
}
\caption{Illustration of the hysteretic restoring force [see Eq. \eqref{eq:hyst:motion}] used in our hysteretic test problem. The terms $r_i$, $k_i$, and $u_i$ denote model parameters.}
\label{fig:hyst_rest_force}
\end{figure}
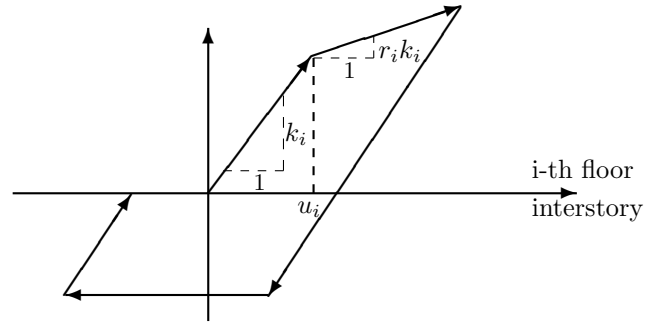
\begin{figure}[hptb]
\centering
\includegraphics[scale=0.6]{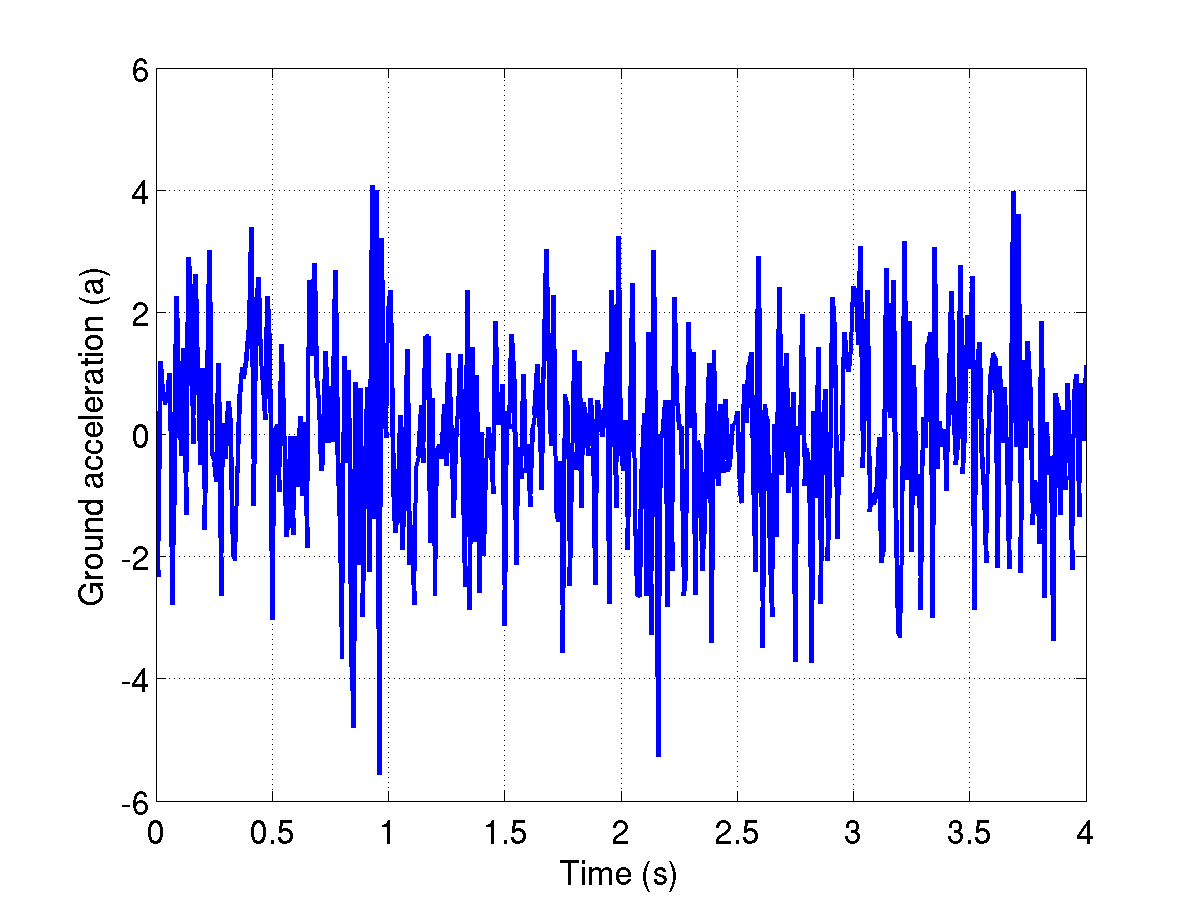}
\vspace{-8pt}
\caption{Horizontal base acceleration (input data) used in the hysteretic test problem~\cite{CheungPrudencio2012}.}
\label{hig:hist_base_acceleration}
\end{figure}

We also define the mass matrix $\bv{M}$ and the stiffness matrix $\bv{K}$:
\begin{equation*}
\bv{M}= 
\begin{bmatrix}
  m_1 & 0   & 0   & 0 \\
  0   & m_2 & 0   & 0 \\
  0   & 0   & m_3 & 0 \\
  0   & 0   & 0   & m_4 \\
\end{bmatrix}
\quad \text{and} \quad
\bv{K}= 
\begin{bmatrix}
  k_1 + k_2 & -k_2    & 0       & 0 \\
  -k_2      & k_2+k_3 & -k_3    & 0 \\
  0         & -k_3    & k_3+k_4 & -k4 \\
  0         &   0     & -k_4    & k_4 \\ 
\end{bmatrix}
\end{equation*}
and the Rayleigh damping matrix
\begin{equation*}
\bv{C} = \rho \bv{M} + \gamma \bv{K} 
\end{equation*}
for given positive scalar parameters $\rho$ and $\gamma$. The response $\bv{q}(t) \equiv [q_1(t),q_2(t),q_3(t),q_4(t)]$ is modeled as satisfying the equation of motion:
\begin{equation}\label{eq:hyst:motion}
\bv{M} \ddot{\bv{q}}(t) + \bv{C} \dot{\bv{q}}(t) + \bv{F}(t) = -\bv{M}\cdot 
\begin{bmatrix} 
1 \\ 
\vdots \\
1
\end{bmatrix}_{4\times1} \cdot a_g(t),
\end{equation}
where $\bv{F}(t) \equiv [F_1(t), F_2(t), F_3(t), F_4(t)]$. In this model, the hysteretic restoring force $\bv{F}(t)$ depends on the whole
time history $[\bv{q}(t),\dot{\bv{q}}(t)]$ of responses from the initial instant until time~$t$.

\begin{figure}[hptb]
\centering
\includegraphics[scale=0.6]{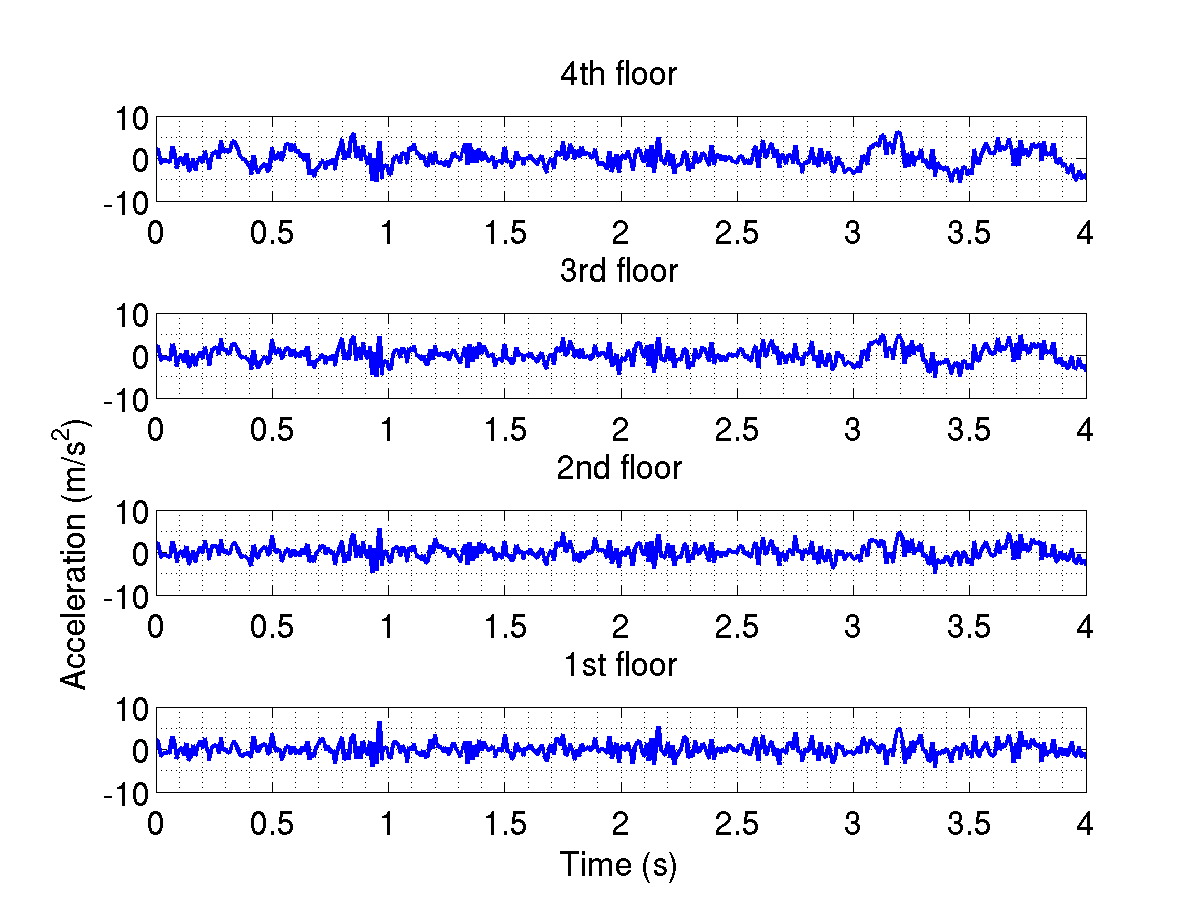}
\vspace{-8pt}
\caption{Horizontal acceleration of each of the four floors (measured data aimed for calibration) used in our hysteretic test problem.}
\label{hig:hist_acceleration_4_floors}
\end{figure}

The (noisy) measured data $y = (y_1, y_2, y_3, y_4)$ available for model calibration consists of 4 s of accelerometer data at each floor (refer to Fig. \ref{hig:hist_acceleration_4_floors}), with a sample interval $\Delta t = 0.01$ s. The simulated dynamic data was obtained by adding Gaussian white noise to the output simulation of the hysteretic model with the following input values:
\begin{align*}
m_1 & = m_2 = m_3 = m_4 = 2\times 10^4 \,kg,\\
k_1 &=  2.2 \times 10^7 \,Nm^{-1},\\
k_2 &=  2.0 \times 10^7 \,Nm^{-1},\\
k_3 &=  1.7 \times 10^7 \,Nm^{-1},\\
k_4 &=  1.45 \times 10^7 \,Nm^{-1},\\
r_1 & = r_2 = r_3 = r_4 = 0.1,\\
u_1 & = u_2 = 8 \times 10^{-3} \,m, \\
u_3 & = u_4 = 7 \times 10^{-3} \,m, \\
\rho &= 7.959 \times 10^{-1}, \\
\gamma &= 2.5 \times 10^{-3},\\
\sigma^2 &= 0.6^2,
\end{align*}
where for $i=1,2,3,4$, $k_i$ is the initial stiffness, $r_i$ is the post-yield stiffness reduction factor, and $u_i$ is  yield displacement.

According to Cheung and Prudencio~\cite{CheungPrudencio2012}, these input values were chosen deliberately so that the excitation $a_g$ did not cause some of the upper floors to enter the nonlinear regime; that is, so that our test inverse problem did not become globally identifiable. 

In this section, 400 time-steps are used, as the data is  available  at instants  
$$t_n = (n - 1) \times \Delta t, \quad 1 \leq n \leq N_T \equiv 401, \quad \Delta t = 0.01$$
however, Cheung and Prudencio used only 250 time steps~\cite{CheungPrudencio2012}. 

An additive noise is assumed to be present in the measurements; i.e.,
\begin{equation*}
y_i(n)=q_i(n)+\varepsilon_i(n), \quad 1 \leq i \leq N_o, \quad  1 \leq n \leq N_T \equiv 401, 
\end{equation*}
where $q_i(n; \theta_2 ,...,\theta_{15})$ denotes the output at time $t_n= n\Delta t \, (\Delta t=0.01s)$ at the $i$-th observed degree of freedom predicted by the proposed structural model, and $y_i(n)$ denotes the corresponding
measured output.

They considered a total of 15 unknown parameters $\bv{\theta} = (\theta_1 , . . . , \theta_{15})$ and modeled the variables $\varepsilon_i$ as independently and identically distributed Gaussian variables with mean zero and some unknown prediction-error variance $\sigma^2$. The variance $\sigma^2$ is assumed to be the same for all $N_o = 4$ floors.  The first component $\theta_1$ is equal to the prediction error variance $\sigma^2$ and the other 14 parameters are related to the four triples $(k_i, r_i, u_i),\, 1 \leq i \leq N_o$ (see Fig. \ref{fig:hyst_rest_force}), to $\rho$, and to $\gamma$. The likelihood function is given by:
\begin{equation}\label{eq:hyst:like}
 f(\bv{y}|\bv{\theta}) = \dfrac{1}{(2 \pi \sigma^2)^{N_oN_T/2}}\exp \left( -\dfrac{1}{2\sigma^2} \displaystyle \sum_{i=1}^{N_o} \sum_{n=1}^{N_T} [y_i(t_n) - {q}_i(t_n; \theta_2,\ldots,\theta_{15})]^2 \right).
\end{equation}
An inverse gamma  prior was used for $\theta_1=\sigma^2$, and a 14-dimensional Gaussian prior was used for $\theta_2 , ..., \theta_{15}$ with zero mean and diagonal covariance matrix equal to a scaled identity matrix.

\subsection{Running the Example}\label{sec:hysteretic-run}

To run the executable provided (available after QUESO installation), and generate figures for the chains, KDEs, CDFs, autocorrelation and scatter plots, enter the following commands:
\begin{lstlisting}[label={},caption={}]
$ cd $HOME/LIBRARIES/QUESO-0.51.0/examples/example
$ rm outputData/*
$ ./example_gsl example_1chain.inp    
$ matlab
   $ plot_all.m	                          # inside matlab   
$ ls -l outputData/*.png
hysteretic_autocorr_thetas.png  hysteretic_kde_theta8.png 
hysteretic_cdf_thetas.png       hysteretic_kde_theta9.png
hysteretic_kde_theta1.png       hysteretic_kde_theta10.png
hysteretic_kde_theta2.png       hysteretic_kde_theta11.png
hysteretic_kde_theta3.png       hysteretic_kde_theta12.png
hysteretic_kde_theta4.png       hysteretic_kde_theta13.png
hysteretic_kde_theta5.png       hysteretic_kde_theta14.png
hysteretic_kde_theta6.png       hysteretic_kde_theta15.png
hysteretic_kde_theta7.png       hysteretic_scatter_thetas.png
\end{lstlisting}

As a result, the user should have created several of PNG figures containing kernel density estimate of the 15 parameters, cumulative density distribution, autocorrelation and scatter plots. The name of the figure files have been chosen to be informative, as shown in the Listing above.

Additional figures may be generated if the user allows the procedure \texttt{debug\_hyst(} be called by the compiler in Line 11 of file \texttt{example\_main.C}; in that case, call the function \texttt{cpp\_gen.m} inside Matlab/Octave.

\subsection{Example Code}\label{sec:hysteretic-code}

The source code for the example is composed of 5 files:
\texttt{example\_main.C} (Listing \ref{code:hysteretic-main-c}), \linebreak
\texttt{example\_likelihood.h} and \texttt{example\_likelihood.C} (Listings \ref{fig-like-example-h} and \ref{fig-like-example-c}),
\texttt{example\_compute.h} and \texttt{example\_compute.C} (Listings \ref{code:hysteretic-compute-h} and \ref{code:hysteretic-compute-c}), and finally \texttt{example\_hyst.h} and \texttt{example\_hyst.C}, which contain the Hysteretic model properly said.

Note that in line 11 of Listings \ref{code:hysteretic-main-c} the `\verb+#if 1+' directive tells the compiler that the application will call \texttt{compute()}, which internally uses QUESO and the Multilevel algorithm. 
On the contrary, the user may calculate the hysteretic force without uncertainty by changing the directive to `\verb+#if 0+', which can assist the analysis of the resulting data.

\lstinputlisting[caption=File \texttt{example\_main.C.}, label={code:hysteretic-main-c}, linerange={33-1000},numbers=left]{hys_example_main.C}

\lstinputlisting[caption=File \texttt{example\_likelihood.h}., label={fig-like-example-h}, linerange={33-1000}]{hys_example_likelihood.h}

\lstinputlisting[caption=File \texttt{example\_likelihood.C}., label={fig-like-example-c}, linerange={33-1000}]{hys_example_likelihood.C}

\lstinputlisting[caption=File \texttt{example\_compute.h.}, label={code:hysteretic-compute-h}, linerange={33-1000}]{hys_example_compute.h}

\lstinputlisting[caption={File \texttt{example\_compute.C}.}, label={code:hysteretic-compute-c}, linerange={33-1000}]{hys_example_compute.C}

\subsection{Input File}\label{sec:hysteretic-input-file}

The options used for solving this example are displayed in Listing \ref{code:hysteretic-input-file}. 

\lstinputlisting[caption={Options for QUESO library used in application code (Listings \ref{code:hysteretic-main-c}-\ref{code:hysteretic-compute-c}})., 
label={code:hysteretic-input-file},]{hys_example.inp}

\subsection{Create your own Makefile}\label{sec:hysteretic-makefile}

Similarly to the other examples presented in this user's manual and also available with QUESO distribution, a user-created makefile is available: `\texttt{Makefile\_hysteretic\_violeta}' which may personalized to each user's computer settings and used to compile the code and create the executable \verb+hysteretic_gsl+. 

Thus to compile, build and execute the code,  commands similar to the following should be entered:
\begin{lstlisting}
$ cd $HOME/LIBRARIES/QUESO-0.51.0/examples/hysteretic/
$ export LD_LIBRARY_PATH=$LD_LIBRARY_PATH:\
  $HOME/LIBRARIES/gsl-1.15/lib/:\
  $HOME/LIBRARIES/boost-1.53.0/lib/:\
  $HOME/LIBRARIES/hdf5-1.8.10/lib:\
  $HOME/LIBRARIES/QUESO-0.51.0/lib 
$ make -f Makefile_hysteretic_violeta 
$ ./hysteretic_gsl example.inp
\end{lstlisting}

Again, the `\verb+export+' instruction above is only necessary if the user has not saved the path for the libraries used during QUESO installation in his/her \verb+.bashrc+ file.

\subsection{Data Post-Processing and Visualization}\label{sec:hysteretic-results}

According to the specifications of the input file in Listing~\ref{code:hysteretic-input-file}, both a folder named \verb+outputData+ and a the following files should be generated:
\begin{verbatim}
rawChain_ml.m 
display_sub0.txt    
\end{verbatim}

Note that in this hysteretic problem a total of 13 levels are required for the Multilevel method (e.g. see the contents of file \texttt{rawChain\_ml.m}).

The sequence of Matlab commands is identical to the ones presented in Sections
\ref{sec:sip-results}, \ref{sec:sfp-results}, \ref{sec:gravity-results} and \ref{sec:tga-results};
therefore, are omitted here. The reader is invited to explore the Matlab files
\texttt{plot\_all.m} and/or  \texttt{cpp\_gen.m},  for details of how the figures have been generated.

\subsubsection{KDE Plots}

\begin{figure}[hptb]
\centering
\subfloat{\includegraphics[scale=0.25]{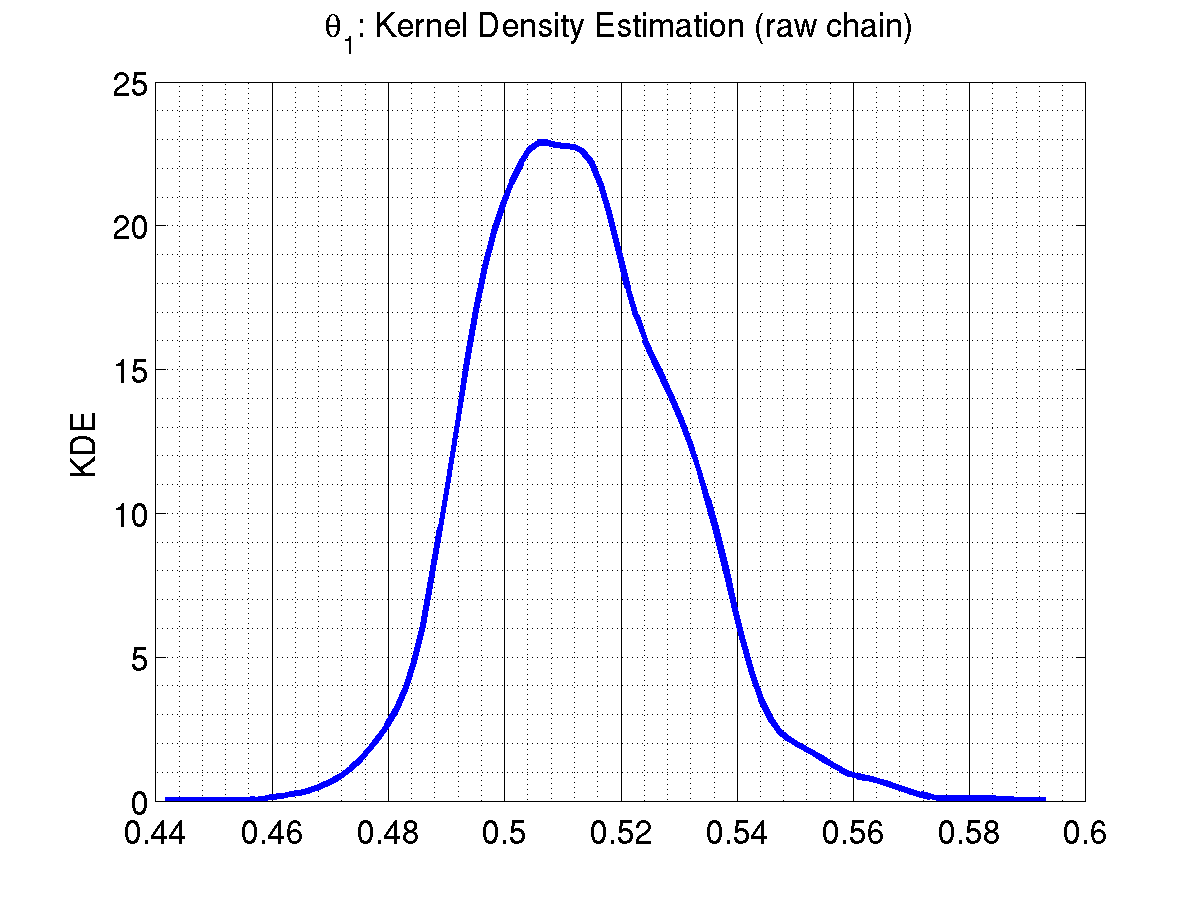}}
\subfloat{\includegraphics[scale=0.25]{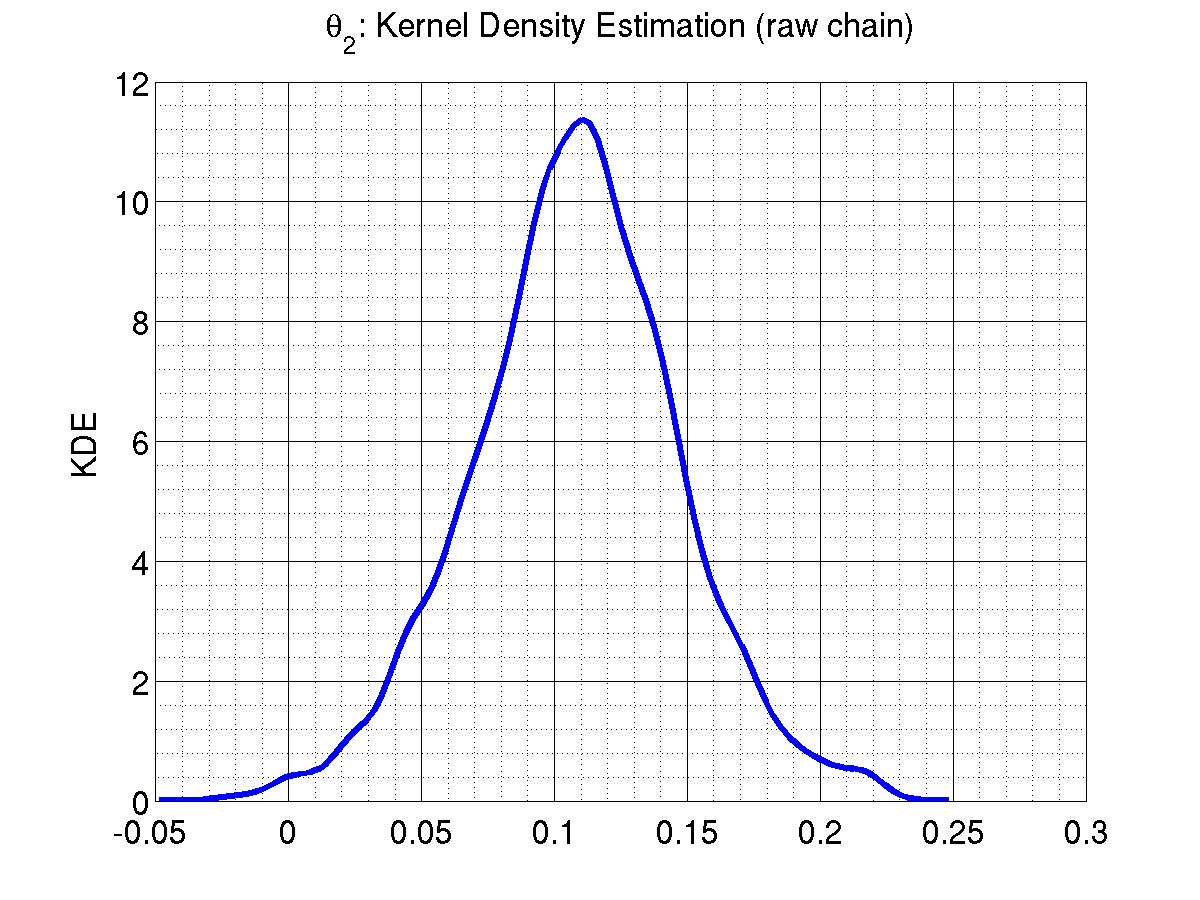}}
\subfloat{\includegraphics[scale=0.25]{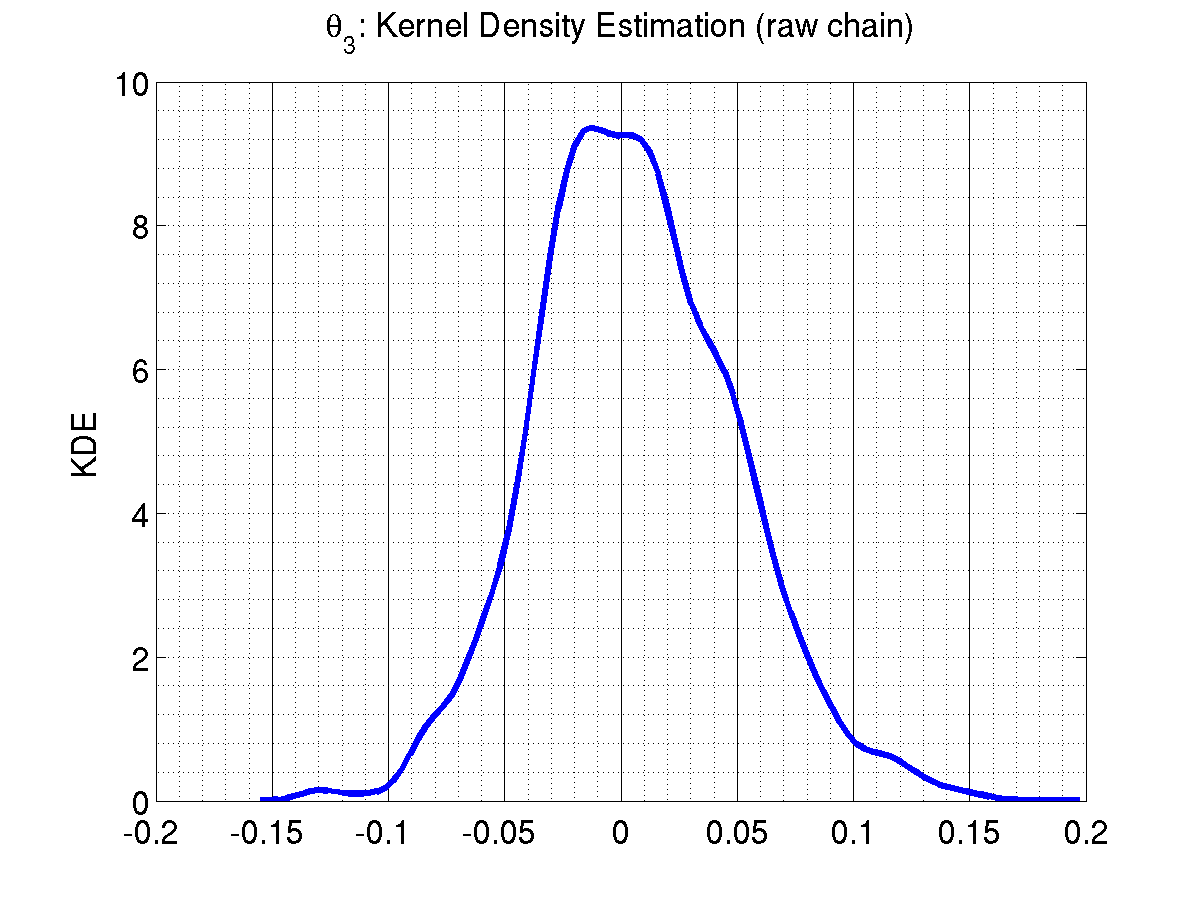}}\\
\subfloat{\includegraphics[scale=0.25]{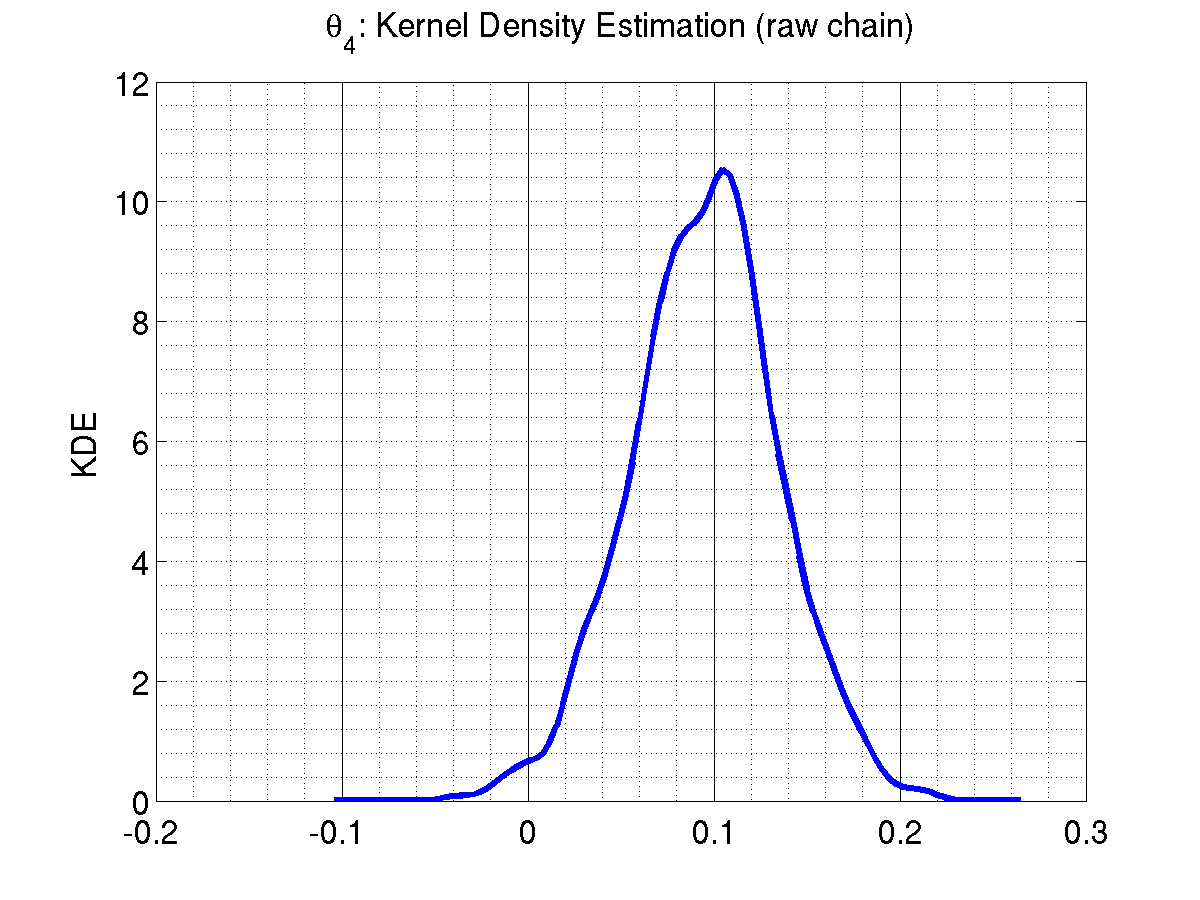}}
\subfloat{\includegraphics[scale=0.25]{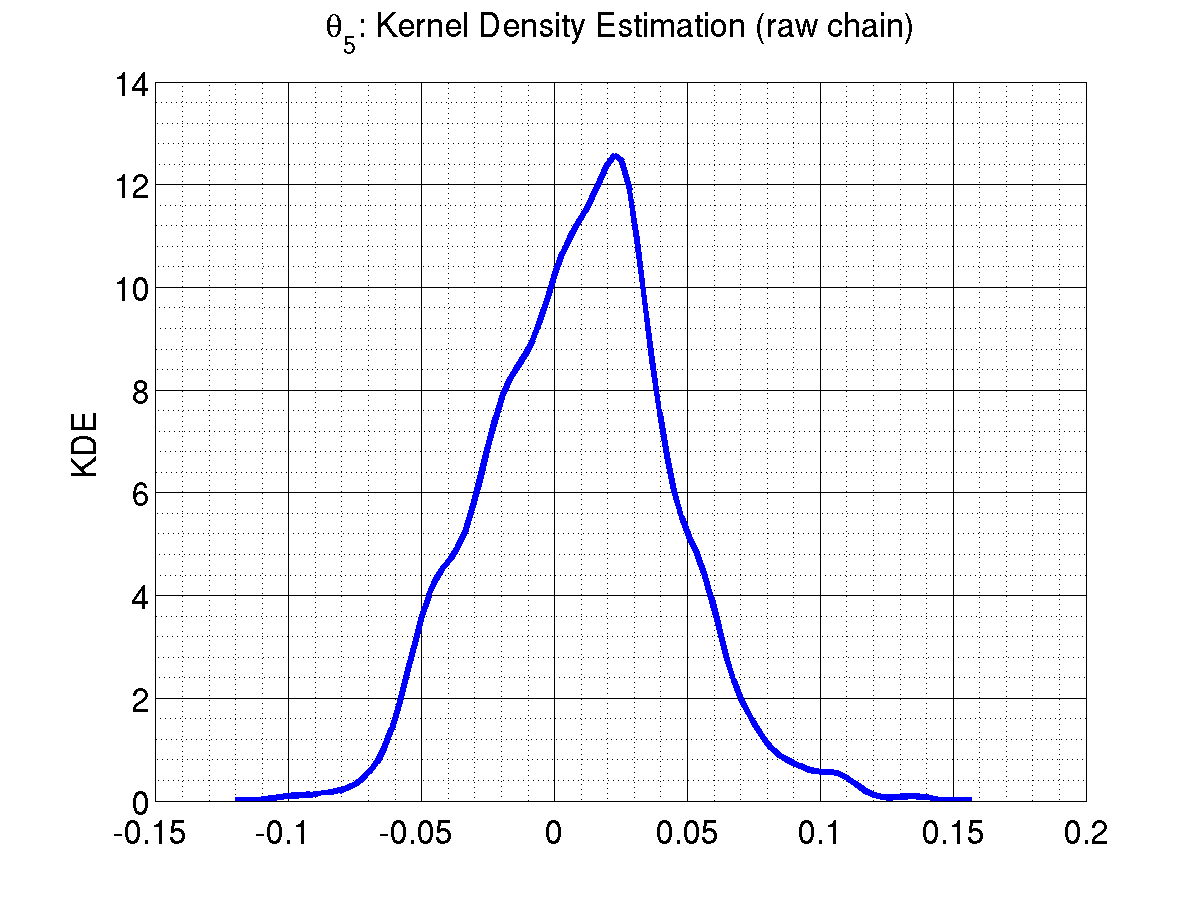}}
\subfloat{\includegraphics[scale=0.25]{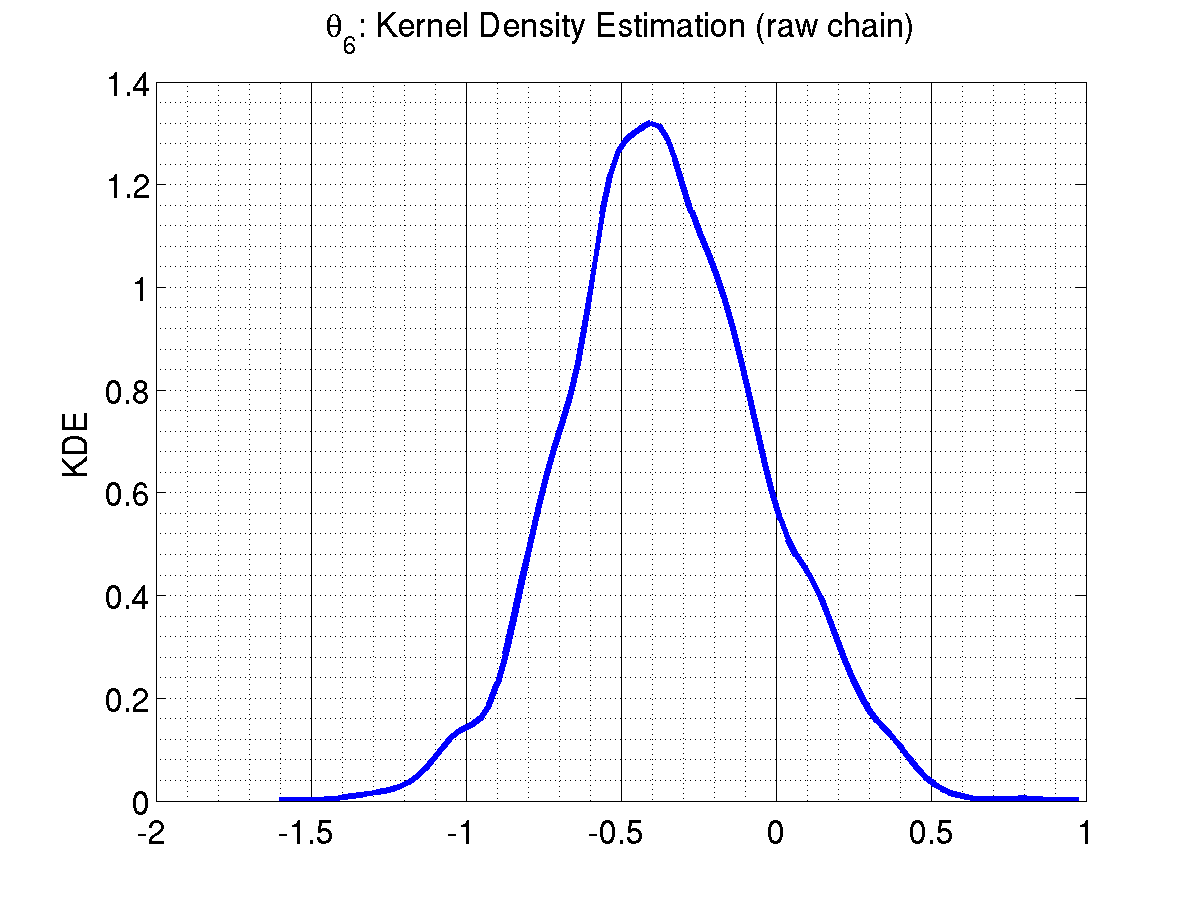}}\\
\subfloat{\includegraphics[scale=0.25]{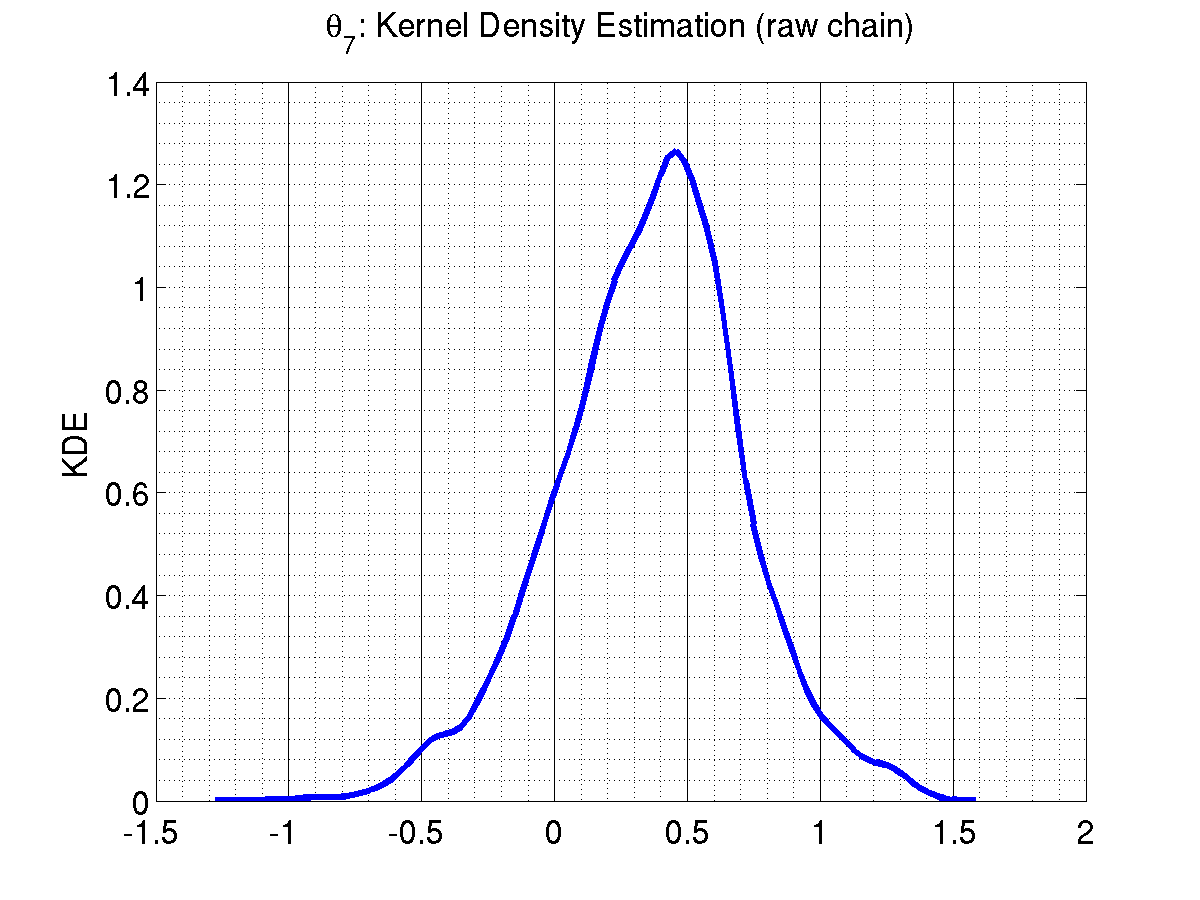}}
\subfloat{\includegraphics[scale=0.25]{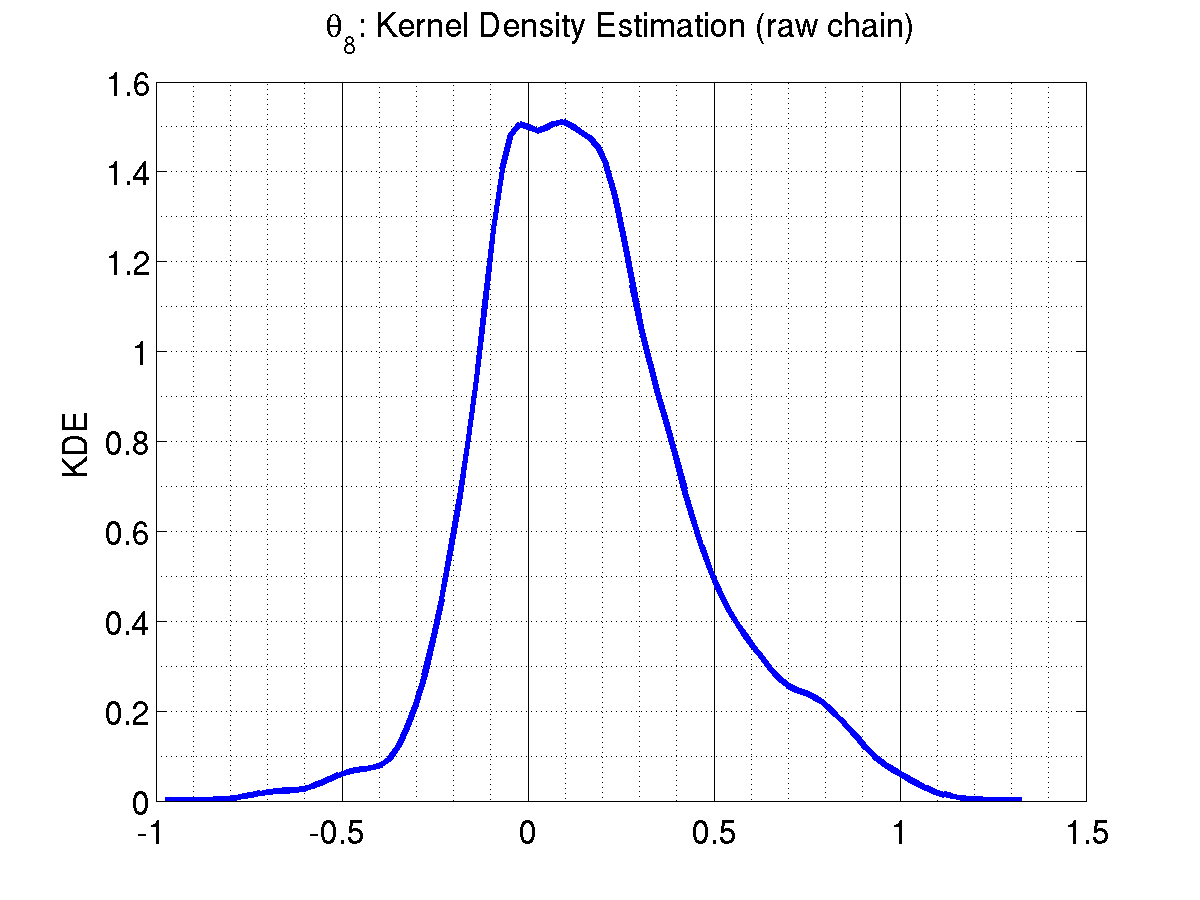}}
\subfloat{\includegraphics[scale=0.25]{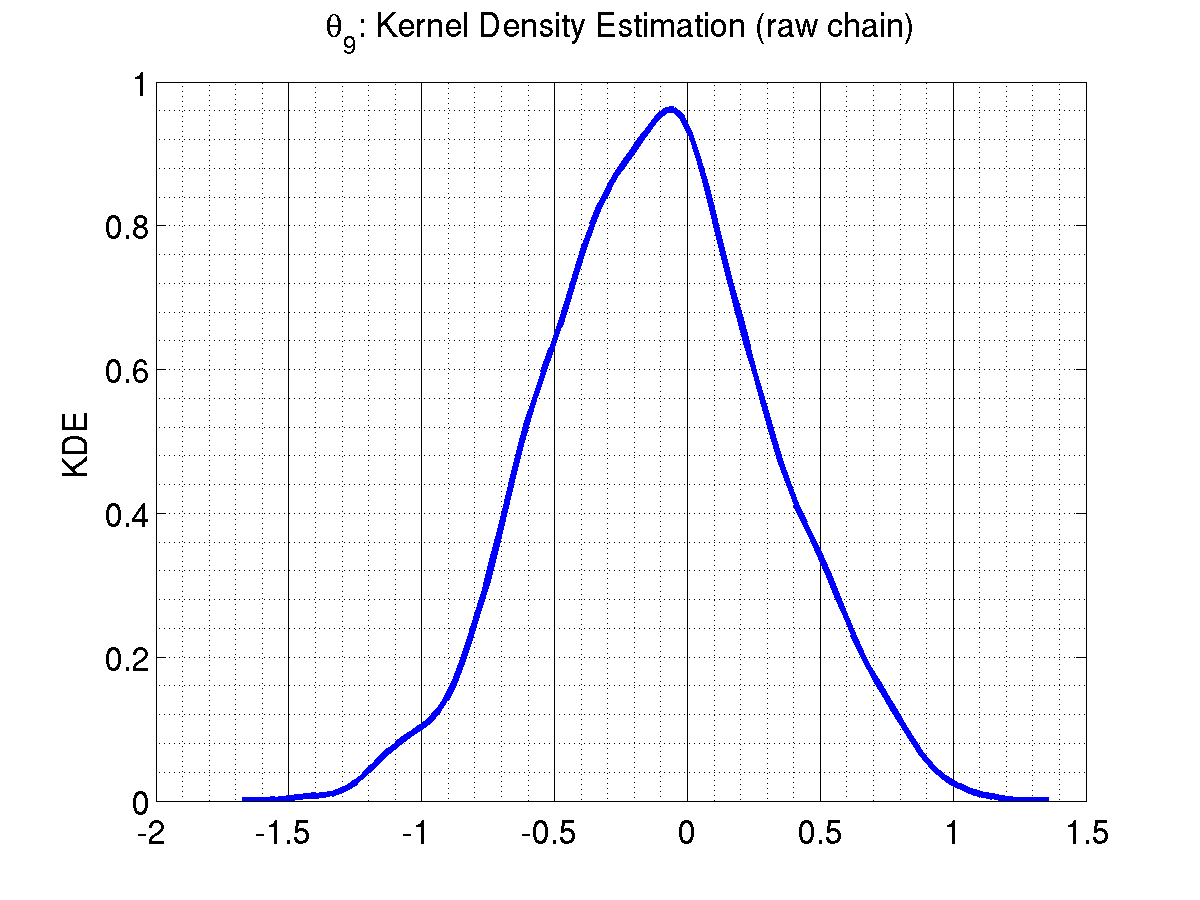}}\\
\subfloat{\includegraphics[scale=0.25]{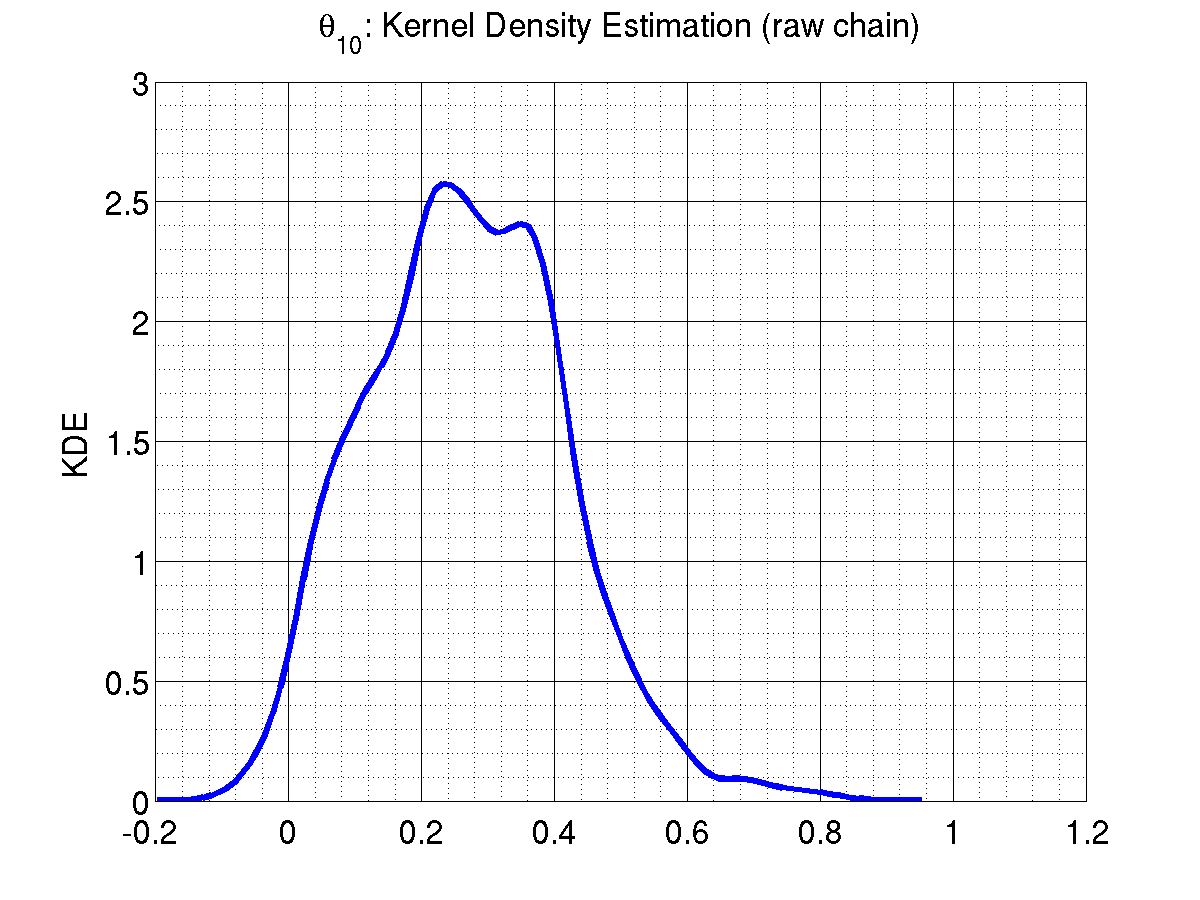}}
\subfloat{\includegraphics[scale=0.25]{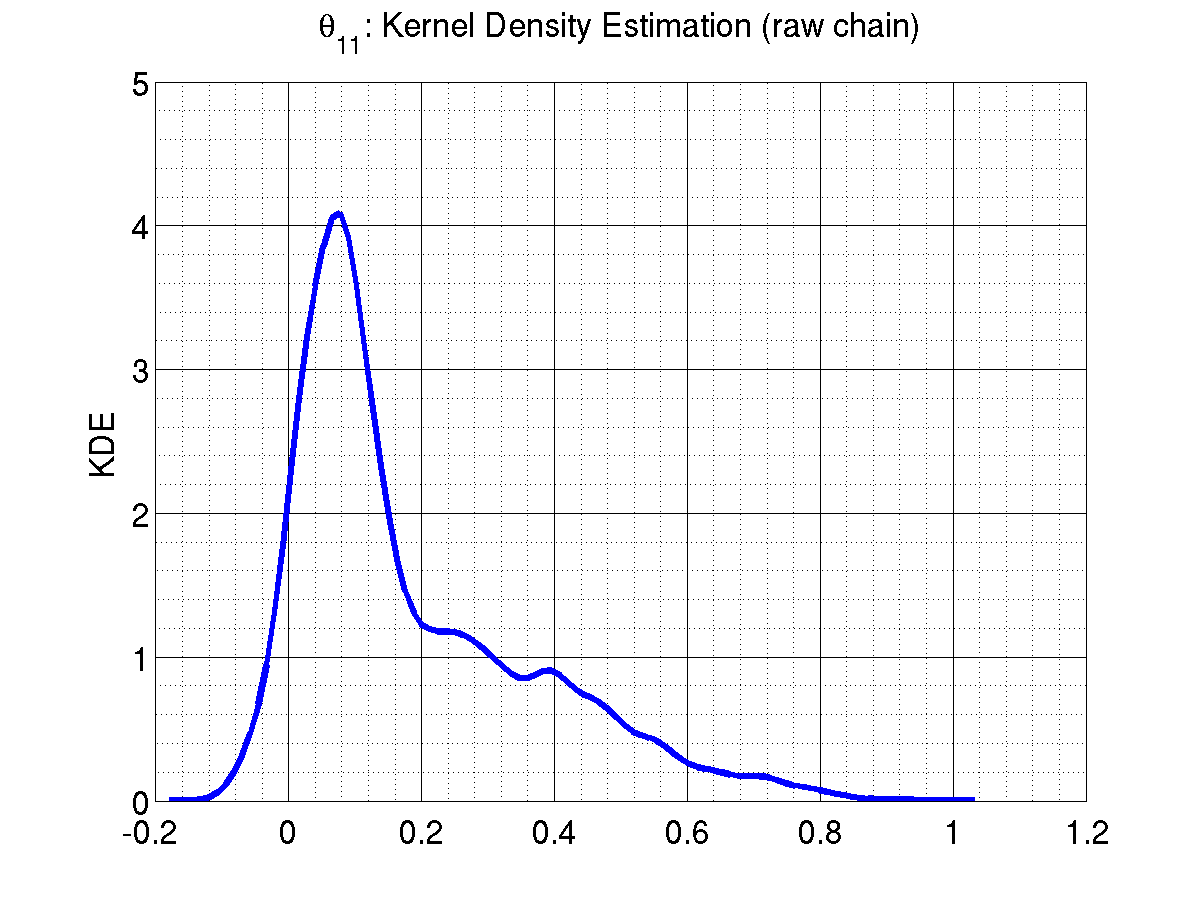}}
\subfloat{\includegraphics[scale=0.25]{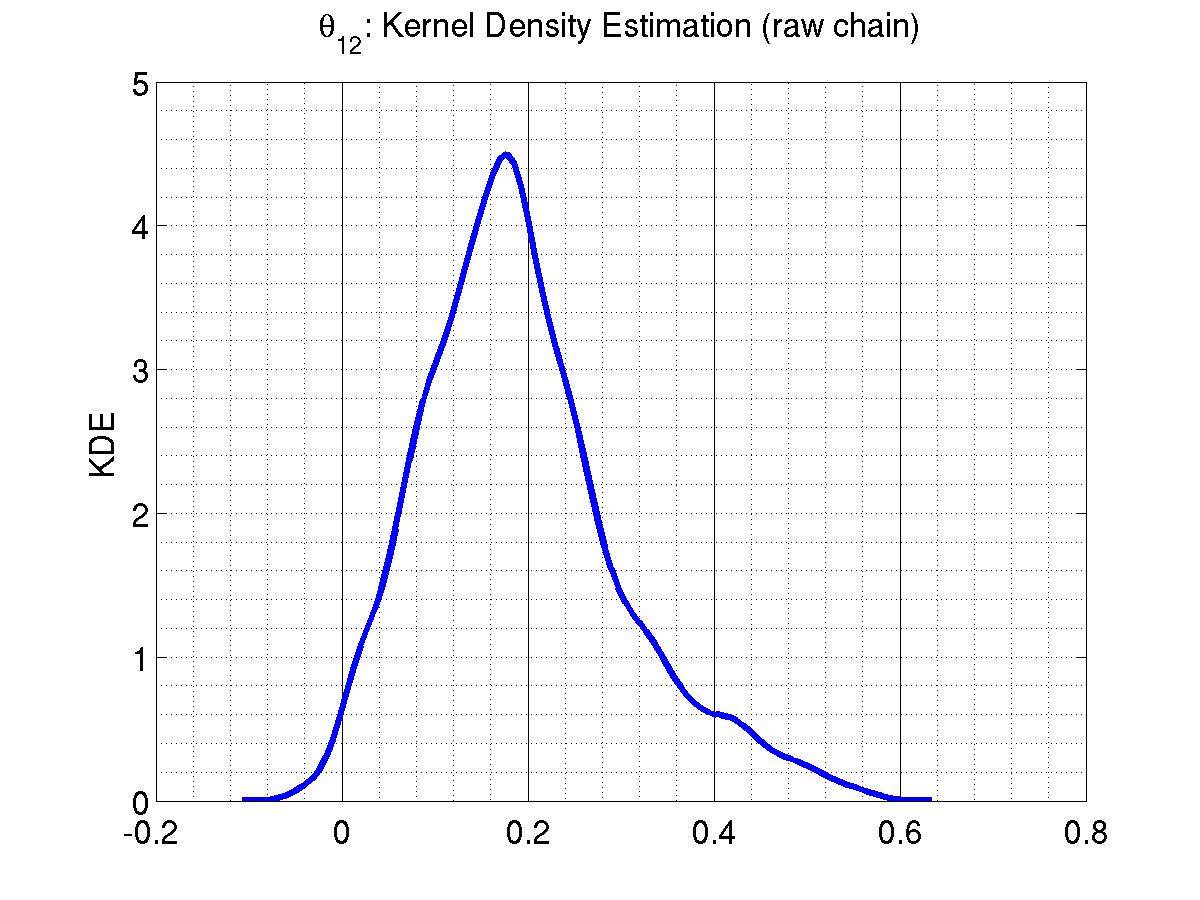}}\\
\subfloat{\includegraphics[scale=0.25]{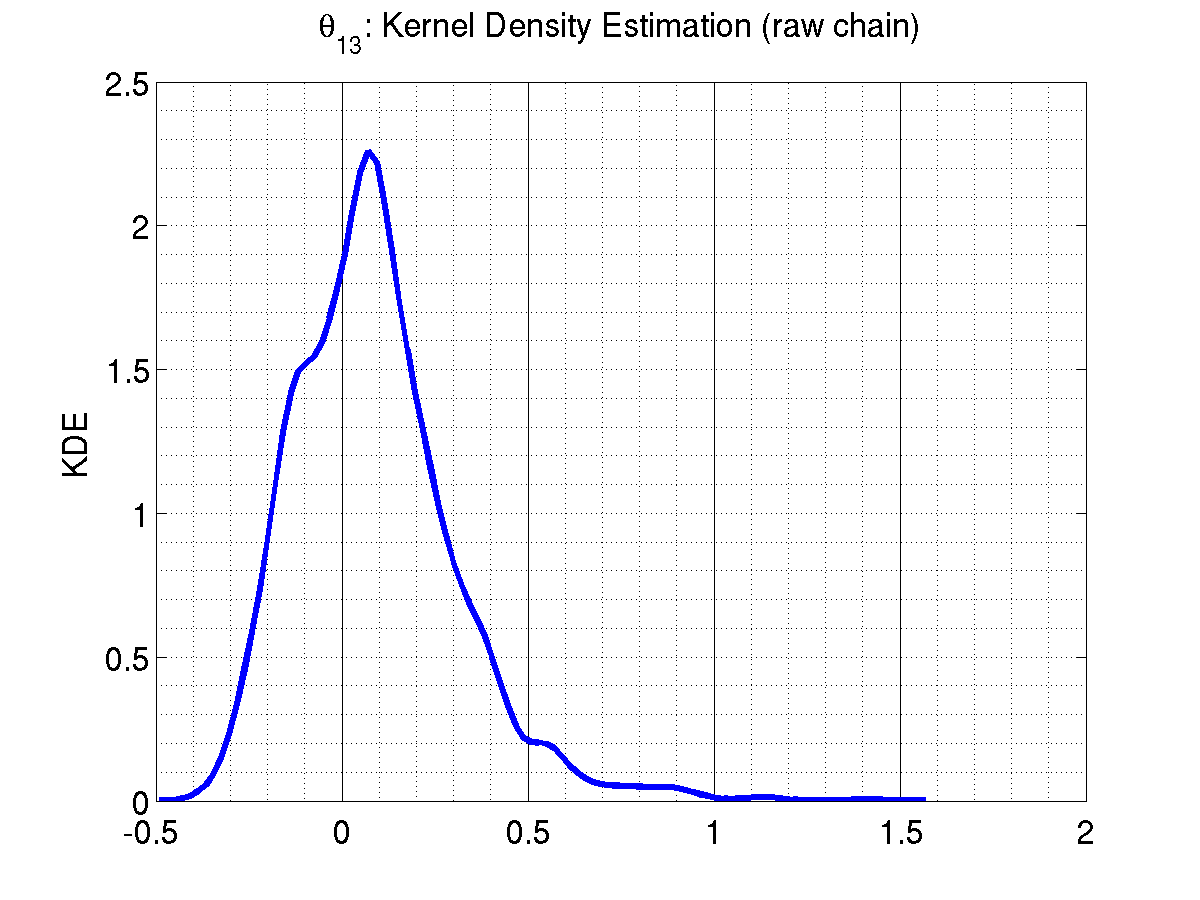}}
\subfloat{\includegraphics[scale=0.25]{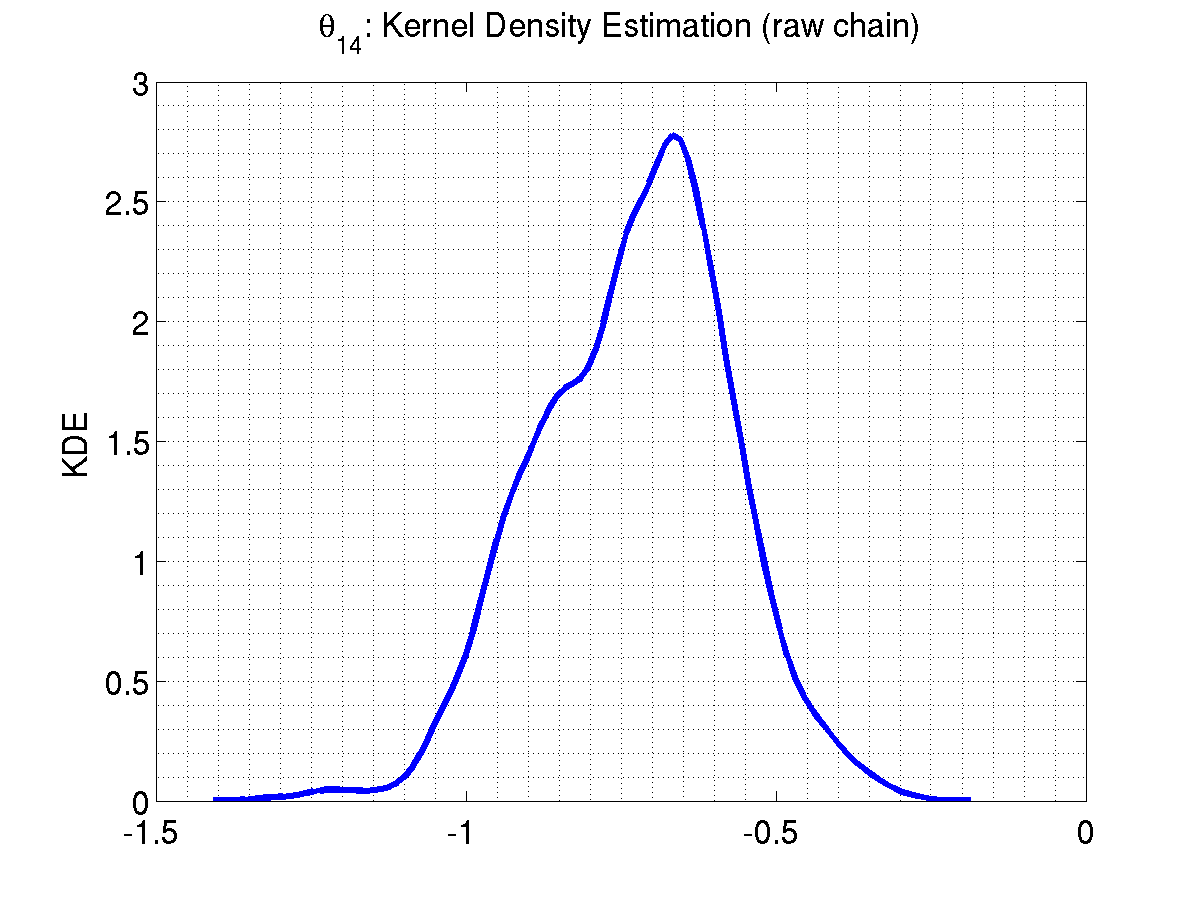}}
\subfloat{\includegraphics[scale=0.25]{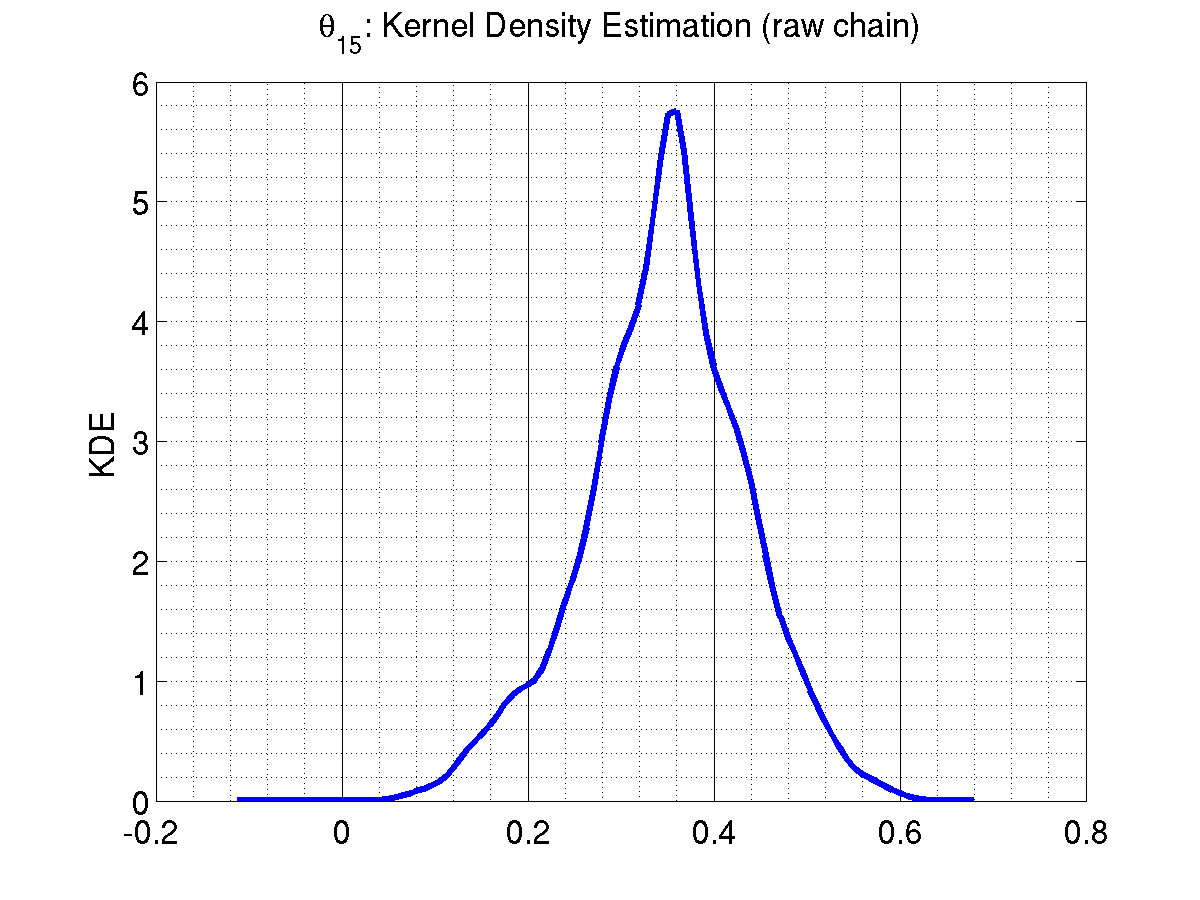}}
\caption{KDE plots of parameter $\bv{\theta}$ at the last level.}
\label{fig:hysteretic_kde}
\end{figure}

\begin{figure}[hptb]
\centering
\subfloat[$\log( f(\bv{y}|\bv{\theta}) )$]{\includegraphics[scale=0.4]{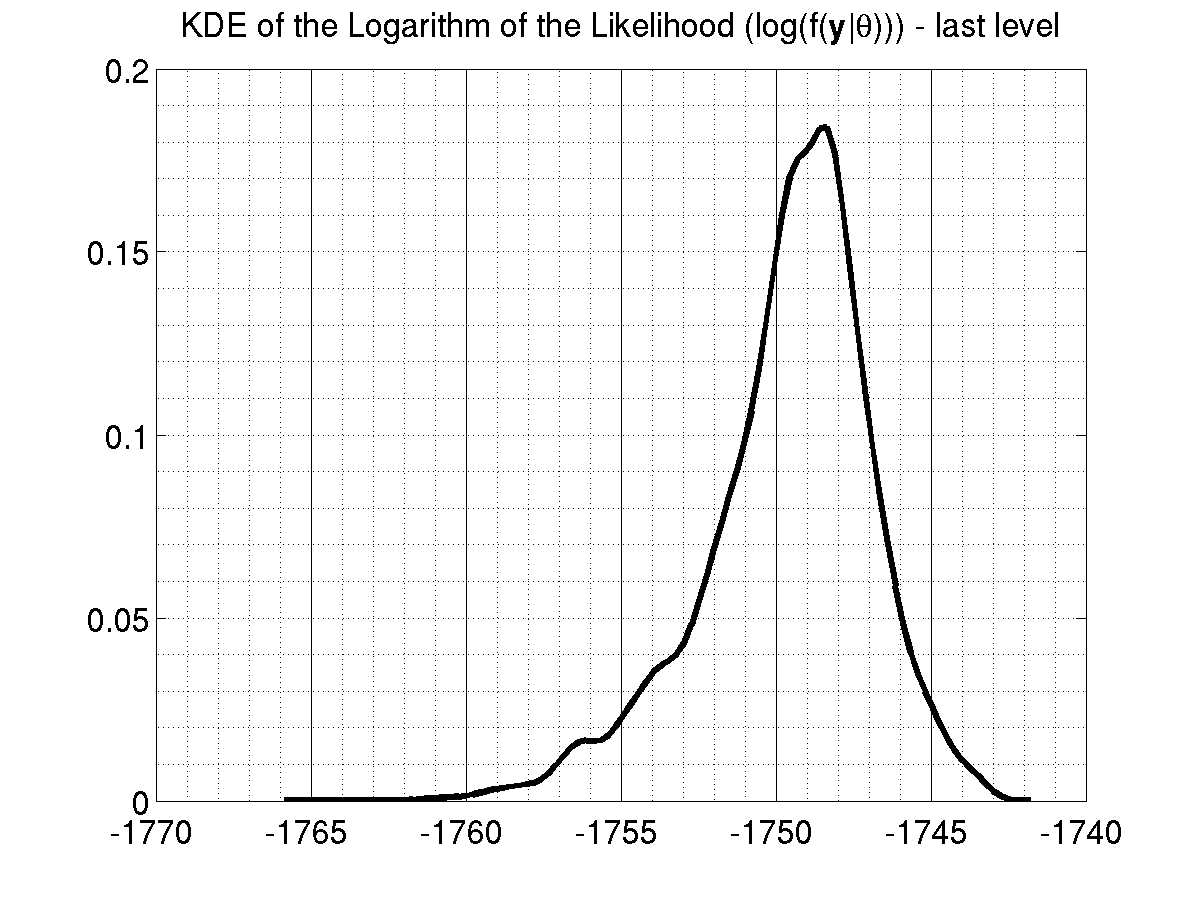}}
\subfloat[$f(\bv{y}|\bv{\theta})$]{\includegraphics[scale=0.4]{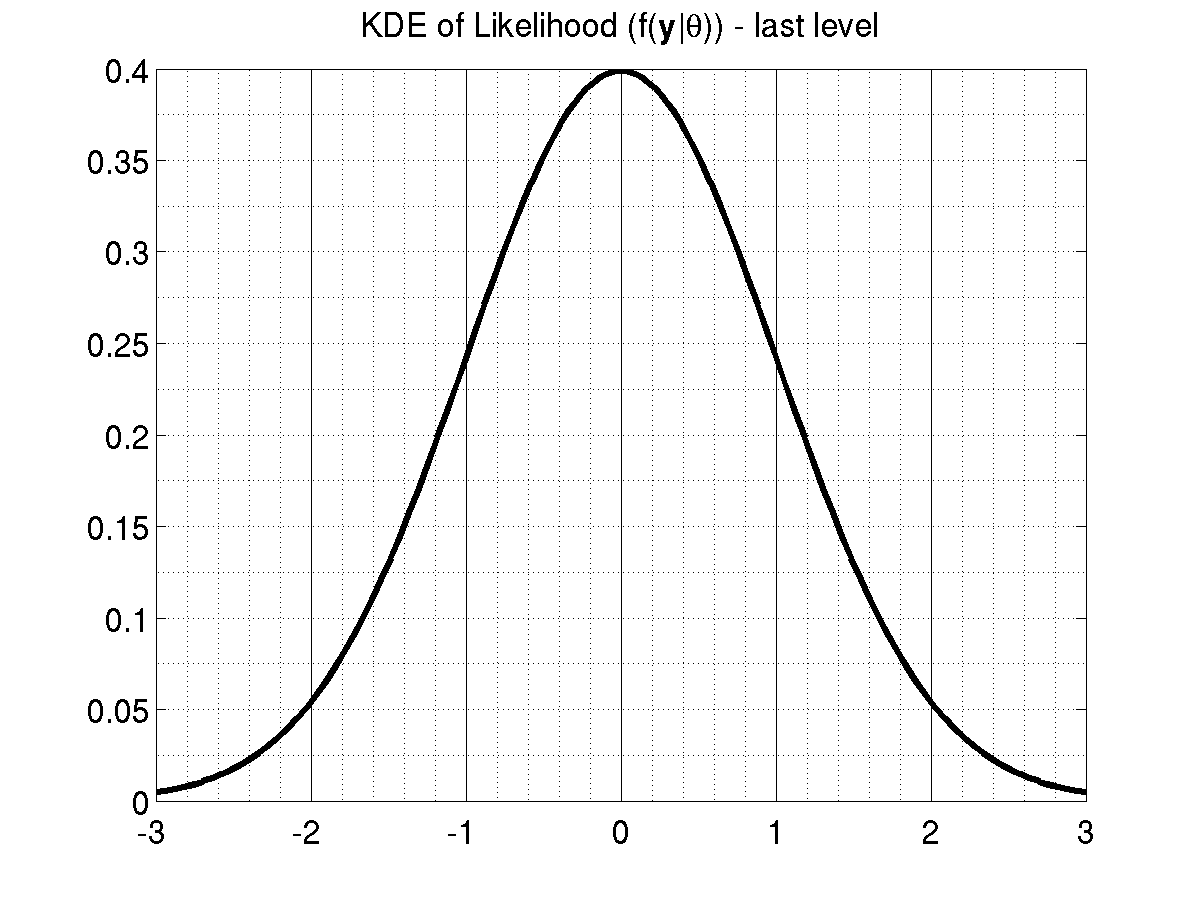}}
\vspace{-8pt}
\caption{KDE plots of the likelihood function, given by Eq. \eqref{eq:hyst:like}, and of its logarithm, at the last level.}
\label{fig:hysteretic_kde_like}
\end{figure}

Figure \ref{fig:hysteretic_kde} presents the KDE plots of each parameter $\theta_i,\, i=1,\ldots,15$.
The Multilevel method also provides data about the logarithm of the likelihood function as well as of the target PDF.
Figure \ref{fig:hysteretic_kde_like} presents the KDE plots of both the likelihood function and of its logarithm.

\subsubsection{Autocorrelation and CDF Plots}

\begin{figure}[hptb]
\centering
\hspace{-10pt}
\subfloat[CDF]{
  \includegraphics[scale=0.45]{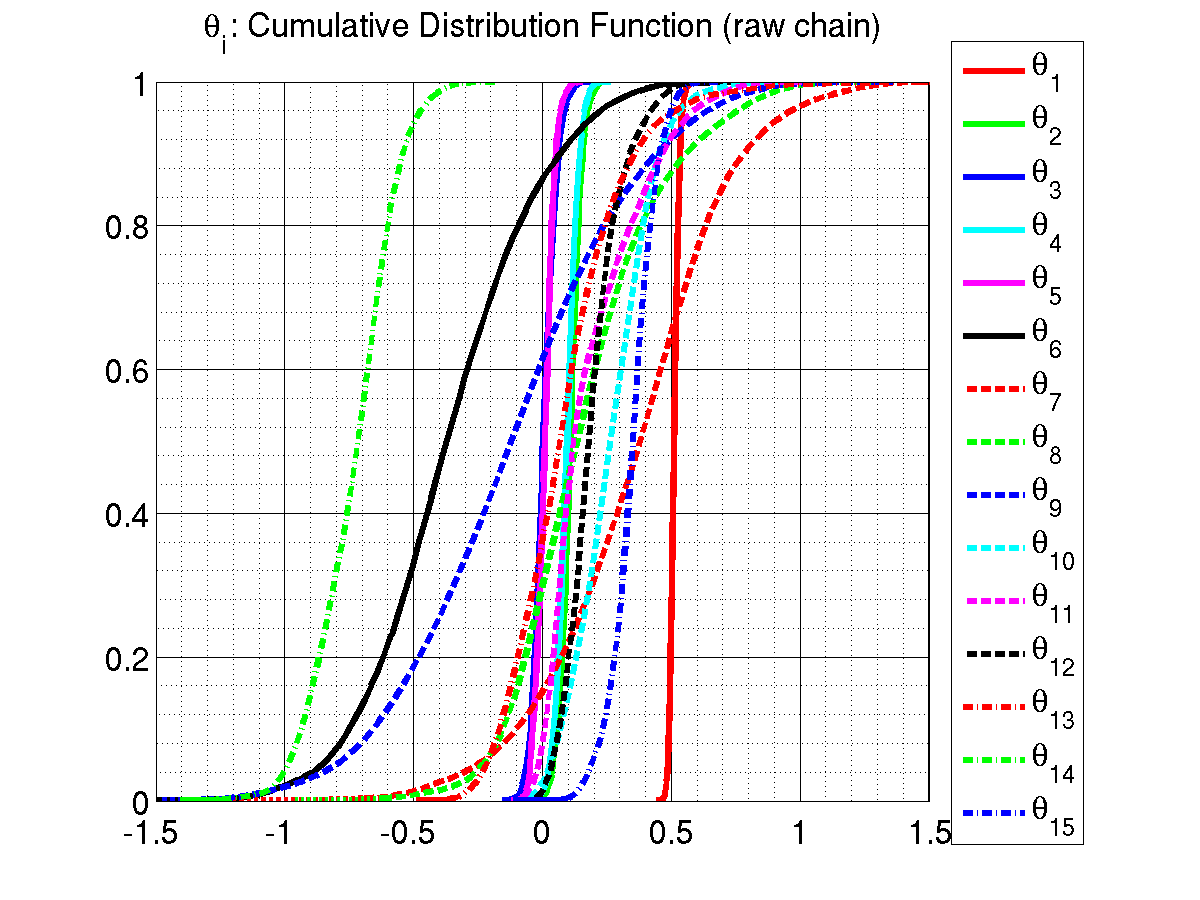}\label{fig:hysteretic_cdf}\hspace{-20pt}}
\subfloat[Autocorrelation]{
  \includegraphics[scale=0.45]{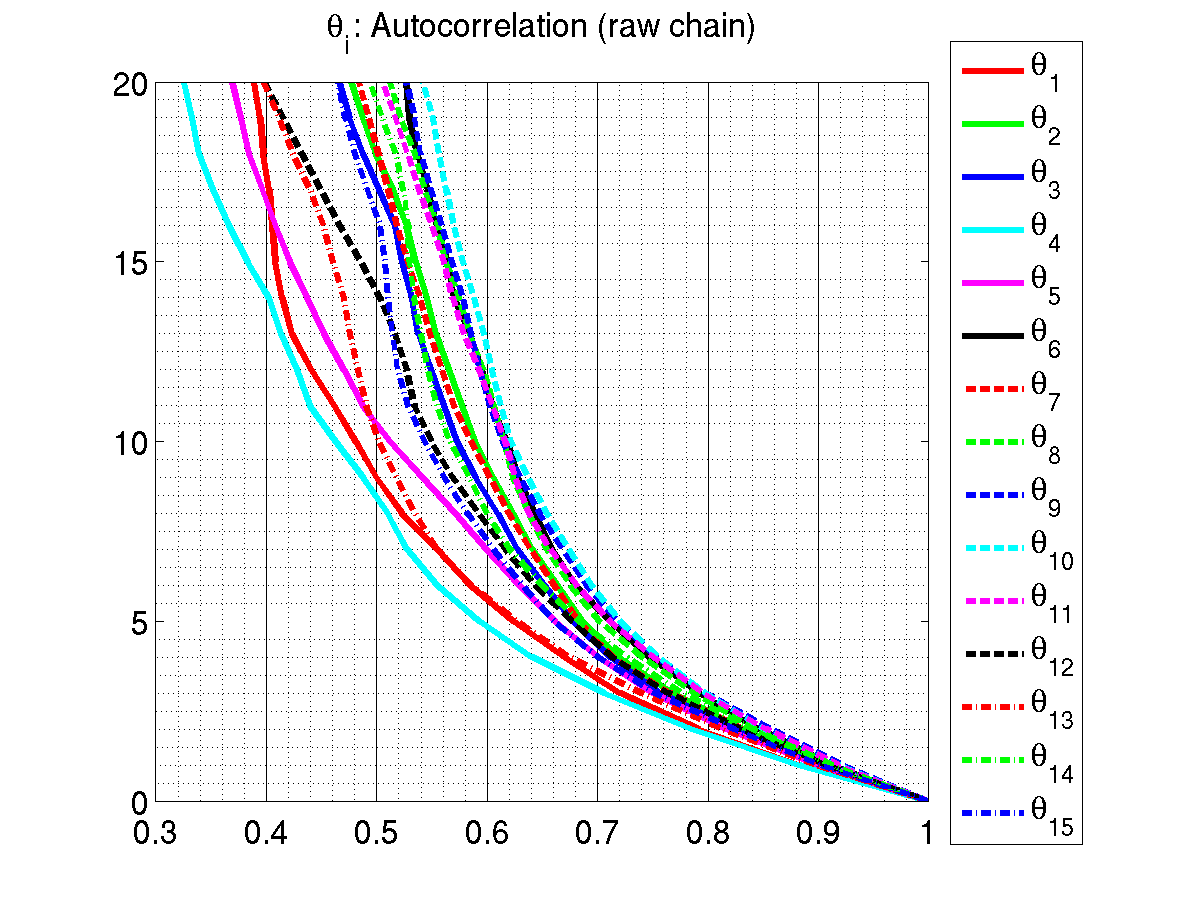}\label{fig:hysteretic_autocorr}}
\vspace{-8pt}
\caption{CDF and autocorrelation plots of parameter $\bv{\theta}$ at the last level.}
\end{figure}

Figure \ref{fig:hysteretic_cdf} combines the CDF of all parameters $\theta_i,\, i=1,\ldots,15$ into a single plot. 
Figure~\ref{fig:hysteretic_autocorr} presents their autocorrelations.

\addcontentsline{toc}{chapter}{References}

\bibliography{users.bbl}

\begin{appendix}



\chapter{Free Software Needs Free Documentation}\label{ch-fsnfd}


\begin{center}
{\it The following article was written by Richard Stallman, founder of the GNU Project.}\\
\end{center}

The biggest deficiency in the free software community today is not in the software$-$it is the lack of good free documentation that we can include with the free software.
Many of our most important programs do not come with free reference manuals and free introductory texts. Documentation is an essential part of any software package; when
an important free software package does not come with a free manual and a free tutorial, that is a major gap. We have many such gaps today.

Consider Perl, for instance. The tutorial manuals that people normally use are non-free. How did this come about? Because the authors of those manuals published them
with restrictive terms$-$no copying, no modification, source files not available$-$which exclude them from the free software world.

That wasn't the first time this sort of thing happened, and it was far from the last. Many times we have heard a GNU user eagerly describe a manual that he is writing,
his intended contribution to the community, only to learn that he had ruined everything by signing a publication contract to make it non-free.

Free documentation, like free software, is a matter of freedom, not price. The problem with the non-free manual is not that publishers charge a price for printed copies$-$that
in itself is fine. (The Free Software Foundation sells printed copies of manuals, too.) The problem is the restrictions on the use of the manual. Free manuals are available
in source code form, and give you permission to copy and modify. Non-free manuals do not allow this.

The criteria of freedom for a free manual are roughly the same as for free software. Redistribution (including the normal kinds of commercial redistribution) must be
permitted, so that the manual can accompany every copy of the program, both on-line and on paper.

Permission for modification of the technical content is crucial too. When people modify the software, adding or changing features, if they are conscientious they will
change the manual too$-$so they can provide accurate and clear documentation for the modified program. A manual that leaves you no choice but to write a new manual to
document a changed version of the program is not really available to our community.

Some kinds of limits on the way modification is handled are acceptable. For example, requirements to preserve the original author's copyright notice, the distribution
terms, or the list of authors, are ok. It is also no problem to require modified versions to include notice that they were modified. Even entire sections that may not
be deleted or changed are acceptable, as long as they deal with nontechnical topics (like this one). These kinds of restrictions are acceptable because they don't
obstruct the community's normal use of the manual.

However, it must be possible to modify all the technical content of the manual, and then distribute the result in all the usual media, through all the usual channels.
Otherwise, the restrictions obstruct the use of the manual, it is not free, and we need another manual to replace it.

Please spread the word about this issue. Our community continues to lose manuals to proprietary publishing. If we spread the word that free software needs free reference
manuals and free tutorials, perhaps the next person who wants to contribute by writing documentation will realize, before it is too late, that only free manuals contribute
to the free software community.

If you are writing documentation, please insist on publishing it under the GNU Free Documentation License or another free documentation license. Remember that this decision
requires your approval$-$you don't have to let the publisher decide. Some commercial publishers will use a free license if you insist, but they will not propose the option;
it is up to you to raise the issue and say firmly that this is what you want. If the publisher you are dealing with refuses, please try other publishers. If you're not sure
whether a proposed license is free, write to lice



\chapter{GNU General Public License}
\label{ch-gpl}


\date{Version 2.1, February 1999}

\begin{center}
Copyright \copyright\ 1991, 1999 Free Software Foundation, Inc.
51 Franklin Street, Fifth Floor, Boston, MA  02110-1301  USA
Everyone is permitted to copy and distribute verbatim copies
of this license document, but changing it is not allowed.
\end{center}

\section*{Preamble}

The licenses for most software are designed to take away your freedom to
share and change it.  By contrast, the GNU General Public Licenses are
intended to guarantee your freedom to share and change free software--to make
sure the software is free for all its users.

This license, the Lesser General Public License, applies to some specially
designated software packages--typically libraries--of the Free Software
Foundation and other authors who decide to use it.  You can use it too, but
we suggest you first think carefully about whether this license or the
ordinary General Public License is the better strategy to use in any
particular case, based on the explanations below.

When we speak of free software, we are referring to freedom of use, not
price.  Our General Public Licenses are designed to make sure that you have
the freedom to distribute copies of free software (and charge for this
service if you wish); that you receive source code or can get it if you want
it; that you can change the software and use pieces of it in new free
programs; and that you are informed that you can do these things.

To protect your rights, we need to make restrictions that forbid distributors
to deny you these rights or to ask you to surrender these rights.  These
restrictions translate to certain responsibilities for you if you distribute
copies of the library or if you modify it.

For example, if you distribute copies of the library, whether gratis or for a
fee, you must give the recipients all the rights that we gave you.  You must
make sure that they, too, receive or can get the source code.  If you link
other code with the library, you must provide complete object files to the
recipients, so that they can relink them with the library after making
changes to the library and recompiling it.  And you must show them these
terms so they know their rights.

We protect your rights with a two-step method: (1) we copyright the library,
and (2) we offer you this license, which gives you legal permission to copy,
distribute and/or modify the library.

To protect each distributor, we want to make it very clear that there is no
warranty for the free library.  Also, if the library is modified by someone
else and passed on, the recipients should know that what they have is not the
original version, so that the original author's reputation will not be
affected by problems that might be introduced by others.

Finally, software patents pose a constant threat to the existence of any free
program.  We wish to make sure that a company cannot effectively restrict the
users of a free program by obtaining a restrictive license from a patent
holder.  Therefore, we insist that any patent license obtained for a version
of the library must be consistent with the full freedom of use specified in
this license.

Most GNU software, including some libraries, is covered by the ordinary GNU
General Public License.  This license, the GNU Lesser General Public License,
applies to certain designated libraries, and is quite different from the
ordinary General Public License.  We use this license for certain libraries
in order to permit linking those libraries into non-free programs.

When a program is linked with a library, whether statically or using a shared
library, the combination of the two is legally speaking a combined work, a
derivative of the original library.  The ordinary General Public License
therefore permits such linking only if the entire combination fits its
criteria of freedom.  The Lesser General Public License permits more lax
criteria for linking other code with the library.

We call this license the "Lesser" General Public License because it does Less
to protect the user's freedom than the ordinary General Public License.  It
also provides other free software developers Less of an advantage over
competing non-free programs.  These disadvantages are the reason we use the
ordinary General Public License for many libraries.  However, the Lesser
license provides advantages in certain special circumstances.

For example, on rare occasions, there may be a special need to encourage the
widest possible use of a certain library, so that it becomes a de-facto
standard.  To achieve this, non-free programs must be allowed to use the
library.  A more frequent case is that a free library does the same job as
widely used non-free libraries.  In this case, there is little to gain by
limiting the free library to free software only, so we use the Lesser General
Public License.

In other cases, permission to use a particular library in non-free programs
enables a greater number of people to use a large body of free software.  For
example, permission to use the GNU C Library in non-free programs enables
many more people to use the whole GNU operating system, as well as its
variant, the GNU/Linux operating system.

Although the Lesser General Public License is Less protective of the users'
freedom, it does ensure that the user of a program that is linked with the
Library has the freedom and the wherewithal to run that program using a
modified version of the Library.

The precise terms and conditions for copying, distribution and modification
follow.  Pay close attention to the difference between a "work based on the
library" and a "work that uses the library".  The former contains code
derived from the library, whereas the latter must be combined with the
library in order to run.

\section*{TERMS AND CONDITIONS FOR COPYING, DISTRIBUTION AND MODIFICATION}

0. This License Agreement applies to any software library or other program
which contains a notice placed by the copyright holder or other authorized
party saying it may be distributed under the terms of this Lesser General
Public License (also called "this License").  Each licensee is addressed as
"you".

A "library" means a collection of software functions and/or data prepared so
as to be conveniently linked with application programs (which use some of
those functions and data) to form executables.

The "Library", below, refers to any such software library or work which has
been distributed under these terms.  A "work based on the Library" means
either the Library or any derivative work under copyright law: that is to
say, a work containing the Library or a portion of it, either verbatim or
with modifications and/or translated straightforwardly into another language.
(Hereinafter, translation is included without limitation in the term
"modification".)

"Source code" for a work means the preferred form of the work for making
modifications to it.  For a library, complete source code means all the
source code for all modules it contains, plus any associated interface
definition files, plus the scripts used to control compilation and
installation of the library.

Activities other than copying, distribution and modification are not covered
by this License; they are outside its scope.  The act of running a program
using the Library is not restricted, and output from such a program is
covered only if its contents constitute a work based on the Library
(independent of the use of the Library in a tool for writing it).  Whether
that is true depends on what the Library does and what the program that uses
the Library does.

1. You may copy and distribute verbatim copies of the Library's complete
source code as you receive it, in any medium, provided that you conspicuously
and appropriately publish on each copy an appropriate copyright notice and
disclaimer of warranty; keep intact all the notices that refer to this
License and to the absence of any warranty; and distribute a copy of this
License along with the Library.

You may charge a fee for the physical act of transferring a copy, and you may
at your option offer warranty protection in exchange for a fee.

2. You may modify your copy or copies of the Library or any portion of it,
thus forming a work based on the Library, and copy and distribute such
modifications or work under the terms of Section 1 above, provided that you
also meet all of these conditions:

  a) The modified work must itself be a software library.

  b) You must cause the files modified to carry prominent notices
  stating that you changed the files and the date of any change.

  c) You must cause the whole of the work to be licensed at no
  charge to all third parties under the terms of this License.

  d) If a facility in the modified Library refers to a function or a
  table of data to be supplied by an application program that uses
  the facility, other than as an argument passed when the facility
  is invoked, then you must make a good faith effort to ensure that,
  in the event an application does not supply such function or
  table, the facility still operates, and performs whatever part of
  its purpose remains meaningful.

  (For example, a function in a library to compute square roots has
  a purpose that is entirely well-defined independent of the
  application.  Therefore, Subsection 2d requires that any
  application-supplied function or table used by this function must
  be optional: if the application does not supply it, the square
  root function must still compute square roots.)

These requirements apply to the modified work as a whole.  If
identifiable sections of that work are not derived from the Library,
and can be reasonably considered independent and separate works in
themselves, then this License, and its terms, do not apply to those
sections when you distribute them as separate works.  But when you
distribute the same sections as part of a whole which is a work based
on the Library, the distribution of the whole must be on the terms of
this License, whose permissions for other licensees extend to the
entire whole, and thus to each and every part regardless of who wrote
it.

Thus, it is not the intent of this section to claim rights or contest
your rights to work written entirely by you; rather, the intent is to
exercise the right to control the distribution of derivative or
collective works based on the Library.

In addition, mere aggregation of another work not based on the Library
with the Library (or with a work based on the Library) on a volume of
a storage or distribution medium does not bring the other work under
the scope of this License.

3. You may opt to apply the terms of the ordinary GNU General Public License
instead of this License to a given copy of the Library.  To do this, you must
alter all the notices that refer to this License, so that they refer to the
ordinary GNU General Public License, version 2, instead of to this License.
(If a newer version than version 2 of the ordinary GNU General Public License
has appeared, then you can specify that version instead if you wish.)  Do not
make any other change in these notices.

Once this change is made in a given copy, it is irreversible for that copy,
so the ordinary GNU General Public License applies to all subsequent copies
and derivative works made from that copy.

This option is useful when you wish to copy part of the code of the Library
into a program that is not a library.

4. You may copy and distribute the Library (or a portion or derivative of it,
under Section 2) in object code or executable form under the terms of
Sections 1 and 2 above provided that you accompany it with the complete
corresponding machine-readable source code, which must be distributed under
the terms of Sections 1 and 2 above on a medium customarily used for software
interchange.

If distribution of object code is made by offering access to copy from a
designated place, then offering equivalent access to copy the source code
from the same place satisfies the requirement to distribute the source code,
even though third parties are not compelled to copy the source along with the
object code.

5. A program that contains no derivative of any portion of the Library, but
is designed to work with the Library by being compiled or linked with it, is
called a "work that uses the Library".  Such a work, in isolation, is not a
derivative work of the Library, and therefore falls outside the scope of this
License.

However, linking a "work that uses the Library" with the Library creates an
executable that is a derivative of the Library (because it contains portions
of the Library), rather than a "work that uses the library".  The executable
is therefore covered by this License.  Section 6 states terms for
distribution of such executables.

When a "work that uses the Library" uses material from a header file that is
part of the Library, the object code for the work may be a derivative work of
the Library even though the source code is not.  Whether this is true is
especially significant if the work can be linked without the Library, or if
the work is itself a library.  The threshold for this to be true is not
precisely defined by law.

If such an object file uses only numerical parameters, data structure layouts
and accessors, and small macros and small inline functions (ten lines or less
in length), then the use of the object file is unrestricted, regardless of
whether it is legally a derivative work.  (Executables containing this object
code plus portions of the Library will still fall under Section 6.)

Otherwise, if the work is a derivative of the Library, you may distribute the
object code for the work under the terms of Section 6.  Any executables
containing that work also fall under Section 6, whether or not they are
linked directly with the Library itself.

6. As an exception to the Sections above, you may also combine or link a
"work that uses the Library" with the Library to produce a work containing
portions of the Library, and distribute that work under terms of your choice,
provided that the terms permit modification of the work for the customer's
own use and reverse engineering for debugging such modifications.

You must give prominent notice with each copy of the work that the Library is
used in it and that the Library and its use are covered by this License.  You
must supply a copy of this License.  If the work during execution displays
copyright notices, you must include the copyright notice for the Library
among them, as well as a reference directing the user to the copy of this
License.  Also, you must do one of these things:

  a) Accompany the work with the complete corresponding machine-readable
  source code for the Library including whatever changes were used in the
  work (which must be distributed under Sections 1 and 2 above); and, if the
  work is an executable linked with the Library, with the complete
  machine-readable "work that uses the Library", as object code and/or source
  code, so that the user can modify the Library and then relink to produce a
  modified executable containing the modified Library.  (It is understood
  that the user who changes the contents of definitions files in the Library
  will not necessarily be able to recompile the application to use the
  modified definitions.)

  b) Use a suitable shared library mechanism for linking with the Library.  A
  suitable mechanism is one that (1) uses at run time a copy of the library
  already present on the user's computer system, rather than copying library
  functions into the executable, and (2) will operate properly with a
  modified version of the library, if the user installs one, as long as the
  modified version is interface-compatible with the version that the work was
  made with.

  c) Accompany the work with a written offer, valid for at least three years,
  to give the same user the materials specified in Subsection 6a, above, for
  a charge no more than the cost of performing this distribution.

  d) If distribution of the work is made by offering access to copy from a
  designated place, offer equivalent access to copy the above specified
  materials from the same place.

  e) Verify that the user has already received a copy of these materials or
  that you have already sent this user a copy.

For an executable, the required form of the "work that uses the Library" must
include any data and utility programs needed for reproducing the executable
from it.  However, as a special exception, the materials to be distributed
need not include anything that is normally distributed (in either source or
binary form) with the major components (compiler, kernel, and so on) of the
operating system on which the executable runs, unless that component itself
accompanies the executable.

It may happen that this requirement contradicts the license restrictions of
other proprietary libraries that do not normally accompany the operating
system.  Such a contradiction means you cannot use both them and the Library
together in an executable that you distribute.

7. You may place library facilities that are a work based on the Library
side-by-side in a single library together with other library facilities not
covered by this License, and distribute such a combined library, provided
that the separate distribution of the work based on the Library and of the
other library facilities is otherwise permitted, and provided that you do
these two things:

  a) Accompany the combined library with a copy of the same work based on the
  Library, uncombined with any other library facilities.  This must be
  distributed under the terms of the Sections above.

  b) Give prominent notice with the combined library of the fact that part of
  it is a work based on the Library, and explaining where to find the
  accompanying uncombined form of the same work.

8. You may not copy, modify, sublicense, link with, or distribute the Library
except as expressly provided under this License.  Any attempt otherwise to
copy, modify, sublicense, link with, or distribute the Library is void, and
will automatically terminate your rights under this License.  However,
parties who have received copies, or rights, from you under this License will
not have their licenses terminated so long as such parties remain in full
compliance.

9. You are not required to accept this License, since you have not signed it.
However, nothing else grants you permission to modify or distribute the
Library or its derivative works.  These actions are prohibited by law if you
do not accept this License.  Therefore, by modifying or distributing the
Library (or any work based on the Library), you indicate your acceptance of
this License to do so, and all its terms and conditions for copying,
distributing or modifying the Library or works based on it.

10. Each time you redistribute the Library (or any work based on the
Library), the recipient automatically receives a license from the original
licensor to copy, distribute, link with or modify the Library subject to
these terms and conditions.  You may not impose any further restrictions on
the recipients' exercise of the rights granted herein.  You are not
responsible for enforcing compliance by third parties with this License.

11. If, as a consequence of a court judgment or allegation of patent
infringement or for any other reason (not limited to patent issues),
conditions are imposed on you (whether by court order, agreement or
otherwise) that contradict the conditions of this License, they do not excuse
you from the conditions of this License.  If you cannot distribute so as to
satisfy simultaneously your obligations under this License and any other
pertinent obligations, then as a consequence you may not distribute the
Library at all.  For example, if a patent license would not permit
royalty-free redistribution of the Library by all those who receive copies
directly or indirectly through you, then the only way you could satisfy both
it and this License would be to refrain entirely from distribution of the
Library.

If any portion of this section is held invalid or unenforceable under any
particular circumstance, the balance of the section is intended to apply, and
the section as a whole is intended to apply in other circumstances.

It is not the purpose of this section to induce you to infringe any patents or
other property right claims or to contest validity of any such claims; this
section has the sole purpose of protecting the integrity of the free software
distribution system which is implemented by public license practices.  Many
people have made generous contributions to the wide range of software
distributed through that system in reliance on consistent application of that
system; it is up to the author/donor to decide if he or she is willing to
distribute software through any other system and a licensee cannot impose that
choice.

This section is intended to make thoroughly clear what is believed to be a
consequence of the rest of this License.

12. If the distribution and/or use of the Library is restricted in certain
countries either by patents or by copyrighted interfaces, the original
copyright holder who places the Library under this License may add an
explicit geographical distribution limitation excluding those countries, so
that distribution is permitted only in or among countries not thus excluded.
In such case, this License incorporates the limitation as if written in the
body of this License.

13. The Free Software Foundation may publish revised and/or new versions of
the Lesser General Public License from time to time.  Such new versions will
be similar in spirit to the present version, but may differ in detail to
address new problems or concerns.

Each version is given a distinguishing version number.  If the Library
specifies a version number of this License which applies to it and "any later
version", you have the option of following the terms and conditions either of
that version or of any later version published by the Free Software Foundation.
If the Library does not specify a license version number, you may choose any
version ever published by the Free Software Foundation.

14. If you wish to incorporate parts of the Library into other free programs
whose distribution conditions are incompatible with these, write to the
author to ask for permission.  For software which is copyrighted by the Free
Software Foundation, write to the Free Software Foundation; we sometimes make
exceptions for this.  Our decision will be guided by the two goals of
preserving the free status of all derivatives of our free software and of
promoting the sharing and reuse of software generally.

\begin{center}
{\sc NO WARRANTY}
\end{center}

15. BECAUSE THE LIBRARY IS LICENSED FREE OF CHARGE, THERE IS NO WARRANTY FOR
THE LIBRARY, TO THE EXTENT PERMITTED BY APPLICABLE LAW.  EXCEPT WHEN
OTHERWISE STATED IN WRITING THE COPYRIGHT HOLDERS AND/OR OTHER PARTIES
PROVIDE THE LIBRARY "AS IS" WITHOUT WARRANTY OF ANY KIND, EITHER EXPRESSED OR
IMPLIED, INCLUDING, BUT NOT LIMITED TO, THE IMPLIED WARRANTIES OF
MERCHANTABILITY AND FITNESS FOR A PARTICULAR PURPOSE.  THE ENTIRE RISK AS TO
THE QUALITY AND PERFORMANCE OF THE LIBRARY IS WITH YOU.  SHOULD THE LIBRARY
PROVE DEFECTIVE, YOU ASSUME THE COST OF ALL NECESSARY SERVICING, REPAIR OR
CORRECTION.

16. IN NO EVENT UNLESS REQUIRED BY APPLICABLE LAW OR AGREED TO IN WRITING
WILL ANY COPYRIGHT HOLDER, OR ANY OTHER PARTY WHO MAY MODIFY AND/OR
REDISTRIBUTE THE LIBRARY AS PERMITTED ABOVE, BE LIABLE TO YOU FOR DAMAGES,
INCLUDING ANY GENERAL, SPECIAL, INCIDENTAL OR CONSEQUENTIAL DAMAGES ARISING
OUT OF THE USE OR INABILITY TO USE THE LIBRARY (INCLUDING BUT NOT LIMITED TO
LOSS OF DATA OR DATA BEING RENDERED INACCURATE OR LOSSES SUSTAINED BY YOU OR
THIRD PARTIES OR A FAILURE OF THE LIBRARY TO OPERATE WITH ANY OTHER
SOFTWARE), EVEN IF SUCH HOLDER OR OTHER PARTY HAS BEEN ADVISED OF THE
POSSIBILITY OF SUCH DAMAGES.

\begin{center}
{\Large\sc END OF TERMS AND CONDITIONS}
\end{center}

\section*{How to Apply These Terms to Your New Programs}

If you develop a new library, and you want it to be of the greatest possible
use to the public, we recommend making it free software that everyone can
redistribute and change. You can do so by permitting redistribution under these
terms (or, alternatively, under the terms of the ordinary General Public
License).

To apply these terms, attach the following notices to the library. It is safest
to attach them to the start of each source file to most effectively convey the
exclusion of warranty; and each file should have at least the ``copyright''
line and a pointer to where the full notice is found.

\begin{verbatim}
one line to give the library's name and an idea of what it does.
Copyright (C) year  name of author

This library is free software; you can redistribute it and/or
modify it under the terms of the GNU Lesser General Public
License as published by the Free Software Foundation; either
version 2.1 of the License, or (at your option) any later version.

This library is distributed in the hope that it will be useful,
but WITHOUT ANY WARRANTY; without even the implied warranty of
MERCHANTABILITY or FITNESS FOR A PARTICULAR PURPOSE.  See the GNU
Lesser General Public License for more details.

You should have received a copy of the GNU Lesser General Public
License along with this library; if not, write to the Free Software
Foundation, Inc., 51 Franklin Street, Fifth Floor, Boston, MA  02110-1301  USA
\end{verbatim}

Also add information on how to contact you by electronic and paper mail.

You should also get your employer (if you work as a programmer) or your school,
if any, to sign a ``copyright disclaimer'' for the library, if necessary. Here
is a sample; alter the names:

\begin{verbatim}
Yoyodyne, Inc., hereby disclaims all copyright interest in
the library `Frob' (a library for tweaking knobs) written
by James Random Hacker.

signature of Ty Coon, 1 April 1990
Ty Coon, President of Vice
\end{verbatim}

That's all there is to it!



\chapter{GNU Free Documentation License}\label{ch-fdl}



\begin{center}
Version 1.3, 3 November 2008 
\end{center}

{\small 
Copyright\textcopyright ~2000, 2001, 2002, 2007, 2008 Free Software Foundation, Inc. \url{http://fsf.org/}
}\\

Everyone is permitted to copy and distribute verbatim copies of this license document, but changing it is not allowed.

\section*{0. Preamble}

The purpose of this License is to make a manual, textbook, or other functional and useful document ``free'' in the sense of freedom: to assure everyone the effective freedom to copy and redistribute it, with or without modifying it, either commercially or noncommercially. Secondarily, this License preserves for the author and publisher a way to get credit for their work, while not being considered responsible for modifications made by others.

This License is a kind of ``copyleft'', which means that derivative works of the document must themselves be free in the same sense. It complements the GNU General Public License, which is a copyleft license designed for free software.

We have designed this License in order to use it for manuals for free software, because free software needs free documentation: a free program should come with manuals providing the same freedoms that the software does. But this License is not limited to software manuals; it can be used for any textual work, regardless of subject matter or whether it is published as a printed book. We recommend this License principally for works whose purpose is instruction or reference.

\section*{1. Applicability and Definitions}

This License applies to any manual or other work, in any medium, that contains a notice placed by the copyright holder saying it can be distributed under the terms of this License. Such a notice grants a world-wide, royalty-free license, unlimited in duration, to use that work under the conditions stated herein. The ``Document'', below, refers to any such manual or work. Any member of the public is a licensee, and is addressed as ``you''. You accept the license if you copy, modify or distribute the work in a way requiring permission under copyright law.

A ``Modified Version'' of the Document means any work containing the Document or a portion of it, either copied verbatim, or with modifications and/or translated into another language.

A ``Secondary Section'' is a named appendix or a front-matter section of the Document that deals exclusively with the relationship of the publishers or authors of the Document to the Document's overall subject (or to related matters) and contains nothing that could fall directly within that overall subject. (Thus, if the Document is in part a textbook of mathematics, a Secondary Section may not explain any mathematics.) The relationship could be a matter of historical connection with the subject or with related matters, or of legal, commercial, philosophical, ethical or political position regarding them.

The ``Invariant Sections'' are certain Secondary Sections whose titles are designated, as being those of Invariant Sections, in the notice that says that the Document is released under this License. If a section does not fit the above definition of Secondary then it is not allowed to be designated as Invariant. The Document may contain zero Invariant Sections. If the Document does not identify any Invariant Sections then there are none.

The ``Cover Texts'' are certain short passages of text that are listed, as Front-Cover Texts or Back-Cover Texts, in the notice that says that the Document is released under this License. A Front-Cover Text may be at most 5 words, and a Back-Cover Text may be at most 25 words.

A ``Transparent'' copy of the Document means a machine-readable copy, represented in a format whose specification is available to the general public, that is suitable for revising the document straightforwardly with generic text editors or (for images composed of pixels) generic paint programs or (for drawings) some widely available drawing editor, and that is suitable for input to text formatters or for automatic translation to a variety of formats suitable for input to text formatters. A copy made in an otherwise Transparent file format whose markup, or absence of markup, has been arranged to thwart or discourage subsequent modification by readers is not Transparent. An image format is not Transparent if used for any substantial amount of text. A copy that is not ``Transparent'' is called ``Opaque''.

Examples of suitable formats for Transparent copies include plain ASCII without markup, Texinfo input format, LaTeX input format, SGML or XML using a publicly available DTD, and standard-conforming simple HTML, PostScript or PDF designed for human modification. Examples of transparent image formats include PNG, XCF and JPG. Opaque formats include proprietary formats that can be read and edited only by proprietary word processors, SGML or XML for which the DTD and/or processing tools are not generally available, and the machine-generated HTML, PostScript or PDF produced by some word processors for output purposes only.

The ``Title Page'' means, for a printed book, the title page itself, plus such following pages as are needed to hold, legibly, the material this License requires to appear in the title page. For works in formats which do not have any title page as such, ``Title Page'' means the text near the most prominent appearance of the work's title, preceding the beginning of the body of the text.

The ``publisher'' means any person or entity that distributes copies of the Document to the public.

A section ``Entitled XYZ'' means a named subunit of the Document whose title either is precisely XYZ or contains XYZ in parentheses following text that translates XYZ in another language. (Here XYZ stands for a specific section name mentioned below, such as ``Acknowledgements'', ``Dedications'', ``Endorsements'', or ``History''.) To ``Preserve the Title'' of such a section when you modify the Document means that it remains a section ``Entitled XYZ'' according to this definition.

The Document may include Warranty Disclaimers next to the notice which states that this License applies to the Document. These Warranty Disclaimers are considered to be included by reference in this License, but only as regards disclaiming warranties: any other implication that these Warranty Disclaimers may have is void and has no effect on the meaning of this License.

\section*{2. Verbatim Copying}

You may copy and distribute the Document in any medium, either commercially or noncommercially, provided that this License, the copyright notices, and the license notice saying this License applies to the Document are reproduced in all copies, and that you add no other conditions whatsoever to those of this License. You may not use technical measures to obstruct or control the reading or further copying of the copies you make or distribute. However, you may accept compensation in exchange for copies. If you distribute a large enough number of copies you must also follow the conditions in section 3.

You may also lend copies, under the same conditions stated above, and you may publicly display copies.

\section*{3. Copying in Quantity}

If you publish printed copies (or copies in media that commonly have printed covers) of the Document, numbering more than 100, and the Document's license notice requires Cover Texts, you must enclose the copies in covers that carry, clearly and legibly, all these Cover Texts: Front-Cover Texts on the front cover, and Back-Cover Texts on the back cover. Both covers must also clearly and legibly identify you as the publisher of these copies. The front cover must present the full title with all words of the title equally prominent and visible. You may add other material on the covers in addition. Copying with changes limited to the covers, as long as they preserve the title of the Document and satisfy these conditions, can be treated as verbatim copying in other respects.

If the required texts for either cover are too voluminous to fit legibly, you should put the first ones listed (as many as fit reasonably) on the actual cover, and continue the rest onto adjacent pages.

If you publish or distribute Opaque copies of the Document numbering more than 100, you must either include a machine-readable Transparent copy along with each Opaque copy, or state in or with each Opaque copy a computer-network location from which the general network-using public has access to download using public-standard network protocols a complete Transparent copy of the Document, free of added material. If you use the latter option, you must take reasonably prudent steps, when you begin distribution of Opaque copies in quantity, to ensure that this Transparent copy will remain thus accessible at the stated location until at least one year after the last time you distribute an Opaque copy (directly or through your agents or retailers) of that edition to the public.

It is requested, but not required, that you contact the authors of the Document well before redistributing any large number of copies, to give them a chance to provide you with an updated version of the Document.

\section*{4. Modifications}

You may copy and distribute a Modified Version of the Document under the conditions of sections 2 and 3 above, provided that you release the Modified Version under precisely this License, with the Modified Version filling the role of the Document, thus licensing distribution and modification of the Modified Version to whoever possesses a copy of it. In addition, you must do these things in the Modified Version:

\begin{enumerate}[A.]
\item  Use in the Title Page (and on the covers, if any) a title distinct from that of the Document, and from those of previous versions (which should, if there were any, be listed in the History section of the Document). You may use the same title as a previous version if the original publisher of that version gives permission.
\item  List on the Title Page, as authors, one or more persons or entities responsible for authorship of the modifications in the Modified Version, together with at least five of the principal authors of the Document (all of its principal authors, if it has fewer than five), unless they release you from this requirement.
\item  State on the Title page the name of the publisher of the Modified Version, as the publisher.
\item  Preserve all the copyright notices of the Document.
\item  Add an appropriate copyright notice for your modifications adjacent to the other copyright notices.
\item  Include, immediately after the copyright notices, a license notice giving the public permission to use the Modified Version under the terms of this License, in the form shown in the Addendum below.
\item  Preserve in that license notice the full lists of Invariant Sections and required Cover Texts given in the Document's license notice.
\item  Include an unaltered copy of this License.
\item  Preserve the section Entitled ``History'', Preserve its Title, and add to it an item stating at least the title, year, new authors, and publisher of the Modified Version as given on the Title Page. If there is no section Entitled ``History'' in the Document, create one stating the title, year, authors, and publisher of the Document as given on its Title Page, then add an item describing the Modified Version as stated in the previous sentence.
\item  Preserve the network location, if any, given in the Document for public access to a Transparent copy of the Document, and likewise the network locations given in the Document for previous versions it was based on. These may be placed in the ``History'' section. You may omit a network location for a work that was published at least four years before the Document itself, or if the original publisher of the version it refers to gives permission.
\item  For any section Entitled ``Acknowledgements'' or ``Dedications'', Preserve the Title of the section, and preserve in the section all the substance and tone of each of the contributor acknowledgements and/or dedications given therein.
\item  Preserve all the Invariant Sections of the Document, unaltered in their text and in their titles. Section numbers or the equivalent are not considered part of the section titles.
\item  Delete any section Entitled ``Endorsements''. Such a section may not be included in the Modified Version.
\item  Do not retitle any existing section to be Entitled ``Endorsements'' or to conflict in title with any Invariant Section.
\item  Preserve any Warranty Disclaimers.
\end{enumerate}
If the Modified Version includes new front-matter sections or appendices that qualify as Secondary Sections and contain no material copied from the Document, you may at your option designate some or all of these sections as invariant. To do this, add their titles to the list of Invariant Sections in the Modified Version's license notice. These titles must be distinct from any other section titles.

You may add a section Entitled ``Endorsements'', provided it contains nothing but endorsements of your Modified Version by various parties--for example, statements of peer review or that the text has been approved by an organization as the authoritative definition of a standard.

You may add a passage of up to five words as a Front-Cover Text, and a passage of up to 25 words as a Back-Cover Text, to the end of the list of Cover Texts in the Modified Version. Only one passage of Front-Cover Text and one of Back-Cover Text may be added by (or through arrangements made by) any one entity. If the Document already includes a cover text for the same cover, previously added by you or by arrangement made by the same entity you are acting on behalf of, you may not add another; but you may replace the old one, on explicit permission from the previous publisher that added the old one.

The author(s) and publisher(s) of the Document do not by this License give permission to use their names for publicity for or to assert or imply endorsement of any Modified Version.

\section*{5. Combining Documents}

You may combine the Document with other documents released under this License, under the terms defined in section 4 above for modified versions, provided that you include in the combination all of the Invariant Sections of all of the original documents, unmodified, and list them all as Invariant Sections of your combined work in its license notice, and that you preserve all their Warranty Disclaimers.

The combined work need only contain one copy of this License, and multiple identical Invariant Sections may be replaced with a single copy. If there are multiple Invariant Sections with the same name but different contents, make the title of each such section unique by adding at the end of it, in parentheses, the name of the original author or publisher of that section if known, or else a unique number. Make the same adjustment to the section titles in the list of Invariant Sections in the license notice of the combined work.

In the combination, you must combine any sections Entitled ``History'' in the various original documents, forming one section Entitled ``History''; likewise combine any sections Entitled ``Acknowledgements'', and any sections Entitled ``Dedications''. You must delete all sections Entitled ``Endorsements''.

\section*{6. Collections of Documents}

You may make a collection consisting of the Document and other documents released under this License, and replace the individual copies of this License in the various documents with a single copy that is included in the collection, provided that you follow the rules of this License for verbatim copying of each of the documents in all other respects.

You may extract a single document from such a collection, and distribute it individually under this License, provided you insert a copy of this License into the extracted document, and follow this License in all other respects regarding verbatim copying of that document.

\section*{7. Aggregation with Independent Works}

A compilation of the Document or its derivatives with other separate and independent documents or works, in or on a volume of a storage or distribution medium, is called an ``aggregate'' if the copyright resulting from the compilation is not used to limit the legal rights of the compilation's users beyond what the individual works permit. When the Document is included in an aggregate, this License does not apply to the other works in the aggregate which are not themselves derivative works of the Document.

If the Cover Text requirement of section 3 is applicable to these copies of the Document, then if the Document is less than one half of the entire aggregate, the Document's Cover Texts may be placed on covers that bracket the Document within the aggregate, or the electronic equivalent of covers if the Document is in electronic form. Otherwise they must appear on printed covers that bracket the whole aggregate.

\section*{8. Translation}

Translation is considered a kind of modification, so you may distribute translations of the Document under the terms of section 4. Replacing Invariant Sections with translations requires special permission from their copyright holders, but you may include translations of some or all Invariant Sections in addition to the original versions of these Invariant Sections. You may include a translation of this License, and all the license notices in the Document, and any Warranty Disclaimers, provided that you also include the original English version of this License and the original versions of those notices and disclaimers. In case of a disagreement between the translation and the original version of this License or a notice or disclaimer, the original version will prevail.

If a section in the Document is Entitled ``Acknowledgements'', ``Dedications'', or ``History'', the requirement (section 4) to Preserve its Title (section 1) will typically require changing the actual title.

\section*{9. Termination}

You may not copy, modify, sublicense, or distribute the Document except as expressly provided under this License. Any attempt otherwise to copy, modify, sublicense, or distribute it is void, and will automatically terminate your rights under this License.

However, if you cease all violation of this License, then your license from a particular copyright holder is reinstated (a) provisionally, unless and until the copyright holder explicitly and finally terminates your license, and (b) permanently, if the copyright holder fails to notify you of the violation by some reasonable means prior to 60 days after the cessation.

Moreover, your license from a particular copyright holder is reinstated permanently if the copyright holder notifies you of the violation by some reasonable means, this is the first time you have received notice of violation of this License (for any work) from that copyright holder, and you cure the violation prior to 30 days after your receipt of the notice.

Termination of your rights under this section does not terminate the licenses of parties who have received copies or rights from you under this License. If your rights have been terminated and not permanently reinstated, receipt of a copy of some or all of the same material does not give you any rights to use it.

\section*{10. Future Revisions of This License}

The Free Software Foundation may publish new, revised versions of the GNU Free Documentation License from time to time. Such new versions will be similar in spirit to the present version, but may differ in detail to address new problems or concerns. See \url{http://www.gnu.org/copyleft/}.

Each version of the License is given a distinguishing version number. If the Document specifies that a particular numbered version of this License ``or any later version'' applies to it, you have the option of following the terms and conditions either of that specified version or of any later version that has been published (not as a draft) by the Free Software Foundation. If the Document does not specify a version number of this License, you may choose any version ever published (not as a draft) by the Free Software Foundation. If the Document specifies that a proxy can decide which future versions of this License can be used, that proxy's public statement of acceptance of a version permanently authorizes you to choose that version for the Document.

\section*{11. Relicensing}

``Massive Multiauthor Collaboration Site'' (or ``MMC Site'') means any World Wide Web server that publishes copyrightable works and also provides prominent facilities for anybody to edit those works. A public wiki that anybody can edit is an example of such a server. A ``Massive Multiauthor Collaboration'' (or ``MMC'') contained in the site means any set of copyrightable works thus published on the MMC site.

``CC-BY-SA'' means the Creative Commons Attribution-Share Alike 3.0 license published by Creative Commons Corporation, a not-for-profit corporation with a principal place of business in San Francisco, California, as well as future copyleft versions of that license published by that same organization.

``Incorporate'' means to publish or republish a Document, in whole or in part, as part of another Document.

An MMC is ``eligible for relicensing'' if it is licensed under this License, and if all works that were first published under this License somewhere other than this MMC, and subsequently incorporated in whole or in part into the MMC, (1) had no cover texts or invariant sections, and (2) were thus incorporated prior to November 1, 2008.

The operator of an MMC Site may republish an MMC contained in the site under CC-BY-SA on the same site at any time before August 1, 2009, provided the MMC is eligible for relicensing.

\section*{ADDENDUM: How to use this License for your documents}

To use this License in a document you have written, include a copy of the License in the document and put the following copyright and license notices just after the title page:

\begin{quotation}\footnotesize
\begin{verbatim}
    Copyright (C)  YEAR  YOUR NAME.
    Permission is granted to copy, distribute and/or modify this document
    under the terms of the GNU Free Documentation License, Version 1.3
    or any later version published by the Free Software Foundation;
    with no Invariant Sections, no Front-Cover Texts, and no Back-Cover Texts.
    A copy of the license is included in the section entitled ``GNU
    Free Documentation License''.
\end{verbatim}
\end{quotation}

If you have Invariant Sections, Front-Cover Texts and Back-Cover Texts, replace the ``with ... Texts.'' line with this:

\begin{quote}\footnotesize
\begin{verbatim}
    with the Invariant Sections being LIST THEIR TITLES, with the
    Front-Cover Texts being LIST, and with the Back-Cover Texts being LIST.
\end{verbatim}
\end{quote}

If you have Invariant Sections without Cover Texts, or some other combination of the three, merge those two alternatives to suit the situation.

If your document contains nontrivial examples of program code, we recommend releasing these examples in parallel under your choice of free software license, such as the GNU General Public License, to permit their use in free software.

\end{appendix}

\end{document}